\documentclass[11pt,twoside]{report}
\usepackage{a4,latexsym}
\usepackage[xdvi]{epsfig}
\begin{document}

\begin{titlepage}
\begin{center}
\bfseries \Large 
DISSERTATION \\[3ex] 
\normalsize \mdseries
submitted to the \\
Combined Faculties for the Natural Sciences and for Mathematics\\
of the Ruperto-Carola University of Heidelberg, Germany\\
for the degree of\\
Doctor of Natural Sciences\\
\vfill
presented by\\[2ex]
Diplom-Physiker Bj{\"o}rn Feuerbacher,\\
born in Heidenheim\\[3ex]
Oral examination: June 25, 2003
\end{center}
\end{titlepage}
\thispagestyle{empty}

\cleardoublepage

\begin{titlepage}
\begin{center}
\bfseries \LARGE Perturbative Check of the Action and Energy Lattice Sum
Rules\\ \vfill
\normalsize \mdseries Referees: Prof. Dr. Heinz-J{\"u}rgen Rothe\\
and Prof. Dr. Dieter Gromes
\end{center}
\end{titlepage}
\thispagestyle{empty}

\cleardoublepage

\begin{titlepage}
\begin{center} \bfseries
Perturbative {\"U}berpr{\"u}fung der Wirkungs- und 
Energie-Gittersummenregeln\\[1ex] Zusammenfassung \mdseries
\end{center}
Gittersummenregeln werden mittels St{\"o}rungstheorie auf dem Gitter
{\"u}ber\-pr{\"u}ft. Die Wirkungssummenregel liefert eine Beziehung zwischen
dem Quark-Anti\-quark-Potential, dessen logarithmischer Ableitung nach dem
Abstand und dem Erwartungswert der Wirkung; die Energiesummenregel dr{\"u}ckt
das Potential als Summe der Energie in den Gluonenfeldern und eines anomalen
Terms aus. Zwei verschiedene unabh{\"a}ngige Berechnungen des
Quark-Antiquark-Poten\-tials werden vorgestellt, und die Transversalit{\"a}t
der gluonischen Vakuumpolarisation auf dem Gitter wird bewiesen. Der
wesentliche Teil der Wirkungssummenregel ist eine Identit{\"a}t, deren 
explizite {\"U}berpr{\"u}fung Methoden und Ergebnisse zur Verf{\"u}gung stellt,
die bei der Behandlung der Energiesummenregel n{\"u}tzlich werden.
Zus{\"a}tzlich wird die Eichinvarianz des Erwartungswertes des Wilson-Loops
bis zur n{\"a}chstf{\"u}hrenden Ordnung bewiesen. Die M{\"o}glichkeit, den
Erwartungswert der Wirkung auf den Erwartungswert der Summe der Plaketten zu
einer festen Zeit einzuschr{\"a}nken, wird diskutiert. Die Energiesummenregel
wird st{\"o}rungstheoretisch bis zur n{\"a}chst\-f{\"u}hrenden Ordnung
{\"u}berpr{\"u}ft, und es wird gezeigt, dass sie mit guter numerischer
Genauigkeit erf{\"u}llt ist. Die einzelnen Beitr{\"a}ge zum
Quark-Antiquark-Potential werden analysiert, und die Einschr{\"a}nkung des
Erwartungswertes der Summe {\"u}ber alle r{\"a}umlichen Plaketten (die Energie
in den magnetischen Feldern) auf den Erwartungswert der Summe der
r{\"a}umlichen Plaketten zu einer festen Zeit wird untersucht. 

\begin{center} \bfseries
Perturbative Check of the Action and Energy Lattice Sum Rules\\[1ex]
Abstract \mdseries
\end{center}
Lattice sum rules are checked using lattice perturbation theory. The action
sum rule gives a relation between the quark-antiquark potential, its
logarithmic derivative with respect to distance and the expectation value of
the action; the energy sum rule expresses the potential as the sum of the
energy in the gluon fields and of an anomalous term. Two different independent
calculations of the quark-antiquark potential are presented, and the
transversality of the gluonic vacuum polarization on the lattice is proven.
The crucial part of the action sum rule is an identity whose explicit
check using perturbation theory provides methods and results which are
useful for checking the energy sum rule. Additionally, the gauge invariance of
the expectation value of the Wilson loop up to next-to-leading order is
proven. The possibility of restricting the expectation value of the action to
one fixed time slice is discussed. The energy sum rule is checked
perturbatively up to next-to-leading order and shown to be satisfied with good
numerical accuracy. The various contributions to the quark-antiquark potential
are analyzed, and the restriction of the expectation value of the sum over all
spatial plaquettes (the energy in the magnetic fields) to one fixed time slice
is examined. 

\end{titlepage}

\thispagestyle{empty}

\cleardoublepage

\begin{titlepage}
\parbox[t][16cm][t]{10cm}{}
\begin{quotation}
\emph{There are very beautiful and elegant ways of getting these things these
days; but suppose that you were inventing it, what would you do to find [it]?
You fiddle around. All the elegant stuff is found later; the way to learn is
not to learn elegant things, it's to fiddle around blind and stupid. Later you
see how it works; polish it up; remove the scaffolding and publish the result
for other students to be amazed at your ingenuity.} \\[2ex]
\rightline{Richard Feynman}
\end{quotation}
\end{titlepage}

\thispagestyle{empty}

\cleardoublepage

\setcounter{page}{1}
\pagestyle{empty}

\tableofcontents

\newpage
\cleardoublepage

\setcounter{page}{1}
\pagestyle{headings}

\chapter{Introduction}

\section{Quantum Chromodynamics}

Only 40 years ago, protons and neutrons were still assumed to be elementary
particles. But already in the fifties of the 20th century, more and more new
so-called ''elementary'' particles had been found at high energy colliders
(see, for example, \cite{pre-quarks}); eventually
dozens were known, most of
them with very short
life times.

The turning point came 1964 with the invention of the quark model
\cite{Gell-Mann, Zweig}, which postulates that all of these particles
(nowadays known as \emph{hadrons}) are composed of sub-particles, the
\emph{quarks}. With this model, the variety of hadrons could be sorted in a
systematic way: the particles with integer spin, the \emph{mesons}, consist
of a quark-antiquark-pair, the particles with half-integer spin, the
\emph{baryons}, are bound states of three quarks or three antiquarks. Today
hundreds of hadrons are known, and most of them fit well into the quark
model. The few exceptions can be attributed to so-called \emph{glueballs} and
other exotic states.

Early experimental support came from high-energy scattering experiments, which
showed that there are sub-particles, called \emph{partons}, in the
proton \cite{Panofsky}. The scale invariance which was observed in these
experiments was well explained by the quark model \cite{Bjorken}.
Based on these and many other experimental results, nowadays the quark model is
universally accepted; a good introduction can be found in \cite{Close}, for
instance.

The theoretical description of the quarks and their interactions is provided by
Quantum Chromodynamics (QCD) \cite{FritzschGellMann1,
FritzschGellMann2}. In
this quantum field theory, the quarks are
 described by Dirac spinors with an
additional degree of freedom called
\emph{color}; they are coupled to $SU(3)$
gauge fields. The quantization of
 the bosonic gauge fields leads to the eight
spin-one particles called
\emph{gluons} which mediate the interactions of the
quarks. A prediction of QCD is the existence of bound states from these gluon
fields, the already above mentioned glueballs. An early theoretical review
of QCD can be found in \cite{MarcianoPagels}; for a review of the
experimental
support see, for instance, \cite{exp_evidence_QCD1}
and
\cite{exp_evidence_QCD2}.

A large problem remains, however: no free quarks or gluons ever have been
observed, despite all efforts taken in the years following the
invention of the quark model \cite{Morpurgo}. The standard explanation for
this is that quarks and gluons can only exist in bound states; this phenomenon
is called \emph{confinement}.

It can be explained if one assumes that the force between the quark-anti\-quark
pair does not
decrease with increasing distance as in electrodynamics, but
instead
stays constant---a flux tube or \emph{string} of gluons forms between
the quarks. Therefore, if the distance is increased, a point will be reached
at which the gluon string contains enough energy for the generation of a
new
quark-antiquark pair out of the vacuum, breaking the string. Hence
instead of
separating the original quark-antiquark-pair, one has produced two new pairs
(this effect is called \emph{screening} of the color charge). Recent
 Monte
Carlo simulations on the lattice (see below) confirm this picture
\cite{Slavo}.

\section{Lattice Gauge Theory}

The most commonly used method in Quantum Field Theory is
perturbation
theory. But for many important problems, unfortunately including
 quark
confinement, perturbative expansions fail: In QCD, the coupling constant
is
small at high energies, corresponding to small separations of the quarks
(this is known as \emph{asymptotic freedom} and is based on the non-abelian
nature of the gauge group), but large at the low energies and relatively
large
distances occuring in hadronic bound states. Therefore many
non-perturbative methods have been developed in the past decades in order to
treat
confinement.

Most promising is the formulation of QCD on a space-time lattice: space and
time are discretized, so that the number of degrees of freedom is reduced
drastically. Usually one works in euclidean space-time. If additionally one
only looks at a finite volume and a finite
 time interval, there is only a
finite number of degrees of freedom left, and
 the generating functional for
this discretized version of QCD can in principle
 be calculated on a computer.
In the limit of infinitely small lattice spacing,
 the results one obtains
should reduce to the continuum results.

Lattice formulations of QCD where first suggested 1974 by Wilson
\cite{Wilson}, Kogut and Susskind \cite{KogutSusskind}. In the three decades
since these proposals, lattice gauge theory has become a branch of particle
physics in its own right. Additionally, it is closely connected to
statistical mechanics and therefore of interest not only to particle
physicist, but also to physicists working on solid state physics and other
many-particle problems. Today, there are lots of encouraging results from
lattice calculations, including the quark-antiquark potential, the
 string
tension, hadron masses, the QCD phase transition from the hadronic
 bound
states to a quark-gluon plasma and so on. An introduction to all of
these
concepts and a summary of the results can be found in \cite{RotheBook},
for example.
 
But one still has to use a very large number of degrees of freedom in order
to get a reasonable approximation to continuum, infinite
 volume QCD. Because
of this, very high-dimensional integrals have to be computed. For example, if
one
 uses a lattice with an extension of only 10 in each of the four
directions and
 for simplicity considers only the gauge fields, there are
already 320.000
degrees of freedom ($10^4 \cdot$ 4
(polarization) $\cdot$
 8 (color)). Hence a 320.000-dimensional integral has
to be performed. Using
only 10 points in every direction in this
320.000-dimensional space in order to
 evaluate this integral numerically, one
would have to compute and add
 $10^{320000}$ terms!

Obviously this is not feasible; hence the numerical integrals are usually done
by Monte Carlo methods. Most gauge field configurations have a large action
and
thus only contribute very little to the generating functional.
Therefore
 the functional integral can be approximated by generating a
sequence of gauge
 field configurations with a probability distribution given
by the Boltzmann
 factor; this technique is called \emph{importance sampling}.
Then the
 expectation values of operators can be approximated by their
ensemble average
 over these representative configurations.

Simulating the gauge fields is relatively simple; the quarks pose bigger
problems---for example, \emph{fermion doubling} \cite{fermiondoubling}
and the sign of the fermionic determinant. Therefore one often works in the
so-called \emph{quenched} approximation: one neglects the effects of dynamical
quarks. Obviously the screening and the string breaking due to the creation
of quark-antiquark pairs can not be observed in such simulations, but they
still can give valuable results on the short distance behaviour of bound
states. 

For a description of the commonly used Monte Carlo algorithms and a review of
the most important results, see e.\ g.\ \cite{RotheBook}.

\section{Sum Rules for the Quark-Antiquark-Potential}

\label{SR_derivation}

Although the lattice results reproduce the expected behaviour of the
quark-antiquark potential, including the confinement, quite nicely and give
good estimates for the string tension, an analytic description would be
preferable instead of these solely numerical results. In order to gain some
insight into the quark-antiquark potential and the corresponding energy
densities of the gluon fields, in 1987 C.\ Michael invented \emph{lattice sum
rules} \cite{Michael}. These connect the quark-antiquark potential with the
expectation values of the energy in the gluon fields; even a separation of the
electric and magnetic contributions is possible.

However, Michael's derivation was partly incorrect (see below); this was
pointed out by Dosch, Nachtmann and Rueter in 1995 \cite{Dosch}.
Shortly afterwards H.\ J.\ Rothe published a corrected version
\cite{RotheSumrules}, and when Michael extended his sum rules in 1996
\cite{MichaelCorrection}, he took this correction into account. Additionally,
Rothe discovered that the trace
 anomaly of the energy momentum tensor gives
an important contribution to the
quark-antiquark potential
\cite{RotheTraceanomaly}.

The main goal of this dissertation is a perturbative check of these sum rules.
In the following subsections, their derivations will be presented, based
on \cite{RotheSumrules}, \cite{RotheTraceanomaly} and \cite{RotheBook}; then I
will outline the problems with the sum rules which will be investigated using
perturbation theory and give an overview of my approach to solve these
problems. Some results will be presented for a general gauge group $G$, but
in most cases I will restrict myself to the commonly used group $SU(N)$.

\subsection{The Wilson Loop and Wilson's Action}

\label{SR_intro}

The starting point for the derivation of both sum rules is the standard formula
for the quark-antiquark potential $\hat{V}$:
\begin{equation}
\hat{V}(\hat{R}) = - \lim_{\hat{T} \to \infty} \frac{1}{\hat{T}} \ln
<W(\hat{R}, \hat{T})>.
\label{PotentialWilson}
\end{equation}
The Wilson loop $W(\hat{R}, \hat{T})$ is the path ordered product of the link
variables $U_l$ along a closed rectangle $C$ with spatial and temporal
size $\hat{R}$ and $\hat{T}$, respectively:
\begin{equation}
W(\hat{R},\hat{T}) = \frac{1}{d(R)} \mbox{Tr} \prod_{l \in C} U_l.
\label{Wilsonloop}
\end{equation}
The link variables $U_l$ are elements of the gauge group, and $d(R)$ is the
dimension of the representation $R$ of the group in which the sources of the
gauge field are. Usually the fundamental representation $F$ is used; then
$d(R) = d(F) = N$ for the group $SU(N)$.

Wilson introduced this loop in 1974 as an order parameter for Quantum
Chromodynamics \cite{Wilson}; a similar structure was already used in 1971 by
F.\ J.\ Wegner for the Ising model \cite{Wegner}. The derivation of
the formula (\ref{PotentialWilson}) which connects the
quark-antiquark-potential to the expectation value of the Wilson loop can be
found in \cite{BrownWeisberger} and in the review article \cite{Gromes}, for
example.

The potential and the size parameters of the Wilson loop are measured in
lattice spacings $a$:
\begin{eqnarray}
\hat{R} &=& \frac{R}{a} \nonumber \\
\hat{T} &=& \frac{T}{a} \nonumber \\
\hat{V} &=& V \cdot a.
\label{Scaling}
\end{eqnarray}
The expectation value in (\ref{PotentialWilson}) is calculated with Wilson's
action (the sum over all plaquettes) in the quenched approximation
(no dynamical fermions):
\begin{equation}
S = \hat{\beta} \sum_{n, \mu, \nu}
\left(1-\frac{1}{2 d(F)}\mbox{Tr}(U_{\mu\nu}(n)+ U_{\mu\nu}^{\dagger}(n))
\right) =: \hat{\beta} \mathcal{P}
\label{WilsonAction}
\end{equation}
with the lattice coupling constant
\begin{equation}
\hat{\beta} = \frac{2 d(F)}{g_0^2}
\label{latticecoupling}
\end{equation}
and the plaquette variables
\begin{equation}
U_{\mu\nu}(n) = U_{\mu}(n)U_{\nu}(n+\hat{\mu}) U^{\dagger}_{\mu}(n+\hat{\nu})
U^{\dagger}_{\nu}(n),
\end{equation}
where $\hat{\mu}$ resp.\ $\hat{\nu}$ are unit vectors pointing in the
corresponding direction. The sum in the action runs over all lattice sites $n$
and all possible orientations of the plaquettes.

\subsection{Action Sum Rule}

\label{ASR_derivation}

Taking the derivative of (\ref{PotentialWilson}) with respect to the logarithm
of the coupling constant, one arrives at:
\begin{equation}
\frac{\partial \hat{V}}{\partial \ln \hat{\beta}} =
\hat{\beta} \frac{\partial \hat{V}}{\partial \hat{\beta}} =
\lim_{\hat{T} \to \infty} \frac{1}{\hat{T}} <S>_{q\bar{q}-0},
\label{potentialderivative}
\end{equation}
where for a general operator $O$, the expectation value in the
quark-antiquark state with respect to the vacuum state is defined as
\begin{equation}
<O>_{q\bar{q}-0} = \frac{<OW>}{<W>} - <O>.
\label{EV_QQ}
\end{equation}
Now the scaling behaviour (\ref{Scaling}) of the potential can be used. In the
continuum limit $a \to 0$, the physical potential $V$ obviously should be
independent of the lattice spacing. This gives:
\begin{equation}
\frac{d}{da} \left(\frac{\hat{V}\left(\frac{R}{a},\hat{\beta}\right)}
{a} \right) = 0 \quad \quad \quad \mbox{ for } a \to 0.
\end{equation}
Carrying out the differentation, using the scaling behaviour of $\hat{R}$, the
definition (\ref{latticecoupling}) and introducing the lattice beta function
\begin{equation}
\beta_L(g_0) = - a \frac{\partial g_0}{\partial a},
\label{latticebeta}
\end{equation}
one arrives at
\begin{equation}
\frac{2 \beta_L}{g_0} \hat{\beta} \frac{\partial \hat{V}}{\partial
\hat{\beta}} =
\hat{V} + \hat{R} \frac{\partial \hat{V}}{\partial \hat{R}}.
\label{PotentialScaling}
\end{equation}
Using this in (\ref{potentialderivative}), the result is:
\begin{equation}
\fbox{$\displaystyle
\hat{V} + \hat{R} \frac{\partial \hat{V}}{\partial \hat{R}} =
\frac{2 \beta_L}{g_0} \lim_{\hat{T} \to \infty} \frac{1}{\hat{T}} <S>_{q\bar{q}-0}.
$}
\label{ActionSumRule}
\end{equation}
This is essentially the action sum rule for the quark-antiquark potential
which
was derived by H.\ J.\ Rothe in 1995 \cite{RotheSumrules}.

As already pointed out, Michael derived this sum rule first (in 1987,
\cite{Michael}), but instead of (\ref{PotentialScaling}), he used a wrong
scaling behaviour for the potential, and therefore in his final formula, the
logarithmic derivative of the potential with respect to $\hat{R}$ was
missing.
 For a confining potential, $\hat{V} \equiv \sigma \hat{R}$, this is
crucial:
using the correct sum rule, the left hand side gives $2 \hat{V}$,
whereas
 Michael's wrong sum rule only had $\hat{V}$ and therefore was wrong by
a factor
of 2.

On the other hand, in his 1987 paper, Michael claimed that the
expectation value $<S>_{q\bar{q}-0}$ on the right hand side can be further
simplified in the limit $\hat{T} \to \infty$ (for a more detailed derivation,
using transfer matrix methods, see \cite{Bali}): plaquettes on different time
slices should give the \emph{same} contribution to the expectation value, so
that
\begin{equation}
<S>_{q\bar{q}-0} \to \hat{T} <L(t)>_{q\bar{q}-0},
\end{equation}
where $L(t)$ is the sum over all plaquettes at a \emph{fixed} time slice $t$:
\begin{equation}
L(t) = \hat{\beta} \sum_{\vec{n}, \mu, \nu}
\left(1-\frac{1}{2 d(F)}\mbox{Tr}(U_{\mu\nu}(\vec{n},t)+
U_{\mu\nu}^{\dagger}(\vec{n},t)) \right).
\end{equation}
Obviously this is the Lagrangian at time $t$.

Therefore the action sum rule becomes:
\begin{equation}
\hat{V} + \hat{R} \frac{\partial \hat{V}}{\partial \hat{R}} =
\frac{2 \beta_L}{g_0} \lim_{\hat{T} \to \infty} <L(t)>_{q\bar{q}-0}.
\end{equation}

Rothe noticed 1995 \cite{RotheTraceanomaly} that this can be rewritten using
the trace anomaly of the energy momentum tensor. This was motivated by
Ji's observation that this trace anomaly contributes 1/4 of the hadron
masses \cite{Ji}. Already in 1977 it had been shown by Collins, Dunkan and
Joglekar \cite{Collins} that the trace of the energy momentum tensor in the
continuum formulation of $SU(N)$ gauge theory is given by
\begin{equation}
T^{\mu}_{\;\;\mu} = \frac{\beta(g)}{2g} F_{\mu\nu}^A F^{\mu\nu,A} = \frac{2
\beta(g)}{g} \mathcal{L}
\end{equation}
in the quenched approximation, where $\mathcal{L}$ is the Lagrangian density,
which is exactly the combination appearing in the formula above. In
1992, Caracciolo, Menotti and Pelisetto showed that this is also true in
lattice
 perturbation theory \cite{Caracciolo}. Using this one finally gets
for
the action sum rule:
\begin{equation}
\hat{V} + \hat{R} \frac{\partial \hat{V}}{\partial \hat{R}} =
\lim_{\hat{T} \to \infty}
\sum_{\vec{x}, \mu} <T_{\mu\mu}(\vec{x},t)>_{q\bar{q}-0}.
\label{Spuranomalie}
\end{equation}

\subsection{Energy Sum Rule}

\label{ESR_derivation}

In order to derive the energy sum rule, one has to use an anisotropic lattice
with lattice spacing $a_t$ in the temporal direction. The anisotropy
parameter is defined as
\[
\xi := a_t/a.
\]
Then the action has to be rewritten:
\begin{eqnarray}
S &=& \hat{\beta}_s \sum_{n, j, k}
\left(1-\frac{1}{2 d(F)}\mbox{Tr}(U_{jk}(n)+ U_{jk}^{\dagger}(n))
\right) \nonumber \\
&&+ \hat{\beta}_t \sum_{n, j}
\left(1-\frac{1}{2 d(F)}\mbox{Tr}(U_{j4}(n)+ U_{j4}^{\dagger}(n)
+U_{4j}(n)+ U_{4j}^{\dagger}(n))
\right) \nonumber \\
&=:& \hat{\beta}_s \mathcal{P}_s + \hat{\beta}_t \mathcal{P}_t,
\end{eqnarray}
where one has to use different coupling constants for the spatial and the
temporal plaquettes. 

In the continuum limit, the expectation value of the Wilson loop
should be independent of the asymmetry, as long as the extension in the
temporal direction is the same:
\[
<W(\hat{R}, \hat{T})>_{\xi = 1} = <W(\hat{R}, \xi \hat{T})>_{\xi} \quad \quad
\quad \mbox{for } a \to 0,
\]
and accordingly
\[
\hat{V}(\hat{R},\hat{\beta}) = \xi \tilde{V}(\hat{R}, \hat{\beta}_s(\xi),
\hat{\beta}_t(\xi))
\]
with
\[
\tilde{V}(\hat{R}, \hat{\beta}_s(\xi), \hat{\beta}_t(\xi)) = - \lim_{\xi\hat{T}
\to \infty} \frac{1}{\xi\hat{T}} \ln <W(\hat{R}, \xi\hat{T})>.
\]

\noindent
It follows that in the continuum limit
\begin{equation}
\frac{d}{d\xi} \left[\xi \tilde{V}(\hat{R}, \hat{\beta}_s(\xi),
\hat{\beta}_t(\xi)) \right] = 0
\label{xiderivative}
\end{equation}
should be satisfied. Carrying out the differentation and then returning
to the isotropic lattice by setting $\xi=1$, one arrives at:
\begin{equation}
\hat{V}(\hat{R},\hat{\beta}) = - \lim_{\hat{T} \to \infty} \frac{1}{\hat{T}}
\left[\frac{\partial \hat{\beta}_t}{\partial \xi} <\mathcal{P}_t>_{q\bar{q}-0}
+ \frac{\partial \hat{\beta}_s}{\partial \xi} <\mathcal{P}_s>_{q\bar{q}-0}
\right]_{\xi = 1}.
\label{anisotropicderivative}
\end{equation}
With the abbreviations
\begin{equation}
\eta_{\pm} := \frac{1}{2} \left[\frac{\partial \hat{\beta}_t}{\partial \xi}
\pm  \frac{\partial \hat{\beta}_s}{\partial \xi} \right]_{\xi = 1},
\end{equation}
this can be rewritten as
\[
\hat{V}(\hat{R},\hat{\beta}) = \lim_{\hat{T} \to \infty} \frac{1}{\hat{T}}
\left[\eta_- <-\mathcal{P}_t+\mathcal{P}_s>_{q\bar{q}-0} - \eta_+
<\mathcal{P}_t+\mathcal{P}_s>_{q\bar{q}-0} \right].
\]
In \cite{Karsch2}, Karsch had shown that
\begin{equation}
\eta_+ = - \frac{1}{4} \frac{2\beta_L(g_0)}{g_0} \hat{\beta}
\end{equation}
by requiring that the string tension calculated from space-time-like Wilson
loops should be identical to the one calculated from purely spatial ones.
Hence the potential becomes:
\[
\hat{V}(\hat{R},\hat{\beta}) = \lim_{\hat{T} \to \infty} \frac{1}{\hat{T}}
\left[\eta_- <-\mathcal{P}_t+\mathcal{P}_s>_{q\bar{q}-0}
+ \frac{1}{4} \frac{2\beta_L(g_0)}{g_0} \hat{\beta}
<\mathcal{P}_t+\mathcal{P}_s>_{q\bar{q}-0} \right],
\]
where obviously the second term contains the expectation value of the
action:
\begin{equation}
\fbox{$\displaystyle
\hat{V}(\hat{R},\hat{\beta}) = \lim_{\hat{T} \to \infty} \frac{1}{\hat{T}}
\left[\eta_- <-\mathcal{P}_t+\mathcal{P}_s>_{q\bar{q}-0}
+ \frac{1}{4} \frac{2\beta_L(g_0)}{g_0} <S>_{q\bar{q}-0} \right].
$}
\label{EnergySumRule}
\end{equation}
This is essentially the energy sum rule which Rothe derived in
\cite{RotheSumrules}. Using the relation between the action and the trace of
the energy momentum tensor, it can be rewritten as
\[
\hat{V}(\hat{R},\hat{\beta}) = \lim_{\hat{T} \to \infty}
\left[\eta_- \frac{1}{\hat{T}} <-\mathcal{P}_t+\mathcal{P}_s>_{q\bar{q}-0}
+ \frac{1}{4} \sum_{\vec{x}, \mu} <T_{\mu\mu}(\vec{x},t)>_{q\bar{q}-0}
\right].
\]
On the other hand, Karsch \cite{Karsch2} showed that in the
continuum limit ($g_0 \to 0$):
\[
\eta_- \to \hat{\beta},
\]
and if one again assumes that the expectation value for
plaquettes on different time slices is identical for large $\hat{T}$, one can
rewrite the first
expectation value:
\[
\lim_{\hat{T} \to \infty} \frac{1}{\hat{T}} <-\mathcal{P}_t+\mathcal{P}_s>_{q\bar{q}-0}
= \lim_{\hat{T} \to \infty}
<-\mathcal{P}'_t(t)+\mathcal{P}'_s(t)>_{q\bar{q}-0},
\]
where $\mathcal{P}'_{s,t}(t)$ is the sum over all spatial
respectively temporal plaquettes on the \emph{fixed} time slice $t$. Therefore
in the continuum limit, the first term on the right hand side of
(\ref{EnergySumRule}) reduces to:
\begin{equation}
\lim_{\hat{T} \to \infty}
\eta_- \frac{1}{\hat{T}} <-\mathcal{P}_t+\mathcal{P}_s>_{q\bar{q}-0}
\to a \sum_{\vec{x}} a^3 \frac{1}{2} \lim_{\hat{T} \to \infty}
<-\vec{E}^2(\vec{x})+\vec{B}^2(\vec{x})>_{q\bar{q}-0},
\end{equation}
which is just the euclidean version of the energy in the gluon fields. Hence
the energy sum rule tells us that the potential energy of the quark-antiquark
pair is given by:
\[
\mbox{energy in the gluon fields} + \frac{1}{4} \mbox{trace anomaly
of the energy momentum tensor}.
\]

\noindent
An especially interesting case to look at is a purely confining potential:
\[
\hat{V}(\hat{R}) = \sigma \hat{R}.
\]
From the action sum rule, one can deduce that
\[
\lim_{\hat{T} \to \infty} \sum_{\vec{x},\mu}
<T_{\mu\mu}(\vec{x},t)>_{q\bar{q}-0}
= \hat{V} + \hat{R} \frac{\partial \hat{V}}{\partial \hat{R}} = 2 \hat{V};
\]
hence the energy sum rule gives:
\[
\frac{1}{2} \hat{V} = - \lim_{\hat{T} \to \infty}
\eta_- \frac{1}{\hat{T}} <-\mathcal{P}_t+\mathcal{P}_s>_{q\bar{q}-0}.
\]
Therefore for a confining potential, the contributions from the energy in the
gluon fields and from the trace anomaly of the energy-momentum-tensor have the
same magnitude: both account for exactly one half of the total potential
energy.

\section{Problems with the Sum Rules}

The main goal of this work is to check the lattice sum rules. This is necessary because of
several reasons: first, neither for the argument leading to the restriction to one fixed time slice,
nor for the scaling arguments on the anisotropic lattice, it is known if they
are justified. Hence their validity should be checked.

But beyond these validity doubts, there are also other reasons to take a
closer look at the sum rules. Because of the opposite signs of the
magnetic and electric field energies in the euclidean formulation,
cancellations between these two contributions can occur when their
expectation values
are calculated in Monte Carlo simulations. These
cancellations are a possible source of errors; a more thorough investigation of
what happens there exactly should be helpful. Additionally, it would be
interesting to
 see how do the contributions of the trace anomaly of the
energy-momentum tensor look like.

There are in principle two ways to check the sum rules and to examine these
interesting details. On the
 one hand, one could perform Monte Carlo
simulations. Using these, one could investigate the physically
interesting
region where the coupling constant is large. But there are some drawbacks,
too: much
computer power and time is needed for these simulations, and
the results are not very
enlightening in general; numerical results do not
tell as much about the underlying physics than
analytical calculations.

Because of these reasons, in this work lattice perturbation theory is
used to study the sum rules. Obviously
this is only possible for weak
coupling, hence the physically interesting and relevant region is missed. 
But this is countered by the fact that much less computer power and time
is needed (only some
low-dimensional numerical integrals in momentum space
have to be done). Additionally, one gets analytical
results, which are
expected to shed some light on the problems mentioned above.

The Feynman rules for lattice perturbation theory are well-known and can be
taken, for example,
 from \cite{RotheBook}; there the necessary weak coupling
expansion of the action is also explained in
detail. The consistency of these
perturbative calculations, a general power counting theorem, and
considerations about the continuum limit can be found in \cite{Reisz}. The
last necessary
 ingredient for the check of the sum rules in the weak coupling
approximation is an expression for the
 potential up to next-to-leading order,
which is provided in \cite{Karsch}.

The outline of the check is as follows: first, I will explain how one obtains a
weak coupling expansion
 of the quark-antiquark potential in lattice
perturbation theory; I will outline an approach by Kovacs
 to this problem
\cite{Kovacs} as well as the methods used by Heller and Karsch, already
mentioned
 above \cite{Karsch}. In the same chapter, I will prove
additionally the transversality of the gluonic
vacuum polarization, which
appears in the calculation of the potential.

Then I will turn to the check of the action sum rule. In principle the
validity of this sum rule is clear - after
 all, it is more or less an
identity. Nevertheless, it seemed to be an easier task to check it rather than
the energy sum rule,
and it provides an opportunity to demonstrate some of the
necessary methods. Additionally, it will turn out that this sum rule
is closely connected with the gauge invariance of the Wilson loop---only some
small additional calculations will be needed to check this gauge invariance,
too. And there is one part
of the action sum rule for which a check is really
needed: the validity of the restriction to one time slice,
 which was already
mentioned above, is not clear at all even for the action sum rule.

Finally the energy sum rule also has to be checked. Many results of the
check of the action sum rule
 will become useful there (a further reason to
check that sum rule first), hence this check will be easier than the
check of the action sum rule, although it looks more complicated at first
sight.
After the check, I will explain how one can derive
results about the sizes of the  various contributions to the potential
(electric, magnetic, anomal) from my calculations and the energy
sum rule.
I will close that chapter again with a section on the restriction to one time
slice.

The last chapter contains a summary and the conclusions. In the appendix,
useful formulas are
 summarized, and some basic calculations are presented.
A detailed account of
 the Fourier transformation needed in chapter 2 for
calculating the potential can also be found there.

\newpage
\thispagestyle{empty}
\cleardoublepage
\chapter{The Quark-Antiquark Potential}

\label{potential}

In this chapter, I will line out how one can calculate the static
quark-antiquark potential on the lattice, which is needed in order to check
the sum rules. First I will give a general derivation, based on the
running of the coupling constant and renormalization group arguments (for more
details, see \cite{Kovacs}), then I will discuss how one can derive these
results directly by a perturbative lattice calculation. Additionally, I will
prove that the gluonic vacuum polarization is transversal in leading order of
the lattice perturbation theory.

\section{The running coupling constant}

\label{runningcoupling}

In leading order, the quark-antiquark potential is given by:
\begin{equation}
V(R) = -\frac{g^2}{4 \pi R} C_2(F),
\end{equation}
where $g$ is the coupling constant of the strong interaction and $C_2(F)$
the quadratic Casimir operator of fundamental representation $F$; for $SU(N)$:
\begin{equation}
C_2(F) = \frac{N^2-1}{2N}.
\end{equation}

If one takes into account that the coupling constant depends also on the
distance between the quark and the anti-quark (''running coupling
constant''), then the formula above needs corrections:
\begin{equation}
V(R) = -\frac{g^2(R)}{4 \pi R} C_2(F) = -\frac{g^2}{4 \pi R} C_2(F) + O(g^4).
\end{equation}
There are various methods to calculate these corrections. In \cite{Fischler},
they were determined in momentum space, using dimensional regularization and
the $\bar{MS}$ renormalization scheme (modified minimial subtraction: not only
the pole in $\epsilon$ is subtracted, but also terms of $\ln 4 \pi$). The
result is:
\begin{equation}
V(\vec{q}^2) = -
\frac{g_{\bar{MS}}^2(\mu^2)}{\vec{q}^2} C_2(F) \left[ 1 +
g_{\bar{MS}}^2(\mu^2) \beta_0 \left(\ln \frac{\mu^2}{\vec{q}^2} - \gamma +
\frac{31}{33} \right) \right].
\end{equation}
Here the Feynman gauge (gauge parameter $\alpha = 1$) and the quenched
approximation (number of dynamic fermions $n_f = 0$) was used. $g_{\bar{MS}}$
is the coupling constant defined in the $\bar{MS}$ scheme, $\mu$ a mass
parameter which has to be introduced during renormalization, $\gamma \approx
0.577216$ is the Euler-Mascheroni constant and $\beta_0$ the first coefficient
in the expansion of the beta function:
\begin{equation}
\beta_0 = \frac{11 C_2(G)}{48\pi^2},
\end{equation}
where $C_2(G) = N$ is the quadratic Casimir operator of $SU(N)$ in the
adjoint representation.

From this one can get an expression for the potential which should result from
lattice calculations by using the relation between $g_{\bar{MS}}$ and the
coupling constant $g_0(a)$ appearing in perturbative calculations on
the lattice. This is done in two steps; first the following relation can
be used:
\[
g_{MS}^2(\mu^2) = g_{MOM}^2(\mu^2) \left[1 - \frac{A(\alpha, n_f, N)}{4 \pi}
g_{MOM}^2(\mu^2) + O(g^4) \right],
\]
where $g_{MS}$ and $g_{MOM}$ are the coupling constants defined in the 
$MS$ and $MOM$ (momentum-space subtraction) renormalization scheme,
respectively. $A(\alpha, n_f, N)$ was determined in \cite{CelmasterGonsalves}
for various values of the gauge parameter $\alpha$ and numbers of
dynamical fermions $n_f$; the value $A(1, 0, N)$ which is relevant here 
is given by:
\[
A(1, 0, N) = \frac{C_2(G)}{2\pi} \left(-\frac{11}{6} \gamma + \frac{11}{6}
\ln 4 \pi + \frac{23}{6} + \frac{1}{36} I \right)
\]
with
\[
I = -2 \int\limits_0^1 \frac{\ln x}{x^2-x+1} \approx 2.3439072.
\]
But here $g_{\bar{MS}}$ is needed; for this one obtains:
\begin{equation}
g_{\bar{MS}}^2(\mu^2) = g_{MOM}^2(\mu^2) \left[1 - \frac{\bar{A}(1, 0,
N)}{4 \pi} g_{MOM}^2(\mu^2) + O(g^4) \right]
\end{equation}
with
\begin{equation}
\bar{A}(1, 0, N) = A(1, 0, N) - \frac{11 C_2(G)}{12\pi} \ln 4 \pi =
\frac{C_2(G)}{2\pi} \left(-\frac{11}{6} \gamma + \frac{23}{6} + \frac{1}{36} I
\right).
\end{equation}

The second step is to connect $g_{MOM}(\mu^2)$ with $g_0(a)$ in the 
continuum limit $a \to 0$ \cite{Hasenfratz, Weisz}:
\begin{equation}
g_{\bar{MOM}}^2(\mu^2) = g_0^2(a) \left[1 + g_0^2(a)
\left(\beta_0 \ln \frac{\pi^2}{a^2 \mu^2} + R(N)\right) +
O(g^4) \right],
\end{equation}
with
\begin{eqnarray}
R(N) &=& C_2(G) \left[\frac{1}{48\pi^2} \left(23 - 22 \ln \pi +
\frac{1}{6} I \right) + 2 P \right] - \frac{1}{8N} \nonumber \\
&=& C_2(G) \left[\frac{1}{48\pi^2} \left(23 - 22 \ln \pi +
\frac{1}{6} I \right) + 2 P  - \frac{1}{8}\right] + \frac{1}{4} C_2(F),
\end{eqnarray}
where $I$ is defined as above and $P \approx 0.0849780$.

Putting everything together, the result for the potential in momentum space is:
\begin{eqnarray*}
V(\vec{q}^2) &=& - \frac{g_0^2(a)}{\vec{q}^2} C_2(F) \\
&& \cdot \left[ 1 + g_0^2(a) \left[\beta_0 \left(\ln
\frac{\pi^2}{a^2\vec{q}^2} - \gamma + \frac{31}{33} \right)
 - \frac{\bar{A}(1, 0, N)}{4 \pi} + R(N) \right] \right].
\end{eqnarray*}
In order to get the potential in coordinate space, one has to do a Fourier
transformation (for details, see appendix \ref{potentialfourier}):
\begin{eqnarray*}
V(R) &=& - \frac{g_0^2(a)}{4 \pi R} C_2(F) \\
&& \cdot \left[ 1 + g_0^2(a) \left[\beta_0 \left(\ln
\frac{\pi^2 R^2}{a^2} + \gamma + \frac{31}{33} \right)
 - \frac{\bar{A}(1, 0, N)}{4 \pi} + R(N) \right] \right];
\end{eqnarray*}
hence the potential, measured in lattice spacings, is:
\begin{eqnarray}
\hat{V}(\hat{R}) &=& - \frac{g_0^2(a)}{4 \pi \hat{R}} C_2(F) \\
&& \cdot \left[ 1 + g_0^2(a) \left[2 \beta_0 \ln
\left(\pi e^{\gamma/2 + 31/66} \hat{R} \right)
 - \frac{\bar{A}(1, 0, N)}{4 \pi} + R(N) \right] \right]. \nonumber
\end{eqnarray}
Inserting the explicit expressions for $\bar{A}(1, 0, N)$ and $R(N)$ given
above, one finally arrives at:
\begin{eqnarray}
\hat{V}(\hat{R}) &=& - \frac{g_0^2(a)}{4 \pi \hat{R}} C_2(F) \\
&& \cdot \left[ 1 + g_0^2(a) \left[2 \beta_0 \ln
\left(e^{\gamma + 31/66 + 48 \pi^2 P / 11 - 3 \pi^2/11} \hat{R} \right)
 + \frac{1}{4} C_2(F) \right] \right] \nonumber \\
&\approx& - \frac{g_0^2(a)}{4 \pi \hat{R}} C_2(F)
\left[ 1 + g_0^2(a) \left[\frac{11}{24 \pi^2} C_2(G) \ln
\left(7.501 \hat{R}\right) + \frac{1}{4} C_2(F) \right] \right] \nonumber
\end{eqnarray}
or
\begin{eqnarray}
\hat{V}(\hat{R}) &\approx& - \frac{g_0^2(a)}{4 \pi \hat{R}} C_2(F)
\left[ 1 + 2 g_0^2(a) \beta_0 \ln
\left(7.501 e^{C_2(F)/8 \beta_0} \hat{R} \right) \right] \nonumber \\
&=& - \frac{g_0^2(a)}{4 \pi \hat{R}} C_2(F)
\left[ 1 + 2 g_0^2(a) \beta_0 \ln \left(7.501
e^{3 \pi^2 (N^2-1)/11 N^2} \hat{R} \right) \right] \nonumber \\
&=& - \frac{g_0^2(a)}{4 \pi \hat{R}}
\left\{ \begin{array}{cc}
\frac{3}{4}
\left[ 1 + 2 g_0^2(a) \beta_0 \ln \left(56.47 \hat{R} \right) \right] &
SU(2) \\
\frac{4}{3}
\left[ 1 + 2 g_0^2(a) \beta_0 \ln \left(82.07 \hat{R} \right) \right] & SU(3)
\\
\end{array} \right.
\label{Potential}
\end{eqnarray}

Alternatively, one can express these results by using the QCD scale
parameters $\Lambda$ of the different renormalization schemes; the following
relations hold \cite{CelmasterGonsalves, Hasenfratz, Weisz}:
\begin{eqnarray}
\frac{\bar{A}(1, 0, N)}{4 \pi} &=& 2 \beta_0 \ln
\frac{\Lambda_{MOM}}{\Lambda_{\bar{MS}}} \nonumber \\
R(N) &=& 2 \beta_0 \ln \frac{\Lambda_{MOM}}{\pi \Lambda_{L}}.
\end{eqnarray}
With this, one can write for the potential
\begin{eqnarray}
\hat{V}(\hat{R}) &=& - \frac{g_0^2(a)}{4 \pi \hat{R}} C_2(F)
\left[ 1 + 2 g_0^2(a) \beta_0 \ln \left(
\frac{e^{\gamma/2 + 31/66} \Lambda_{\bar{MS}}}{\Lambda_{L}}
\hat{R} \right) \right] \nonumber \\
&\approx& - \frac{g_0^2(a)}{4 \pi \hat{R}} C_2(F)
\left[ 1 + 2 g_0^2(a) \beta_0 \ln \left(2.135
\frac{\Lambda_{\bar{MS}}}{\Lambda_{L}} \hat{R} \right) \right].
\end{eqnarray}
The ratio of the scale parameters is:
\begin{equation}
\frac{\Lambda_{\bar{MS}}}{\Lambda_{L}} \approx
\left\{ \begin{array}{cc}
26.45 & SU(2) \\
38.45 & SU(3) \\
\end{array} \right.
\end{equation}

These expressions for the potential agree with results from Monte Carlo
simulations \cite{Kovacs}, as well as with a perturbative calculation on the
lattice \cite{Karsch}. Additionally, the potential shows the right scaling
behaviour, compared with (\ref{PotentialScaling}).

In some older publications, one can find the following simpler expression
for the potential instead of the one derived above \cite{wrongpotential}:
\begin{equation}
\hat{V}(\hat{R}) = - \frac{g_0^2(a)}{4 \pi \hat{R}} C_2(F)
\left[ 1 + 2 g_0^2(a) \beta_0 \ln \hat{R} \right],
\end{equation}
the (big) factor $2.135 \frac{\Lambda_{\bar{MS}}}{\Lambda_{L}}$ in the
logarithm is missing there completely! There are two reasons for this:
\begin{enumerate}
\item The contribution from the tadpole graph in the gluonic vacuum
polarization was left out by these authors; therefore the term
proportional to $C_2(F)^2$ is missing. That term contributes a factor $7.529$
($SU(2)$) or $10.942$ ($SU(3)$), respectively, in the logarithm.
\item For the ultraviolet and infrared cutoffs, they used only rough
approximations, choosing them idential to $a$ and $R$, respectively; hence
the ratio of the cutoffs is identical to $\hat{R}$ in their results. In
contrast, the correct calculation yields an additional factor of 7.501 in the
logarithm.
\end{enumerate}

\section{Weak coupling expansion on the lattice}

\label{potential_lattice}

A weak coupling expansion of the quark-antiquark potential on the lattice was
done first by Heller and Karsch in 1985 \cite{Karsch}. Their methods and
results are summarized in this section for future reference.

As already mentioned in the introduction, the quark-antiquark potential can be
derived from the expectation value of the Wilson loop:
\begin{equation}
\hat{V}(\hat{R},\hat{\beta}) = -\lim_{\hat{T} \to \infty} \frac{1}{\hat{T}}
\ln <W(\hat{R}, \hat{T})>_{subtr},
\end{equation}
with
\begin{equation}
W = \frac{1}{d(F)} \mbox{Tr} \prod_{l \in C} U_l. \quad \quad \mbox{(path
ordered)}
\end{equation}
The subscript ''subtr'' means that in order to get the correct potential from
the formula above, one has to subtract the self energy contributions to the
quarks. This subscript will be omitted in most places in the following.

The rectangular curve $C$ is chosen to have the corners $n_0$,
$n_0+\hat{R}\hat{\mu}$, $n_0+\hat{R}\hat{\mu}+\hat{T}\hat{\nu}$,
$n_0+\hat{T}\hat{\nu}$, where $\hat{\mu}$ and $\hat{\nu}$ are the unit vectors
in one spatial and the temporal direction of the lattice, respectively. The
link variables are connected with the vector potential by the following
relation:
\begin{equation}
U_l = e^{i g_0 A_l}.
\end{equation}

Using the Baker-Hausdorf formula repeatedly, one then arrives at the
following expression for $W$:
\begin{eqnarray*}
W \!\!\! &=& \!\! \frac{1}{d(F)} \mbox{Tr} \exp\left[i g_0 \sum_l A_l -
\frac{1}{2} g_0^2 \sum_{l_1 < l_2} [A_{l_1}, A_{l_2}] - \frac{1}{4} i g_0^3
\sum_{l_1 < l_2 < l_3} [[A_{l_1}, A_{l_2}], A_{l_3}] \right. \\
&&\left. - \frac{1}{12} i g_0^3 \sum_{(l_1, l_2) < l_3} [A_{l_1}, [A_{l_2},
A_{l_3}]] - \frac{1}{12} i g_0^3 \sum_{l_1 < l_2} [[A_{l_1}, A_{l_2}], A_{l_2}]
+ O(g_0^4) \right],  
\end{eqnarray*}
where $A_l$ is given by:
\begin{eqnarray*}
A_l &=& A_{\mu}(n) \quad \quad \mbox{for } l=(n,n+\hat{\mu}) \\
A_l &=& -A_{\mu}(n-\hat{\mu}) \quad \quad \mbox{for } l=(n,n-\hat{\mu}).
\end{eqnarray*}
Expanding the exponential and taking the trace, one obtains the perturbative
expansion for the Wilson loop:
\begin{equation}
W = 1 - g_0^2 \omega^{(2)} - g_0^3 \omega^{(3)} - g_0^4 \omega^{(4)} + O(g_0^5)
\end{equation}
with
\begin{eqnarray}
\omega^{(2)} &=& \frac{1}{4d(F)} \left( \sum_l A_l^A \right)^2 \\
\omega^{(3)} &=& \frac{i}{6d(F)} \mbox{Tr} \left( \sum_l A_l \right)^3
+ \frac{i}{2d(F)} \mbox{Tr} \left( \sum_l A_l \sum_{l_1 < l_2} [A_{l_1},
A_{l_2}] \right) \\
\omega^{(4)} &=& - \frac{1}{24d(F)} \mbox{Tr} \left( \sum_l A_l \right)^4
- \frac{1}{8d(F)} \mbox{Tr} \left(\sum_{l_1 < l_2} [A_{l_1}, A_{l_2}] \right)^2
\nonumber \\ &&- \frac{1}{4d(F)} \mbox{Tr} \left( \sum_l A_l \sum_{l_1 < l_2 <
l_3} [[A_{l_1}, A_{l_2}], A_{l_3}] \right) \nonumber \\
&&- \frac{1}{12d(F)} \mbox{Tr} \left( \sum_l A_l \sum_{(l_1, l_2) < l_3}
[A_{l_1}, [A_{l_2}, A_{l_3}]] \right) \nonumber \\
&&- \frac{1}{12d(F)} \mbox{Tr} \left( \sum_l A_l \sum_{l_1 < l_2} [[A_{l_1},
A_{l_2}], A_{l_2}] \right) \nonumber \\ &&- \frac{1}{4d(F)} \mbox{Tr} \left(
\left(\sum_l A_l\right)^2 \sum_{l_1 < l_2} [A_{l_1}, A_{l_2}] \right).
\end{eqnarray}
For doing explicit calculations with these expressions, it is convenient to
split $\omega^{(4)}$ up into its parts in the following way:
\begin{equation}
\omega^{(4)} = -\omega^{(4A)} - \ldots - \omega^{(4F)}.
\label{omega4split}
\end{equation}

Additionally, in order to calculate the expectation value of the Wilson loop,
one needs a perturbative expansion for the action. Expectation values are
calculated in the usual way:
\begin{equation}
<O> = \frac{\int DU O e^{-S}}{\int DU e^{-S}},
\end{equation}
where $S$ is Wilson's action (\ref{WilsonAction}) and the integral runs over
all elements $U$ of the gauge group. Carrying out the Faddeev-Popov gauge
fixing and rewriting the integration measure using the vector potential
$A^A_{\mu}$, one arrives at the following perturbative expansion (see,
for example, \cite{RotheBook}):
\begin{equation}
S_{eff} = S^{(0)} + g_0 S^{(1)} + g_0^2 S^{(2)} + g_0^2 S^{(2)}_{FP}
+ g_0^2 S^{(2)}_{meas} + O(g_0^3).
\end{equation}
$S^{(1)}$ gives the three-gluon, $S^{(2)}$ the four-gluon vertex; $S^{(FP)}$
comes from the Fadeev-Popov determinant, and $S^{(meas)}$ from the
transformation of the integration measure. $S^{(0)}$ is the term of
order $g_0^0$, which is quadratic in the gauge fields:
\begin{equation}
S^{(0)} = -\frac{1}{2} \int\limits_{BZ} \sum_{\mu, \nu}
\frac{d^4p}{(2\pi)^4} \hat{A}^{A}_{\mu}(-p)
\left(\hat{p}^2 \delta_{\mu\nu} -
\left(1 - \frac{1}{\alpha}\right) \hat{p}_{\mu} \hat{p}_{\nu} \right)
\hat{A}^{A}_{\nu}(p).
\end{equation}
Here, $\alpha$ is the usual gauge parameter; for all the results summarized
below, the Feynman gauge $\alpha = 1$ had been used. The subscript
\emph{BZ} denotes the region of integration: the first Brillouin
zone, all components of $p$ run from $-\pi$ to $+\pi$. Additionally, the
following standard abbreviations were introduced above:
\begin{eqnarray}
\hat{p}_{\mu} &=& 2 \sin(p_{\mu}/2) \nonumber \\ 
\hat{p}^2 &=& \sum_{\mu=1}^4
\hat{p}_{\mu}^2.
\end{eqnarray}
Note also that the sum over the Lorentz indices $\mu$ and $\nu$ was explicitly
written out, i.\ e.\ the usual sum convention is not used in this work:
\emph{Repeated indices only have to be summed if this is explicitly denoted.}

Now the expectation value of an operator becomes:
\begin{eqnarray}
<O> &=& \frac{\int DA \; O \; e^{-S_{eff}} }{\int DA \; e^{-S_{eff}}}
= \frac{\int DA \; O \; e^{-S^{(0)}}}{\int DA \; e^{-S^{(0)}}}(1+O(g_0))
\nonumber \\ &=:& <O>_0 + O(g_0),
\end{eqnarray}
and the free gluon propagator is given by
\begin{eqnarray}
<A^A_{\mu}(p)A^B_{\nu}(q)>_0 &=& \delta^{AB}
(2 \pi)^4 \delta(p+q) \frac{\delta_{\mu\nu} - \left(1 -
\frac{1}{\alpha} \right) \frac{\hat{p}_{\mu} \hat{p}_{\nu}}{\hat{p}^2}}
{\hat{p}^2} \label{Propagator} \\
&=& \delta^{AB} (2 \pi)^4 \delta(p+q) \frac{\delta_{\mu\nu}}{\hat{p}^2} \quad
\mbox{(Feynman gauge)}. \nonumber
\end{eqnarray}

The explicit expressions for the other coefficients in the expansion of the
action can also be found partly in \cite{Karsch} (but notice that the
action is defined there with just the opposite sign in comparison to the
convention here!); for the three- and four-gluon vertices, see \cite{vertices}
or \cite{Kawai}. A more modern version, better suited for calculations, is
given in \cite{RotheBook}.

\subsection{Leading order}

\label{potential_LO}

Using the formula (\ref{PotentialWilson}) and the expansion of the Wilson loop
outlined above, one gets in leading order:
\begin{equation}
\hat{V} = g_0^2 \lim_{\hat{T} \to \infty} \frac{1}{\hat{T}}
<\omega^{(2)}>_{0,subtr},
\end{equation}
with
\begin{eqnarray}
&& <\omega^{(2)}>_0 \\ &=& 2 C_2(F) \int\limits_{BZ}
\frac{d^4p}{(2\pi)^4}
\frac{\sin^2(p_{\mu}\hat{R}/2)\sin^2(p_{\nu}\hat{T}/2)}
{\hat{p}^2} \left(\frac{1}{\sin^2(p_{\mu}/2)} +
\frac{1}{\sin^2(p_{\nu}/2)} \right). \nonumber
\end{eqnarray}
For a detailed derivation of this expression, see appendix \ref{sums}. As
mentioned above, $subtr$ means that the self-energy contributions have to be
subtracted. $\mu$ and $\nu$ represent here the spatial respectively temporal
direction of the Wilson loop; these are different, \emph{fixed} indizes and
are \emph{not} summed over. Because of the symmetry of the lattice, these two
indices can be chosen arbitrarily without changing the result. 

In this order, the potential can be calculated exactly. First, perform the
limit $\hat{T} \to \infty$. The following relations hold:
\begin{eqnarray*}
\lim_{\hat{T} \to \infty} \frac{1}{\hat{T}}
\frac{\sin^2(p_{\mu}\hat{R}/2)\sin^2(p_{\nu}\hat{T}/2)}
{\sin^2(p_{\mu}/2)}
&=& 0 \\
\lim_{\hat{T} \to \infty} \frac{1}{\hat{T}}
\frac{\sin^2(p_{\mu}\hat{R}/2)\sin^2(p_{\nu}\hat{T}/2)}
{\sin^2(p_{\nu}/2)}
&=& 2 \pi \delta(p_{\nu}).
\end{eqnarray*}
With this one gets:
\begin{equation}
\hat{V} = 2 C_2(F) g_0^2 \int\limits_{BZ}
\frac{d^3p}{(2\pi)^3}
\frac{\sin^2(p_{\mu}\hat{R}/2)} {\hat{\vec{p}}^2},
\end{equation}
where
\[
\hat{\vec{p}}^2 = \sum_{\lambda \ne \nu} \hat{p}_{\lambda}^2;
\]
hence
\begin{eqnarray*}
V &=& \frac{2 C_2(F) g_0^2}{a} \int\limits_{BZ} \frac{d^3p}{(2\pi)^3}
\frac{\sin^2(p_{\mu}\hat{R}/2)} {\hat{\vec{p}}^2} \\
&=& 2 C_2(F) g_0^2 \int\limits_{-\pi/a}^{\pi/a} \frac{d^3p'}{(2\pi)^3}
\frac{\sin^2(p'_{\mu} R/2)} {\frac{4}{a^2} \sum_{\lambda \ne \nu}
\sin^2(p'_{\lambda} a/2)}.
\end{eqnarray*}

In the continuum limit $a \to 0$, the integral gets its main contributions
for  $p'_{\lambda}a \approx 0$. Hence one can expand the sines in the
denominator: $\sin^2\left(p'_{\lambda}a/2)\right) \approx \frac{1}{4}
p'^2_{\lambda} a^2$. The integration limits $\pm\pi/a$ go to $\pm\infty$ in the
limit $a \to 0$. Therefore one gets:
\begin{equation}
V = 2 C_2(F) g_0^2 \int\limits_{-\infty}^{\infty} \frac{d^3p'}{(2\pi)^3}
\frac{\sin^2(p'_{\mu} R/2)} {\sum_{\lambda \ne \nu} p'^2_{\lambda}}.
\end{equation}

\noindent
Now use the relation
\[
\int_0^{\infty} dt e^{-tx} = \frac{1}{x};
\]
with this one can write:
\begin{eqnarray*}
V &=& 2 C_2(F) g_0^2  \int\limits_{-\infty}^{\infty} \frac{d^3p'}{(2\pi)^3}
\int_0^{\infty} dt \sin^2(p'_{\mu} R/2) e^{-t \sum_{\lambda \ne \nu}
p'^2_{\lambda}} \\
&=& 2 C_2(F) g_0^2  \int_0^{\infty} dt \int\limits_{-\infty}^{\infty}
\frac{dp'_{\mu}}{2\pi} \sin^2(p'_{\mu}R/2) e^{-t p'^2_{\mu}}
\left(\int\limits_{-\infty}^{\infty} \frac{dx}{2\pi} e^{-t x^2} \right)^2
\\
&=& \frac{1}{2\pi} C_2(F) g_0^2  \int_0^{\infty} \frac{dt}{t}
\int\limits_{-\infty}^{\infty} \frac{dp'_{\mu}}{2\pi}
\sin^2(p'_{\mu}R/2) e^{-t p'^2_{\mu}} \\
&=& \frac{1}{4\pi} C_2(F) g_0^2 \int_0^{\infty} \frac{dt}{t}
\int\limits_{-\infty}^{\infty} \frac{dp'_{\mu}}{2\pi}
(1 - e^{i p'_{\mu} R}) e^{-t p'^2_{\mu}}.
\end{eqnarray*}
After completing the square in the exponent, the integral over $p'_{\mu}$ can
be done:
\[
V = \frac{1}{8\pi^{3/2}} C_2(F) g_0^2  \int_0^{\infty} \frac{dt}{t^{3/2}}
(1 - e^{-R^2/4t}).
\]

Only the second term depends on $R$; the first gives an infinite contribution
which is independent of $R$---hence it is the self energy contribution (this
can also be seen by a close inspection of the graphs contributing to
$<\omega^{(2)}>_0$). After subtracting it, what remains is:
\begin{eqnarray}
V &=& -\frac{g_0^2}{8\pi^{3/2}} C_2(F) \int_0^{\infty}
\frac{dt}{t^{3/2}} e^{-R^2/4t} \nonumber \\
&=& -\frac{g_0^2}{4\pi^{3/2}R} C_2(F) \int_0^{\infty} dx
e^{-x^2}
= -\frac{g_0^2}{4\pi^{3/2}R} C_2(F) \int_{-\infty}^{\infty} dx
e^{-x^2} \nonumber \\
&=& -\frac{g_0^2}{4 \pi R} C_2(F).
\end{eqnarray}

Hence indeed one recovers the correct formula for the potential in leading
order from lattice perturbation theory.

\subsection{Next-to-leading order}

\label{potential_NLO}

In next-to-leading order, there are several different contributions to the
expectation value of the Wilson loop:
\begin{equation}
<W> = 1 - g_0^2 \ldots + g_0^4 <\omega^{(3)}S^{(1)}>_{conn} - g_0^4
<\omega^{(4)}>_0 - g_0^4 W_{VP} + O(g_0^6),
\end{equation}
(note the different sign for the $\omega^{(3)}$-term, compared to
\cite{Karsch}) where the subscript ''conn'' means that one has to consider
only the connected graphs contributing to the expectation value, and $W_{VP}$
stands for the contributions to $<W>$ coming from the vacuum polarization
graphs:
\begin{eqnarray}
W_{VP} &=& -<\omega^{(2)}S^{(2)}>_{conn} +
<\omega^{(2)}\frac{1}{2}\left(S^{(1)}\right)^2>_{conn} -
<\omega^{(2)}S^{(2)}_{FP}>_{conn} \nonumber \\ && -
<\omega^{(2)}S^{(2)}_{meas}>_{conn} \\ &=& 2 C_2(F) \int\limits_{BZ}
\frac{d^4p}{(2 \pi)^4} \frac{\sin^2(p_{\mu} \hat{R}/2) \sin^2(p_{\nu}
\hat{T}/2)}{(\hat{p}^2)^2} \nonumber \\ && \cdot
\left(\frac{\Pi_{\mu\mu}(p)}{\sin^2(p_{\mu}/2)} - 2
\frac{\Pi_{\mu\nu}(p)}{\sin(p_{\mu}/2)\sin(p_{\nu}/2)} +
\frac{\Pi_{\nu\nu}(p)}{\sin^2(p_{\nu}/2)} \right).
\end{eqnarray}
Again, the signs of three of the terms are opposite to the ones in
\cite{Karsch} because of the different sign convention for the action.

\vspace{2ex}

\noindent
The explicit contributions are, given here for future reference:
\begin{eqnarray}
&&<\omega^{(3)}S^{(1)}>_{conn} \nonumber \\ &=& -C_2(G) C_2(F) \int\limits_{BZ}
\frac{d^4p}{(2 \pi)^4} \int\limits_{BZ} \frac{d^4k}{(2 \pi)^4}
\frac{1}{\hat{p}^2 \hat{k}^2 (\widehat{p+k})^2} \nonumber \\
&& \Bigg[ \Bigg\{ \frac{\sin^2(p_{\mu}\hat{R}/2)}{\sin(p_{\mu/2})}
\sin(p_{\nu}\hat{T}/2) \cos(p_{\nu}/2) \sin((2p+k)_{\mu}/2) \nonumber \\
&&\cdot \Bigg(\frac{\sin(p_{\nu}\hat{T}/2)
\sin((2p+k)_{\nu}/2)}{\sin(p_{\nu}/2) \sin(k_{\nu}/2) \sin((p+k)_{\nu}/2)} -
\frac{\sin((2p+k)_{\nu}\hat{T}/2)}{\sin(k_{\nu}/2) \sin((p+k)_{\nu}/2)}\Bigg)
\nonumber \\ &&+ 4 \frac{\sin((k-p)_{\nu}/2) \cos((p+k)_{\mu}/2)}
{\sin(p_{\mu}/2) \sin(k_{\mu}/2) \sin((p+k)_{\nu}/2)} \nonumber \\ && \cdot
\sin(p_{\mu}\hat{R}/2) \sin(p_{\nu}\hat{T}/2) \sin((p+k)_{\nu}\hat{T}/2)
\sin(k_{\mu}\hat{R}/2) \nonumber \\ && \cdot \cos((p+k)_{\mu}\hat{R}/2)
\cos(k_{\nu}\hat{T}/2) \Bigg\} + \Bigg\{ (\mu,\hat{R}) \leftrightarrow
(\nu,\hat{T}) \Bigg\} \Bigg]
\end{eqnarray}
Here again the same comment with respect to the sign as above applies.

\vspace{2ex}

For the calculation of $<\omega^{(4)}>$, use the decomposition
(\ref{omega4split}) and determine each of the six contributions separately
\cite{Heller}:
\begin{eqnarray}
<\omega^{(4A)}>_0 &=& \left(2 C_2(F)^2 - \frac{1}{3} C_2(G) C_2(F) \right)
\\ &&
\Bigg[\int\limits_{BZ} \frac{d^4p}{(2 \pi)^4}  \frac{\sin^2(p_{\mu}\hat{R}/2)
\sin^2(k_{\nu}\hat{T}/2)}{\hat{p}^2} \left(\frac{1}{\sin^2(p_{\mu})}
+ \frac{1}{\sin^2(p_{\nu})}\right) \Bigg]^2 \nonumber
\label{omega4start}
\end{eqnarray}
\begin{eqnarray}
<\omega^{(4B)}>_0 &=& -\frac{1}{2} C_2(G) C_2(F) g_0^4 \int\limits_{BZ}
\frac{d^4p}{(2 \pi)^4} \int\limits_{BZ} \frac{d^4k}{(2 \pi)^4} \nonumber \\ &&
\frac{1}{\hat{p}^2 \hat{k}^2}  \Bigg[ \left\{ \frac{\sin^2(p_{\mu}\hat{R}/2)
\sin^2(k_{\nu}\hat{T}/2)} {\sin^2(p_{\mu/2}) \sin^2(k_{\nu}/2)} \right\}
+ \left\{ (\mu,\hat{R}) \leftrightarrow (\nu,\hat{T}) \right\} \Bigg]
\nonumber \\ && -\frac{N^2-1}{16} g_0^4 \int\limits_{BZ} \frac{d^4p}{(2 \pi)^4}
\int\limits_{BZ} \frac{d^4k}{(2 \pi)^4} \frac{1}{\hat{p}^2 \hat{k}^2} 
\Bigg[\Bigg\{ \frac{1}{2} \frac{\sin^2((p_{\mu}+k_{\mu})\hat{R}/2)}
{\sin^2(p_{\mu/2})} \nonumber \\
&&\frac{\sin^2((p_{\nu}+k_{\nu})\hat{T}/2)}{\sin^2(k_{\mu}/2)}
\left[\frac{\sin((p_{\mu}-k_{\mu})\hat{R}/2)}{\sin((p_{\mu}+k_{\mu})\hat{R}/2)}
 - \frac{\sin((p_{\mu}-k_{\mu})/2)}{\sin((p_{\mu}+k_{\mu})/2)} \right]^2
\nonumber \\
&&+2\frac{\sin^2((k_{\nu}-p_{\nu})\hat{T}/2)\sin^2(p_{\mu}\hat{R}/2)\sin^2(k_{\mu}\hat{R}/2)}
{\sin^2(p_{\mu}/2)\sin^2(k_{\mu}/2)} \Bigg\} \nonumber \\ && + \left\{
(\mu,\hat{R}) \leftrightarrow (\nu,\hat{T}) \right\} \Bigg]
\end{eqnarray}
\begin{eqnarray}
<\omega^{(4C)}>_0 \! &=& \frac{1}{2} C_2(G) C_2(F) g_0^4 \int\limits_{BZ}
\frac{d^4p}{(2 \pi)^4} \int\limits_{BZ} \frac{d^4k}{(2 \pi)^4}
\frac{1}{\hat{p}^2 \hat{k}^2} \nonumber \\ && \Bigg[ \Bigg\{
\frac{\sin(p_{\mu}\hat{R}/2)}{\sin(p_{\mu}/2)} 2 \sin^2(p_{\nu}\hat{T}/2)
(\Sigma_1 - \Sigma_2) \nonumber \\ &&- \frac{\sin(p_{\mu}\hat{R}/2)
\sin^2(p_{\nu}\hat{T}/2)}{\sin(p_{\mu}/2)} \frac{\sin(k_{\mu}\hat{R}/2)
\cos(k_{\nu}\hat{T})}{\sin(k_{\mu}/2)} \nonumber \\ && \cdot
\left(\Sigma_R(-p_{\mu},k_{\mu}) - \Sigma_R(k_{\mu},-p_{\mu})\right)
\nonumber \\ &&+ \frac{\sin^2(p_{\mu}\hat{R}/2)
\sin^2(p_{\nu}\hat{T}/2)}{\sin^2(p_{\mu}/2)} \Sigma_R(k_{\mu},-k_{\mu})
\nonumber \\ &&+ \frac{\sin(p_{\mu}\hat{R}/2)
\sin^2(p_{\nu}\hat{T}/2)}{\sin(p_{\mu}/2)} \frac{\sin(k_{\mu}\hat{R}/2)
\cos(k_{\nu}\hat{T})}{\sin(k_{\mu}/2)} \Sigma_R(-k_{\mu},-p_{\mu}) \nonumber \\
&&+ \frac{\sin^2(p_{\mu}\hat{R}/2) \sin^2(p_{\nu}\hat{T}/2)}{\sin^2(p_{\nu}/2)}
\Sigma_R(k_{\mu},-k_{\mu}) \Bigg\} + \left\{(\mu,\hat{R}) \leftrightarrow
(\nu,\hat{T}) \right \} \nonumber \\ &&+ 2 \frac{\sin^2(p_{\mu}\hat{R}/2)
\sin^2(p_{\nu}\hat{T}/2)}{\sin^2(p_{\mu}/2)} \Sigma_T(k_{\nu},-k_{\nu})
\nonumber \\ &&+ \frac{\sin^2(p_{\mu}\hat{R}/2)
\sin^2(p_{\nu}\hat{T}/2)}{\sin^2(p_{\nu}/2)} \frac{\sin^2(k_{\mu}\hat{R}/2)
\cos(k_{\nu} \hat{T})}{\sin^2(k_{\mu}/2)} \Bigg]
\end{eqnarray}
\begin{eqnarray}
<\omega^{(4D)}>_0 \! &=& \frac{1}{6} C_2(G) C_2(F) g_0^4 \int\limits_{BZ}
\frac{d^4p}{(2 \pi)^4} \int\limits_{BZ} \frac{d^4k}{(2 \pi)^4}
\frac{1}{\hat{p}^2 \hat{k}^2} \nonumber \\ && \Bigg[ \Bigg\{
\frac{\sin(p_{\mu}\hat{R}/2)}{\sin(p_{\mu}/2)} \sin^2(p_{\nu}\hat{T}/2)
\nonumber \\ && \cdot \left(\Sigma_R(0,-p_{\mu}) + \Sigma_1 - \Sigma_2 -
\Sigma_R(k_{\mu}-p_{\mu}, -k_{\mu}) \right) \nonumber \\
&&+ \frac{\sin(p_{\mu}\hat{R}/2)}{\sin(p_{\mu}/2)} \sin^2(p_{\nu}T/2)
\nonumber \\ && \cdot \left(\Sigma_R(-p_{\mu},0) + \Sigma_1 - \Sigma_2 -
\Sigma_R(-k_{\mu}, k_{\mu}-p_{\mu}) \right) \nonumber \\
&& + \frac{\sin^2(p_{\mu}\hat{R}/2)}{\sin^2(p_{\mu}/2)}
\sin^2(p_{\nu}\hat{T}/2) \frac{\sin^2(k_{\nu}\hat{T}/2)}{\sin(k_{\nu}/2)}
\nonumber \\ && + \frac{\sin(p_{\mu}\hat{R}/2)}{\sin^2(p_{\mu}/2)}
\sin^2(p_{\nu}\hat{T}/2) \frac{\sin^2(k_{\mu}\hat{R}/2)}{\sin^2(k_{\mu}/2)}
\left(1 + \cos(k_{\nu}\hat{T}) \right) \nonumber \\ &&-
\frac{\sin(p_{\mu}\hat{R}/2)}{\sin(p_{\mu}/2)} \sin^2(p_{\nu}\hat{T}/2) 
\frac{\sin(k_{\mu}\hat{R}/2)}{\sin(k_{\mu}/2)} \cos(k_{\nu}\hat{T}) \nonumber
\\ &&\left(\Sigma_R(-p_{\mu},-k_{\mu}) - \Sigma_R(-k_{\mu},-p_{\mu}) \right)
\nonumber \\ &&- \frac{\sin(p_{\mu}\hat{R}/2)}{\sin(p_{\mu}/2)}
\sin^2(p_{\nu}\hat{T}/2)  \frac{\sin(k_{\mu}\hat{R}/2)}{\sin(k_{\mu}/2)}
\cos(k_{\nu}\hat{T}) \Sigma_R(-p_{\mu},k_{\mu}) \nonumber \\ &&-
\frac{\sin(p^2_{\mu}\hat{R}/2)}{\sin^2(p_{\mu}/2)} \sin^2(p_{\nu}\hat{T}/2)
\Sigma_R(k_{\mu},-k_{\mu}) \nonumber \\ &&-
\frac{\sin^2(p_{\mu}\hat{R}/2)}{\sin^2(p_{\nu}/2)} \sin^2(p_{\nu}\hat{T}/2)
\Sigma_R(k_{\mu},-k_{\mu}) \Bigg\} + \left\{(\mu,\hat{R}) \leftrightarrow
(\nu,\hat{T}) \right \} \nonumber \\ &&+ 2
\frac{\sin^2(p_{\mu}\hat{R}/2)}{\sin^2(p_{\nu}/2)} \sin^2(p_{\nu}\hat{T}/2)
\frac{\sin^2(k_{\mu}\hat{R}/2)}{\sin^2(k_{\mu}/2)} \nonumber \\ &&- 2
\frac{\sin^2(p_{\mu}\hat{R}/2)}{\sin^2(p_{\mu}/2)} \sin^2(p_{\nu}\hat{T}/2)
\Sigma_T(k_{\nu},-k_{\nu}) \nonumber \\ &&-
\frac{\sin^2(p_{\mu}\hat{R}/2)}{\sin^2(p_{\nu}/2)} \sin^2(p_{\nu}\hat{T}/2)
\frac{\sin^2(k_{\mu}\hat{R}/2)}{\sin^2(k_{\mu}/2)} \cos(k_{\nu} \hat{T})
\nonumber \\ &&+ 2 \frac{\sin^2(p_{\mu}\hat{R}/2)}{\sin^2(p_{\mu}/2)}
\sin^2(p_{\nu}\hat{T}/2) \frac{\sin^2(k_{\nu}\hat{T}/2)}{\sin^2(k_{\nu}/2)}
\cos(k_{\mu} \hat{R}) \Bigg]
\end{eqnarray}
\begin{eqnarray}
<\omega^{(4E)}>_0 &=& \frac{1}{6} C_2(G) C_2(F) g_0^4 \int\limits_{BZ} \frac{d^4p}{(2
\pi)^4} \int\limits_{BZ} \frac{d^4k}{(2 \pi)^4} \frac{1}{\hat{p}^2 \hat{k}^2}
\nonumber \\ && \Bigg[ \Bigg\{ \frac{\sin(p_{\mu}\hat{R}/2)}{\sin(p_{\mu}/2)}
\sin^2(p_{\nu}\hat{T}/2) \Sigma_R(-p_{\mu},0) \nonumber \\
&& -\frac{\sin(p_{\mu}\hat{R}/2)}{\sin(p_{\mu}/2)} \sin^2(p_{\nu}\hat{T}/2)
\Sigma_R(k_{\mu}, -p_{\mu}-k_{\mu}) \nonumber \\
&& +\frac{\sin(p_{\mu}\hat{R}/2)}{\sin(p_{\mu}/2)} \sin^2(p_{\nu}\hat{T}/2)
\Sigma_R(0,-p_{\mu}) \nonumber \\
&& -\frac{\sin(p_{\mu}\hat{R}/2)}{\sin(p_{\mu}/2)} \sin^2(p_{\nu}\hat{T}/2)
\Sigma_R(-p_{\mu}-k_{\mu},k_{\mu}) \nonumber \\
&& +\frac{\sin^2(p_{\mu}\hat{R}/2) \sin^2(p_{\nu}\hat{T}/2)}{\sin^2(p_{\mu}/2)}
 \hat{R} +\frac{\sin^2(p_{\mu}\hat{R}/2)
\sin^2(p_{\nu}\hat{T}/2)}{\sin^2(p_{\nu}/2)} \hat{R} \nonumber \\ && +
\frac{\sin(p_{\mu}\hat{R}/2)}{\sin(p_{\mu}/2)} \sin^2(p_{\nu}\hat{T}/2)
\frac{\sin(k_{\mu}\hat{R}/2)}{\sin(k_{\mu}/2)}
\frac{\sin((p_{\mu}+k_{\mu})\hat{R}/2)}{\sin((p_{\mu}+k_{\mu})/2)} \nonumber \\
&& \cdot \cos(k_{\nu}\hat{T}) \Bigg\} \nonumber \\ &&+ \Bigg\{(\mu,\hat{R})
\leftrightarrow (\nu,\hat{T}) \Bigg\} +
2\frac{\sin^2(p_{\mu}\hat{R}/2)}{\sin^2(p_{\mu}/2)} \sin^2(p_{\nu}\hat{T}/2)
\hat{T}\Bigg]
\label{omega4end}
\end{eqnarray}
and finally, due to symmetry properties in the colour indices
\begin{equation}
<\omega^{(4F)}>_0 = 0.
\end{equation}
Here the functions $\Sigma$ are the parts of the following functions
$\tilde{\Sigma}$ which are even in $p_{\mu} \to -p_{\mu}, k_{\mu} \to
-k_{\mu}$:
\begin{eqnarray*}
\tilde{\Sigma}_1 &=& e^{i p_{\mu} (\hat{R}-1)/2} \sum_{x_1=0}^{\hat{R}-3}
\sum_{x_2=x_1+1}^{\hat{R}-2} \sum_{x_3=x_2+1}^{\hat{R}-1} e^{-i p_{\mu} x_1}
e^{i k_{\mu} (x_2-x_3)} \\
\tilde{\Sigma}_2 &=& e^{i p_{\mu} (\hat{R}-1)/2} \sum_{x_1=0}^{\hat{R}-3}
\sum_{x_2=x_1+1}^{\hat{R}-2} \sum_{x_3=x_2+1}^{\hat{R}-1} e^{-i p_{\mu} x_2}
e^{i k_{\mu} (x_1-x_3)} \\
\tilde{\Sigma}_R(p_{\mu}, k_{\mu}) &=& e^{-i p_{\mu} (\hat{R}-1)/2} e^{-i
k_{\mu} (\hat{R}-1)/2} \sum_{x_1=0}^{\hat{R}-2} \sum_{x_2=x_1+1}^{\hat{R}-1}
e^{i p_{\mu} x_1} e^{i k_{\mu} x_2}.
\end{eqnarray*}
The explicit expressions are for $\hat{R} > 2$ \cite{Heller}:
\begin{eqnarray}
\Sigma_1 &=& -\frac{1}{4} (\hat{R}-2)
\frac{\sin(p_{\mu}\hat{R}/2)}{\sin(p_{\mu}/2)} + \frac{1}{4}
\frac{\sin(p_{\mu}(\hat{R}-2)/2)}{\sin(p_{\mu}/2)} \nonumber \\ &&- \frac{1}{4}
\frac{\sin((k_{\mu}-p_{\mu})(\hat{R}-2)/2) \cos(k_{\mu} (\hat{R}+1)/2 -
p_{\mu})}{\sin((k_{\mu}-p_{\mu})/2) \sin^2(k_{\mu}/2)} \nonumber \\ &&+
\frac{1}{4} \frac{\sin(p_{\mu}(\hat{R}-2)/2) \cos(p_{\mu}-k_{\mu})}
{\sin(p_{\mu}/2) \sin^2(k_{\mu}/2)} \\
\Sigma_2 &=& - \frac{1}{4}
\frac{\sin(p_{\mu}(\hat{R}-2)/2)}{\sin(p_{\mu}/2)} \frac{\cos(k_{\mu} -
p_{\mu}/2)}{\sin(k_{\mu}/2) \sin((p_{\mu}+k_{\mu})/2)} \nonumber \\ && +
\frac{1}{4} \frac{\sin(k_{\mu}(\hat{R}-2)/2)}{\sin^2(k_{\mu}/2)}
\frac{\cos((p_{\mu}+k_{\mu})\hat{R}/2 + k_{\mu}/2 - p_{\mu})}
{\sin((p_{\mu}+k_{\mu})/2)} \nonumber \\ && + \frac{1}{4}
\frac{\sin((k_{\mu}-p_{\mu})(\hat{R}-2)/2) \cos(k_{\mu}(\hat{R}+2)/2 -
p_{\mu}/2)} {\sin(p_{\mu}/2) \sin(k_{\mu}/2) \sin((k_{\mu}-p_{\mu})/2)}
\nonumber \\ && - \frac{1}{4} \frac{\sin(k_{\mu}(\hat{R}-2)/2)
\cos((p_{\mu}+k_{\mu})\hat{R}/2 + k_{\mu} - p_{\mu})} {\sin(p_{\mu}/2)
\sin^2(k_{\mu}/2)}
\end{eqnarray}
For $\hat{R} \le 2$, both of these functions vanish because of the constraints
in the sums. Additionally for $\hat{R} > 1$ one has:
\begin{eqnarray}
\Sigma_R(p_{\mu}, k_{\mu}) &=& \mbox{Re} \, \tilde{\Sigma}_R(p_{\mu}, k_{\mu})
\nonumber \\ &=& \frac{1}{2} \frac{\sin(p_{\mu}(\hat{R}-1)/2)
\sin(k_{\mu} \hat{R}/2 - p_{\mu}/2)}{\sin(p_{\mu}/2) \sin(k_{\mu}/2)}
\nonumber \\ &&+ \frac{1}{2} \frac{\sin((p_{\mu}+k_{\mu})(\hat{R}-1)/2)
\sin(p_{\mu}/2)} {\sin((p_{\mu}+k_{\mu})/2) \sin(k_{\mu}/2)} \nonumber \\
&=& \frac{1}{2} \frac{\sin(p_{\mu}\hat{R}/2)}{\sin(p_{\mu}/2)}
\frac{\sin(k_{\mu}\hat{R}/2)}{\sin(k_{\mu}/2)} - \frac{1}{2}
\frac{\sin((p_{\mu}+k_{\mu})\hat{R}/2)}{\sin((p_{\mu}+k_{\mu})/2)}
\end{eqnarray}
For $\hat{R} = 1$, this function also vanishes, again because of the
constraints in the sums. Obviously the function is also even under
$p_{\mu} \leftrightarrow k_{\mu}$; therefore terms like
\[
\Sigma_R(p_{\mu}, k_{\mu}) - \Sigma_R(k_{\mu}, p_{\mu})
\]
in the expressions (\ref{omega4start}) to (\ref{omega4end}) vanish.

Additionally, the part of $\tilde{\Sigma}_R$ which is odd under $p_{\mu} \to
-p_{\mu}, k_{\mu} \to -k_{\mu}$ appears also several times in the expressions
above (in $<\omega^{4B)}>_0$ and in $<\omega^{(3)}S^{(1)}>$). It is given by:
\begin{eqnarray}
&&O_R(p_{\mu}, k_{\mu}) = \mbox{Im} \, \tilde{\Sigma}_R(p_{\mu}, k_{\mu})
\\ &=& \frac{1}{2} \left[
\frac{\sin((p_{\mu}+k_{\mu})(\hat{R}-1)/2)}{\sin((p_{\mu} + k_{\mu})/2)}
\frac{\cos(p_{\mu}/2)}{\sin(k_{\mu}/2)} \right. \nonumber \\ && \left.
- \frac{\cos(k_{\mu} \hat{R}/2 - p_{\mu}/2)}{\sin(k_{\mu}/2)}
\frac{\sin(p_{\mu}(\hat{R}-1)/2)}{\sin(p_{\mu}/2)}
\right] \nonumber \\
&=& \frac{1}{4}
\frac{\sin((p_{\mu}-k_{\mu})/2)}{\sin(p_{\mu}/2)\sin(k_{\mu}/2)}
\left[
\frac{\sin((p_{\mu}+k_{\mu}) \hat{R}/2)}{\sin((p_{\mu} + k_{\mu})/2)}
- \frac{\sin((p_{\mu}-k_{\mu}) \hat{R}/2)}{\sin((p_{\mu} - k_{\mu})/2)}
\right]. \nonumber
\end{eqnarray}
Obviously, $O_R(p_{\mu}, k_{\mu})$ is also odd under $p_{\mu} \leftrightarrow
k_{\mu}$.

\vspace{2ex}

\noindent
Finally, the vacuum polarization is given by the sum of the following five
contributions, where the fourth and the fifth are unique to lattice
perturbation theory:
\begin{eqnarray}
\Pi_{\mu\nu}^{gluonloop} &=& \frac{1}{2} C_2(G) \int\limits_{BZ}
\frac{d^dk}{(2\pi)^d} \frac{1}{\hat{p}^2 \left(\widehat{p+k}\right)^2}
\left[ 2 \delta_{\mu\nu} \cos^2(k_{\nu}/2) \left(\widehat{2p+k}\right)^2
\right. \nonumber \\
&& + \left(\widehat{2k+p})\right)_{\mu}
\left(\widehat{2k+p}\right)_{\nu} \sum_{\rho \ne \mu, \nu} \cos^2(p_{\rho}/2)
\nonumber \\ && \left. + 2 \left(\widehat{2p+k}\right)_{\mu}
\left(\widehat{k-p}\right)_{\nu} \cos((p_{\mu}+k_{\mu})/2) \cos(k_{\nu}/2)
\right] \\
\Pi_{\mu\nu}^{gluontadpole} &=& \frac{1}{2d} \left(2 C_2(F) - \frac{1}{3}
C_2(G) \right) \left(\delta_{\mu\nu} \hat{p}^2 - \hat{p}_{\mu} \hat{p}_{\nu}
\right) \nonumber \\
&&+ \frac{1}{6} C_2(G) \delta_{\mu\nu} \left[
(d+3) \Delta_0 + 1 - \frac{2}{d} + 3 \left(\Delta_0 -
\frac{1}{2d}\right) \cos(p_{\mu}) \right. \nonumber \\ && \left. - \left(7
\Delta_0 - \frac{5}{2d}\right) \sum_{\rho} \cos(p_{\rho}) \right] \\
\Pi_{\mu\nu}^{ghostloop} &=& -\frac{1}{2} C_2(G) \int\limits_{BZ}
\frac{d^dk}{(2\pi)^d} \frac{1}{\hat{p}^2 \left(\widehat{p+k}\right)^2}
\nonumber \\ && \left[ 2 \left(\widehat{p+k}\right)_{\mu} \hat{k}_{\nu}
\cos(k_{\mu}/2) \cos((p_{\nu}+k_{\nu})/2) \right]
\\
\Pi_{\mu\nu}^{ghosttadpole} &=& - \frac{1}{4d} C_2(G) \delta_{\mu\nu} \\
\Pi_{\mu\nu}^{measure} &=& - \frac{1}{12} C_2(G) \delta_{\mu\nu}.
\end{eqnarray}
For later convenience, the results are given here for an arbitrary numbers of
dimensions $d$. Additonally, the abbreviation
\begin{equation}
\Delta_0 = \int\limits_{BZ} \frac{d^dk}{(2\pi)^d} \frac{1}{\hat{k}^2}
\end{equation}
has been introduced (compare appendix \ref{commonintegrals}).

\vspace{3ex}

It is not obvious how one could get an explicit expression for the potential
from these eight-dimensional integrals in momentum space, reproducing the
result (\ref{Potential}). One way is to evaluate the integrals numerically;
Heller and Karsch did this already in their original paper \cite{Karsch} and
fitted their results to the following formula:
\[
\hat{V}(\hat{R}) = V_{self} - \frac{4}{3} \frac{g_0^2}{4 \pi \hat{R}}
\left(1 + \frac{11}{16\pi^2} g_0^2 \ln(\hat{R} M)^2 \right).
\]
Here the factor $\frac{4}{3}$ comes from the quadratic Casimir operator
$C_2(F)$ for $SU(3)$ in the fundamental representation, the factor
$\frac{11}{16\pi^2}$ comes from the first coefficient of the beta function,
and $V_{self}$ denotes the contributions from the self energy. Comparing with
(\ref{Potential}), one should get:
\[
M = 82.07
\]
respectively
\[
\ln M = 4.408.
\]
The results Heller and Karsch obtained for $\ln M$ for $\hat{R} \gg 1$ are
consistent with this prediction - a further verification that
(\ref{Potential}) indeed gives the right result for the quark-antiquark
potential.

\subsection{Transversality of the gluonic vacuum polarization}

Using the results Heller and Karsch obtained for the vacuum polarization
\cite{Karsch}, one can show that it is transversal even for finite lattice
spacing. This is not crucial for this work, but an interesting
result in itself. So far, the transversality of the vacuum polarization,
calculated in lattice perturbation theory, only had been checked in the
continuum limit \cite{Kawai}.

The contribution proportional to $C_2(F)$ of the vacuum polarization
is obviously transversal. For the contributions proportional to
$C_2(G)$ it has to be shown that
\begin{equation}
\sum_{\mu, \nu} \hat{p}_{\mu} \hat{p}_{\nu} \Pi_{\mu\nu} = 0,
\end{equation}
which proves the transversality. In order to show this, look at the various
contributions separately:
\begin{eqnarray}
\sum_{\mu, \nu}  \hat{p}_{\mu} \hat{p}_{\nu} \Pi_{\mu\nu}^{gluonloop} &=&
C_2(G) g_0^2 \left[ \hat{p}^2 \left( \left(d - \frac{1}{2}\right) \Delta_0 -
\frac{1}{2} + \frac{1}{2d} \right) \right. 
\nonumber \\ && \left.+ \left(\hat{p}^2\right)^2 \left(-\frac{1}{4}
\left(\Delta_0 - \frac{1}{2d}\right) - \frac{1}{4} I \right)
+ \hat{p}^4 \frac{1}{4} \left(\Delta_0 - \frac{1}{2d}\right)
\right] \nonumber \\ \\
\sum_{\mu, \nu}  \hat{p}_{\mu} \hat{p}_{\nu} \Pi_{\mu\nu}^{gluontadpole} &=&
C_2(G) g_0^2 \left[ \hat{p}^2 \left((1-d) \Delta_0 + \frac{7}{12} -
\frac{7}{12d} \right) \right. \nonumber \\ && \left. +
\left(\hat{p}^2\right)^2 \frac{1}{4} \left(\Delta_0 - \frac{1}{2d}\right) -
\hat{p}^4 \frac{1}{4} \left(\Delta_0 - \frac{1}{2d}\right) \right] \\
\sum_{\mu, \nu}  \hat{p}_{\mu} \hat{p}_{\nu} \Pi_{\mu\nu}^{ghostloop} &=&
C_2(G) g_0^2 \left[\hat{p}^2 \left(- \frac{1}{2} \Delta_0 + \frac{1}{4d}
\right) + \left(\hat{p}^2\right)^2 \frac{1}{4} I \right] \\
\sum_{\mu, \nu}  \hat{p}_{\mu} \hat{p}_{\nu} \Pi_{\mu\nu}^{ghosttadpole} &=&
C_2(G) g_0^2 \left[ \hat{p}^2 \left( -\frac{1}{6d} \right) \right] \\
\sum_{\mu, \nu}  \hat{p}_{\mu} \hat{p}_{\nu} \Pi_{\mu\nu}^{measure} &=&
C_2(G) g_0^2  \left[ \hat{p}^2 \left( -\frac{1}{12} \right) \right].
\end{eqnarray}
Here the abbreviation
\[
I := \int\limits_{BZ} \frac{d^dk}{(2\pi)^d} \int\limits_{BZ} \frac{d^dq}{(2\pi)^d}
\frac{\delta^{(d)}(p+k+q)}{\hat{q}^2 \hat{k}^2}
\]
was introduced. In the calculation, the following Ward identity for the
three-gluon vertex was used:
\begin{eqnarray}
\sum_{\mu} \hat{p}_{\mu} \Gamma_{\mu\nu\lambda}(p,q,k)
&=& i g_0 (2\pi)^d \delta^{(d)}(p+q+k) \\
&&\cdot \left[\cos(p_{\lambda}/2) (\hat{q}^2 \delta_{\nu\lambda} -
\hat{q}_{\nu} \hat{q}_{\lambda}) - \cos(p_{\nu}/2) (\hat{k}^2
\delta_{\nu\lambda} - \hat{k}_{\nu} \hat{k}_{\lambda}) \right]. \nonumber
\end{eqnarray}
Another useful formula which was needed here is:
\begin{equation}
\hat{p} \, \hat{k} \, \cos((p+k)/2) = \frac{1}{2}
\left(\widehat{(p+k})^2 - \hat{p}^2 - \hat{k}^2 \right).
\end{equation}
For the evaluation of the other integrals appearing in the derivation, see
appendix C.

Adding up all these contributions, everything cancels, and one arrives indeed
at the result that the vacuum polarization is transversal. Hence it can be
written in the following form: 
\begin{equation}
\Pi_{\mu\nu}(p) = \left(\delta_{\mu\nu} \hat{p}^2 - \hat{p}_{\mu} \hat{p}_{\nu}
\right) \Pi(p).
\end{equation}

Additionally, one should show that the function $\Pi(p)$ is regular at $p=0$
or alternatively that $\Pi_{\mu\nu}(0) = 0$. For  the contribution proportional
to $C_2(F)$ this is again obvious; and again the contributions proportional
to $C_2(G)$ are determined separately:
\begin{eqnarray}
\Pi_{\mu\nu}^{gluonloop}(0) &=& C_2(G) g_0^2 \delta_{\mu\nu} \left[
\left(d-\frac{1}{2}\right) \Delta_0 + \frac{1}{2d} - \frac{1}{2} \right] \\
\Pi_{\mu\nu}^{gluontadpole}(0) &=& C_2(G) g_0^2 \delta_{\mu\nu} \left[
(d-1) \left(-\Delta_0 + \frac{7}{12d} \right) \right] \\
\Pi_{\mu\nu}^{ghostloop}(0) &=& C_2(G) g_0^2 \delta_{\mu\nu}
\left[-\frac{1}{2} \Delta_0 + \frac{1}{4d} \right] \\
\Pi_{\mu\nu}^{ghosttadpole}(0) &=& C_2(G) g_0^2 \delta_{\mu\nu}
\left[-\frac{1}{6d}\right] \\
\Pi_{\mu\nu}^{measure}(0) &=& C_2(G) g_0^2 \delta_{\mu\nu}
\left[-\frac{1}{12}\right].
\end{eqnarray}
Again, some of the integrals listed in appendix C were used. Adding
everything, one indeed arrives at the claimed result
\begin{equation}
\Pi_{\mu\nu}(0) = 0.
\end{equation}

\newpage
\thispagestyle{empty}
\cleardoublepage
\def\w{
  \fmfstraight
  \fmfleft{w1,w2,w3}
  \fmftop{w4,w5,w6}
  \fmfright{w9,w8,w7}
  \fmfbottom{w12,w11,w10}
  \fmf{plain}{w1,w3}
  \fmf{plain}{w3,w7}
  \fmf{plain}{w7,w9}
  \fmf{plain}{w9,w1}
}

\def\p{
  \begin{fmfsubgraph}(0.375w, 0.4h)(1cm, 1cm)
  \fmfleft{p1,p2,p3}
  \fmftop{p4,p5,p6}
  \fmfright{p9,p8,p7}
  \fmfbottom{p12,p11,p10}
  \fmf{plain}{p1,p3}
  \fmf{plain}{p3,p7}
  \fmf{plain}{p7,p9}
  \fmf{plain}{p9,p1}
  \end{fmfsubgraph}
}

\def\pw{
  \p
  \w
}

\chapter{Action Sum Rule}

\label{ASR_chapter}

In order to derive the action sum rule
\begin{equation}
\hat{V} + \hat{R} \frac{\partial \hat{V}}{\partial \hat{R}} = 
\frac{2 \beta_L}{g_0} \lim_{\hat{T} \to \infty} <L(t)>_{q\bar{q}-0},
\end{equation}
essentially three steps were needed:
\begin{enumerate}
\item Taking the logarithmic derivative of (\ref{PotentialWilson}); this led
to the identity (\ref{potentialderivative}).
\item Using the scaling behaviour of the potential; the result was 
(\ref{ActionSumRule}).
\item Taking the limit of large $\hat{T}$ and thereby restricting the sum over
all plaquettes to one fixed time slice.
\end{enumerate}

The explicit form of the potential up to next-to-leading order was already
derived in chapter 2; it´s easy to see that it has indeed the right scaling
behaviour. Hence only the first and the third step remain to be checked. As
already pointed out, the first step gives essentially an identity, so that in
principle a check for this is not needed.

Nevertheless, it is instructive to see how this identity looks like on the
level of Feynman graphs, and the results derived here will be helpful
later. Additionally, it will turn out that this identity is closely connected
with the gauge invariance of the Wilson loop, so that this invariance can
also be checked perturbatively.

The first section will give an introduction and contains the necessary
expansions; the second section will deal with the perturbative check of
(\ref{potentialderivative}) up to next-to-leading order. Then I will point out
the connections to the gauge invariance of the Wilson loop, and how this can
be checked. Finally the restriction to one fixed time slice will be examined.

\section{Preliminaries}

\label{ASR_prelim}

The identity (\ref{potentialderivative}), when inserting the definition
(\ref{EV_QQ}) and the formula (\ref{PotentialWilson}), becomes
\begin{equation}
- \hat{\beta} \frac{\partial}{\partial \hat{\beta}} \lim_{\hat{T} \to \infty}
\frac{1}{\hat{T}} \ln  <W(\hat{R}, \hat{T})>
= \lim_{\hat{T} \to \infty} \frac{1}{\hat{T}} \left(\frac{<SW>}{<W>} -
<S>\right).
\end{equation}
Obviously, this identity has to be satisfied even for all finite $\hat{T}$:
\begin{equation}
- \hat{\beta} \frac{\partial}{\partial \hat{\beta}} \ln  <W(\hat{R}, \hat{T})>
= \frac{<SW>}{<W>} - <S>.
\end{equation}
Carrying out the derivative and multiplying by $<W>$ gives:
\begin{equation}
- \hat{\beta} \frac{\partial}{\partial \hat{\beta}} <W> = <SW> - <S><W>.
\end{equation}
Now insert the relation (\ref{latticecoupling}) between the lattice coupling
constant $\hat{\beta}$ and the coupling constant $g_0(a)$ appearing in
perturbative lattice calculations; this finally leads to:
\begin{equation}
g_0^2 \frac{\partial}{\partial g_0^2} <W> = <SW> - <S><W> = <SW>_{conn}.
\label{actionidentity}
\end{equation}
That is the formula which will be checked perturbatively in the next section.

\section{Perturbative check}

\label{ASR_check}

\subsection{Leading order}

\label{ASR_check_LO}

Using the expansions
\begin{equation}
S = S^{(0)} + O(g_0).
\end{equation}
and
\begin{equation}
W = 1 - g_0^2 \omega^{(2)} + O(g_0^3),
\end{equation}
(\ref{actionidentity}) becomes
\begin{equation}
<\omega^{(2)}>_0 = <\omega^{(2)}S^{(0)}>_{conn}.
\label{ASR_LO}
\end{equation}
The expectation value on the left hand side was already given in section
\ref{potential_LO}:
\begin{eqnarray}
&&<\omega^{(2)}>_0 \\ &=& 2 C_2(F) \int\limits_{BZ}
\frac{d^4p}{(2\pi)^4}
\frac{\sin^2(\frac{1}{2}p_{\mu}\hat{R})\sin^2(\frac{1}{2}p_{\nu}\hat{T})}
{\hat{p}^2} \left(\frac{1}{\sin^2(\frac{1}{2}p_{\mu})} +
\frac{1}{\sin^2(\frac{1}{2}p_{\nu})} \right); \nonumber
\end{eqnarray}
this result was derived using Feynman gauge $\alpha = 1$. In the following, if
it is not explicitly indicated otherwise, all expectation values will be
calculated in that gauge.

The easiest way to calculate the correlator on the right hand side of
(\ref{ASR_LO}) is to compute the correlator between two gauge fields and the
action first; this is equivalent to the insertion of the action into a gluon
line:
\begin{equation}
<A^A_{\mu}(p)A^B_{\nu}(q) S>_{conn} =
\delta^{AB} \frac{\delta_{\mu\nu}
- \frac{\hat{p}_{\mu} \hat{p}_{\nu}}{\hat{p}^2}}
{\hat{p}^2} (2 \pi)^4 \delta(p+q).
\label{Sinsertion}
\end{equation}
This results hold for every gauge parameter $\alpha$---hence inserting the
action into a gluon line with arbitrary gauge parameter simply gives a
propagator in Landau gauge $\alpha=\infty$! Therefore the calculation of the
correlator $<\omega^{(2)}S^{(0)}>_{conn}$ gives exactly the same result as if
one would calculate the expectation value $<\omega^{(2)}>_0$ using the
propagator in Landau gauge. For convenience, this expectation value will now
be calculated in an arbitrary gauge. 

First, $\omega^{(2)}$ is given by:
\begin{equation}
\omega^{(2)} = \frac{1}{4d(F)} \left( \sum_l A_l^A \right)^2,
\end{equation}
with the sum along the Wilson loop:
\begin{eqnarray}
\sum_l A_l^A &=& \sum_{l=0}^{\hat{R}-1} A^A_{\mu}(n_0+l \hat{\mu})
+ \sum_{l=0}^{\hat{T}-1} A^A_{\nu}(n_0+ \hat{\mu} \hat{R} + l \hat{\nu}) \\
&&- \sum_{l=0}^{\hat{R}-1} A^A_{\mu}(n_0+ \hat{\mu} \hat{R} + \hat{\nu} \hat{T}
- l \hat{\mu}) - \sum_{l=0}^{\hat{T}-1} A^A_{\nu}(n_0 + \hat{\nu} \hat{T}
- l \hat{\nu}). \nonumber
\end{eqnarray}
Expressing this in Fourier space, one gets:
\begin{equation}
\sum_l A_l^A =
-2i \int\limits_{BZ} \frac{d^4p}{(2\pi)^4}
\sin(p_{\mu}\hat{R}/2) \sin(p_{\nu}\hat{T}/2)
e^{i p n_c} \frac{A^A_{\alpha}(p)(\delta_{\mu\alpha} - \delta_{\nu\alpha})}
{\sin(p_{\alpha}/2)},
\end{equation}
where
\[
n_c = n_0+\hat{\mu}\hat{R}/2+\hat{\nu}\hat{T}/2
\]
is the center of the Wilson loop.

Hence calculating the expectation value, using the propagator
(\ref{Propagator}), the result is:
\begin{eqnarray}
&&<\omega^{(2)}>_{0,\alpha} \nonumber \\ &=& -\frac{1}{d(F)} \sum_{\alpha,
\beta} \int\limits_{BZ} \frac{d^4p}{(2\pi)^4} \int\limits_{BZ}
\frac{d^4q}{(2\pi)^4} <A^A_{\alpha}(p) A^A_{\beta}(q)>_{0,\alpha} e^{i(p+q)
n_c} \frac{\delta_{\mu\alpha} - \delta_{\nu\alpha}} {\sin(p_{\alpha}/2)}
\nonumber \\ && \frac{\delta_{\mu\beta} - \delta_{\nu\beta}}
{\sin(q_{\beta}/2)} \sin(p_{\mu}\hat{R}/2) \sin(p_{\nu}\hat{T}/2)
\sin(q_{\mu}\hat{R}/2) \sin(q_{\nu}\hat{T}/2) \nonumber \\
&=& 2 C_2(F) \sum_{\alpha, \beta} \int\limits_{BZ} \frac{d^4p}{(2\pi)^4}
\frac{\sin^2(p_{\mu}\hat{R}/2) \sin^2(p_{\nu}\hat{T}/2)}{\hat{p}^2} \nonumber
\\ && \frac{\delta_{\mu\alpha} - \delta_{\nu\alpha}} {\sin(p_{\alpha}/2)}
\frac{\delta_{\mu\beta} - \delta_{\nu\beta}} {\sin(p_{\beta}/2)}
\left(\delta_{\mu\nu} - \left(1 - \frac{1}{\alpha} \right)
\frac{\hat{p}_{\mu}\hat{p}_{\nu}}{\hat{p}^2} \right) \nonumber \\
&=& 2 C_2(F) \int\limits_{BZ}
\frac{d^4p}{(2\pi)^4}
\frac{\sin^2(\frac{1}{2}p_{\mu}\hat{R})\sin^2(\frac{1}{2}p_{\nu}\hat{T})}
{\hat{p}^2} \nonumber \\ && \cdot \left(\frac{1}{\sin^2(\frac{1}{2}p_{\mu})} +
\frac{1}{\sin^2(\frac{1}{2}p_{\nu})} \right) \nonumber \\
&=& <\omega^{(2)}>_{0,\alpha=1}.
\end{eqnarray}
Therefore the expectation value $<\omega^{(2)}>_0$ does not depend on the
gauge parameter $\alpha$, i.\ e., it is gauge invariant! Naively one could have
expected this result, because the Wilson loop is defined in a gauge-invariant
way. But gauge transformations on the lattice are defined with respect to the
link variables, and it is not entirely clear what effects this can have in a
weak-coupling expansion. Hence it is nice to see that even in the limit of
small coupling, the gauge invariance is preserved. The close connection
between the action sum rule and the gauge invariance of the Wilson loop which
becomes apparent here will be discussed further in section
\ref{gauge_invariance}.

\noindent
Additionally, using this gauge invariance, it is obvious that indeed
\[
<\omega^{(2)}>_0 = <\omega^{(2)}S^{(0)}>_{conn}
\]
holds---hence the action sum rule is valid in leading order.

\vspace{2ex}

This validity as well as the gauge invariance of the Wilson loop in leading
order can also be expressed in diagrammatical form in the following way:
\\[1ex]
\begin{picture}(12,4.5)
\put(-3.4,-21.6)
{\epsfig{file=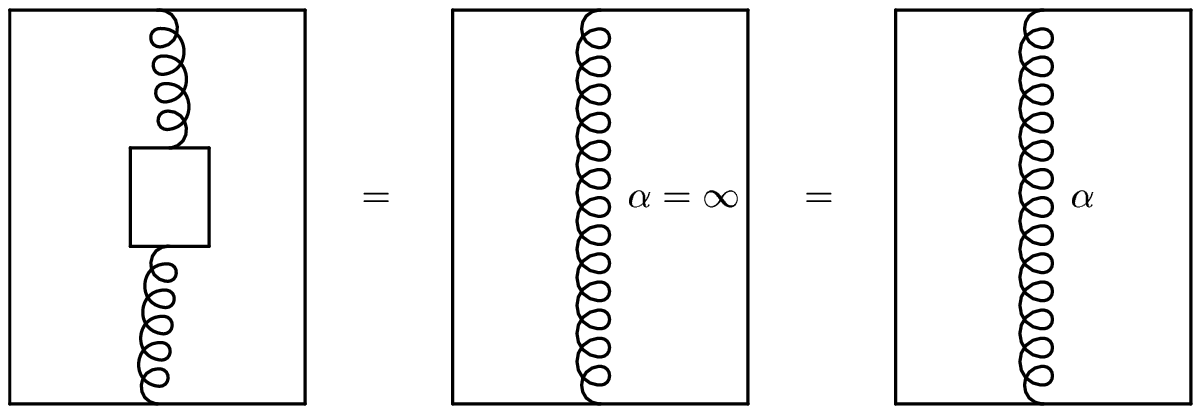, height=30cm, width=20cm}}  
\end{picture}\\[3ex]
In these diagrams, a sum of the end points of the gluon line along the Wilson
loop as well as along the plaquette, and a sum over all possible positions
and orientations of the plaquette is to be understood. In the third graph, the
gauge parameter $\alpha$ can be arbitrary. Gluon lines without an explicit
index $\alpha$ are in Feynman gauge.

\vspace{3ex}

\noindent
Another way to express this more concisely is to introduce the operator
\begin{equation}
G := \frac{1}{2} \int\limits_{BZ} A^A_{\mu}(p) \hat{p}_{\mu} \hat{p}_{\nu}
A^A_{\nu}(-p);
\label{gaugeoperator}
\end{equation}
then one can write:
\begin{equation}
<A^A(p)A^B(q)>_{0, \alpha} =  <A^A(p)A^B(q)>_{0} - \left(1 - \frac{1}{\alpha}\right)
<A^A(p)A^B(q)G>_{conn}.
\label{arbitrarygauge}
\end{equation}
If one denotes the insertion of this operator into a gluon line by a cross,
one can also express this diagrammatically:
\\[1ex]
\begin{picture}(12,2)
\put(-3.6,-23.7)
{
\epsfig{file=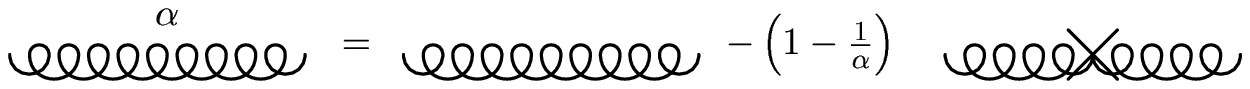, height=30cm, width=20cm}
}
\end{picture}\\
where again a gluon line without an explicit index represents a propagator in
Feynman gauge.

Now, using this notation, the gauge invariance of the Wilson loop in leading
order can be expressed in the following concise way:
\begin{equation}
<\omega^{(2)}G>_{conn} = 0,
\end{equation}
or again diagrammatically:
\\[1ex]
\begin{picture}(12,4.5)
\put(-3.5,-21.5)
{
\epsfig{file=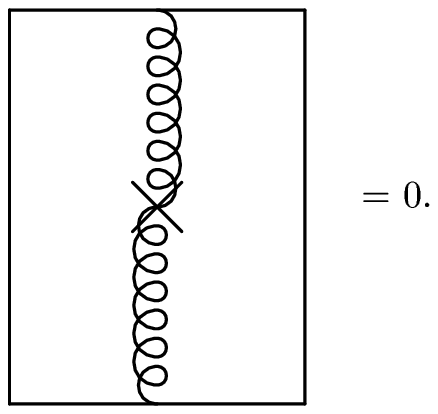, height=30cm, width=20cm}
}
\end{picture}

\subsection{Next-to-leading order}

\label{ASR_check_NLO}

Taking the higher terms in the expansions of the Wilson loop
\begin{equation}
W = 1 - g_0^2 \omega^{(2)} - g_0^3 \omega^{(3)} - g_0^4 \omega^{(4)} + O(g_0^5)
\end{equation}
into account, one gets:
\begin{eqnarray}
&& g_0^2 \frac{\partial}{\partial g_0^2}<W> \nonumber \\
&=& - g_0^2 <\omega^{(2)}>_0
+ 2 g_0^4 <\omega^{(2)}S^{(2)}>_{conn}
- g_0^4 <\omega^{(2)} \left(S^{(1)}\right)^2>_{conn} \nonumber \\ &&
+ 2 g_0^4 <\omega^{(2)}S^{(2)}_{FP}>_{conn} + 2 g_0^4
<\omega^{(2)}S^{(2)}_{meas}>_{conn} - 2 g_0^4 <\omega^{(4)}>_0 \nonumber \\ &&
+ 2 g_0^4 <\omega^{(3)}S^{(1)}>_0 + O(g_0^5).
\label{Wderivative}
\end{eqnarray}

For the right side of (\ref{actionidentity}), the higher terms in the
expansion of the action have also to be used:
\[
S = S^{(0)} + g_0 S^{(1)} + g_0^2 S^{(2)} + O(g_0^3).
\]
The result is:
\begin{eqnarray}
&& <WS>-<W><S> \nonumber \\
&=& - g_0^2 <S^{(0)}\omega^{(2)}>_{conn} + g_0^4
<S^{(0)}S^{(2)}\omega^{(2)}>_{conn} \nonumber \\ && - \frac{1}{2} g_0^4
<S^{(0)}\left(S^{(1)}\right)^2\omega^{(2)}>_{conn}
+ g_0^4 <S^{(0)}S^{(2)}_{FP}\omega^{(2)}>_{conn} \nonumber \\ && + g_0^4
<S^{(0)}S^{(2)}_{meas}\omega^{(2)}>_{conn} - g_0^4
<S^{(2)}\omega^{(2)}>_{conn} \nonumber \\ &&
+ g_0^4 <\left(S^{(1)}\right)^2\omega^{(2)}>_{conn} - g_0^4
<S^{(1)}\omega^{(3)}>_0 + g_0^4 <S^{(0)}S^{(1)}\omega^{(3)}>_{conn}
\nonumber \\
&&- g_0^4 <S^{(0)}\omega^{(4)}>_{conn} + O(g_0^6).
\label{SW_expansion}
\end{eqnarray}

Inserting this into the formula (\ref{actionidentity}) and using the validity
of the action sum rule in leading order, one arrives at the following formula
which has to be shown to be true:
\begin{eqnarray}
&& <S^{(0)}S^{(2)}\omega^{(2)}>_{conn} - \frac{1}{2} 
<S^{(0)}\left(S^{(1)}\right)^2\omega^{(2)}>_{conn}
+  <S^{(0)}S^{(2)}_{FP}\omega^{(2)}>_{conn} \nonumber \\
&& + <S^{(0)}S^{(2)}_{meas}\omega^{(2)}>_{conn}
+ <S^{(0)}S^{(1)}\omega^{(3)}>_{conn}
- <S^{(0)}\omega^{(4)}>_{conn} \nonumber \\
&&= \nonumber \\
&& + 3 <\omega^{(2)}S^{(2)}>_{conn}
- 2 <\omega^{(2)} \left(S^{(1)}\right)^2>_{conn} 
+ 2 <\omega^{(2)}S^{(2)}_{FP}>_{conn} \nonumber \\ && + 2
<\omega^{(2)}S^{(2)}_{meas}>_{conn} + 3 <\omega^{(3)}S^{(1)}>_0
- 2<\omega^{(4)}>_0.
\label{actionidentity_nlo}
\end{eqnarray}

\begin{figure}[t]
\begin{picture}(12,13)
\put(-3.5,-13)
{
\epsfig{file=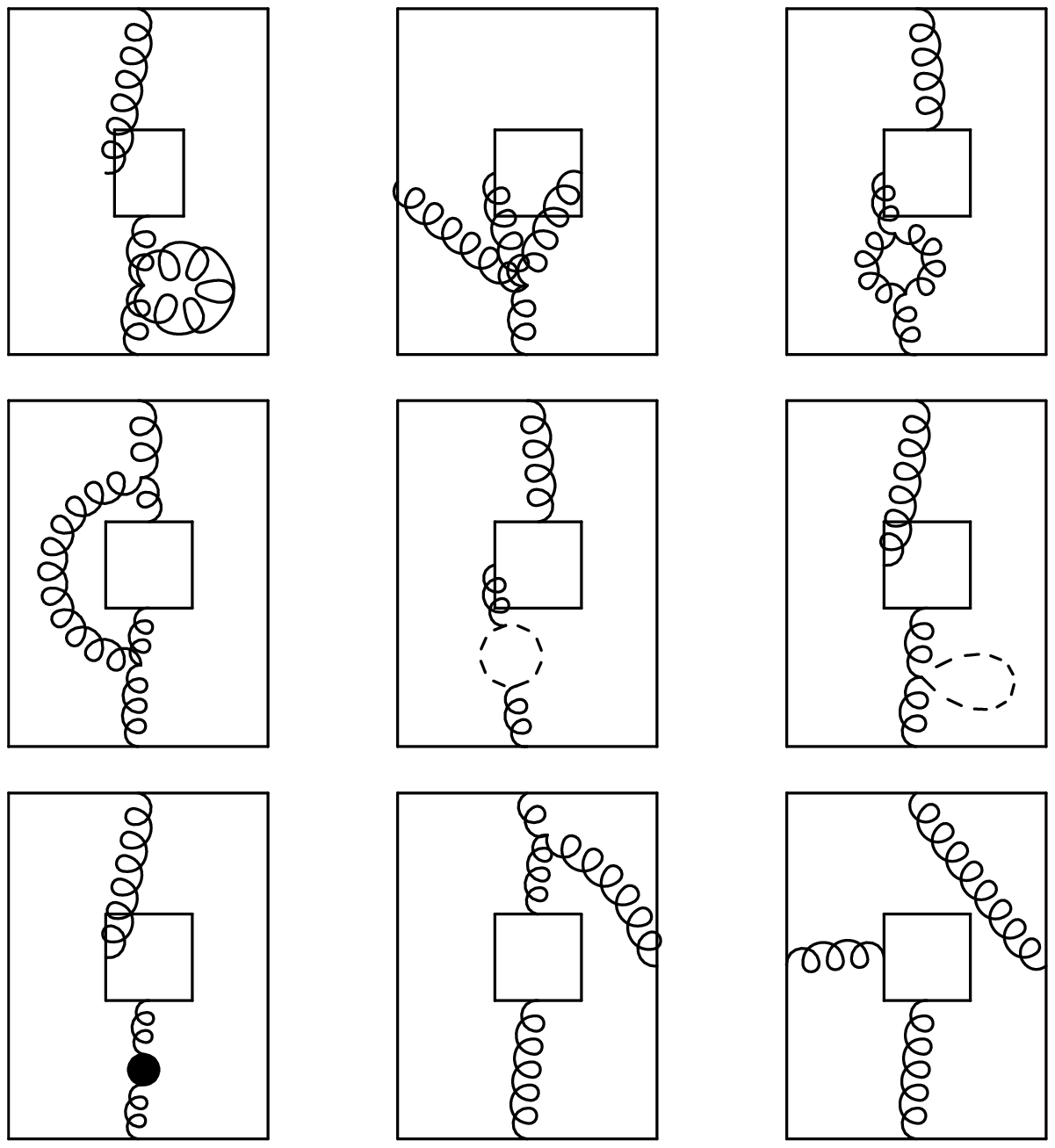, height=30cm, width=20cm}
}
\end{picture}

\caption{Contributions to the l.h.s.\ of (\ref{actionidentity_nlo})}
\label{actionidentity_NLO_l}
\end{figure}

\begin{figure}[t]
\begin{picture}(12,13)
\put(-3.5,-13)
{
\epsfig{file=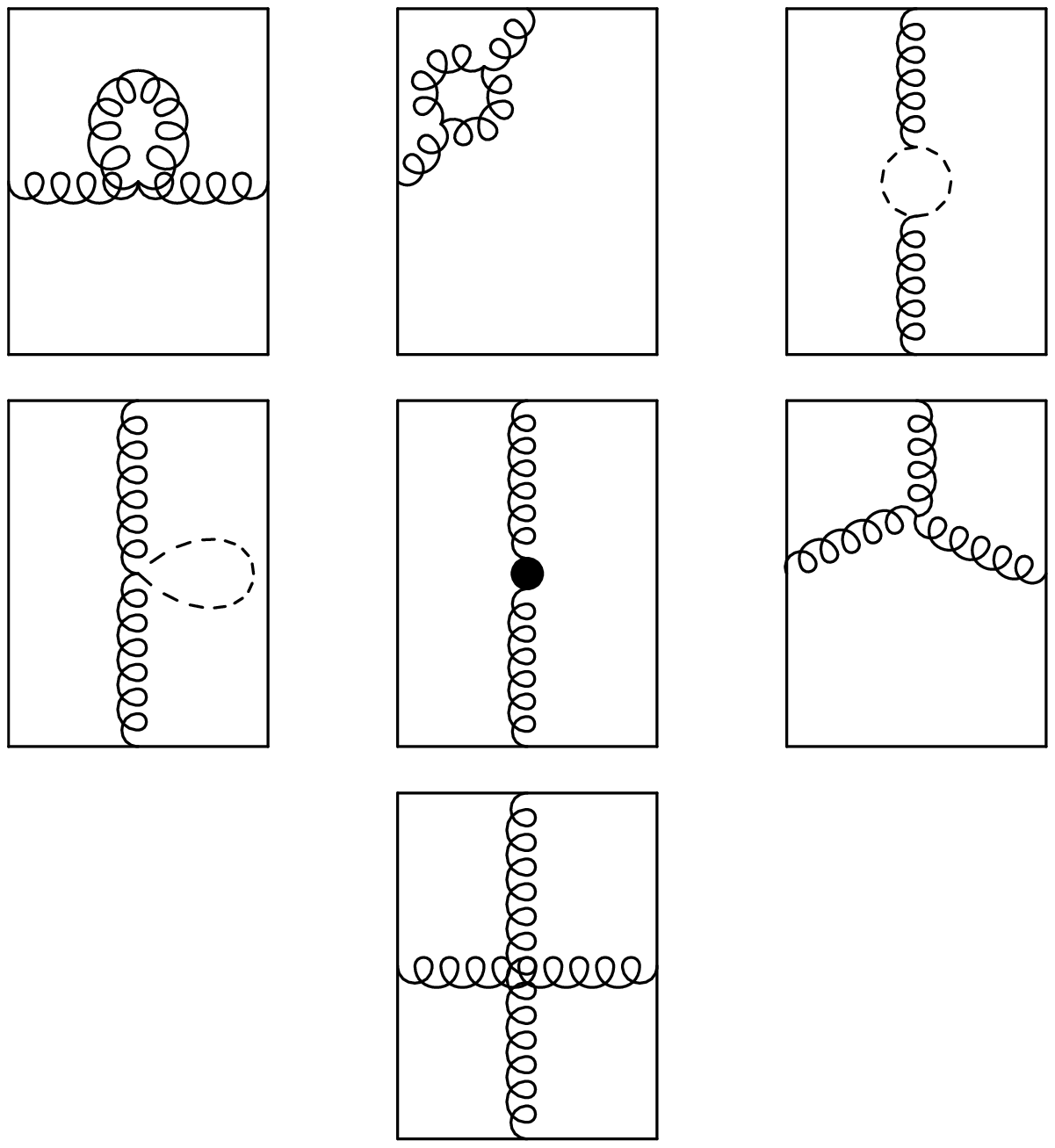, height=30cm, width=20cm}
}
\end{picture}
\caption{Contributions to the r.h.s.\ of (\ref{actionidentity_nlo})}
\label{actionidentity_NLO_r}
\end{figure}

Now one can use that the insertion of $S^{(0)}$ into a gluon line transforms
the propagator into Landau gauge. For every graph, one has to count the
numbers of gluons lines into which $S^{(0)}$ can be inserted; then one
gets the following simple results:
\begin{eqnarray}
&&<S^{(0)}\omega^{(2)}S^{(2)}>_{conn, \alpha=\infty}
= 3 <\omega^{(2)}S^{(2)}>_{conn, \alpha=\infty} \nonumber \\
&&<S^{(0)}\omega^{(2)}(S^{(1)})^2>_{conn, \alpha=\infty}
= 4 <\omega^{(2)}(S^{(1)})^2>_{conn, \alpha=\infty} \nonumber \\
&&<S^{(0)}\omega^{(2)}S^{(2)}_{FP}>_{conn, \alpha=\infty} = 2 
<\omega^{(2)}S^{(2)}_{FP}>_{conn, \alpha=\infty} \nonumber \\
&&<S^{(0)}\omega^{(2)}S^{(2)}_{meas}>_{conn, \alpha=\infty} = 2
<\omega^{(2)}S^{(2)}_{meas}>_{conn, \alpha=\infty} \nonumber \\
&&<S^{(0)}\omega^{(3)}S^{(1)}>_{conn, \alpha=\infty}
= 3 <\omega^{(3)}S^{(1)}>_{conn, \alpha=\infty} \nonumber \\
&& <S^{(0)}\omega^{(4)}>_{conn, \alpha=\infty} = 2
<\omega^{(4)}>_{\alpha=\infty}.
\end{eqnarray}

Inserting these expressions into (\ref{actionidentity_nlo}), one sees that
indeed the left hand side is equal to the right hand side. Hence if one
calculates every graph in Landau gauge, the action sum rule is obviously valid
up to next-to-leading order. But this is not entirely satisfying, because the
results for the potential in chapter 2, taken from \cite{Karsch}, were
calculated in Feynman gauge instead of Landau gauge; and it will turn out
later that checking the energy sum rule can also be done much more easily in
Feynman gauge.

Hence what is needed is either a check that the expectation value of the
Wilson loop is gauge invariant up to next-to-leading order, or an explicit
check of (\ref{actionidentity_nlo}) in the Feynman gauge. As will be explained
in section \ref{gauge_invariance}, these two are closely related; checking
(\ref{actionidentity_nlo}) will provide more than half of the check of the
gauge invariance already.

Therefore now (\ref{actionidentity_nlo}) has to be checked. The relevant
Feynman diagrams contributing to the left and right hand side, respectively,
are given in the figures \ref{actionidentity_NLO_l} and
\ref{actionidentity_NLO_r}. There are essentially three types of graphs:
vacuum polarization graphs, one with a three-gluon vertex and one with two
independent gluon lines. For the last two, the end points of the gluon lines
are not summed over the entire Wilson loop; there are constraints: see the
explicit form of $\omega^{(3)}$ and $\omega^{(4)}$ given in section
\ref{potential_lattice}. In contrast, in the vacuum polarization graphs both
gluon end points are summed over the complete Wilson loop. Additionally, in
all of the graphs, a sum over all possible positions and orientations of the
plaquette is to be understood.

The vacuum polarization graphs can be grouped into two types again: the ones
where the action is inserted into an \emph{internal} gluon line (the second
and fourth graph in figure \ref{actionidentity_NLO_l}), and the ones where it
is inserted into an \emph{external} line (the first, third, fifth, sixth and
seventh graph). The second class includes a gluon loop, a gluon tadpole, a
ghost loop, a ghost tadpole and a contribution coming from the integration
measure; the latter two are special for lattice perturbation theory. These
graphs can be dealt with easily if one uses the following observation:
\begin{equation}
\sum_l <A_l A^A_{\mu}(x) S^{(0)}>_{conn} = \sum_l <A_l A^A_{\mu}(x)>_{0,
\alpha}.
\label{one_endpoint}
\end{equation}
This means that even if only one end of the gluon line into which the action
is inserted is summed over the whole Wilson loop, while the other end is
entirely arbitrary, the result will be the same as if one would have used a
gluon line in an arbitrary gauge, but without the insertion. This can also be
expressed diagrammatically:\\
\begin{picture}(12,4.5)
\put(-3.5,-21.4)
{
\epsfig{file=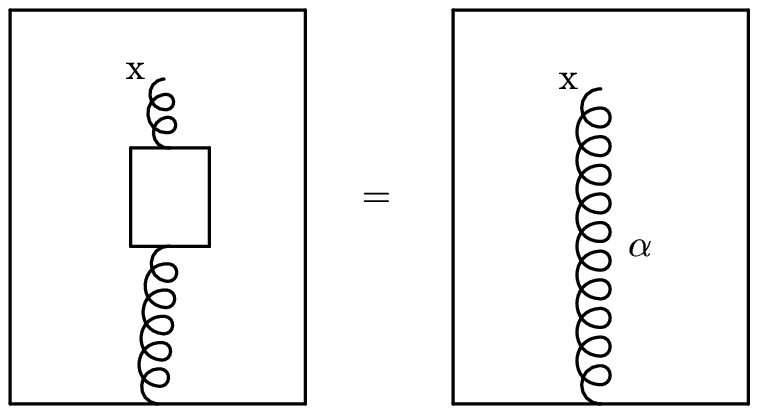, height=30cm, width=20cm}
}
\end{picture}\\
Using this, it becomes obvious that
\begin{eqnarray}
<S^{(0)}\omega^{(2)}S^{(2)}>_{conn}
&=& 2 <\omega^{(2)}S^{(2)}>_{conn} + \mbox{ insertions into internal lines}
\nonumber \\
<S^{(0)}\omega^{(2)}(S^{(1)})^2>_{conn}
&=& 2 <\omega^{(2)}(S^{(1)})^2>_{conn} + \mbox{ ins.\ into internal lines}
\nonumber \\
<S^{(0)}\omega^{(2)}S^{(2)}_{FP}>_{conn}
&=& 2 <\omega^{(2)}S^{(2)}_{FP}>_{conn} \nonumber \\
<S^{(0)}\omega^{(2)}S^{(2)}_{meas}>_{conn}
&=& 2 <\omega^{(2)}S^{(2)}_{meas}>_{conn}
\label{VP_external}
\end{eqnarray}
or, again with diagrams:\\
\begin{picture}(12,9.5)
\put(-3.5,-16.7)
{
\epsfig{file=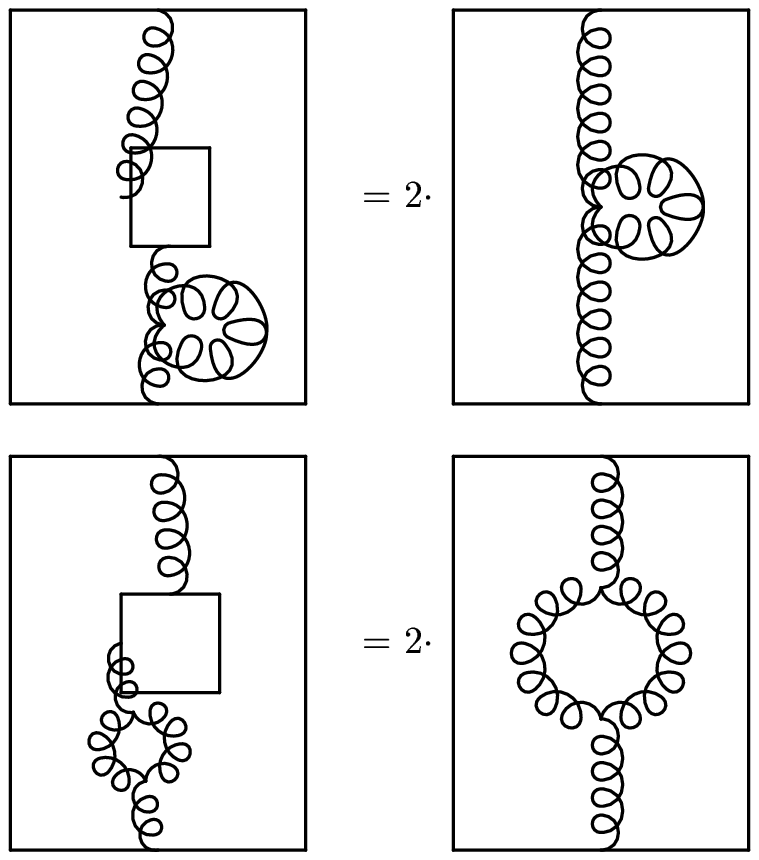, height=30cm, width=20cm}
}
\end{picture}\\
and the same for the graphs with ghost lines and the graph with the insertion
of the contribution of the integration measure.

Then there are the graphs where the action is inserted into \emph{internal}
gluon lines of the vacuum polarization. Taking them also into account, one can
write:
\begin{eqnarray}
&&<S^{(0)}\omega^{(2)}(S^{(1)})^2>_{conn} = 4 
<(S^{(1)})^2\omega^{(2)}>_{conn} + \Delta <(S^{(1)})^2\omega^{(2)}> \nonumber
\\ && <S^{(0)}\omega^{(2)}S^{(2)}>_{conn} = 3 <S^{(2)}\omega^{(2)}>_{conn} +
\Delta <S^{(2)}\omega^{(2)}>,
\end{eqnarray}
where the $\Delta$-terms come from the operator $G$ contained in the action.
The explicit results are:
\begin{eqnarray}
&&\Delta <(S^{(1)})^2\omega^{(2)}> \nonumber \\ &=& -4 C_2(G) C_2(F)
\sum_{\alpha, \beta} \int\limits_{BZ} \frac{d^4p}{(2\pi)^4}
\frac{\sin^2(p_{\mu}\hat{R}/2) \sin^2(p_{\nu}\hat{T}/2)} {\hat{p}^4}
\frac{\delta_{\mu\alpha}-\delta_{\nu\alpha}}{\sin(p_{\alpha}/2)}
\frac{\delta_{\mu\beta}-\delta_{\nu\beta}}{\sin(p_{\beta}/2)} \nonumber \\ &&
\int\limits_{BZ} \frac{d^4k}{(2\pi)^4} \frac{1}{\hat{k}^4(\widehat{p+k})^2}
\Bigg[ (\widehat{p+2k})_{\alpha} (\widehat{p+2k})_{\beta}
\sum_{\gamma} \hat{k}^2_{\gamma} \cos^2(p_{\gamma} / 2)
\nonumber \\ &+& \delta_{\alpha\beta} \cos(k_{\alpha} / 2)
\cos(k_{\beta} / 2) ((\widehat{2p+k})\hat{k})^2 \nonumber \\
&+& \hat{k}_{\alpha} \hat{k}_{\beta} \cos((k+p)_{\alpha} / 2)
\cos((p+k)_{\beta} / 2) (\widehat{k-p})^2 \nonumber \\
&-& \! \! \sum_{\gamma} \hat{k}_{\gamma} (\widehat{k-p})_{\gamma}
\cos(p_{\gamma} / 2) [(\widehat{p+2k})_{\alpha} \hat{k}_{\beta}
\cos((p+k)_{\beta} / 2) + \alpha \leftrightarrow \beta] \nonumber \\
&-& \hat{k}(\widehat{2p+k}) [(\widehat{p+2k})_{\alpha}
 \hat{k}_{\beta} \cos(k_{\beta} / 2) \cos(p_{\beta}/2) + \alpha \leftrightarrow \beta]
\nonumber \\ &+& \hat{k}(\widehat{2p+k}) [(\widehat{k-p})_{\alpha}
\cos(k_{\alpha} / 2) \hat{k}_{\beta} \cos((p+k)_{\beta} / 2)
+ \alpha \leftrightarrow \beta] \Bigg]
\end{eqnarray}
and
\begin{eqnarray}
&&\Delta <\omega^{(2)}S^{(2)}> \nonumber \\ &=& - C_2(G)C_2(F) \sum_{\alpha,
\beta} \int\limits_{BZ} \frac{d^4p}{(2\pi)^4} \frac{\sin^2(p_{\mu}\hat{R}/2)
\sin^2(p_{\nu}\hat{T}/2)} {\hat{p}^4}
\frac{\delta_{\mu\alpha}-\delta_{\nu\alpha}}{\sin(p_{\alpha}/2)}
\frac{\delta_{\mu\beta}-\delta_{\nu\beta}}{\sin(p_{\beta}/2)} \nonumber \\ &&
\int\limits_{BZ} \frac{d^4k}{(2\pi)^4} \frac{1}{\hat{k}^4} \Bigg[
\delta_{\alpha\beta} \bigg\{2 \sum_{\rho \ne \alpha} \hat{k}_{\rho}^2
\cos^2(k_{\alpha}/2) - 4 \cos^2(k_{\alpha}/2) \sum_{\rho} \sin^2(p_{\rho}/2)
\hat{k}^2_{\rho} \nonumber \\ &-& \frac{2}{3} \sin^2(k_{\alpha}/2)
\hat{p}^2 + 2 \hat{k}_{\alpha}^2 \cos^2(k_{\alpha}/2) \sin^2(p_{\alpha}/2)
\bigg\} \nonumber \\
&+& \! \hat{p}_{\alpha} \hat{p}_{\beta} \bigg\{ \frac{4}{3}
\sin^2(k_{\alpha}/2) + \frac{4}{3} \sin^2(k_{\beta}/2)
- 2 \sin^2(k_{\alpha}/2) \sin^2(k_{\beta}/2) \bigg\} \Bigg]
\end{eqnarray}

For the graph with the three-gluon vertex (the so-called ''spider graph''), one
can make use of the fact that in $\omega^{(3)}$, one of the end points of
the three gluon lines is summed over the entire gluon loop, without any
constraints. Therefore, according to (\ref{one_endpoint}), an insertion of the
action into this line just reproduces the normal spider graph without the
insertion, and one has to consider only the effect of inserting the action
into the other two lines. The result is, the $\Delta$-term again coming from
the operator $G$ contained in the action:
\begin{equation}
<S^{(0)}\omega^{(3)}S^{(1)}>_{conn} = 3
<\omega^{(3)}S^{(1)}>_{conn} + \Delta<\omega^{(3)}S^{(1)}>,
\end{equation}
with
\begin{eqnarray}
&& \Delta<\omega^{(3)}S^{(1)}> \nonumber \\
&=& -2 C_2(G) C_2(F) \int\limits_{BZ} \frac{d^4p}{(2\pi)^4} \int\limits_{BZ} \frac{d^4q}{(2\pi)^4}
\frac{\sin(p_{\mu}R/2)\sin(p_{\nu}T/2)}{\hat{p}^2\hat{q}^4
(\widehat{p+q})^2} \nonumber \\ && \cdot \Bigg[ \Bigg\{
\frac{1}{\sin((p_{\mu}+q_{\mu})/2)} \Bigg(
\frac{\cos(q_{\mu}/2)}{\sin(p_{\mu}/2)} \sum_{\rho \ne \mu}
(\widehat{2p+q})_{\rho} \hat{q}_{\rho} \nonumber \\ && +
\frac{\hat{q}_{\mu} (\widehat{q+p})_{\nu} \cos(q_{\nu}/2)
\cos(p_{\mu}/2)}{\sin(p_{\nu}/2)} + \frac{\cos(q_{\mu}/2) \cos((q_{\nu}
+p_{\nu})/2) \hat{q}_{\nu} \hat{p}_{\mu}} {\sin(p_{\nu}/2)} \Bigg)
\nonumber \\ &&  \cdot \Bigg(\sin(p_{\nu}T/2) \sin((p_{\mu}+2q_{\mu})/2) \Bigg[
\frac{\sin((p_{\mu}+2q_{\mu})R/2)}{\sin((p_{\mu}+2q_{\mu})/2)}  -
\frac{\sin(p_{\mu}R/2)}{\sin(p_{\mu}/2)} \Bigg] \nonumber \\  && + 2
\sin(q_{\nu}T/2) \sin((p_{\mu}+q_{\mu})R/2) \cos((p_{\nu}+q_{\nu})T/2) 
\cos(q_{\mu}R/2) \Bigg) \Bigg\} \nonumber \\  && + \Bigg\{(\mu,R)
\leftrightarrow (\nu,T)\Bigg\} \Bigg].
\end{eqnarray}

For calculating the effect of inserting the action into the graph with two
independent gluon lines, it is convenient to split up $\omega^{(4)}$ as
shown already in (\ref{omega4split}) and treat the various terms
separately. In $\omega^{(4A)}$, all end points of gluon lines are summed over
the entire Wilson loop, hence using (\ref{one_endpoint}), one gets:
\begin{equation}
<\omega^{(4A)}S^{(0)}>_{conn} = 2 <\omega^{(4A)}>_0.
\end{equation}
The other contributions are much more complicated. In $\omega^{(4C)}$ to
$\omega^{(4F)}$, at least one of the end points of the gluon lines is summed
over the entire Wilson loop, and therefore (\ref{one_endpoint}) can be used to
simplify the calculation: insertions of the action in that gluon line simply
reproduce the same result as for the graph without any insertions. In
$\omega^{(4B)}$, even this simplification is not possible.

The end result is, where again the $\Delta$-terms come from the insertion of
$G$
\begin{equation}
<\omega^{(4)}S^{(0)}>_{conn} = 2 <\omega^{(4)}> + \Delta <\omega^{(4)}>,
\end{equation}
respectively
\begin{eqnarray}
<\omega^{(4B)}S^{(0)}>_{conn} &=& 2 <\omega^{(4B)}>_0 + \Delta
<\omega^{(4B)}>; \nonumber \\
<\omega^{(4C)}S^{(0)}>_{conn} &=& 2 <\omega^{(4C)}>_0 +
\Delta <\omega^{(4C)}>; \nonumber \\ \ldots
\end{eqnarray}
with:
\begin{eqnarray}
&& \Delta <\omega^{(4B)}> \nonumber \\ &=& -\frac{C_2(G) C_2(F)}{8} g_0^4
\int\limits_{BZ} \frac{d^4p}{(2 \pi)^4} \int\limits_{BZ}
\frac{d^4k}{(2 \pi)^4} \frac{1}{\hat{p}^2 \hat{k}^2} \nonumber \\ &&
\cdot \Bigg[ \Bigg\{ -4 \frac{\hat{k}^2_{\mu}}{\hat{k}^2} \Bigg(
\Big(\sin^2(p_{\nu}T/2) \cos^2(k_{\nu}T/2) + \sin^2(k_{\nu}T/2)
\cos^2(p_{\nu}T/2) \Big) \nonumber \\ && \,\,\, \cdot \frac{\sin^2(p_{\mu}R/2)}
{\sin^2(p_{\mu}/2)} \frac{\sin^2(k_{\mu}R/2)} {\sin^2(k_{\mu}/2)} \nonumber \\
&&+ \frac{\sin^2(k_{\mu}R/2)} {\sin^2(k_{\mu}/2)} \frac{\sin^2(p_{\nu}T/2)}
{\sin^2(p_{\nu}/2)} + \frac{1}{4} \frac{\sin^2((p_{\nu}+k_{\nu})T/2)
\sin^2((p_{\mu}-k_{\mu})/2)} {\sin^2(p_{\mu}/2) \sin^2(k_{\mu}/2)} \nonumber \\
&& \,\,\, \cdot
\left[\frac{\sin((p_{\mu}+k_{\mu})R/2)}{\sin((p_{\mu}+k_{\mu})/2)} -
\frac{\sin((p_{\mu}-k_{\mu})R/2)}{\sin((p_{\mu}-k_{\mu})/2)}\right]^2 \Bigg)
\nonumber \\ &&+ 8 \frac{\hat{k}_{\mu}\hat{k}_{\nu}}{\hat{k}^2}
\Bigg(\frac{\sin^2(p_{\mu}R/2)}{\sin^2(p_{\mu}/2)} \frac{\sin^2(k_{\mu}R/2)}
{\sin(k_{\mu}/2)} \frac{\sin^2(k_{\nu}T/2)} {\sin(k_{\nu}/2)}
\cos^2(p_{\nu}T/2) \nonumber \\ && - \frac{1}{2}
\frac{\sin((p_{\nu}+k_{\nu})T/2) \sin((p_{\mu}-k_{\mu})/2)} {\sin(p_{\mu}/2)
\sin(k_{\mu}/2)} \nonumber \\ && \cdot
\left[\frac{\sin((p_{\mu}+k_{\mu})R/2)}{\sin((p_{\mu}+k_{\mu})/2)} -
\frac{\sin((p_{\mu}-k_{\mu})R/2)}{\sin((p_{\mu}-k_{\mu})/2)}\right] \Bigg)
\Bigg\} + \Bigg\{(\mu,R) \leftrightarrow (\nu,T)\Bigg\} \Bigg] \nonumber \\ \\
&&\Delta <\omega^{(4C)}> \nonumber \\ &=& -\frac{C_2(G) C_2(F)}{4} g_0^4
\int\limits_{BZ} \frac{d^4p}{(2 \pi)^4} \int\limits_{BZ} \frac{d^4k}{(2 \pi)^4}
\frac{1}{\hat{p}^2 \hat{k}^2} \nonumber \\ && \cdot \Bigg[ \Bigg\{4
\frac{\hat{k}^2_{\mu}}{\hat{k}^2} \frac{\sin(p_{\mu}R/2)}
{\sin(p_{\mu}/2)} \sin^2(p_{\nu}T/2) (\Sigma_1 - \Sigma_2) \nonumber \\ && +
\frac{\sin^2(p_{\mu}R/2) \sin^2(p_{\nu}T/2)}{\sin^2(p_{\alpha}/2)}
(\delta_{\alpha\mu} + \delta_{\alpha\nu}) \frac{\hat{k}^2_{\mu}}{\hat{k}^2}
\left(\frac{\sin^2(k_{\mu}R/2)}{\sin^2(k_{\mu}/2)} - R\right) \nonumber \\ &&
+ 2 \frac{\sin(p_{\mu}R/2)}{\sin(p_{\mu}/2)} \sin^2(p_{\nu}T/2)
\frac{\sin(k_{\mu}R/2)}{\sin(k_{\mu}/2)} \cos(k_{\nu}T)
\frac{\hat{k}^2_{\mu}}{\hat{k}^2} \Sigma_R(p_{\mu}, k_{\mu}) \nonumber \\
&& - \frac{\sin(p_{\mu}R/2)}{\sin(p_{\mu}/2)} \sin^2(p_{\nu}T/2)
\frac{\sin^2(k_{\nu}T/2)}{\sin(k_{\nu}/2)}
\frac{\hat{k}_{\mu}\hat{k}_{\nu}}{\hat{k}^2} \nonumber \\ && \cdot
\Bigg(\frac{\sin(p_{\mu}(R-1)/2)}{\sin(p_{\mu}/2)}
\frac{\cos(p_{\mu}/2)}{\sin(k_{\mu}/2)} \nonumber \\ && -
\frac{\cos(k_{\mu}R/2+p_{\mu}/2)}{\sin(k_{\mu}/2)}
\frac{\sin((p_{\mu}+k_{\mu})(R-1)/2)}{\sin((p_{\mu}+k_{\mu})/2)} \nonumber \\
&& - 2 \frac{\sin(k_{\mu}(R-1)/2)\sin(k_{\mu}(R+1)/2)}{\sin(k_{\mu}/2)}
\frac{\sin(p_{\mu}R/2)}{\sin(p_{\mu}/2)} \nonumber \\ &&+ 2
\frac{\cos(k_{\mu}(R+1)/2)} {\sin(p_{\mu}/2)}
\frac{\sin((p_{\mu}+k_{\mu})(R-1)/2)} {\sin((p_{\mu}+k_{\mu})/2)} \Bigg)
\Bigg\} + \Bigg\{(\mu,R) \leftrightarrow (\nu,T)\Bigg\} \nonumber \\ && + 2
\sin^2(p_{\mu}R/2) \frac{\sin^2(p_{\nu}T/2)}{\sin^2(p_{\nu}/2)}
\frac{\hat{k}^2_{\mu}}{\hat{k}^2} \frac{\sin^2(k_{\mu}R/2)}
{\sin^2(k_{\mu}/2)} \cos(k_{\nu} T) \nonumber \\ && + 2 \sin^2(p_{\nu}T/2)
\frac{\sin^2(p_{\mu}R/2)}{\sin^2(p_{\mu}/2)} \frac{\hat{k}^2_{\nu}}
{\hat{k}^2} \left(\frac{\sin^2(k_{\nu}T/2)}{\sin^2(k_{\nu}/2)} - T\right)
\nonumber \\ && - 2 \frac{\sin^2(p_{\mu}R/2)}{\sin^2(p_{\mu}/2)}
\sin^2(p_{\nu}T/2) \frac{\hat{k}_{\mu}\hat{k}_{\nu}}{\hat{k}^2}
\frac{\sin^2(k_{\mu}R/2)} {\sin(k_{\mu}/2)}
\frac{\sin^2(k_{\nu}T/2)}{\sin(k_{\nu}/2)} \nonumber \\ && + 2
\sin^2(p_{\mu}R/2) \frac{\sin(p_{\nu}T/2)}{\sin(p_{\nu}/2)}
\frac{\sin^2(k_{\mu}R/2)}{\sin(k_{\mu}/2)} \sin(k_{\nu}T/2)
\frac{\hat{k}_{\mu}\hat{k}_{\nu}}{\hat{k}^2} \nonumber \\ && \cdot
\left(\frac{\sin(p_{\nu}T/2)}{\sin(p_{\nu}/2)}
\frac{\sin(k_{\nu}T/2)}{\sin(k_{\nu}/2)} -
\frac{\sin((p_{\nu}+k_{\nu})T/2)}{\sin((p_{\nu}+k_{\nu})/2)} \right) \Bigg] \\
&& \Delta <\omega^{(4D)}> \nonumber \\ &=& -\frac{C_2(G)C_2(F)}{12} g_0^4
\int\limits_{BZ} \frac{d^4p}{(2 \pi)^4} \int\limits_{BZ}
\frac{d^4k}{(2 \pi)^4} \frac{1}{\hat{p}^2 \hat{k}^2} \Bigg[ \Bigg\{4
\frac{\hat{k}^2_{\mu}}{\hat{k}^2} \frac{\sin(p_{\mu}R/2)}
{\sin(p_{\mu}/2)} \sin^2(p_{\nu}T/2) \nonumber \\ && \cdot \Bigg(\Sigma_1 -
\Sigma_2 + \frac{1}{2} \frac{\sin(p_{\mu}R/2)} {\sin(p_{\mu}/2)} R -
\frac{1}{2} \frac{\sin(k_{\mu}R/2)}{\sin(k_{\mu}/2)}
\frac{\sin((p_{\mu}+k_{\mu})R/2)}{\sin((p_{\mu}+k_{\mu})/2)} \Bigg) \nonumber
\\ && - \frac{\sin^2(p_{\mu}R/2) \sin^2(p_{\nu}T/2)}{\sin^2(p_{\alpha}/2)}
(\delta_{\alpha\mu} + \delta_{\alpha\nu}) \frac{\hat{k}^2_{\mu}}{\hat{k}^2}
\left(\frac{\sin^2(k_{\mu}R/2)}{\sin^2(k_{\mu}/2)} - R\right) \nonumber \\ &&
+ 2 \sin^2(p_{\mu}R/2) \frac{\sin^2(p_{\nu}T/2)}{\sin^2(p_{\nu}/2)}
\frac{\sin^2(k_{\mu}R/2)}{\sin^2(k_{\mu}/2)} \frac{\hat{k}^2_{\mu}}
{\hat{k}^2} \nonumber \\ && +2 \sin^2(p_{\nu}T/2)
\frac{\sin^2(p_{\mu}R/2)}{\sin^2(p_{\mu}/2)}
\frac{\sin^2(k_{\mu}R/2)}{\sin^2(k_{\mu}/2)} \frac{\hat{k}^2_{\mu}}
{\hat{k}^2} (1+\cos(k_{\nu}T)) \nonumber \\ && - 2 \sin^2(p_{\nu}T/2)
\frac{\sin(p_{\mu}R/2)}{\sin(p_{\mu}/2)}
\frac{\sin(k_{\mu}R/2)}{\sin(k_{\mu}/2)} \frac{\hat{k}^2_{\mu}}
{\hat{k}^2} \cos(k_{\nu}T) \Sigma_R(p_{\mu},k_{\mu}) \nonumber \\ && + 2
\sin^2(p_{\nu}T/2) \frac{\sin^2(p_{\mu}R/2)}{\sin^2(p_{\mu}/2)}
\frac{\sin^2(k_{\mu}R/2)}{\sin(k_{\mu}/2)} \frac{\sin^2(k_{\nu}T/2)}
{\sin(k_{\nu}/2)} \frac{\hat{k}_{\mu}\hat{k}_{\nu}}{\hat{k}^2}
\nonumber \\ && - \sin^2(p_{\nu}T/2) \frac{\sin(p_{\mu}R/2)}{\sin(p_{\mu}/2)}
\frac{\sin^2(k_{\nu}T/2)}{\sin(k_{\nu}/2)}
\frac{\hat{k}_{\mu}\hat{k}_{\nu}}{\hat{k}^2}
\Bigg(\!2\frac{\sin(p_{\mu}(R-1)/2)}{\sin(p_{\mu}/2)}
\frac{\cos(p_{\mu}/2)}{\sin(k_{\mu}/2)} \nonumber \\ && -
2\frac{\cos(k_{\mu}R/2+p_{\mu}/2)}{\sin(k_{\mu}/2)}
\frac{\sin((p_{\mu}+k_{\mu})(R-1)/2)}{\sin((p_{\mu}+k_{\mu})/2)} \nonumber \\
&& - \frac{\sin(k_{\mu}(R-1)/2)\sin(k_{\mu}(R+1)/2)}{\sin(k_{\mu}/2)}
\frac{\sin(p_{\mu}R/2)}{\sin(p_{\mu}/2)} \nonumber \\ &&+
\frac{\cos(k_{\mu}(R+1)/2)} {\sin(p_{\mu}/2)}
\frac{\sin((p_{\mu}+k_{\mu})(R-1)/2)} {\sin((p_{\mu}+k_{\mu})/2)} \Bigg)
\Bigg\} + \Bigg\{(\mu,R) \leftrightarrow (\nu,T)\Bigg\} \nonumber \\ && -2
\sin^2(p_{\mu}R/2) \frac{\sin(p_{\nu}T/2)}{\sin(p_{\nu}/2)}
\frac{\sin^2(k_{\mu}R/2)}{\sin(k_{\mu}/2)} \sin(k_{\nu}T/2)
\frac{\hat{k}_{\mu}\hat{k}_{\nu}}{\hat{k}^2} \nonumber \\ && \cdot
\Bigg(\frac{\sin(p_{\nu}T/2)}{\sin(p_{\nu}/2)}
\frac{\sin(k_{\nu}T/2)}{\sin(k_{\nu}/2)} - \frac{\sin((p_{\nu}+k_{\nu})T/2)}
{\sin((p_{\nu}+k_{\nu})/2)} \Bigg) \nonumber \\ && -2 \sin^2(p_{\nu}T/2)
\frac{\sin^2(p_{\mu}R/2)}{\sin^2(p_{\mu}/2)}
\frac{\hat{k}^2_{\nu}}{\hat{k}^2}
\left(\frac{\sin^2(k_{\nu}T/2)}{\sin^2(k_{\nu}/2)} - T\right) \nonumber \\ &&
+ 4 \frac{\sin^2(p_{\nu}T/2)\sin^2(p_{\mu}R/2)}{\sin^2(p_{\alpha}/2)}
(\delta_{\alpha\mu} - \delta_{\alpha\nu})
\frac{\sin^2(k_{\mu}R/2)}{\sin(k_{\mu}/2)} \frac{\sin^2(k_{\nu}T/2)}
{\sin(k_{\nu}/2)} \frac{\hat{k}_{\mu}\hat{k}_{\nu}} {\hat{k}^2}
\nonumber \\ && + 4 \sin^2(p_{\mu}R/2)
\frac{\sin^2(p_{\nu}T/2)}{\sin^2(p_{\nu}/2)}
\frac{\sin^2(k_{\mu}R/2)}{\sin^2(k_{\mu}/2)} \frac{\hat{k}_{\mu}^2}
{\hat{k}^2} \nonumber \\ && + 4 \sin^2(p_{\nu}T/2)
\frac{\sin^2(p_{\mu}R/2)}{\sin^2(p_{\mu}/2)}
\frac{\sin^2(k_{\nu}T/2)}{\sin^2(k_{\nu}/2)} \cos(k_{\mu} R)
\frac{\hat{k}_{\nu}^2} {\hat{k}^2} \nonumber \\ && - 2 \sin^2(p_{\mu}R/2)
\frac{\sin^2(p_{\nu}T/2)}{\sin^2(p_{\nu}/2)}
\frac{\sin^2(k_{\mu}R/2)}{\sin^2(k_{\mu}/2)} \cos(k_{\nu}T)
\frac{\hat{k}_{\mu}^2} {\hat{k}^2} \nonumber \\ && + 2 \sin^2(p_{\nu}T/2)
\frac{\sin^2(p_{\mu}R/2)}{\sin^2(p_{\mu}/2)}
\frac{\sin^2(k_{\mu}R/2)}{\sin(k_{\mu}/2)} \frac{\sin^2(k_{\nu}T/2)}
{\sin(k_{\nu}/2)} \frac{\hat{k}_{\mu}\hat{k}_{\nu}} {\hat{k}^2} \Bigg]
\\ &&\Delta <\omega^{(4E)}> \nonumber \\ &=& -\frac{C_2(G)C_2(F)}{6} g_0^4
\int\limits_{BZ} \frac{d^4p}{(2 \pi)^4} \int\limits_{BZ}
\frac{d^4k}{(2 \pi)^4} \frac{1}{\hat{p}^2 \hat{k}^2} \Bigg[ \Bigg\{
\frac{\sin(p_{\mu}R/2)}{\sin(p_{\mu}/2)} \sin^2(p_{\nu}T/2) \nonumber \\ &&
\cdot \Bigg(\frac{1}{4} \frac{\sin(p_{\mu}R/2)}{\sin(p_{\mu}/2)} (2R+T) +
\frac{\hat{k}^2_{\mu}}{\hat{k}^2} \frac{\sin(k_{\mu}R/2)}
{\sin(k_{\mu}/2)} \frac{\sin((p_{\mu}+k_{\mu})R/2)}{\sin((p_{\mu}+k_{\mu})/2)}
\nonumber \\ &&\cdot (\cos(k_{\nu}T)-1) + \frac{\hat{k}_{\mu}\hat{k}_{\nu}}
{\hat{k}^2} \frac{\sin^2(k_{\nu}T/2)}{\sin(k_{\nu}/2)} \sin(k_{\mu}R/2)
\frac{\sin((p_{\mu}+k_{\mu})R/2)}{\sin((p_{\mu}+k_{\mu})/2)} \Bigg) \Bigg\}
\nonumber \\ && + \Bigg\{(\mu,R) \leftrightarrow (\nu,T)\Bigg\} + \frac{1}{2}
\sin^2(p_{\nu}T/2) \frac{\sin^2(p_{\mu}R/2)}{\sin^2(p_{\mu}/2)} T + 2
\frac{\hat{k}_{\mu}\hat{k}_{\nu}}{\hat{k}^2}
\frac{\sin(p_{\nu}T/2)}{\sin(p_{\nu}/2)} \nonumber \\ && \cdot
\sin^2(p_{\mu}R/2) \frac{\sin^2(k_{\mu}R/2)}{\sin(k_{\mu}/2)} \sin(k_{\nu}T/2)
\frac{\sin((p_{\nu}+k_{\nu})T/2)}{\sin((p_{\nu}+k_{\nu})/2)} \Bigg]
\end{eqnarray}
and finally
\begin{equation}
\Delta <\omega^{(4F)}> = 0,
\end{equation}
again because of symmetry properties in the colour indices. The
functions $\Sigma_1$, $\Sigma_2$ and $\Sigma_R$ can be found in section
\ref{potential_NLO}.\\
Putting everything together, the result is:
\begin{eqnarray}
&& \Delta <\omega^{(4)}> \nonumber \\
&=& \Bigg\{\frac{C_2(G)C_2(F)}{6} \Delta_0 \int\limits_{BZ} \frac{d^4p}{(2\pi)^4}
\frac{1}{\hat{p}^2} \sin^2(p_{\nu}T/2)
\frac{\sin(p_{\mu}R/2)}{\sin(p_{\mu}/2)} \Bigg(\frac{\sin(p_{\mu}R/2)}
{\sin(p_{\mu}/2)} \nonumber \\ && + \frac{1}{2}
\frac{\sin(p_{\mu}(R-2)/2)}{\sin(p_{\mu}/2)} \Bigg) \nonumber \\
&&+ C_2(G)C_2(F) \int\limits_{BZ} \frac{d^4p}{(2\pi)^4} \int\limits_{BZ} \frac{d^4k}{(2\pi)^4}
\frac{1}{\hat{p}^2\hat{k}^2}
\frac{\sin^2(p_{\mu}R/2)\sin^2(p_{\nu}T/2)}{\sin^2(p_{\mu}/2)} \nonumber \\ &&
\cdot \Bigg( \frac{\hat{k}^2_{\nu}}{\hat{k}^2} \frac{\sin^2(k_{\nu}T/2)}
{\sin^2(k_{\nu}/2)} + \frac{1}{6} \frac{\hat{k}^2_{\mu}}{\hat{k}^2}
\frac{\sin^2(k_{\mu}R/2)} {\sin^2(k_{\mu}/2)} \Bigg) \nonumber \\
&&- \frac{2 C_2(G) C_2(F)}{3} \int\limits_{BZ} \frac{d^4p}{(2\pi)^4} \int\limits_{BZ}
\frac{d^4k}{(2\pi)^4} \frac{1}{\hat{p}^2\hat{k}^4} \frac{\sin^2(p_{\mu}R/2)
\sin^2(p_{\nu}T/2)}{\sin^2(p_{\mu}/2)} \nonumber \\ && \cdot
\sin^2(k_{\mu}R/2) \sin^2(k_{\nu}T/2) \left(\frac{\hat{k}^2_{\nu}}
{\sin^2(k_{\nu}/2)} - \frac{1}{2} \frac{\hat{k}^2_{\mu}}
{\sin^2(k_{\mu}/2)} \right) \nonumber \\
&&- \frac{C_2(G) C_2(F)}{2} \int\limits_{BZ} \frac{d^4p}{(2\pi)^4)} \int\limits_{BZ}
\frac{d^4k}{(2\pi)^4)} \frac{1}{\hat{p}^2\hat{k}^2}
\frac{\hat{k}^2_{\mu}}{\hat{k}^2} \frac{\sin^2(p_{\mu}R/2)
\sin^2(k_{\mu}R/2)} {\sin^2(p_{\mu}/2) \sin^2(k_{\mu}/2)} \nonumber \\
&& \,\,\, \cdot \sin^2(k_{\nu}T/2) \nonumber \\
&&- \frac{C_2(G) C_2(F)}{9} \int\limits_{BZ} \frac{d^4p}{(2\pi)^4} \int\limits_{BZ}
\frac{d^4k}{(2\pi)^4} \frac{1}{\hat{p}^2\hat{k}^2}
\frac{\hat{k}^2_{\mu}}{\hat{k}^2} \frac{\sin^2((p_{\mu}-k_{\mu})/2)}
{\sin^2(p_{\mu}/2) \sin^2(k_{\mu}/2)} \nonumber \\ && \,\,\, \cdot
\sin^2((p_{\nu}+k_{\nu})T/2)
\left(\frac{\sin((p_{\mu}+k_{\mu})R/2)}{\sin((p_{\mu}+k_{\mu})/2)} -
\frac{\sin((p_{\mu}-k_{\mu})R/2)}{\sin((p_{\mu}-k_{\mu})/2)}\right)^2
\nonumber \\ &&- \frac{C_2(G) C_2(F)}{2} \int\limits_{BZ} \frac{d^4p}{(2\pi)^4}
\int\limits_{BZ} \frac{d^4k}{(2\pi)^4} \frac{1}{\hat{p}^2\hat{k}^2}
\frac{\hat{k}^2_{\mu}}{\hat{k}^2}
\frac{\sin^2(p_{\nu}T/2)}{\sin^2(p_{\nu}/2)}
\frac{\sin^2(k_{\mu}R/2)}{\sin^2(k_{\mu}/2)} \nonumber \\ &&- \frac{C_2(G)
C_2(F)}{3} \int\limits_{BZ} \frac{d^4p}{(2\pi)^4)} \int\limits_{BZ}
\frac{d^4k}{(2\pi)^4)} \frac{1}{\hat{p}^2\hat{k}^2}
\frac{\hat{k}^2_{\mu}}{\hat{k}^2} \sin^2(p_{\nu}T/2)
\frac{\sin(p_{\mu}R/2)}{\sin(p_{\mu}/2)} \nonumber \\ && \,\,\, \cdot
\frac{\sin(k_{\mu}R/2)}{\sin(k_{\mu}/2)} 
\frac{\sin((p_{\mu}+k_{\mu})R/2)}{\sin((p_{\mu}+k_{\mu})/2)} \nonumber \\ &&+
\frac{C_2(G) C_2(F)}{3} \int\limits_{BZ} \frac{d^4p}{(2\pi)^4}
\int\limits_{BZ} \frac{d^4k}{(2\pi)^4} \frac{1}{\hat{p}^2\hat{k}^2}
\frac{\hat{k}^2_{\mu}}{\hat{k}^2} \sin^2(p_{\nu}T/2)
\frac{\sin(p_{\mu}R/2)}{\sin(p_{\mu}/2)} \nonumber \\ && \cdot \Bigg(
\frac{\sin(p_{\mu}(R-2)/2) \cos(p_{\mu}+k_{\mu})}{\sin(p_{\mu}/2)
\sin^2(k_{\mu}/2)} \nonumber \\ && -  \frac{\sin((p_{\mu}+k_{\mu})(R-2)/2)
\cos(k_{\mu}(R+1)/2+p_{\mu})}{\sin((p_{\mu}+k_{\mu})/2) \sin^2(k_{\mu}/2)}
\nonumber \\ && + \frac{\sin(p_{\mu}(R-2)/2)
\cos(p_{\mu}/2-k_{\mu})}{\sin(p_{\mu}/2) \sin(k_{\mu}/2)
\sin((p_{\mu}+k_{\mu})/2)} - \frac{\cos((p_{\mu}+k_{\mu})R/2 + k_{\mu}/2 -
p_{\mu})}{\sin((p_{\mu}+k_{\mu})/2)} \nonumber \\ && \cdot
\frac{\sin(k_{\mu}(R-2)/2)} {\sin^2(k_{\mu}/2)} +
\frac{\sin((p_{\mu}+k_{\mu})(R-2)/2) \cos(k_{\mu}(R+2)/2+p_{\mu}/2)}
{\sin(p_{\mu}/2) \sin(k_{\mu}/2) \sin((p_{\mu}+k_{\mu})/2)} \nonumber \\ && -
\frac{\sin(k_{\mu}(R-2)/2) \cos((p_{\mu}-k_{\mu})R/2-k_{\mu}-p_{\mu})}
{\sin(p_{\mu}/2) \sin^2(k_{\mu}/2)} \Bigg) \nonumber \\ &&+ C_2(G) C_2(F)
\int\limits_{BZ} \frac{d^4p}{(2\pi)^4} \int\limits_{BZ} \frac{d^4k}{(2\pi)^4}
\frac{1}{\hat{p}^2\hat{k}^2} \frac{\hat{k}_{\mu}\hat{k}_{\nu}}
{\hat{k}^2} \Bigg[\frac{\sin^2(p_{\mu}R/2)}{\sin^2(p_{\mu}/2)}
\cos^2(p_{\nu}T/2) \nonumber \\ && \cdot
\frac{\sin^2(k_{\mu}R/2)}{\sin(k_{\mu}/2)}
\frac{\sin^2(k_{\nu}T/2)}{\sin(k_{\nu}/2)} - \frac{1}{2}
\frac{\sin((p_{\nu}+k_{\nu})T/2) \sin((p_{\mu}-k_{\mu})/2)} {\sin(p_{\mu}/2)
\sin(k_{\mu}/2)} \nonumber \\ && \cdot
\left(\frac{\sin((p_{\mu}+k_{\mu})R/2)}{\sin((p_{\mu}+k_{\mu})/2)} -
\frac{\sin((p_{\mu}-k_{\mu})R/2)}{\sin((p_{\mu}-k_{\mu})/2)}\right) \nonumber
\\ && \cdot \frac{\sin(p_{\mu}R/2)}{\sin(p_{\mu}/2)} \frac{\sin(k_{\nu}T/2)}
{\sin(k_{\nu}/2)} \cos(p_{\nu}T/2) \cos(k_{\mu}R/2) \Bigg] \nonumber \\ && +
\frac{C_2(G) C_2(F)}{6} \int\limits_{BZ} \frac{d^4p}{(2\pi)^4}
\int\limits_{BZ} \frac{d^4k}{(2\pi)^4} \frac{1}{\hat{p}^2\hat{k}^2}
\frac{\hat{k}_{\mu}\hat{k}_{\nu}} {\hat{k}^2} \sin^2(p_{\nu}T/2)
\frac{\sin(p_{\mu}R/2)}{\sin(p_{\mu}/2)} \nonumber \\ && \,\,\, \cdot
\sin(k_{\mu}R/2) \frac{\sin^2(k_{\nu}T/2)}{\sin(k_{\nu}/2)}
\left(\frac{\sin(p_{\mu}R/2)}{\sin(p_{\mu}/2)} \frac{\sin(k_{\mu}R/2)}
{\sin(k_{\mu}/2)} + \frac{\sin((p_{\mu}+k_{\mu})R/2)}
{\sin((p_{\mu}+k_{\mu})/2)} \right) \nonumber \\ && - \frac{C_2(G) C_2(F)}{12}
\int\limits_{BZ} \frac{d^4p}{(2\pi)^4} \int\limits_{BZ} \frac{d^4k}{(2\pi)^4}
\frac{1}{\hat{p}^2\hat{k}^2} \frac{\hat{k}_{\mu}\hat{k}_{\nu}}
{\hat{k}^2} \sin^2(p_{\nu}T/2) \frac{\sin(p_{\mu}R/2)}{\sin(p_{\mu}/2)}
\nonumber \\ && \,\,\, \cdot \frac{\sin^2(k_{\nu}T/2)}{\sin(k_{\nu}/2)} 
\Bigg(5 \frac{\sin(p_{\mu}(R-1)/2)}{\sin(p_{\mu}/2)} \frac{\cos(p_{\mu}/2)}
{\sin(k_{\mu}/2)} \nonumber \\ && - 5
\frac{\cos(k_{\mu}R/2+p_{\mu}/2)}{\sin(k_{\mu}/2)}
\frac{\sin((p_{\mu}+k_{\mu})(R-1)/2)}{\sin((p_{\mu}+k_{\mu})/2)} \nonumber \\
&& - 7 \frac{\sin(k_{\mu}(R-1)/2)\sin(k_{\mu}(R+1)/2)}{\sin(k_{\mu}/2)}
\frac{\sin(p_{\mu}R/2)}{\sin(p_{\mu}/2)} \nonumber \\ && + 7
\frac{\cos(k_{\mu}(R+1)/2)} {\sin(p_{\mu}/2)}
\frac{\sin((p_{\mu}+k_{\mu})(R-1)/2)} {\sin((p_{\mu}+k_{\mu})/2)} \Bigg)
\Bigg\} + \Bigg\{(\mu,R) \leftrightarrow (\nu,T)\Bigg\}. \nonumber \\
\end{eqnarray}
\\[1ex]
Looking again at (\ref{actionidentity_nlo}), one sees now that
\begin{equation} \Delta <S^{(2)}\omega^{(2)}> - \frac{1}{2}  \Delta
<\left(S^{(1)}\right)^2\omega^{(2)}> + \Delta <S^{(1)}\omega^{(3)}> - \Delta
<\omega^{(4)}> = 0. \label{oneGinsertion} \end{equation} has to be satisfied
in order to show that (\ref{actionidentity}) is true in next-to-leading order.

The only way to show this is to numerically evaluate the various
eight-dimensional integrals in momentum space for arbitrary $\hat{R}$ and
$\hat{T}$. Obviously this would take a lot of computer time; in the light of
the fact that (\ref{actionidentity}) is an identity and therefore in
principle has not to be checked (as explained at the beginning of the chapter,
these calculations are mainly done for illustrative purposes, and
because the results will become useful later) this is not justifiable.

Therefore only a check for the very special case $\hat{R}=\hat{T}=1$ was done.
Most of the results above simplify significantly then (for example, the
functions $\Sigma_1$, $\Sigma_2$, $\Sigma_R$ and $O_R$ vanish); most of the
integrations can even be done almost exactly, with only the one numerical
constant $\Delta_0$ remaining. 

But the contributions from $\Delta <\left(S^{(1)}\right)^2\omega^{(2)}>$ and
$\Delta <S^{(1)}\omega^{(3)}>$, which are now given by the following integrals
\begin{eqnarray}
&& \Delta
\left.<\left(S^{(1)}\right)^2\omega^{(2)}>\right|_{\hat{R}=\hat{T}=1}
\nonumber \\ &=& - 2 \int\limits_{BZ} \frac{d^4p}{(2\pi)^4} \int\limits_{BZ}
\frac{d^4k}{(2\pi)^4} \frac{1}{\hat{p}^2)^2 (\hat{k}^2)^2 (\widehat{p+k})^2}
\nonumber \\ && \Bigg[\Bigg((\widehat{p+2k})_{\mu}^2 \sum_{\rho=1}^4
\hat{k}_{\rho}^2 \cos^2(p_{\rho}/2)
+ \cos^2(k_{\mu}/2) \left((\widehat{2p+k}) \hat{k} \right)^2 \nonumber \\
&& + \hat{k}_{\mu}^2 \cos^2((p + k)_{\mu}/2) (\widehat{k-p})^2 \nonumber \\
&&- 2 (\widehat{p + 2k})_{\mu} \hat{k}_{\mu} \cos((p + k)_{\mu}/2) 
\sum_{\rho=1}^4 \hat{k}_{\rho} (\widehat{k-p})_{\rho} \cos(p_{\rho}/2)
\nonumber \\  && - 2 \Big[(\widehat{2p+k}) \hat{k} ((\widehat{p+2k})_{\mu}
\hat{k}_{\mu} \cos(k_{\mu}/2) \cos(p_{\mu}/2) \nonumber \\
&& - (\widehat{k-p})_{\mu} \cos(k_{\mu}/2) \hat{k}_{\mu} 
\cos((p + k)_{\mu}/2)\Big] \Bigg) \hat{p}_{\nu}^2 \nonumber \\
&& - \Bigg( (\widehat{p+2k})_{\mu} (\widehat{p+2k})_{\nu}
\sum_{\rho=1}^4 \hat{k}_{\rho}^2 \cos^2(p_{\rho}/2)  \nonumber \\
&& + \hat{k}_{\mu} \hat{k}_{\nu} \cos((p + k)_{\mu}/2) \cos((p + k)_{\nu}/2) 
(\widehat{k-p})^2 \nonumber \\
&& - \sum_{\rho=1}^4 \hat{k}_{\rho} (\widehat{k-p})_{\rho} \cos(p_{\rho}/2)
\nonumber \\ && \cdot \Big[(\widehat{p+2k})_{\mu} \hat{k}_{\nu} \cos((p +
k)_{\nu}/2)  + (\widehat{p+2k})_{\nu} \hat{k}_{\mu} \cos((p + k)_{\mu}/2)
\Big] \nonumber \\ && - (\widehat{2p+k}) \hat{k}
\Big[ (\widehat{p+2k})_{\mu} \hat{k}_{\nu} \cos(k_{\nu}/2) \cos(p_{\nu}/2)
\nonumber \\ && + (\widehat{p+2k})_{\nu} \hat{k}_{\mu} \cos(k_{\mu}/2)
\cos(p_{\mu}/2) - (\widehat{k-p})_{\mu} \cos(k_{\mu}/2) \hat{k}_{\nu} \cos((p
+ k)_{\nu}/2) \nonumber \\ && - (\widehat{k-p})_{\nu} \cos(k_{\nu}/2)
\hat{k}_{\nu} \cos((p + k)_{\mu}/2) \Big] \Bigg) \hat{p}_{\mu} \hat{p}_{\nu} \\
&& \Delta \left.<S^{(1)}\omega^{(3)}>\right|_{\hat{R}=\hat{T}=1} \nonumber \\
&=& - \int\limits_{BZ} \frac{d^4p}{(2\pi)^4} \int\limits_{BZ}
\frac{d^4k}{(2\pi)^4} \frac{1}{\hat{p}^2 (\hat{k}^2)^2 (\widehat{p+k})^2}   
\Bigg( \Bigg\{ \Big[\hat{p}_{\nu} \cos(k_{\mu}/2) \sum_{\rho \ne \mu}
(\widehat{p+k})_{\rho} \hat{k}_{\rho} \nonumber \\
&& +\hat{p}_{\mu} \hat{k}_{\mu} (\widehat{p+k})_{\nu} \cos(k_{\nu}/2)
\cos(p_{\mu}/2)
+\hat{p}^2_{\mu} \hat{k}_{\nu} \cos(k_{\mu}/2) \cos((p+k)_{\nu}/2)
\Big] \nonumber \\ && \cdot \hat{k}_{\nu} \cos((p+k)_{\nu}/2) \cos(k_{\mu}/2)
\Bigg\} + \Bigg\{ \hat{\mu} \leftrightarrow \hat{\nu} \Bigg\} \Bigg),
\end{eqnarray}
still have to evaluated completely numerically. This was done using
the standard routine \emph{Vegas} from the Numerical Recipes
\cite{NumericalRecipes}. On the other hand, the integrals which can be done
by hand give:
\begin{eqnarray}
\Delta \left.<S^{(2)}\omega^{(2)}>\right|_{\hat{R}=\hat{T}=1} &=&
\frac{5}{192} \Delta_0 - \frac{5}{36} \Delta_0^2 \\
\Delta \left.<\omega^{(4)}>\right|_{\hat{R}=\hat{T}=1} &=& \frac{5}{192}
\Delta_0 - \frac{1}{6} \Delta_0^2
\end{eqnarray}

Using the result of a numerical integration for $\Delta_0$, one finally
gets the following results:
\begin{eqnarray}
\Delta <S^{(2)}\omega^{(2)}> &=& 0.00070077 +- 0.00000005 \nonumber \\
- \frac{1}{2}  \Delta <\left(S^{(1)}\right)^2\omega^{(2)}> &=&
0.00187522 +- 0.00000145 \nonumber \\
\Delta <S^{(1)}\omega^{(3)}> &=&  -0.00254364 +- 0.00000265 \nonumber \\
- \Delta <\omega^{(4)}> &=& -0.00003397 +- 0.00000062
\end{eqnarray}
Add everything up:
\begin{eqnarray}
&&\Delta <S^{(2)}\omega^{(2)}> - \frac{1}{2}
\Delta <\left(S^{(1)}\right)^2\omega^{(2)}>
+ \Delta <S^{(1)}\omega^{(3)}> - \Delta <\omega^{(4)}> \nonumber \\
&=&  -0.00000162 +- 0.00000478,
\end{eqnarray}
hence in the range of the numerical error, (\ref{oneGinsertion}) is indeed
satisfied---or, in other words, the action sum rule respectively the identity
(\ref{actionidentity}) is valid up to next-to-leading order (for
$\hat{R}=\hat{T}=1$).

\section{Gauge invariance of the Wilson Loop}

\label{gauge_invariance}

The gauge invariance of the expectation value of the Wilson loop in leading
order has already been checked in section \ref{ASR_check_LO}. For checking this
in next-to-leading order, one makes use again of the relation
(\ref{arbitrarygauge}) which connects the propagator in Feynman gauge with the
propagator in an arbitrary gauge by using the insertion of the operator $G$.

If one applies this relation to an arbitrary graph with $n$ gluon lines in an
arbitrary gauge $\alpha$, it follows that one can write this graph as a sum of
the following contributions: the same graph with all propagators in Feynman
gauge, $n$ graphs with $n-1$ propagators in Feynman gauge and the insertion of
$G$ in the remaining line, $n(n-1)/2$ graphs with $n-2$ propagators in Feynman
gauge and insertions of $G$ into the two remaining lines, and so on; the last
one is a graph with insertions of $G$ into \emph{all} lines. The coefficients
of the terms in this sum are $(-1)^m \left(1 - \frac{1}{\alpha}\right)^m$,
where $m$ is the number of insertions of $G$. 

Here is an example to illustrate this ($n=3$); as usual a gluon line without
an explicit index $\alpha$ represents a propagator in Feynman gauge, and a
cross denotes the insertion of the operator $G$ (\ref{gaugeoperator}):\\
\begin{picture}(12,10)
\put(-3.5,-15.9)
{
\epsfig{file=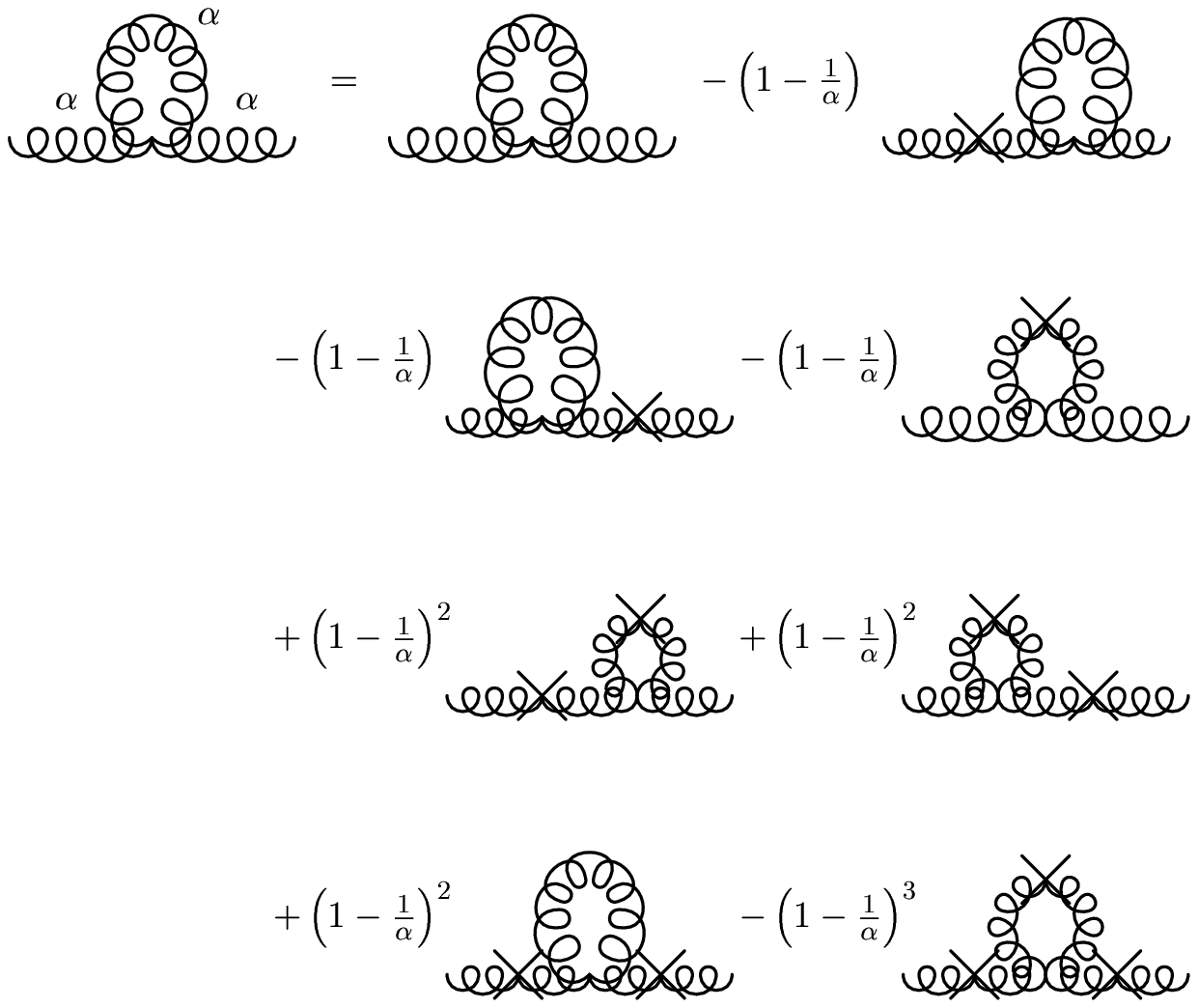, height=30cm, width=20cm}
}
\end{picture}
\\

Now looking at the graphs contributing to $<W>$ in next-to-leading order (see
figure \ref{actionidentity_NLO_r}), ones sees that there are up to four gluon
lines in them. Hence one would expect that if one calculates $<W>$ in an
arbitrary gauge, one gets a polynomial in $\left(1 - \frac{1}{\alpha}\right)$
of degree four. If one can show that all terms in this polynomial with the
exception of one of order zero (which is independent of $\alpha$) vanish, then
the gauge invariance of the expectation value of the Wilson loop is proven.

\subsubsection{Terms of order three and four}

The only contributions with a power of $\left(1 - \frac{1}{\alpha}\right)$
greater than two come from the two vacuum polarization graphs (gluon tadpole
and gluon loop) where $G$ can be inserted into external as well as into
internal lines, and from the spider graph. One can easily show that these
contributions give zero. For this, use (\ref{one_endpoint}); introducing the
operator $G$ again, that formula is equivalent to:
\begin{equation}
\sum_l <A_l A^A_{\mu}(x) G>_{conn} = 0;
\label{oneendpoint_G}
\end{equation}
This means that even if only one end of the gluon line into which $G$ is
inserted is summed over the whole Wilson loop, while the other end is entirely
arbitrary, the contribution of the graph will vanish. Diagrammatically:\\
\begin{picture}(12,4.5)
\put(-3.5,-21.2)
{
\epsfig{file=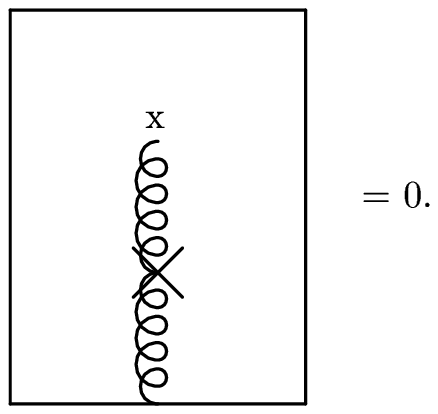, height=30cm, width=20cm}
}
\end{picture}\\
So, obviously the contributions from the vacuum polarization graphs where $G$
is inserted into an external line, and from the spider graph where $G$ is
inserted into the line whose endpoints is summed over the whole loop, vanish,
and only terms up to $\left(1 - \frac{1}{\alpha}\right)^2$ remain in the
polynomial which results from $<W>$.

\subsubsection{Terms of order one}

The contributions of order $\left(1 - \frac{1}{\alpha}\right)^1$ were already
given in section \ref{ASR_check_NLO} - the $\Delta$ terms were just the
contributions to $<SW>_{conn}$ which resulted from exactly one insertion of
$G$ into a gluon line. It also has been shown in that section that the
condition that all of these terms add up to zero (\ref{oneGinsertion}) is
equivalent to the validity of (\ref{actionidentity}) in next-to-leading order.
But as already stressed several times, (\ref{actionidentity}) is an identity,
hence one can conclude that the contributions of the graphs with exactly one
$G$ insertion indeed add up to zero.

\subsubsection{Terms of order two}

The only contributions which are left are the ones with two insertions of $G$.
Using (\ref{oneendpoint_G}), one sees that such contributions can only come
from $<\omega^{(3)}S^{(1)}>_0$, $<\omega^{(4B)}>_0$ and $<\omega^{(2)}
\left(S^{(1)}\right)^2>_{conn}$. Fortunately, the explicit calculation of
these three contributions is not too hard; the end result is:
\begin{eqnarray}
<\omega^{(3)}S^{(1)}>_0 &=& <\omega^{(2)} \left(S^{(1)}\right)^2>_{conn}
= 2 <\omega^{(4B)}>_0 \nonumber \\
&=& 8 \left(1 - \frac{1}{\alpha}\right)^2 C_2(G) C_2(F) \!
\int\limits_{BZ} \frac{d^4p}{(2\pi)^4} \int\limits_{BZ} \frac{d^4k}{(2\pi)^4} \frac{1}{(\hat{k}^2)^2
((\widehat{p+k})^2)^2} \nonumber \\
&& \cdot \sin^2(p_{\mu} \hat{R}/2) \sin^2(p_{\nu} \hat{T}/2) \nonumber \\
&& \cdot \left(\frac{\hat{k}_{\mu}}{\hat{p}_{\mu}} \cos((p+k)_{\mu}/2)
- \frac{\hat{k}_{\nu}}{\hat{p}_{\nu}} \cos((p+k)_{\nu}/2) \right)^2.
\end{eqnarray}
Considering now that $<W>$ is given by
\begin{equation}
<W> = - \frac{1}{2} g_0^3 <\omega^{(2)} \left(S^{(1)}\right)^2>_{conn}
- g_0^4 <\omega^{(4)}>_0 + g_0^4 <\omega^{(3)} S^{(1)}>_0 + \ldots,
\end{equation}
one sees that the three contributions of order $\left(1 - \frac{1}{\alpha}
\right)^2$ indeed add up to zero.

\vspace{3ex}

Hence it has been shown that in the polynomial in $\left(1 -
\frac{1}{\alpha}\right)$ of degree 4 one gets from calculating $<W>$ in an
arbitrary gauge, only the first term (of order $\left(1 -
\frac{1}{\alpha}\right)^0$ and hence independent of $\alpha$) survives: The
term of order 1 gives zero because of (\ref{oneGinsertion}), which is
equivalent to the identity (\ref{actionidentity}); the term of order 2 has
been shown to vanish exactly above, and the terms of order 3 and 4 vanish
because of (\ref{oneendpoint_G}). Therefore the proof is complete: up to
next-to-leading order, the expectation value of the Wilson loop is indeed
gauge invariant.

\section{Restriction to a fixed time slice}

\label{ASR_restriction}

What remains to be checked is the restriction to one fixed time slice,
expressed by the equation
\begin{equation}
\lim_{\hat{T} \to \infty} \frac{1}{\hat{T}} <S>_{q\bar{q}-0} = 
\lim_{\hat{T} \to \infty} <L(t)>_{q\bar{q}-0}.
\end{equation}
Using (\ref{EV_QQ}), this is equivalent to:
\begin{equation}
\lim_{\hat{T} \to \infty} \frac{1}{\hat{T}} <SW>_{conn} = 
\lim_{\hat{T} \to \infty} <LW>_{conn}.
\end{equation}
Alternatively, with (\ref{actionidentity}), it can also be written as:
\begin{equation}
\lim_{\hat{T} \to \infty} <LW>_{conn} = g_0^2 \frac{\partial}{\partial
g_0^2} \lim_{\hat{T} \to \infty} \frac{1}{\hat{T}} <W>
\label{lagrange}
\end{equation}
\\[4ex]
The expansion of $<LW>_{conn}$ is very similar to the one for $<SW>_{conn}$
(\ref{SW_expansion}):
\begin{eqnarray}
&&<LW>- <L><W> \nonumber \nonumber \\
&=& -g_0^2 <L^{(0)} \omega^{(2)}>_{conn}
+ g_0^4 <L^{(0)} S^{(2)} \omega^{(2)}>_{conn} \nonumber \\
&&- \frac{1}{2} g_0^4 <L^{(0)} (S^{(1)})^2 \omega^{(2)}>_{conn}
+ g_0^4 <L^{(0)} S^{(2)}_{FP} \omega^{(2)}>_{conn}  \nonumber \\
&& + g_0^4 <L^{(0)} S^{(2)}_{meas} \omega^{(2)}>_{conn}
- g_0^4 <L^{(2)} \omega^{(2)}>_{conn} \nonumber \\
&&+ g_0^4 <L^{(1)} S^{(1)} \omega^{(2)}>_{conn}
- g_0^4 <L^{(1)} \omega^{(3)}>_{conn} 
\nonumber \\
&& + g_0^4 <L^{(0)} S^{(1)} \omega^{(3)}>_{conn} - g_0^4 <L^{(0)}
\omega^{(4)}>_{conn} + O(g_0^6)
\label{LW_expansion}
\end{eqnarray}
For the right hand side of (\ref{lagrange}), one can again use
(\ref{Wderivative}). 

\subsection{Leading order}

\label{ASR_restriction_LO}

On the right hand side of (\ref{lagrange}), in leading order the only
contribution comes from
\begin{eqnarray}
<\omega^{(2)}>_0
&=&  2 C_2(F) \int\limits_{BZ} \frac{d^4p}{(2\pi)^4} \frac{\sin^2(p_3
\hat{R}/2) \sin^2(p_4 \hat{T}/2)}{\hat{p}^2} \nonumber \\ &&
\cdot \left(\frac{1}{\sin^2(p_3/2)} + \frac{1}{\sin^2(p_4/2)}
\right).
\end{eqnarray}
For convenience, the spatial direction of the Wilson loop, previously denoted
simply by $\mu$, has been chosen to be the $x_3$ direction here, and
euclidean time, previously denoted by $\nu$, has been identified with $x_4$.

\noindent
Now the limit of large $\hat{T}$ has to be considered. Obviously
\[
\lim_{\hat{T} \to \infty} \frac{1}{\hat{T}} \int\limits_{BZ}
\frac{d^4p}{(2\pi)^4} \frac{\sin^2(p_4 \hat{T}/2)}{\hat{p}^2}
\frac{\sin^2(p_3 \hat{R}/2) }{\sin^2(p_3/2)} = 0.
\]
Only the second term can give a non-vanishing contribution; using
\[
\lim_{\hat{T} \to \infty} \frac{1}{\hat{T}}
\frac{\sin^2(p_4 \hat{T}/2)}{\sin^2(p_4/2)} = 2 \pi \delta(p_4)
\]
(compare with section \ref{potential_LO}), what remains is:
\begin{equation}
\lim_{\hat{T} \to \infty} \frac{1}{\hat{T}} <\omega^{(2)}>_0
= 2 C_2(F) \int\limits_{BZ} \frac{d^3p}{(2\pi)^3}
\frac{\sin^2(p_3 \hat{R}/2)}{\hat{\vec{p}}^2}
\end{equation}
with
\[
\hat{\vec{p}}^2 = \sum_{j=1}^3 \hat{p}_j^2.
\]
\\[3ex]
On the other hand, an insertion of $L^{(0)}$ leads to non-conservation of the
fourth component of the momentum vector:
\begin{eqnarray}
<A^A_{\mu}(p)A^B_{\nu}(q) L^{(0)}>_{conn} &=& \! \delta^{AB}
\frac{1}{2} \frac{(2 \pi)^3 \delta^3(\vec{p}+\vec{q})}{\hat{p}^2 \hat{q}^2}
\Bigg[\left(\delta_{\mu\nu} \hat{p}^2 - \hat{p}_{\mu}\hat{p}_{\nu}\right)
e^{-i(p+q)_{\nu}/2}  \nonumber \\
&& + \left(\delta_{\mu\nu} \hat{q}^2 - \hat{q}_{\mu} \hat{q}_{\nu}\right)
e^{-i(p+q)_{\mu}/2} \Bigg].
\label{L_propagator}
\end{eqnarray}
Making use of this, an explicit calculation yields the following result:
\begin{eqnarray}
&&<L^{(0)} \omega^{(2)}>_{conn} \nonumber \\
&=& 2 C_2(F) \! \int\limits_{BZ} \! \frac{d^4p}{(2\pi)^4} \!
\int\limits_{-\pi}^{\pi} \! \frac{dq_4}{2\pi} \frac{\hat{\vec{p}}^2 +
\frac{1}{2} \left(\hat{p}_4^2 + \hat{q}_4^2 \right)}{\hat{p}^2
\left(\hat{\vec{p}}^2 + \hat{q}_4^2 \right)} 
\sin^2(p_3 \hat{R}/2) \sin(p_4 \hat{T}/2) \sin(q_4 \hat{T}/2) \nonumber \\ &&
\left(-\frac{\cos((p_4+q_4)(n_{c,4}-t))}{\sin^2(p_3/2)} +
\frac{\cos((p_4+q_4)(n_{c,4}-t-1/2))}{\sin(p_4/2)\sin(q_4/2)} \right),
\end{eqnarray}
where again $n_c$ is the center of the Wilson loop and hence
\[
n_{c,4} = n_{0,4} + \frac{1}{2} \hat{T}.
\]
In the limit of large $\hat{T}$, the first term vanishes due to
the fast oscillations of the two sines depending on $\hat{T}$; the second gives
only a non-vanishing contribution for $p_4 = q_4 = 0$:
\begin{equation}
\lim_{\hat{T} \to \infty} <L^{(0)} \omega^{(2)}>_{conn}
= 2 C_2(F) \int\limits_{BZ} \frac{d^3p}{(2\pi)^3}
\frac{\sin^2(p_3 \hat{R}/2)}{\hat{\vec{p}}^2},
\end{equation}
which is identical to the result above. Hence in leading order the
restriction to the fixed time slice $t$ indeed works:
\begin{equation}
\lim_{\hat{T} \to \infty} \frac{1}{\hat{T}} <S>_{q\bar{q}-0} =
\lim_{\hat{T} \to \infty} <L(t)>_{q\bar{q}-0} + O(g_0^4).
\end{equation}

\vspace{2ex}

For the special case $n_{c,4} = t$ (the fixed time slice lying in the middle
of the Wilson loop), the result above simplifies to
\begin{eqnarray}
&&<L^{(0)} \omega^{(2)}>_{conn} \nonumber \\
&=& 2 C_2(F) \! \int\limits_{BZ} \! \frac{d^4p}{(2\pi)^4} \!
\int\limits_{-\pi}^{\pi} \! \frac{dq_4}{2\pi} \frac{\hat{\vec{p}}^2 +
\frac{1}{2} \left(\hat{p}_4^2 + \hat{q}_4^2 \right)}{\hat{p}^2
\left(\hat{\vec{p}}^2 + \hat{q}_4^2 \right)}
\sin^2(p_3 \hat{R}/2) \sin(p_4 \hat{T}/2) \sin(q_4 \hat{T}/2) \nonumber \\ &&
\left(-\frac{1}{\sin^2(p_3/2)} +
\frac{\cos((p_4+q_4)/2))}{\sin(p_4/2)\sin(q_4/2)} \right),
\end{eqnarray}
and it is obvious that the first term vanishes for all $\hat{T}$, because the
integrand is odd in $p_4$ as well as in $q_4$. For the cosine in the numerator
of the second term, use the trigonometric relation $\cos((p_4+q_4)/2))
= \cos(p_4/2) \cos(q_4/2) - \sin(p_4/2) \sin(q_4/2)$; the sines do not
contribute because they would again give a function odd in $p_4$ as well as in
$q_4$. Hence what remains is:
\begin{eqnarray}
&&<L^{(0)} \omega^{(2)}>_{conn} \\
&=& \! \! 2 C_2(F) \! \int\limits_{BZ} \! \frac{d^4p}{(2\pi)^4} \!
\int\limits_{-\pi}^{\pi} \! \frac{dq_4}{2\pi}  \frac{\hat{\vec{p}}^2 +
\frac{1}{2} \left(\hat{p}_4^2 + \hat{q}_4^2 \right)}{\hat{p}^2
\left(\hat{\vec{p}}^2 + \hat{q}_4^2 \right)}
\sin^2(p_3 \hat{R}/2) \frac{\sin(p_4 \hat{T}/2)}{\tan(p_4/2)} \frac{\sin(q_4
\hat{T}/2)} {\tan(q_4/2)}, \nonumber
\end{eqnarray}
an expression which in the limit $\hat{T} \to \infty$ again gives
\begin{equation}
\lim_{\hat{T} \to \infty} <L^{(0)} \omega^{(2)}>_{conn}
= 2 C_2(F) \int\limits_{BZ} \frac{d^3p}{(2\pi)^3}
\frac{\sin^2(p_3 \hat{R}/2)}{\hat{\vec{p}}^2}.
\end{equation}

\subsection{Next-to-leading order}

\label{ASR_restriction_NLO}

\begin{figure}[b]
\begin{picture}(12,4.5)
\put(-3.5,-22)
{
\epsfig{file=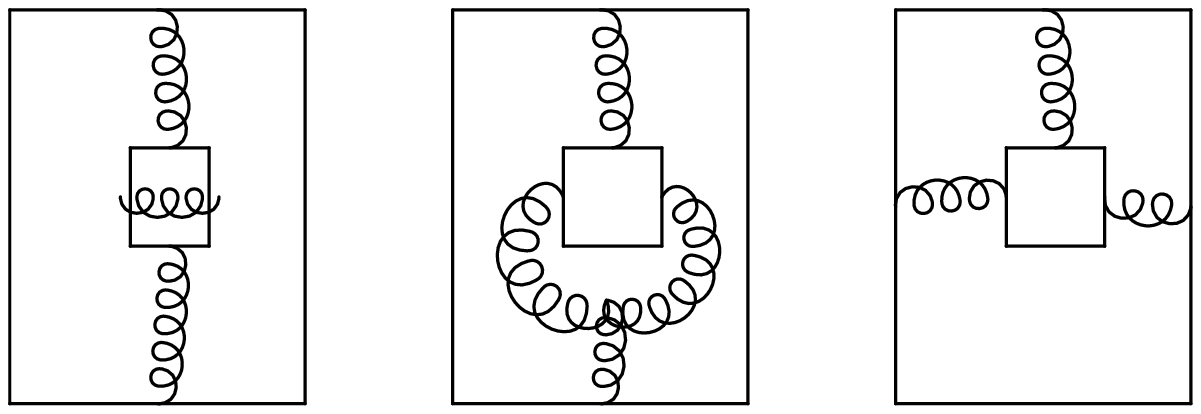, height=30cm, width=20cm}
}
\end{picture}
\caption{Additional contributions to $<LW>_{conn}$}
\label{L_graphs}
\end{figure}

The graphs corresponding to the next-to-leading order contributions were mostly
given in the figures \ref{actionidentity_NLO_l} and \ref{actionidentity_NLO_r}
already, but now the plaquettes are not summed over \emph{all} possible
positions, but only over the ones with a fixed time. Additionally, three
extra graphs have to be considered now, corresponding to $<L^{(2)}
\omega^{(2)}>_{conn}$, $<L^{(1)} S^{(1)} \omega^{(2)}>_{conn}$ and $<L^{(1)}
\omega^{(3)}>_{conn}$ (see figure \ref{L_graphs}).
There again the possible positions of the plaquette lie only on the
fixed time slice.

By inspecting the calculation of $S^{(1)}$ and $S^{(2)}$ in
\cite{RotheBook} and modifying it to give explicit results for $L^{(1)}$ and 
$L^{(2)}$, one sees that essentially the contributions of the three graphs
above can be obtained by using the usual four-gluon vertex in $<L^{(2)}
\omega^{(2)}>_{conn}$ and the usual three-gluon vertex in the two other
expectation values. The only crucial differences are that one has to replace
the usual four momentum conservation at these vertices with conservation of
only the spatial components of the momentum, and that some additional phase
factors appear. This will be studied in more detail below.

Now, as usual, consider the different types of graphs separately. There are:
\begin{itemize}
\item vacuum polarization graphs
\item spider graphs
\item graphs with two independent gluon lines
\end{itemize}
For simplicity, the Wilson loop will now be chosen to lie symmetrically to
the fixed time slice, and only the special case $t=0$ will be treated. Thus
one has $n_{0,4} = -\frac{1}{2}\hat{T}$.

\subsubsection{The vacuum polarization graphs}

In contrast to the situation in section \ref{ASR_check}, there are now two
additional vacuum polarization graphs (the first two displayed in figure
\ref{L_graphs}); the other five are very similar to the ones already
encountered in section \ref{ASR_check}, but now instead of calculating the
effect of inserting the operator $G$, one has to use (\ref{L_propagator}) in
the gluon lines. 

First, look at the cases where $L$ is inserted into an external line
(represented by the first, third, fifth, sixth and seventh graph in figure
\ref{actionidentity_NLO_l}). Because of the non-conservation of the time
component of the momentum, the relation (\ref{one_endpoint}) can not be used
any more; one has to calculate the contributions explicitly. The result is:
\begin{eqnarray}
&&4 C_2(F) \sum_{\alpha, \beta, \gamma} \int\limits_{BZ} \frac{d^4p}{(2 \pi)^4}
\int\limits_{BZ} \frac{d^4q}{(2 \pi)^4}
\frac{\sin^2(p_3 \hat{R}/2) \sin(p_4 \hat{T}/2) \sin(q_4
\hat{T}/2)}{(\hat{p}^2)^2 \hat{q}^2} \Pi_{\alpha\beta}(p) \nonumber
\\ && \frac{1}{2} \Bigg[\left(\delta_{\beta\gamma} \hat{p}^2
 - \hat{p}_{\beta}\hat{p}_{\gamma} \right) e^{-i(p+q)_{\gamma}/2}
+ \left(\delta_{\beta\gamma} \hat{q}^2- \hat{q}_{\beta}\hat{q}_{\gamma}\right)
e^{-i(p+q)_{\beta}/2} \Bigg]
\nonumber \\ &&
\frac{\delta_{3\alpha} - \delta_{4\alpha}}{\sin(p_{\alpha}/2)}
\frac{\delta_{3\gamma} - \delta_{4\gamma}}{\sin(q_{\gamma}/2)}
(2\pi)^3 \delta(\vec{p}+\vec{q}).
\end{eqnarray}
In the limit $\hat{T} \to \infty$, only the terms with $\alpha=\gamma=4$ give
a non-vanishing contribution, and these only for $p_4 = q_4 = 0$. Hence the
limit is, carrying out the integration over $q_1$, $q_2$ and $q_3$ using the
Delta-function:
\begin{equation}
4 C_2(F) \sum_{\alpha, \beta, \gamma} \int\limits_{BZ} \frac{d^3p}{(2 \pi)^3}
\frac{\sin^2(p_3 \hat{R}/2)}{\left(\hat{\vec{p}}^2\right)^2}
\Pi_{44}(\vec{p},0).
\end{equation}
This is identical to
\[
2 \lim_{\hat{T} \to \infty} \frac{1}{\hat{T}} W_{VP},
\]
where $W_{VP}$ can be found in section \ref{potential_NLO}, and, using
(\ref{VP_external}), therefore also identical to
\begin{eqnarray}
&&\lim_{\hat{T} \to \infty} \frac{1}{T}
\Bigg(-<S^{(0)}\omega^{(2)}S^{(2)}>_{conn,ext} +
<S^{(0)}\omega^{(2)}\frac{1}{2}\left(S^{(1)}\right)^2>_{conn,ext} 
\nonumber \\ && - <S^{(0)}\omega^{(2)}S^{(2)}_{FP}>_{conn,ext} -
<S^{(0)}\omega^{(2)}S^{(2)}_{meas}>_{conn,ext} \Bigg),
\end{eqnarray}
where only the insertions of $S^{(0)}$ into the \emph{external} gluon
lines are considered.

Next, look at the graphs where the sum over the plaquettes on the time slice
$t=0$ is inserted into an \emph{internal} line, represented by the second and
fourth graph in figure \ref{actionidentity_NLO_l}, and at the two graphs with
insertions of $L^{(1)}$ respectively $L^{(2)}$ depicted in figure
\ref{L_graphs}. They give the following contribution:
\begin{eqnarray}
&&2 C_2(F) \sum_{\alpha, \beta} \int\limits_{BZ} \frac{d^4p}{(2 \pi)^4}
\int\limits_{-\pi}^{\pi} \frac{dq_4}{2 \pi}
\frac{\sin^2(p_3 \hat{R}/2) \sin(p_4 \hat{T}/2) \sin(q_4
\hat{T}/2)}{\hat{p}^2 \left(\hat{\vec{p}}^2+\hat{q}_4^2\right)} \nonumber
\\ &&
\frac{\delta_{3\alpha} - \delta_{4\alpha}}{\sin(p_{\alpha}/2)}
\frac{\delta_{3\beta} - \delta_{4\beta}}{\sin(q_{\beta}/2)}
\Pi^L_{\alpha\beta}(\vec{p},p_4,q_4),
\end{eqnarray}
where $\Pi^L$ represents the vacuum polarization tensor with $L^{(0)}$
inserted into one of its internal lines respectively the two contributions
with $L^{(1)}$ or $L^{(2)}$. In the limit $\hat{T} \to \infty$, this reduces
to:
\begin{equation}
2 C_2(F) \sum_{\alpha, \beta} \int\limits_{BZ} \frac{d^3p}{(2 \pi)^3}
\frac{\sin^2(p_3 \hat{R}/2)}{\left(\hat{\vec{p}}^2\right)^2}
\Pi^L_{44}(\vec{p},0,0).
\end{equation}

Now first consider the contribution from the second graph in figure
\ref{actionidentity_NLO_l}, where $L^{(0)}$ is inserted into the internal line
of the gluon tadpole. The four-gluon vertex appearing there is denoted by
$\Gamma_{\mu\nu\rho\lambda}^{ABCD}(k,q,r,s)$. Its explicit form can be found in
\cite{RotheBook}, for example; it is not important here. Then the contribution
coming from this graph is proportional to:
\begin{eqnarray}
&& \sum_{\rho, \lambda}
\int\limits_{BZ} \frac{d^4r}{(2\pi)^4} \int\limits_{BZ} \frac{d^4s}{(2\pi)^4}
(2\pi)^4 \delta(p+q+r+s)
\Gamma_{\mu\nu\rho\lambda}^{ABCD}(p,q,r,s) \frac{(2\pi)^3
\delta(\vec{r}+\vec{s})}{\hat{r}^2\hat{s}^2} \nonumber \\ && \frac{1}{2}
\Bigg[ \left(\delta_{\rho\lambda} \hat{r}^2 - \hat{r}_{\rho} \hat{r}_{\lambda} 
\right) e^{-i(r+s)_{\lambda}/2}
+ \left(\delta_{\rho\lambda} \hat{s}^2 - \hat{s}_{\rho}
\hat{s}_{\lambda}\right)  e^{-i(r+s)_{\rho}/2} \Bigg],
\end{eqnarray}
where the Delta-function coming from the four-gluon vertex has been extraced
explicitly from $\Gamma$. Splitting this four-dimensional Delta-function up
into the spatial and temporal components and looking now only at the relevant
component where $p_4=q_4=0$, this reduces to:
\begin{eqnarray*}
&&\int\limits_{BZ} \frac{d^4r}{(2\pi)^4} \int\limits_{BZ} \frac{d^4s}{(2\pi)^4}
(2\pi)^3\delta(\vec{p}+\vec{q}+\vec{r}+\vec{s}) (2\pi) \delta(r_4+s_4)  
\frac{(2\pi)^3 \delta(\vec{r}+\vec{s})}{\hat{r}^2\hat{s}^2} \\ &&
\sum_{\rho, \lambda}
\Gamma_{\mu\nu\rho\lambda}^{ABCD}((\vec{p},0),(\vec{q},0),r,s) \\ &&
\frac{1}{2} \Bigg[\left(\delta_{\rho\lambda} \hat{r}^2 - \hat{r}_{\rho} \hat{r}_{\lambda} 
\right) e^{-i(r+s)_{\lambda}/2}
+ \left(\delta_{\rho\lambda} \hat{s}^2 - \hat{s}_{\rho}
\hat{s}_{\lambda}\right)  e^{-i(r+s)_{\rho}/2} \Bigg].
\end{eqnarray*}
Carrying out the four $s$-integrations, using the second and third
Delta-function, yields:
\begin{equation}
\sum_{\rho, \lambda}
\int\limits_{BZ} \frac{d^4r}{(2\pi)^4} (2\pi)^3(\vec{p}+\vec{q})
\Gamma_{\mu\nu\rho\lambda}^{ABCD}((\vec{p},0),(\vec{q},0),r,-r)
\frac{\delta_{\rho\lambda} \hat{r}^2 - \hat{r}_{\rho}
\hat{r}_{\lambda}}{\left(\hat{r}^2\right)^2}.
\end{equation}
Using (\ref{Sinsertion}), one sees that this is the same result which one
would have obtained if one would have inserted $S^{(0)}$ into the internal
line of the gluon tadpole diagram and then looked only at the contribution for
$p_4=q_4=0$. Hence the result is:
\begin{equation}
\lim_{\hat{T} \to \infty} <L^{(0)}\omega^{(2)}S^{(2)}>_{conn}
= \lim_{\hat{T} \to \infty} \frac{1}{\hat{T}}
<S^{(0)}\omega^{(2)}S^{(2)}>_{conn},
\end{equation}
if only the insertion into the internal line is considered. But looking at the
results for insertions into the external lines obtained above, one sees that
this formula is true even if \emph{all} possible insertions are considered.

Exactly the same arguments can be made for the fourth graph in
\ref{actionidentity_NLO_l}, where $L^{(0)}$ is inserted into internal lines of
the gluon loop. Here, too, one obtains
\begin{equation}
\lim_{\hat{T} \to \infty} <L^{(0)}\omega^{(2)}\left(S^{(1)}\right)^2>_{conn}
= \lim_{\hat{T} \to \infty} \frac{1}{\hat{T}}
<S^{(0)}\omega^{(2)}\left(S^{(1)}\right)^2>_{conn},
\end{equation}
using the Delta-functions from the two three-gluon vertices. Using the results
obtained above for insertions into external lines, the end result is:
\begin{eqnarray}
&&\lim_{\hat{T} \to \infty}
\Bigg(-<L^{(0)}\omega^{(2)}S^{(2)}>_{conn} +
<L^{(0)}\omega^{(2)}\frac{1}{2}\left(S^{(1)}\right)^2>_{conn} 
\nonumber \\ && - <L^{(0)}\omega^{(2)}S^{(2)}_{FP}>_{conn} -
<L^{(0)}\omega^{(2)}S^{(2)}_{meas}>_{conn} \Bigg) \nonumber \\
&=&\lim_{\hat{T} \to \infty} \frac{1}{\hat{T}}
\Bigg(-<S^{(0)}\omega^{(2)}S^{(2)}>_{conn} +
<S^{(0)}\omega^{(2)}\frac{1}{2}\left(S^{(1)}\right)^2>_{conn} 
\nonumber \\ && - <S^{(0)}\omega^{(2)}S^{(2)}_{FP}>_{conn} -
<S^{(0)}\omega^{(2)}S^{(2)}_{meas}>_{conn} \Bigg),
\end{eqnarray}
where now insertions into \emph{all} gluon lines are allowed.

\vspace{3ex}

What remains are the two additional vacuum polarization graphs which were
depicted in figure \ref{L_graphs}, incorporating $L^{(1)}$ and $L^{(2)}$.
First look at the second of these graphs, which involves a gluon loop where one
of the two three-gluon vertices is replaced by $L^{(1)}$. As already
mentioned above, the insertion of $L^{(1)}$ at this place amounts to using
a slightly altered version of the three-gluon vertex
$\Gamma^{ABC}_{\mu\nu\lambda}(p,k,q)$. The usual form can be found again in
\cite{RotheBook}, for example; it comes from $S^{(1)}$. Doing the same
derivation as outlined there, but for $L(0)$ instead for $S$, one gets:
\begin{eqnarray}
\Gamma^{(L),ABC}_{\mu\nu\lambda}(p,q,k) &=& i g_0 (2\pi)^3
\delta(\vec{p}+\vec{q}+\vec{k}) f_{ABC} \nonumber \\ &&
\Big[e^{i(p+q+k)(\hat{\mu}+\hat{\nu})/2}
\delta_{\lambda\nu} \cos(p_{\nu}/2) (\widehat{q-k})_{\mu}
\nonumber \\ &&
+ e^{i(p+q+k)(\hat{\nu}+\hat{\lambda})/2}
\delta_{\mu\lambda} \cos(q_{\lambda}/2) (\widehat{k-p})_{\nu}
\nonumber \\ &&
+ e^{i(p+q+k)(\hat{\lambda}+\hat{\mu})/2}
\delta_{\nu\mu} \cos(k_{\mu}/2) (\widehat{p-q})_{\lambda} \Big].
\label{L3_vertex}
\end{eqnarray}
Using this, the graph\\[1ex]
\begin{picture}(12,3)
\put(-3.5,-23.3)
{
\epsfig{file=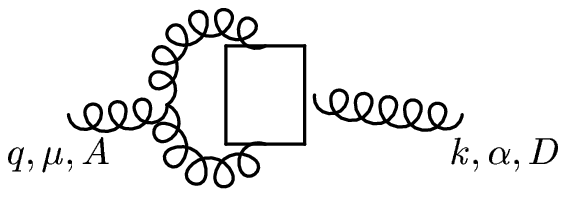, height=30cm, width=20cm}
}
\end{picture}\\
gives a contribution proportional to
\begin{eqnarray}
&&\int\limits_{BZ} \frac{d^4r}{(2\pi)^4} \int\limits_{BZ} \frac{d^4s}{(2\pi)^4}
\frac{1}{\hat{r}^2\hat{s}^2} (2\pi)^4 \delta(q-s-r) (2\pi)^3 \delta(\vec{k}
+ \vec{r} + \vec{s}) \nonumber \\ && \sum_{\nu, \lambda}
\Gamma^{ABC}_{\mu\nu\lambda}(q,-r,-s)
\Gamma^{(L),DBC}_{\alpha\nu\lambda}(k,r,s),
\end{eqnarray}
where the Delta-functions have been extracted from the $\Gamma$s and are
displayed explicitly. Doing the integrations over the three spatial
components of $s$ using the first Delta-function, this gives:
\begin{eqnarray*}
&&\int\limits_{BZ} \frac{d^4r}{(2\pi)^4} \int\limits_{-\pi}^{\pi}
\frac{ds_4}{2\pi}
\frac{1}{\hat{r}^2\left((\widehat{\vec{r}-\vec{q}})^2+\hat{s}_4^2\right)}
(2\pi) \delta(q_4-s_4-r_4) (2\pi)^3 \delta(\vec{k} + \vec{q}) \\ &&
\sum_{\nu, \lambda} \Gamma^{ABC}_{\mu\nu\lambda}(q,-r,(\vec{r},-s_4))
\Gamma^{(L),DBC}_{\alpha\nu\lambda}(k,r,(-\vec{r},s_4)).
\end{eqnarray*}
Now in the limit $\hat{T} \to \infty$, again only the contribution for
$q_4=k_4=0$ is needed. Then the first Delta-function can be used to do the
$s_4$-integration, and one gets:
\[
\sum_{\nu, \lambda} \int\limits_{BZ} \frac{d^4r}{(2\pi)^4}
\frac{(2\pi)^3 \delta(\vec{k} + \vec{q})}
{\hat{r}^2\left((\widehat{\vec{r}-\vec{q}})^2+\hat{r}_4^2\right)}
\Gamma^{ABC}_{\mu\nu\lambda}((\vec{q},0),-r,r)
\Gamma^{(L),DBC}_{\alpha\nu\lambda}((\vec{k},0),r,-r).
\]
Looking now at (\ref{L3_vertex}), one sees that in this special case the phase
factors give only factors of one, so that
\[
\Gamma^{(L),DBC}_{\alpha\nu\lambda}((\vec{k},0),r,-r)
= \Gamma^{DBC}_{\alpha\nu\lambda}((\vec{k},0),r,-r),
\]
and the contribution from the graph is simply
\begin{equation}
\sum_{\nu, \lambda} \int\limits_{BZ} \frac{d^4r}{(2\pi)^4}
\frac{(2\pi)^3 \delta(\vec{k} + \vec{q})}
{\hat{r}^2\left((\widehat{\vec{r}-\vec{q}})^2+\hat{r}_4^2\right)}
\Gamma^{ABC}_{\mu\nu\lambda}((\vec{q},0),-r,r)
\Gamma^{DBC}_{\alpha\nu\lambda}((\vec{k},0),r,-r).
\end{equation}
But that is identical to the contribution which the general gluon loop graph
would give if one would also consider only $q_4=k_4=0$ there! Hence the final
result is:
\begin{equation}
\lim_{\hat{T} \to \infty} <L^{(1)}S^{(1)}\omega^{(2)}>_{conn}
= \lim_{\hat{T} \to \infty}
\frac{1}{\hat{T}} <\left(S^{(1)}\right)^2\omega^{(2)}>_{conn}.
\end{equation}
Exactly the same type of argumentation can be used for the first graph, with
the insertion of $L^{(2)}$, so that one gets also:
\begin{equation}
\lim_{\hat{T} \to \infty} <L^{(2)}\omega^{(2)}>_{conn}
= \lim_{\hat{T} \to \infty}
\frac{1}{\hat{T}} <S^{(2)}\omega^{(2)}>_{conn}.
\end{equation}

Summarizing: in the limit $\hat{T} \to \infty$, all vacuum polarization
graphs contributing to the correlator of $L(0)$ and $W$ give the same
contribution as the vacuum polarization graphs contributing to the
correlator of $S$ and $W$, divided by $\hat{T}$.

\subsubsection{The spider graphs}

There are two spider graphs here: the usual one with the three-gluon vertex,
where $L^{(0)}$ is inserted into a gluon line (the eighth in figure
\ref{actionidentity_NLO_l}), and the one where the three-gluon vertex is
replaced by $L^{(1)}$ (the third in figure \ref{L_graphs}). For the normal
spider graph, one has:
\begin{equation}
\lim_{\hat{T} \to \infty} \frac{1}{\hat{T}} <S^{(1)} \omega^{(3)}>_0 = 0,
\end{equation}
hence in the limit of high $\hat{T}$, the contributions of these two spider
graphs should also vanish.

Unfortunately there is no such elegant argument here as in the case of the
vacuum polarization graphs; the contributions have to be calculated
explicitly. The second one is simpler tban the first; one gets for it:
\begin{eqnarray}
&&<\omega^{(3)}L^{(1)}>_{conn} \nonumber \\
&=& \frac{C_2(F)C_2(G)}{2} \int\limits_{BZ} \frac{d^4p}{(2\pi)^4}
\int\limits_{BZ} \frac{d^4k}{(2\pi)^4} \int\limits_{-\pi}^{\pi}
\frac{dq_4}{2\pi} \, \frac{\sin(p_3 \hat{R}/2) \sin(p_4 \hat{T}/2)}
{\hat{p}^2 \hat{k}^2 \left((\widehat{\vec{p}+\vec{k}})^2+\hat{q}_4^2 \right)}
\nonumber \\ && \left[ 4 \frac{\cos((k+p)_4/2) \cos((p+k)_3/2)
(\widehat{k-p})_4} {\sin(p_3/2)} \cos(k_3 \hat{R}/2) \frac{\sin((p+k)_3
\hat{R}/2)}{\sin((p+k)_3/2)} \right. \nonumber \\ &&
\cdot \frac{\sin(k_4 \hat{T}/2)}{\sin(k_4/2)} \cos(q_4 (\hat{T}-1)/2)
\nonumber \\ && - 2 \frac{\cos(p_4/2) (\widehat{2k+p})_3}{\sin(p_3/2)}
\sin(p_4/2) \cos(q_4/2) \cos(k_4/2) \nonumber \\ && \cdot \frac{\sin(q_4
\hat{T}/2)}{\sin(q_4/2)} \frac{\sin(k_4 \hat{T}/2)}{\sin(k_4/2)} \sin((2k+p)_3
\hat{R}/2) \nonumber \\ &&
- 4 \frac{\cos(p_4/2) (\widehat{2k+p})_3}{\sin(p_3/2)} \sin(p_3 \hat{R}/2)
\sin(p_4/2) \cos((k+q)_4/2) O_T(q_4,k_4) \nonumber \\ &&
- 4 \frac{\cos(p_4/2) \cos^2(k_4/2)(\widehat{2p+k})_3}{\sin(p_4/2)}
\frac{\sin((p+k)_3 \hat{R}/2)}{\sin((p+k)_3/2)}
\frac{\sin(k_4 \hat{T}/2)}{\sin(k_4/2)} \nonumber \\ && \cdot \cos(k_3
\hat{R}/2) \cos(q_4 (\hat{T}-1)/2) \nonumber \\ &&
+ 4 \frac{\cos(p_3/2) (\widehat{q-k})_4}{\sin(p_4/2)} \cos(p_4/2)
\cos((q+k)_4/2) \sin((q+k)_4 \hat{T}/2) \nonumber \\ && \cdot
O_R(k_3,-p_3-k_3).
\end{eqnarray}
In the limit $\hat{T} \to \infty$, terms like
\[
\frac{\sin(q_4 \hat{T}/2)}{\sin(q_4/2)}
\]
give constant contributions, whereas sines and cosines which depend on
$\hat{T}$, but are not divided by the corresponding sines of the coordinates,
give contributions proportional to $\hat{T}^{-1}$ because of the fast
oscillations of these functions. Therefore all of the terms in the expression
above go to zero in the limit:
\begin{equation}
\lim_{\hat{T} \to \infty} <\omega^{(3)}L^{(1)}>_{conn} = 0.
\end{equation}

\noindent
The first spider graph is much more complicated:
\begin{eqnarray}
&&<L^{(0)}\omega^{(3)}S^{(1)}>_{conn}
\sim \sum_{\gamma,\delta,\epsilon,\varphi}
\int\limits_{BZ} \frac{d^4p}{(2\pi)^4}
\int\limits_{BZ} \frac{d^4k}{(2\pi)^4} \int\limits_{-\pi}^{\pi}
\frac{dq_4}{2\pi} \Gamma_{\delta\epsilon\gamma}(p,k,-p-k) \nonumber \\ &&
\frac{1}{2} \left(\frac{\left.\left[\delta_{\gamma\varphi} \left(
\left(\widehat{\vec{p}+\vec{k}}\right)^2 + \hat{q}_4^2\right) -
\hat{q}_{\gamma} \hat{q}_{\varphi} \right] \right|_{\vec{q} =
-\vec{p}-\vec{k}} e^{-i(p+k+q)_{\gamma}/2}}
{\hat{p}^2 \hat{k}^2 \left(\widehat{p+k} \right)^2 \left(
\left(\widehat{\vec{p}+\vec{k}}\right)^2 + \hat{q}_4^2 \right)} \right.
\nonumber \\ && - \left. \frac{\left[\delta_{\gamma\varphi}
\left(\widehat{p+k}\right)^2 - \left(\widehat{p+k}\right)_{\gamma}
\left(\widehat{p+k}\right)_{\varphi} \right] e^{-i(p+k+q)_{\varphi}/2}}
{\hat{p}^2 \hat{k}^2 \left(\widehat{p+k} \right)^2 \left(
\left(\widehat{\vec{p}+\vec{k}}\right)^2 + \hat{q}_4^2 \right)}
\right) \nonumber \\
&& \cdot \Bigg[\frac{\delta_{3\delta}-\delta_{4\delta}}{\sin(p_{\delta}/2)}
\sin(p_3 \hat{R}/2) \sin(p_4 \hat{T}/2) \nonumber \\
&& \cdot \Bigg( 2 \delta_{3\epsilon} \delta_{4\varphi}
\frac{\sin(k_3 \hat{R}/2)}{\sin(k_3/2)} \frac{\sin(q_4 \hat{T}/2)}{\sin(q_4/2)}
\left(\cos((p+k)_3 \hat{R}/2 + k_4 \hat{T}/2) \right. \nonumber \\
&& \left. - i \sin((p+k)_3 \hat{R}/2 + k_4 \hat{T}/2) \right) \nonumber \\
&& + \delta_{3\epsilon} \delta_{3\varphi} \Big(-i \sin((q-k)_4 \hat{T}/2)
\frac{\sin(k_3 \hat{R}/2)}{\sin(k_3/2)} \frac{\sin((p+k)_3
\hat{R}/2)}{\sin((p+k)_3/2)} \nonumber \\
&& + 2 \sin((q+k)_4 \hat{T}/2) O_R(k_3, -p_3-k_3) \Big) \nonumber \\
&& - 2 \delta_{4\epsilon} \delta_{3\varphi}
\frac{\sin(k_4 \hat{T}/2)}{\sin(k_4/2)} \frac{\sin((p+k)_3
\hat{R}/2)}{\sin((p+k)_3/2)} \left(\cos(q_4 \hat{T}/2 - k_3 \hat{R}/2)
\right. \nonumber \\ && \left. + i \sin(q_4 \hat{T}/2 + k_3 \hat{R}/2) \right)
\nonumber \\ && - \delta_{4\epsilon} \delta_{4\varphi} \Big(i \sin((p+2k)_3
\hat{R}/2) \frac{\sin(k_4 \hat{T}/2)}{\sin(k_4/2)} \frac{\sin(q_4
\hat{T}/2)}{\sin(q_4/2)} \nonumber \\
&& - 2 \sin(p_3 \hat{R}/2) O_T(k_4, q_4) \Big) \Bigg) \nonumber \\
&& - \left(\frac{\delta_{3\phi}}{\sin((p+k)_3/2)} +
\frac{\delta_{4\phi}}{\sin(q_4/2)} \right) \sin((p+k)_3 \hat{R}/2) \sin(q_4
\hat{T}/2) \nonumber \\
&& \cdot \Bigg( 2 \delta_{3\epsilon} \delta_{4\delta}
\frac{\sin(k_3 \hat{R}/2)}{\sin(k_3/2)} \frac{\sin(p_4 \hat{T}/2)}{\sin(p_4/2)}
\left(\cos(p_3 \hat{R}/2 - k_4 \hat{T}/2) \right. \nonumber \\
&& \left. + i \sin(p_3 \hat{R}/2 + k_4 \hat{T}/2) \right) \nonumber \\
&& + \delta_{3\epsilon} \delta_{3\delta} \Big(-i \sin((p-k)_4 \hat{T}/2)
\frac{\sin(k_3 \hat{R}/2)}{\sin(k_3/2)} \frac{\sin(p_3
\hat{R}/2)}{\sin(p_3/2)} \nonumber \\
&& + 2 \sin((p+k)_4 \hat{T}/2) O_R(k_3, p_3) \Big) \nonumber\\
&& - 2 \delta_{4\epsilon} \delta_{3\delta}
\frac{\sin(k_4 \hat{T}/2)}{\sin(k_4/2)} \frac{\sin(p_3
\hat{R}/2)}{\sin(p_3/2)} \left(\cos(p_4 \hat{T}/2 - k_3 \hat{R}/2)
\right. \nonumber \\ && \left. + i \sin(p_4 \hat{T}/2 + k_3 \hat{R}/2) \right)
\nonumber \\ && - \delta_{4\epsilon} \delta_{4\delta} \Big(i \sin((p-k)_3
\hat{R}/2) \frac{\sin(k_4 \hat{T}/2)}{\sin(k_4/2)} \frac{\sin(p_4
\hat{T}/2)}{\sin(p_4/2)} \nonumber \\
&& - 2 \sin((p+k)_3 \hat{R}/2) O_T(k_4, p_4) \Big) \Bigg) \Bigg].
\end{eqnarray}
Here the explicit expression for the three-gluon vertex
$\Gamma_{\delta\epsilon\gamma}$ was not inserted and the sums were not carried
out; this would only have given an even more complicated result. This is not
necessary because even here one can already see that all of the terms which
appear vanish in the limit (using the same arguments as above). Hence the
result for this spider graph is also:
\begin{equation}
\lim_{\hat{T} \to \infty} <L^{(0)}\omega^{(3)}S^{(1)}>_{conn} = 0.
\end{equation}
Thus indeed both spider graphs vanish in the limit of large $\hat{T}$.

\subsubsection{The graphs with two independent gluon lines}

As usual, $\omega^{(4)}$ is splitted up into its parts and every
contribution is calculated separately. First one obtains, again using the
symmetry properties of the colour indices:
\begin{equation}
<L^{(0)} \omega^{(4F)}>_{conn} = 0.
\end{equation}

The contribution of $\omega^{(4)}$ which is proportional to
$\left(C_2(F)\right)^2$ only comes from $\omega^{(4A)}$ and can also be easily
calculated:
\begin{equation}
<L^{(0)} \omega^{(4A)}>_{conn} = <\omega^{(2)}> <L^{(0)}
\omega^{(2)}>_{conn} + C_2(G) C_2(F) \cdot \left( \ldots \right)
\end{equation}
In the limit of large $\hat{T}$, $<L^{(0)} \omega^{(2)}>_{conn}$ gives the same
contribution as $<S^{(0)} \omega^{(2)}>_{conn}/\hat{T}$ and therefore
$<L^{(0)} \omega^{(4A)}>_{conn}$ gives the same contribution as
$<S^{(0)} \omega^{(4A)}>_{conn}/\hat{T}$.

The other parts are much more complicated. Because of the
non-conserva\-tion of the momentum when $L^{(0)}$ is inserted,
(\ref{one_endpoint}) no longer applies, so that the insertion into \emph{both}
lines has to be considered now. This leads to a doubling of the number of the
graphs which have to be computed.

It is convenient to split the various contributions up again, classifying them
according to the number of links in spatial and temporal direction. First look
at the graphs where only spatial links appear; denote the relevant parts of
$\omega^{(4)}$ by $\omega^{(4)}_{RR}$:
\begin{eqnarray}
&&<L^{(0)} \omega^{(4A)}_{RR}>_{conn} \nonumber \\ &=& - 2\frac{C_2(G)
C_2(F)}{3} \int\limits_{BZ} \frac{d^4p}{(2\pi)^4} \int\limits_{BZ}
\frac{d^4k}{(2\pi)^4} \int\limits_{-\pi}^{\pi} \frac{dq_4}{2\pi}
\frac{1}{\hat{k}^2} \frac{\hat{\vec{p}}^2 + \frac{1}{2} \hat{p}_4^2 +
\frac{1}{2} \hat{q}_4^2 - \hat{p}_3^2}{\hat{p}^2
\left(\hat{\vec{p}}^2 + \hat{q}_4^2 \right)} \nonumber \\ &&
\frac{\sin^2(p_3 \hat{R}/2)}{\sin^2(p_3/2)} \frac{\sin^2(k_3
\hat{R}/2)}{\sin^2(k_3/2)} \sin^2(k_4 \hat{T}/2) \sin(p_4 \hat{T}/2)
\sin(q_4 \hat{T}/2) \\
&&<L^{(0)} \omega^{(4B)}_{RR}>_{conn} \nonumber \\ &=& -
\frac{C_2(G) C_2(F)}{2} \int\limits_{BZ} \frac{d^4p}{(2\pi)^4}
\int\limits_{BZ} \frac{d^4k}{(2\pi)^4} \int\limits_{-\pi}^{\pi}
\frac{dq_4}{2\pi} \frac{1}{\hat{k}^2} \frac{\hat{\vec{p}}^2 + \frac{1}{2}
\hat{p}_4^2 + \frac{1}{2} \hat{q}_4^2 - \hat{p}_3^2}{\hat{p}^2
\left(\hat{\vec{p}}^2 + \hat{q}_4^2 \right)} \nonumber \\ &&
\cdot \Bigg[ \sin((k-p)_4 \hat{T}/2) \sin((k+q)_4 \hat{T}/2)
\frac{\sin^2(p_3 \hat{R}/2)}{\sin^2(p_3/2)}
\frac{\sin^2(k_3 \hat{R}/2)}{\sin^2(k_3/2)} \nonumber \\
&& + 4 \sin((p+k)_4 \hat{T}/2) \sin((q-k)_4 \hat{T}/2) O_R^2(p_3, k_3) 
\Bigg] \\
&&<L^{(0)} \omega^{(4C)}_{RR}>_{conn} \nonumber \\ &=& \frac{C_2(F) C_2(G)}{2}
\int\limits_{BZ} \frac{d^4p}{(2\pi)^4} \int\limits_{BZ} \frac{d^4k}{(2\pi)^4}
\int\limits_{-\pi}^{\pi} \frac{dq_4}{2\pi} \frac{1}{\hat{k}^2}
\frac{\hat{\vec{p}}^2 + \frac{1}{2} \hat{p}_4^2 + \frac{1}{2} \hat{q}_4^2
- \hat{p}_3^2}{\hat{p}^2 \left(\hat{\vec{p}}^2 +
\hat{q}_4^2 \right)} \nonumber \\ &&
\cdot \sin^2(k_4 \hat{T}/2) \frac{\sin(k_3 \hat{R}/2)}{\sin(k_3/2)}
\cos(p_4 \hat{T}/2) \cos(q_4 \hat{T}/2) \\ && \cdot
\Bigg[2 \left.\left(\Sigma_1 - \Sigma_2 \right)\right|_{p_3 \leftrightarrow
k_3} + \frac{\sin(k_3 \hat{R}/2)}{\sin(k_3/2)} \Sigma_R(p_3, -p_3)
+ \frac{\sin(p_3 \hat{R}/2)}{\sin(p_3/2)} \Sigma_R(p_3,k_3) \Bigg]
\nonumber \\
&&<L^{(0)} \omega^{(4D)}_{RR}>_{conn} \nonumber \\ &=& \frac{C_2(F) C_2(G)}{6}
\int\limits_{BZ} \frac{d^4p}{(2\pi)^4} \int\limits_{BZ} \frac{d^4k}{(2\pi)^4}
\int\limits_{-\pi}^{\pi} \frac{dq_4}{2\pi} \frac{1}{\hat{k}^2}
\frac{\hat{\vec{p}}^2 + \frac{1}{2} \hat{p}_4^2 + \frac{1}{2} \hat{q}_4^2
- \hat{p}_3^2}{\hat{p}^2 \left(\hat{\vec{p}}^2 +
\hat{q}_4^2 \right)} \nonumber \\ &&
\cdot \sin^2(k_4 \hat{T}/2) \frac{\sin(k_3 \hat{R}/2)}{\sin(k_3/2)}
\cos(p_4 \hat{T}/2) \cos(q_4 \hat{T}/2) \nonumber \\ && \cdot
\Bigg[2 \left.\left(\Sigma_1 - \Sigma_2 \right)\right|_{p_3 \leftrightarrow
k_3} + 2 \, \Sigma_R(0, -k_3) - 2 \, \Sigma_R(p_3-k_3,-p_3) \nonumber \\
&& + 2 \frac{\sin(k_3 \hat{R}/2)}{\sin(k_3/2)} \frac{\sin^2(p_3
\hat{R}/2)}{\sin^2(p_3/2)} - \frac{\sin(p_3 \hat{R}/2)}{\sin(p_3/2)}
\Sigma_R(p_3, k_3) \nonumber \\ && + \frac{\sin(k_3 \hat{R}/2)}{\sin(k_3/2)}
\Sigma_R(p_3, -p_3) \Bigg] \\
&&<L^{(0)} \omega^{(4E)}_{RR}>_{conn} \nonumber \\ &=& \frac{C_2(F) C_2(G)}{6}
\int\limits_{BZ} \frac{d^4p}{(2\pi)^4} \int\limits_{BZ} \frac{d^4k}{(2\pi)^4}
\int\limits_{-\pi}^{\pi} \frac{dq_4}{2\pi} \frac{1}{\hat{k}^2}
\frac{\hat{\vec{p}}^2 + \frac{1}{2} \hat{p}_4^2 + \frac{1}{2} \hat{q}_4^2
- \hat{p}_3^2}{\hat{p}^2 \left(\hat{\vec{p}}^2 +
\hat{q}_4^2 \right)} \nonumber \\
&& \cdot \cos(p_4 \hat{T}/2) \cos(q_4 \hat{T}/2) \sin^2(k_4 \hat{T}/2)
\frac{\sin(k_3 \hat{R}/2)}{\sin(k_3/2)} \Bigg[
\frac{\sin(k_3 \hat{R}/2)}{\sin(k_3/2)} \hat{R} \nonumber \\
&& + \frac{\sin(k_3 \hat{R}/2)}{\sin(k_3/2)} \frac{\sin(p_3
\hat{R}/2)}{\sin(p_3/2)} \frac{\sin((p+k)_3 \hat{R}/2)}{\sin((p+k)_3/2)}
\Bigg].
\end{eqnarray}
Because of the fast oscillations of the two cosines for $\hat{T} \to \infty$,
the contributions of these integrals vanish in the limit of large $\hat{T}$.

The next group consists of the graphs with two spatial and two temporal links,
where $L^{(0)}$ is inserted into the line connecting the two spatial links.
The corresponding parts of $\omega^{(4)}$ are denoted by $\omega^{(4)}_{RT1}$:
\begin{eqnarray}
&&<L^{(0)} \omega^{(4A)}_{RT1}>_{conn} \nonumber \\ &=& - 2\frac{C_2(G)
C_2(F)}{3} \int\limits_{BZ} \frac{d^4p}{(2\pi)^4} \int\limits_{BZ}
\frac{d^4k}{(2\pi)^4} \int\limits_{-\pi}^{\pi} \frac{dq_4}{2\pi}
\frac{1}{\hat{k}^2} \frac{\hat{\vec{p}}^2 + \frac{1}{2} \hat{p}_4^2 +
\frac{1}{2} \hat{q}_4^2 - \hat{p}_3^2}{\hat{p}^2 \left(\hat{\vec{p}}^2 +
\hat{q}_4^2 \right)} \nonumber \\ &&
\frac{\sin^2(p_3 \hat{R}/2)}{\sin^2(p_3/2)} \frac{\sin^2(k_4
\hat{T}/2)}{\sin^2(k_4/2)} \sin^2(k_3 \hat{R}/2) \sin(p_4 \hat{T}/2)
\sin(q_4 \hat{T}/2) \\
&=& 0 \nonumber \\
&&<L^{(0)} \omega^{(4B)}_{RT1}>_{conn} \nonumber \\ &=& - \frac{C_2(G)
C_2(F)}{2} \int\limits_{BZ} \frac{d^4p}{(2\pi)^4} \int\limits_{BZ}
\frac{d^4k}{(2\pi)^4} \int\limits_{-\pi}^{\pi} \frac{dq_4}{2\pi}
\frac{1}{\hat{k}^2} \frac{\hat{\vec{p}}^2 + \frac{1}{2} \hat{p}_4^2 +
\frac{1}{2} \hat{q}_4^2 - \hat{p}_3^2}{\hat{p}^2 \left(\hat{\vec{p}}^2 +
\hat{q}_4^2 \right)} \nonumber \\ && \cdot \frac{\sin^2(p_3
\hat{R}/2)}{\sin^2(p_3/2)} \frac{\sin^2(k_4 \hat{T}/2)}{\sin^2(k_4/2)}
\cos(p_4 \hat{T}/2) \cos(q_4 \hat{T}/2) \\
&& <L^{(0)} \omega^{(4C)}_{RT1}>_{conn} \nonumber \\ &=& \frac{C_2(G)
C_2(F)}{2} \int\limits_{BZ} \frac{d^4p}{(2\pi)^4} \int\limits_{BZ}
\frac{d^4k}{(2\pi)^4} \int\limits_{-\pi}^{\pi} \frac{dq_4}{2\pi}
\frac{1}{\hat{k}^2} \frac{\hat{\vec{p}}^2 + \frac{1}{2} \hat{p}_4^2 +
\frac{1}{2} \hat{q}_4^2 - \hat{p}_3^2}{\hat{p}^2 \left(\hat{\vec{p}}^2 +
\hat{q}_4^2 \right)} \nonumber \\
&& \cdot \Bigg[\frac{\sin^2(k_4 \hat{T}/2)}{\sin^2(k_4/2)} \sin^2(k_3
\hat{R}/2) \cos(p_4 \hat{T}/2) \cos(q_4 \hat{T}/2) \Sigma_R(p_3, -p_3)
\nonumber \\ && + \frac{\sin^2(k_4 \hat{T}/2)}{\sin^2(k_4/2)} \sin^2(k_3
\hat{R}/2) \cos(p_4 \hat{T}/2) \cos(q_4 \hat{T}/2) \frac{\sin^2(p_3
\hat{R}/2)}{\sin^2(p_3/2)} \Bigg] \\
&& <L^{(0)} \omega^{(4D)}_{RT1}>_{conn} \nonumber \\ &=& \frac{C_2(G)
C_2(F)}{6} \int\limits_{BZ} \frac{d^4p}{(2\pi)^4} \int\limits_{BZ}
\frac{d^4k}{(2\pi)^4} \int\limits_{-\pi}^{\pi} \frac{dq_4}{2\pi}
\frac{1}{\hat{k}^2} \frac{\hat{\vec{p}}^2 + \frac{1}{2} \hat{p}_4^2 +
\frac{1}{2} \hat{q}_4^2 - \hat{p}_3^2}{\hat{p}^2 \left(\hat{\vec{p}}^2 +
\hat{q}_4^2 \right)} \nonumber \\
&&\cdot \Bigg[-\frac{\sin^2(k_4 \hat{T}/2)}{\sin^2(k_4/2)} \sin^2(k_3
\hat{R}/2) \cos(p_4 \hat{T}/2) \cos(q_4 \hat{T}/2) \Sigma_R(p_3, -p_3)
\nonumber \\ && + 2 \frac{\sin^2(k_4 \hat{T}/2)}{\sin^2(k_4/2)} \sin^2(k_3
\hat{R}/2) \cos(p_4 \hat{T}/2) \cos(q_4 \hat{T}/2) \frac{\sin^2(p_3
\hat{R}/2)}{\sin^2(p_3/2)} \Bigg] \\
&& <L^{(0)} \omega^{(4E)}_{RT1}>_{conn} \nonumber \\ &=& \frac{C_2(G)
C_2(F)}{6} \int\limits_{BZ} \frac{d^4p}{(2\pi)^4} \int\limits_{BZ}
\frac{d^4k}{(2\pi)^4} \int\limits_{-\pi}^{\pi} \frac{dq_4}{2\pi}
\frac{1}{\hat{k}^2} \frac{\hat{\vec{p}}^2 + \frac{1}{2} \hat{p}_4^2 +
\frac{1}{2} \hat{q}_4^2 - \hat{p}_3^2}{\hat{p}^2 \left(\hat{\vec{p}}^2 +
\hat{q}_4^2 \right)} \nonumber \\
&& \cdot \frac{\sin^2(k_4 \hat{T}/2)}{\sin^2(k_4/2)} \sin^2(k_3 \hat{R}/2)
\cos(p_4 \hat{T}/2) \cos(q_4 \hat{T}/2) \hat{R}.
\end{eqnarray}
For $\hat{T} \to \infty$, the factor $\frac{\sin^2(k_4 \hat{T}/2)}{\sin^2(k_4/2)}$
gives a linear dependence on $\hat{T}$, but the two cosines both give
factors of $\hat{T}^{-1}$, so that in total all of these integrals go with
$\hat{T}^{-1}$ in the limit and hence vanish.

Then there are the graphs with two spatial and two temporal links where
$L^{(0)}$ is inserted into the line connecting the two temporal links. The
corresponding parts of $\omega^{(4)}$ are denoted by $\omega^{(4)}_{RT2}$:
\begin{eqnarray}
&&<L^{(0)} \omega^{(4A)}_{RT2}>_{conn} \nonumber \\ &=& - 2\frac{C_2(G)
C_2(F)}{3} \int\limits_{BZ} \frac{d^4p}{(2\pi)^4} \int\limits_{BZ}
\frac{d^4k}{(2\pi)^4} \int\limits_{-\pi}^{\pi} \frac{dq_4}{2\pi}
\frac{1}{\hat{k}^2} \frac{\hat{\vec{p}}^2}{\hat{p}^2 \left(\hat{\vec{p}}^2 +
\hat{q}_4^2 \right)} \nonumber \\ 
&& \cdot \frac{\sin^2(k_3 \hat{R}/2)}{\sin^2(k_3/2)} \sin^2(k_4 \hat{T}/2)
\frac{\sin(p_4 \hat{T}/2)}{\tan(p_4/2)} \frac{\sin(q_4
\hat{T}/2)}{\tan(q_4/2)} \sin^2(p_3 \hat{R}/2) \\
&&<L^{(0)} \omega^{(4B)}_{RT2}>_{conn} \nonumber \\ &=& - \frac{C_2(G)
C_2(F)}{2} \int\limits_{BZ} \frac{d^4p}{(2\pi)^4} \int\limits_{BZ}
\frac{d^4k}{(2\pi)^4} \int\limits_{-\pi}^{\pi} \frac{dq_4}{2\pi}
\frac{1}{\hat{k}^2} \frac{\hat{\vec{p}}^2}{\hat{p}^2 \left(\hat{\vec{p}}^2 +
\hat{q}_4^2 \right)} \nonumber \\
&& \cdot \frac{\sin^2(k_3 \hat{R}/2)}{\sin^2(k_3/2)} \frac{\sin(p_4
\hat{T}/2)}{\tan(p_4/2)} \frac{\sin(q_4 \hat{T}/2)}{\tan(q_4/2)} \\
&& <L^{(0)} \omega^{(4C)}_{RT2}>_{conn} \nonumber \\ &=& \frac{C_2(G)
C_2(F)}{2} \int\limits_{BZ} \frac{d^4p}{(2\pi)^4} \int\limits_{BZ}
\frac{d^4k}{(2\pi)^4} \int\limits_{-\pi}^{\pi} \frac{dq_4}{2\pi}
\frac{1}{\hat{k}^2} \frac{\hat{\vec{p}}^2}{\hat{p}^2 \left(\hat{\vec{p}}^2 +
\hat{q}_4^2 \right)} \nonumber \\
&& \cdot \Bigg[3 \sin^2(k_4 \hat{T}/2) \frac{\sin^2(k_3 \hat{R}/2)}{\sin^2(k_3/2)}
\cos((p+q)_4/2) \Sigma_T(p_4,q_4) \nonumber \\
&&+ \cos(k_4 \hat{T}) \frac{\sin^2(k_3 \hat{R}/2)}{\sin^2(k_3/2)}
\sin^2(p_3 \hat{R}/2) \frac{\sin(p_4
\hat{T}/2)}{\tan(p_4/2)} \frac{\sin(k_4 \hat{T}/2)}{\tan(q_4/2)} \nonumber \\
&&+ \sin^2(p_3 \hat{R}/2) \Sigma_R(k_3, -k_3) \frac{\sin(p_4
\hat{T}/2)}{\tan(p_4/2)} \frac{\sin(q_4 \hat{T}/2)}{\tan(q_4/2)} \Bigg] \\
&& <L^{(0)} \omega^{(4D)}_{RT2}>_{conn} \nonumber \\ &=& \frac{C_2(G)
C_2(F)}{6} \int\limits_{BZ} \frac{d^4p}{(2\pi)^4} \int\limits_{BZ}
\frac{d^4k}{(2\pi)^4} \int\limits_{-\pi}^{\pi} \frac{dq_4}{2\pi}
\frac{1}{\hat{k}^2} \frac{\hat{\vec{p}}^2}{\hat{p}^2 \left(\hat{\vec{p}}^2 +
\hat{q}_4^2 \right)} \nonumber \\
&& \cdot \Bigg[ \left(3 - \cos(k_4 \hat{T}) \right) \sin^2(p_3 \hat{R}/2)
\frac{\sin^2(k_3 \hat{R}/2)}{\sin^2(k_3/2)} \frac{\sin(p_4
\hat{T}/2)}{\tan(p_4/2)} \frac{\sin(q_4 \hat{T}/2)}{\tan(q_4/2)} \nonumber \\
&& + \left(1 + 2 \cos(p_3 \hat{R})\right) \sin^2(k_4 \hat{T}/2)
\frac{\sin^2(k_3 \hat{R}/2)}{\sin^2(k_3/2)} \frac{\sin(p_4
\hat{T}/2)}{\tan(p_4/2)} \frac{\sin(q_4 \hat{T}/2)}{\tan(q_4/2)} \nonumber \\
&&- 3 \sin^2(k_4 \hat{T}/2) \frac{\sin^2(k_3 \hat{R}/2)}{\sin^2(k_3/2)}
\cos((p+q)_4/2) \Sigma_T(p_4,q_4) \nonumber \\
&&- \sin^2(p_3 \hat{R}/2) \Sigma_R(k_3, -k_3) \frac{\sin(p_4
\hat{T}/2)}{\tan(p_4/2)} \frac{\sin(q_4 \hat{T}/2)}{\tan(q_4/2)} \Bigg] \\
&& <L^{(0)} \omega^{(4E)}_{RT2}>_{conn} \nonumber \\ &=& \frac{C_2(G)
C_2(F)}{6} \int\limits_{BZ} \frac{d^4p}{(2\pi)^4} \int\limits_{BZ}
\frac{d^4k}{(2\pi)^4} \int\limits_{-\pi}^{\pi} \frac{dq_4}{2\pi}
\frac{1}{\hat{k}^2} \frac{\hat{\vec{p}}^2}{\hat{p}^2 \left(\hat{\vec{p}}^2 +
\hat{q}_4^2 \right)} \nonumber \\
&& \cdot \Bigg[3 \sin^2(k_4 \hat{T}/2) \frac{\sin^2(k_3 \hat{R}/2)}{\sin^2(k_3/2)}
\frac{\sin((p+q)_4 \hat{T}/2)}{\tan((p+q)_4/2)} \nonumber \\
&& + \hat{R} \sin^2(p_3 \hat{R}) \frac{\sin(p_4
\hat{T}/2)}{\tan(p_4/2)} \frac{\sin(q_4 \hat{T}/2)}{\tan(q_4/2)} \Bigg].
\end{eqnarray}
Not all of these terms vanish seperately, but their sum gives:
\begin{eqnarray}
<L^{(0)} \omega^{(4)}_{RT2}>_{conn} &=& \frac{C_2(G) C_2(F)}{2}
\int\limits_{BZ} \frac{d^4p}{(2\pi)^4} \int\limits_{BZ} \frac{d^4k}{(2\pi)^4}
\int\limits_{-\pi}^{\pi} \frac{dq_4}{2\pi} \frac{1}{\hat{k}^2}
\frac{\hat{\vec{p}}^2}{\hat{p}^2 \left(\hat{\vec{p}}^2 +
\hat{q}_4^2 \right)} \nonumber \\
&& \cdot \frac{\sin^2(k_3 \hat{R}/2)}{\sin^2(k_3/2)} \frac{\sin(p_4
\hat{T}/2)}{\tan(p_4/2)} \frac{\sin(q_4 \hat{T}/2)}{\tan(q_4/2)} \cos(k_4
\hat{T}) \cos(p_3 \hat{R}). \nonumber \\
\end{eqnarray}
This expression vanishes in the limit of large $\hat{T}$, again because of the
fast oscillations of the cosine.

The next large group consists of the graphs with three spatial links and one
temporal link. Denoting the corresponding part with $\omega^{(4)}_{RRRT}$,
they give:
\begin{eqnarray}
&&<L^{(0)} \omega^{(4A)}_{RRRT}>_{conn} \nonumber \\ &=& - \frac{C_2(G)
C_2(F)}{3} \int\limits_{BZ} \frac{d^4p}{(2\pi)^4} \int\limits_{BZ}
\frac{d^4k}{(2\pi)^4} \int\limits_{-\pi}^{\pi} \frac{dq_4}{2\pi}
\frac{1}{\hat{k}^2} \frac{\hat{p}_4^2+\hat{q}_4^2}{\hat{p}^2
\left(\hat{\vec{p}}^2 + \hat{q}_4^2 \right)} \nonumber \\ 
&& \cdot \frac{\sin^2(k_3 \hat{R}/2)}{\sin^2(k_3/2)} \sin^2(k_4 \hat{T}/2)
\frac{\sin(p_4 \hat{T}/2)}{\tan(p_4/2)} \frac{\sin(q_4
\hat{T}/2)}{\tan(q_4/2)} \sin^2(p_3 \hat{R}/2) \\
&&<L^{(0)} \omega^{(4B)}_{RRRT}>_{conn} \nonumber \\ &=& - \frac{C_2(G)
C_2(F)}{2} \int\limits_{BZ} \frac{d^4p}{(2\pi)^4} \int\limits_{BZ}
\frac{d^4k}{(2\pi)^4} \int\limits_{-\pi}^{\pi} \frac{dq_4}{2\pi}
\frac{1}{\hat{k}^2} \frac{\hat{p}_3 \hat{p}_4}{\hat{p}^2 \left(\hat{\vec{p}}^2
+ \hat{q}_4^2 \right)} \frac{\sin(q_4 \hat{T}/2)}{\tan(q_4/2)} \nonumber \\
&& \cdot \frac{\sin(k_3 \hat{R}/2)}{\sin(k_3/2)}
\Bigg[\frac{\sin^2(p_3 \hat{R}/2)}{\sin(p_3/2)} \frac{\sin(k_3
\hat{R}/2)}{\sin(k_3/2)} - 2 \cos(p_3 \hat{R}/2) O_R(p_3, q_3) \Bigg]
\nonumber \\
&&\cdot \left(\cos^2(k_4 \hat{T}/2) \sin(p_4 \hat{T}/2)
\cos(p_4/2) - \sin^2(k_4 \hat{T}/2) \cos(p_4 \hat{T}/2) \sin(p_4/2) \right)
\nonumber \\ \\
&& <L^{(0)} \omega^{(4C)}_{RRRT}>_{conn} \nonumber \\ &=& \frac{C_2(G)
C_2(F)}{2} \int\limits_{BZ} \frac{d^4p}{(2\pi)^4} \int\limits_{BZ}
\frac{d^4k}{(2\pi)^4} \int\limits_{-\pi}^{\pi} \frac{dq_4}{2\pi}
\frac{1}{\hat{k}^2} \frac{\hat{p}_4}{\hat{p}^2 \left(\hat{\vec{p}}^2 +
\hat{q}_4^2 \right)} \frac{\sin(q_4 \hat{T}/2)}{\tan(q_4/2)} \nonumber \\
&& \cdot \Bigg[\left(\sin(p_4 (\hat{T}+1)/2) + \sin(p_4 (\hat{T}-1)/2)
\right) \sin(p_3/2) \sin(p_3 \hat{R}/2)
\left(\Sigma_{1R} - \Sigma_{2R} \right) \nonumber \\
&& + \sin^2(p_3 \hat{R}/2) \sin(p_4 (\hat{T}+1)/2) \Sigma_R(k_3, -k_3)
\nonumber \\
&&+ \frac{\sin(k_3 \hat{R}/2)}{\sin(k_3/2)} \sin(p_3 \hat{R}/2) \sin(p_3/2)
\sin(p_4 (\hat{T}-1)/2) \cos(k_4 \hat{T}) \Sigma_R(p_3, k_3) \nonumber \\
&& + \sin(p_4 \hat{T}/2) \sin^2(p_3 \hat{R}/2) \cos(p_4/2) \Sigma_R(k_3, -k_3)
\nonumber \\
&& - \sin(p_4 (\hat{T}-1)/2) \frac{\sin(k_3 \hat{R}/2)}{\sin(k_3/2)}
\sin(p_3/2) \sin(p_3 \hat{R}/2) \Sigma_R(k_3,p_3) \nonumber \\
&& - \sin(p_4 (\hat{T}-1)/2) \frac{\sin(k_3 \hat{R}/2)}{\sin(k_3/2)}
\sin(p_3/2) \cos(p_3 \hat{R}/2) O_R(k_3, p_3) \nonumber \\
&&+ \sin^2(p_3 \hat{R}/2) \left(\sin^2(k_4 \hat{T}/2) \sin(p_4 (\hat{T}-1)/2)
+ \sin(p_4 \hat{T}/2) \cos(k_4 \hat{T}) \right) \nonumber \\
&&+ 2 \sin(p_4 (\hat{T}-1)/2) \sin^2(k_4 \hat{T}/2) \frac{\sin(k_3
\hat{R}/2)}{\sin(k_3/2)} \cos(p_3 \hat{R}/2) \sin(p_3/2) O_R(k_3, p_3)
\nonumber \\
&&- 2 \cos(p_4 \hat{T}/2) \sin(p_4/2) \sin^2(k_4 \hat{T}/2) \frac{\sin^2(k_3
\hat{R}/2)}{\sin^2(k_3/2)} \sin^2(p_3 \hat{R}/2) \Bigg] \\
&& <L^{(0)} \omega^{(4D)}_{RRRT}>_{conn} \nonumber \\ &=& \frac{C_2(G)
C_2(F)}{12} \int\limits_{BZ} \frac{d^4p}{(2\pi)^4} \int\limits_{BZ}
\frac{d^4k}{(2\pi)^4} \int\limits_{-\pi}^{\pi} \frac{dq_4}{2\pi}
\frac{1}{\hat{k}^2} \frac{\hat{p}_4}{\hat{p}^2 \left(\hat{\vec{p}}^2 +
\hat{q}_4^2 \right)} \frac{\sin(q_4 \hat{T}/2)}{\tan(q_4/2)} \nonumber \\
&& \cdot \Bigg[\left(\sin(p_4 (\hat{T}+1)/2) + \sin(p_4 (\hat{T}-1)/2)
\right) \sin(p_3/2) \sin(p_3 \hat{R}/2) \nonumber \\
&&\cdot \left(\Sigma_{1R} - \Sigma_{2R} + \Sigma_R(0,k_3) -\Sigma_R
(p_3+k_3,-p_3) \right) \nonumber \\
&& + \frac{\sin^2(k_3 \hat{R}/2)}{\sin^2(k_3/2)} \sin^2(p_3 \hat{R}/2)
\left(\sin(p_4 (\hat{T}-1)/2) + \sin(p_4 (\hat{T}+1)/2) \cos(k_4
\hat{T}) \right) \nonumber \\
&& - \sin(p_3/2) \sin(p_3 \hat{R}/2) \sin(p_4 (\hat{T}-1)/2) \cos(k_4 \hat{T})
\frac{\sin(k_3 \hat{R}/2)}{\sin(k_3/2)} \Sigma_R(k_3, p_3) \nonumber \\
&& - \sin^2(p_3 \hat{R}/2) \sin(p_4 (\hat{T}+1)/2) \Sigma_R(k_3, -k_3)
\nonumber \\
&&+ 2 \sin^2(p_3 \hat{R}/2) \frac{\sin^2(k_3 \hat{R}/2)}{\sin^2(k_3/2)}
\left(\sin(p_4 \hat{T}/2) \cos(p_4/2) \right. \nonumber \\
&& \left. \, \, \, - \sin^2(k_4 \hat{T}/2) \sin(p_4 (\hat{T}+1)/2) \right)
\nonumber \\
&& + \sin^2(p_3 \hat{R}/2) \frac{\sin^2(k_3 \hat{R}/2)}{\sin^2(k_3/2)}
\left(\sin(p_4 \hat{T}/2) \cos(p_4/2) \right. \nonumber \\
&& \left. \,\,\, - \sin^2(k_4\hat{T}/2) \sin(p_4 (\hat{T}-1)/2) \right)
\nonumber \\
&& + \sin^2(k_4 \hat{T}/2) \frac{\sin(k_3 \hat{R}/2)}{\sin(k_3/2)}
\sin(p_4 (\hat{T}-1)/2) \sin(p_3/2) \nonumber \\ && \cdot \left(\sin(p_3
\hat{R}/2) \Sigma_R(p_3, k_3) + \cos(p_3 \hat{R}/2) O_R(p_3, k_3) \right)
\nonumber \\ &&- \sin^2(p_3 \hat{R}/2) \sin(p_4 \hat{T}/2) \cos(p_4/2)
\Sigma_R(k_3, -k_3) \nonumber \\
&&+ \frac{\sin^2(k_3 \hat{R}/2)}{\sin^2(k_3/2)} \sin^2(p_3 \hat{R}/2)
\nonumber \\ &&\cdot \left(\sin^2(k_4 \hat{T}/2) \sin(p_4 (\hat{T}-1)/2) +
\cos(k_4 \hat{T}) \sin(p_4 \hat{T}/2) \cos(p_4/2) \right) \Bigg] \\
&& <L^{(0)} \omega^{(4E)}_{RRRT}>_{conn} \nonumber \\ &=&
\frac{C_2(G) C_2(F)}{12} \int\limits_{BZ} \frac{d^4p}{(2\pi)^4}
\int\limits_{BZ} \frac{d^4k}{(2\pi)^4} \int\limits_{-\pi}^{\pi}
\frac{dq_4}{2\pi} \frac{1}{\hat{k}^2} \frac{\hat{p}_4}{\hat{p}^2
\left(\hat{\vec{p}}^2 + \hat{q}_4^2 \right)}
\frac{\sin(q_4 \hat{T}/2)}{\tan(q_4/2)} \nonumber \\
&& \cdot \Bigg[\sin^2(k_3 \hat{R}/2) \sin(p_4 \hat{T}/2) \cos(p_4/2) \hat{R}
\nonumber \\
&&+ \frac{\sin(k_3 \hat{R}/2)}{\sin(k_3/2)}
\frac{\sin((p+k)_3 \hat{R}/2)}{\sin((p+k)_3/2)} 
\sin(p_3 \hat{R}/2) \sin(p_3/2) \nonumber \\
&& \cdot \left(\sin^2(k_4 \hat{T}/2) + \cos(k_4 \hat{T}) \right)
\sin(p_4 (\hat{T}-1)/2) \nonumber \\
&&+ \left(\sin(p_4 (\hat{T}+1)/2) + \sin(p_4 (\hat{T}+1)/2) \right)
\sin(p_3 \hat{R}/2) \sin(p_3/2) \nonumber \\ && \cdot \left(\Sigma_R(p_3,0) -
\Sigma_R(k_3, p_3-k_3) \right) \nonumber \\
&&+ \sin(p_4 (\hat{T}+1)/2) \sin^2(p_3 \hat{R}/2) \hat{R} \Bigg].
\end{eqnarray}
All of these contributions go to zero in the limit of large $\hat{T}$.

What remains are the graphs with four temporal links and the
graphs with three temporal and one spatial link. Only these can give non-vanishing
contributions in the limit of large $\hat{T}$; the first ones will be denoted
by $\omega^{(4)}_{TT}$ and the others with $\omega^{(4)}_{RTTT}$. The explicit
results are:
\begin{eqnarray}
&&<L^{(0)} \omega^{(4A)}_{TT}>_{conn} \nonumber \\ &=& - 2\frac{C_2(G)
C_2(F)}{3} \int\limits_{BZ} \frac{d^4p}{(2\pi)^4} \int\limits_{BZ}
\frac{d^4k}{(2\pi)^4} \int\limits_{-\pi}^{\pi} \frac{dq_4}{2\pi}
\frac{1}{\hat{k}^2} \frac{\hat{\vec{p}}^2}{\hat{p}^2 \left(\hat{\vec{p}}^2 +
\hat{q}_4^2 \right)} \nonumber \\ 
&& \cdot \frac{\sin^2(k_4 \hat{T}/2)}{\sin^2(k_4/2)} \sin^2(k_3 \hat{R}/2)
\frac{\sin(p_4 \hat{T}/2)}{\tan(p_4/2)} \frac{\sin(q_4
\hat{T}/2)}{\tan(q_4/2)} \sin^2(p_3 \hat{R}/2) \label{start} \\
&&<L^{(0)} \omega^{(4B)}_{TT}>_{conn} \nonumber \\ &=& - \frac{C_2(G)
C_2(F)}{2} \int\limits_{BZ} \frac{d^4p}{(2\pi)^4} \int\limits_{BZ}
\frac{d^4k}{(2\pi)^4} \int\limits_{-\pi}^{\pi} \frac{dq_4}{2\pi}
\frac{1}{\hat{k}^2} \frac{\hat{\vec{p}}^2}{\hat{p}^2 \left(\hat{\vec{p}}^2 +
\hat{q}_4^2 \right)} \nonumber \\
&& \cdot \Bigg[ \sin^2((k-p)_3 \hat{R}/2) \frac{\sin(p_4
\hat{T}/2)}{\tan(p_4/2)} \frac{\sin(q_4 \hat{T}/2)}{\tan(q_4/2)} 
\frac{\sin^2(k_4 \hat{T}/2)}{\sin^2(k_4/2)} \nonumber \\
&& + 4 \sin((k-p)_3 \hat{R}/2) \sin((k+p)_3 \hat{R}/2) 
\frac{\sin(q_4 \hat{T}/2)}{\tan(q_4/2)} \frac{\sin^2(k_4 \hat{T}/2)}{\sin^2(k_4/2)}
O_T(p_4,k_4) \nonumber \\
&&- 4 \sin^2((k+p)_3 \hat{R}/2) \cos((p+q)_4/2) O_T(p_4,k_4) O_T(q_4,-k_4)
\Bigg] \\
&& <L^{(0)} \omega^{(4C)}_{TT}>_{conn} \nonumber \\ &=& \frac{C_2(G)
C_2(F)}{2} \int\limits_{BZ} \frac{d^4p}{(2\pi)^4} \int\limits_{BZ}
\frac{d^4k}{(2\pi)^4} \int\limits_{-\pi}^{\pi} \frac{dq_4}{2\pi}
\frac{1}{\hat{k}^2} \frac{\hat{\vec{p}}^2}{\hat{p}^2 \left(\hat{\vec{p}}^2 +
\hat{q}_4^2 \right)} \nonumber \\
&& \cdot \Bigg[2 \frac{\sin(p_4 \hat{T}/2)}{\tan(p_4/2)} \sin^2(p_3 \hat{R}/2)
\left.\left(\Sigma_{1T} - \Sigma_{2T} \right)\right|_{p_4 = - q_4}
\cos(q_4/2) \nonumber \\
&& + 2 \frac{\sin(k_4 \hat{T}/2)}{\sin(k_4/2)} \sin^2(k_3 \hat{R}/2)
\left(\Sigma_{3T} - \Sigma_{4T}\right) \cos((p+q)_4/2) \nonumber \\
&& - 2 \frac{\sin(p_4 \hat{T}/2)}{\tan(p_4/2)} \frac{\sin(k_4
\hat{T}/2)}{\sin(k_4/2)} O_T(q_4, k_4) \sin(q_4/2)  \nonumber \\
&& \,\,\, \cdot \left(\sin^2(k_3 \hat{R}/2)
\cos(p_3 \hat{R}) - \sin^2(p_3 \hat{R}/2) \cos(k_3 \hat{R}) \right) \nonumber
\\ && + \frac{\sin(p_4 \hat{T}/2)}{\tan(p_4/2)} \frac{\sin(q_4
\hat{T}/2)}{\tan(q_4/2)} \sin^2(p_3 \hat{R}/2) \Sigma_T(q_4, -q_4) \nonumber \\
&& + \frac{\sin^2(k_4 \hat{T}/2)}{\sin^2(k_4/2)} \sin^2(k_3 \hat{R}/2)
\nonumber \\
&& \,\,\, \cdot \left(\Sigma_T(p_4,q_4) \cos((p+q)_4/2) - O_T(p_4,q_4)
\sin((p+q)_4/2) \right) \nonumber \\
&& + \sin^2(p_3 \hat{R}/2) \cos(k_3 \hat{T}) \frac{\sin(p_4
\hat{T}/2)}{\tan(p_4/2)} \frac{\sin(k_4 \hat{T}/2)}{\sin(k_4/2)}
\nonumber \\ && \,\,\, \cdot
\left(\Sigma_T(q_4, k_4) \cos(q_4/2) - O_T(q_4, k_4) \sin(q_4/2) \right)
\nonumber \\ && + \sin^2(k_3 \hat{R}/2) \cos(p_3 \hat{T}) \frac{\sin(p_4
\hat{T}/2)}{\tan(p_4/2)} \frac{\sin(k_4 \hat{T}/2)}{\sin(k_4/2)}
\nonumber \\ && \,\,\, \cdot
\left(\Sigma_T(k_4, q_4) \cos(q_4/2) - O_T(k_4, q_4) \sin(q_4/2) \right)
\Bigg] \\
&& <L^{(0)} \omega^{(4D)}_{TT}>_{conn} \nonumber \\ &=& \frac{C_2(G)
C_2(F)}{6} \int\limits_{BZ} \frac{d^4p}{(2\pi)^4} \int\limits_{BZ}
\frac{d^4k}{(2\pi)^4} \int\limits_{-\pi}^{\pi} \frac{dq_4}{2\pi}
\frac{1}{\hat{k}^2} \frac{\hat{\vec{p}}^2}{\hat{p}^2 \left(\hat{\vec{p}}^2 +
\hat{q}_4^2 \right)} \nonumber \\
&& \cdot \Bigg[2 \frac{\sin(p_4 \hat{T}/2)}{\tan(p_4/2)} \sin^2(p_3 \hat{R}/2)
\cos(q_4/2) \nonumber \\ && \,\,\, \cdot
\left[\left.\left(\Sigma_{1T} - \Sigma_{2T} \right)\right|_{p_4 = q_4}
+ \Sigma_T(0,q_4) - \Sigma_T(q_4+k_4,-k_4) \right]  \nonumber \\
&& + 2 \frac{\sin(k_4 \hat{T}/2)}{\sin(k_4/2)} \sin^2(k_3 \hat{R}/2)
\cos((p+q)_4/2) \nonumber \\ && \,\,\, \cdot
\left[\left(\Sigma_{3T} - \Sigma_{4T}\right) + \Sigma_T(p_4+q_4,-k_4)
-\Sigma_T(p_4-k_4,q_4) \right] \nonumber \\
&& + 2 \frac{\sin(p_4 \hat{T}/2)}{\tan(p_4/2)} \frac{\sin(q_4
\hat{T}/2)}{\tan(q_4/2)} \frac{\sin^2(k_4 \hat{T}/2)}{\sin^2(k_4/2)}
\nonumber \\ && \,\,\, \cdot
\left(\sin^2(p_3 \hat{R}/2) \cos^2(k_3 \hat{R}/2)
+ \sin^2(k_3 \hat{R}/2) \cos^2(p_3 \hat{R}/2) \right) \nonumber \\
&& + 2 \frac{\sin(p_4 \hat{T}/2)}{\tan(p_4/2)} \frac{\sin(k_4
\hat{T}/2)}{\sin(k_4/2)} O_T(k_4, q_4) \sin(q_4/2) \nonumber \\
&& \,\,\, \cdot \left( \sin^2(p_3 \hat{R}/2)  \cos(k_3 \hat{R}) - 
\sin^2(k_3 \hat{R}/2) \cos(p_3 \hat{R}) \right)
\nonumber \\
&& - \sin^2(p_3 \hat{R}/2) \cos(k_3 \hat{R}) \frac{\sin(p_4
\hat{T}/2)}{\tan(p_4/2)} \frac{\sin(k_4 \hat{T}/2)}{\sin(k_4/2)}
\nonumber \\ && \,\,\, \cdot
\left(\Sigma_T(k_4, q_4) \cos(q_4/2) - O_T(k_4, q_4) \sin(q_4/2) \right)
\nonumber \\
&& - \sin^2(p_3 \hat{R}/2) \frac{\sin(p_4 \hat{T}/2)}{\tan(p_4/2)} 
\frac{\sin(q_4 \hat{T}/2)}{\tan(q_4/2)}  \Sigma_T(k_4, -k_4) \nonumber \\
&& - \sin^2(k_3 \hat{R}/2) \frac{\sin(k_4 \hat{T}/2)}{\sin(k_4/2)}
\cos(p_3 \hat{R}) \frac{\sin(p_4 \hat{T}/2)}{\tan(p_4/2)}
\nonumber \\ && \,\,\, \cdot
\left(\Sigma_T(k_4, q_4) \cos(q_4/2) + O_T(k_4, q_4) \sin(q_4/2) \right)
\nonumber \\
&& - \sin^2(k_3 \hat{R}/2) \frac{\sin^2(k_4 \hat{T}/2)}{\sin^2(k_4/2)}
\nonumber \\ && \,\,\, \cdot
\left(\Sigma_T(p_4,q_4) \cos((p+q)_4/2) - O_T(p_4,q_4) \sin((p+q)_4/2) \right)
\Bigg] \\
&& <L^{(0)} \omega^{(4E)}_{TT}>_{conn} \nonumber \\ &=&
\frac{C_2(G) C_2(F)}{6} \int\limits_{BZ} \frac{d^4p}{(2\pi)^4} \int\limits_{BZ}
\frac{d^4k}{(2\pi)^4} \int\limits_{-\pi}^{\pi} \frac{dq_4}{2\pi}
\frac{1}{\hat{k}^2} \frac{\hat{\vec{p}}^2}{\hat{p}^2 \left(\hat{\vec{p}}^2 +
\hat{q}_4^2 \right)} \nonumber \\
&& \cdot \Bigg[\sin^2(p_3 \hat{R}/2) \frac{\sin(q_4 \hat{T}/2)}{\tan(q_4/2)}
\left(\Sigma_T(p_4,0) - \Sigma_T(k_4, p_4-k_4) \right) \nonumber \\
&& + \sin^2(k_3 \hat{R}/2) \frac{\sin(k_4 \hat{T}/2)}{\sin(k_4/2)}
\left(\Sigma_T(k_4,p_4+q_4) - \Sigma_T(p_4, q_4+k_4) \right) \nonumber \\
&& + \frac{\sin(p_4 \hat{T}/2)}{\tan(p_4/2)} \frac{\sin(q_4
\hat{T}/2)}{\tan(q_4/2)} \hat{T} \sin^2(p_3 \hat{R}/2) \nonumber \\
&& + \frac{\sin^2(k_4 \hat{T}/2)}{\sin^2(k_4/2)}
\frac{\sin((p+q)_4 \hat{T}/2)}{\tan((p+q)_4/2)} \sin^2(k_3 \hat{R}/2) 
\nonumber \\
&& + \frac{\sin(p_4 \hat{T}/2)}{\tan(p_4/2)} \frac{\sin(k_4
\hat{T}/2)}{\sin(k_4/2)}
\frac{\sin((k+q)_4 \hat{T}/2)}{\sin((k+q)_4/2)} \cos(q_4/2) \nonumber \\
&& \,\,\, \cdot \left(\sin^2(p_3 \hat{R}/2) \cos(k_3 \hat{R})
+ \sin^2(k_3 \hat{R}/2) \cos(p_3 \hat{R}) \right) \Bigg].
\end{eqnarray}
and
\begin{eqnarray}
&&<L^{(0)} \omega^{(4A)}_{RTTT}>_{conn} \nonumber \\ &=& - \frac{C_2(G)
C_2(F)}{3} \int\limits_{BZ} \frac{d^4p}{(2\pi)^4} \int\limits_{BZ}
\frac{d^4k}{(2\pi)^4} \int\limits_{-\pi}^{\pi} \frac{dq_4}{2\pi}
\frac{1}{\hat{k}^2} \frac{\hat{p}_4^2+\hat{q}_4^2}{\hat{p}^2
\left(\hat{\vec{p}}^2 + \hat{q}_4^2 \right)} \nonumber \\ 
&& \cdot \frac{\sin^2(k_4 \hat{T}/2)}{\sin^2(k_4/2)} \sin^2(k_4 \hat{R}/2)
\frac{\sin(p_4 \hat{T}/2)}{\tan(p_4/2)} \frac{\sin(q_4
\hat{T}/2)}{\tan(q_4/2)} \sin^2(p_3 \hat{R}/2) \\
&&<L^{(0)} \omega^{(4B)}_{RTTT}>_{conn} \nonumber \\ &=& - C_2(G) C_2(F)
\int\limits_{BZ} \frac{d^4p}{(2\pi)^4} \int\limits_{BZ}
\frac{d^4k}{(2\pi)^4} \int\limits_{-\pi}^{\pi} \frac{dq_4}{2\pi}
\frac{1}{\hat{k}^2} \frac{1}{\hat{p}^2
\left(\hat{\vec{p}}^2 + \hat{q}_4^2 \right)} \nonumber \\ 
&& \cdot \Bigg[\hat{p}_4 \sin(p_4 (\hat{T}-1)/2) 
\frac{\sin(k_4 \hat{T}/2)}{\sin(k_4/2)} \frac{\sin(q_4
\hat{T}/2)}{\tan(q_4/2)} + 2 \hat{q}_4 O_T(q_4,k_4) \nonumber \\
&& \,\,\,\,\, \cdot \left(\cos(p_4 \hat{T}/2) \cos(p_4/2) \cos(q_4/2) 
- \sin(p_4 (\hat{T}-1)/2) \sin(p_4/2) \right) \Bigg] \nonumber \\ \\
&&<L^{(0)} \omega^{(4C)}_{RTTT}>_{conn} \nonumber \\ &=& - \frac{C_2(G)
C_2(F)}{2} \int\limits_{BZ} \frac{d^4p}{(2\pi)^4} \int\limits_{BZ}
\frac{d^4k}{(2\pi)^4} \int\limits_{-\pi}^{\pi} \frac{dq_4}{2\pi}
\frac{1}{\hat{k}^2} \frac{1}{\hat{p}^2
\left(\hat{\vec{p}}^2 + \hat{q}_4^2 \right)} \nonumber \\ 
&& \cdot \Bigg[2 \hat{p}_4 \cos(p_4/2) \sin(p_4 \hat{T}/2) \cos(q_4/2)
\sin^2(p_3 \hat{R}/2)
\left.\left(\Sigma_{1T}-\Sigma_{2T} \right)\right|_{p_4=q_4} \nonumber \\
&& - 2 \hat{p}_4 \cos(p_4/2) \sin(p_4 \hat{T}/2) \sin^2(p_3 \hat{R}/2)
\cos(k_3 \hat{R}) \sin(q_4/2) O_T(q_4,k_4) \nonumber \\
&& + \hat{p}_4 \cos(p_4/2) \sin(p_4 \hat{T}/2) \sin^2(p_3 \hat{R}/2)
\frac{\sin(q_4 \hat{T}/2)}{\tan(q_4/2)} \Sigma_T(k_4, -k_4) \nonumber \\
&& + \hat{p}_4 \cos(p_4/2) \sin(p_4 \hat{T}/2) \sin^2(p_3 \hat{R}/2) \cos(k_3
\hat{R}) \frac{\sin(k_4 \hat{T}/2)}{\sin(k_4/2)} \nonumber \\
&& \,\,\, \cdot
\left(\Sigma_T(q_4,k_4) \cos(q_4/2) - O_T(q_4, k_4) \sin(q_4/2) \right)
\nonumber \\
&& + 2 \hat{p}_4 \sin(p_4 (\hat{T}+1)/2) \sin^2(p_3 \hat{R}/2)
\frac{\sin(q_4 \hat{T}/2)}{\tan(q_4/2)} \Sigma_T(k_4, -k_4) \nonumber \\
&&- 2 \sin^2(k_3 \hat{R}/2) \frac{\sin(k_4 \hat{T}/2)}{\sin(k_4/2)}
\sin^2(p_3 \hat{R}/2) \hat{p}_4 \cos(p_4/2) \sin(p_4 \hat{T}/2)
\Sigma_T(k_,q_4) \nonumber \\
&& + \sin^2(p_3 \hat{R}/2) \frac{\sin(q_4 \hat{T}/2)}{\tan(q_4/2)} 
\hat{p}_4 \sin(p_4 (\hat{T}-1)/2) \Sigma_T(k_4,-k_4) \nonumber \\
&& + \sin^2(k_3 \hat{R}/2) \frac{\sin(k_4 \hat{T}/2)}{\sin(k_4/2)} 
\nonumber \\ && \,\,\, \cdot
\left(\hat{q}_4 \left(\sin(p_4 (\hat{T}-1)/2) \sin(p_4/2)-\cos(p_4 \hat{T}/2)
\right)O_T(k_4,q_4) \right. \nonumber \\ 
&& \,\,\,\,\, \left. - \hat{p}_4 \sin(p_4 (\hat{T}-1)/2)
\cos(q_4/2) \Sigma_T(k_4,q_4) \right) \nonumber \\
&& + \hat{p}_4 \left(\sin(p_4 (\hat{T}-1)/2)-\sin(p_4 (\hat{T}+1)/2) \right)
\sin^2(p_3 \hat{R}/2) \nonumber \\ && \,\,\, \cdot \frac{\sin^2(k_4
\hat{T}/2)}{\sin^2(k_4/2)} \frac{\sin(q_4 \hat{T}/2)}{\sin(q_4/2)}
\left(\cos(k_3 \hat{R} + \sin^2(k_3 \hat{R}/2) \right) \nonumber \\
&& + 2 \sin^2(k_3 \hat{R}/2) \frac{\sin(k_4 \hat{T}/2)}{\sin(k_4/2)}
\sin^2(p_3 \hat{R}/2) \nonumber \\ && \,\,\,\ \cdot \hat{q}_4 \left(\sin(p_4
(\hat{T}-1) \sin(p_4/2) - \cos(p_4 \hat{T}/2) \right) O_T(k_4,q_4) \Bigg] \\
&&<L^{(0)} \omega^{(4D)}_{RTTT}>_{conn} \nonumber \\ &=& -
\frac{C_2(G) C_2(F)}{3} \int\limits_{BZ} \frac{d^4p}{(2\pi)^4} \int\limits_{BZ}
\frac{d^4k}{(2\pi)^4} \int\limits_{-\pi}^{\pi} \frac{dq_4}{2\pi}
\frac{1}{\hat{k}^2} \frac{1}{\hat{p}^2
\left(\hat{\vec{p}}^2 + \hat{q}_4^2 \right)} \nonumber \\ 
&& \cdot \Bigg[2 \hat{p}_4 \cos(p_4/2) \sin(p_4 \hat{T}/2) \cos(q_4/2)
\sin^2(p_3 \hat{R}/2) \nonumber \\ && \,\,\, \cdot
\left[\left.\left(\Sigma_{1T}-\Sigma_{2T} \right)\right|_{p_4=q_4}
+ \Sigma_T(0,q_4) - \Sigma_T(k_4+q_4,-k_4) \right] \nonumber \\
&& + 2 \hat{p}_4 \cos(p_4/2) \sin(p_4 \hat{T}/2)
\sin^2(p_3 \hat{R}/2) \frac{\sin^2(k_4 \hat{T}/2)}{\sin^2(k_4/2)}
\frac{\sin(q_4 \hat{T}/2)}{\tan(q_4/2)} 
\nonumber \\ && \,\,\, \cdot \cos^2(k_3 \hat{R}/2) \nonumber \\
&& + 2 \hat{p}_4 \cos(p_4/2) \sin(p_4 \hat{T}/2)
\sin^2(p_3 \hat{R}/2) \frac{\sin(k_4 \hat{T}/2)}{\sin(k_4/2)}
\cos(k_3 \hat{R}) \nonumber \\ && \,\,\, \cdot \sin(q_4/2) O_T(k_4,q_4)
\nonumber \\ && - \hat{p}_4 \cos(p_4/2) \sin(p_4 \hat{T}/2) \sin^2(p_3
\hat{R}/2) \frac{\sin(q_4 \hat{T}/2)}{\tan(q_4/2)} \Sigma_T(k_4, -k_4)
\nonumber \\ && - \hat{p}_4 \cos(p_4/2) \sin(p_4 \hat{T}/2) \sin^2(p_3
\hat{R}/2) \cos(k_3 \hat{R}) \frac{\sin(k_4 \hat{T}/2)}{\sin(k_4/2)}
\nonumber \\ && \,\,\, \cdot
\left(\Sigma_T(k_4,q_4) \cos(q_4/2) - O_T(k_4,q_4) \sin(q_4/2) \right)
\nonumber \\
&& - 2 \sin^2(p_3 \hat{R}/2) \sin^2(k_3 \hat{R}/2) \frac{\sin(k_4
\hat{T}/2)}{\sin(k_4/2)} \hat{q}_4 \nonumber \\ && \,\,\, \cdot
\left(\cos(k_4 \hat{T}/2) - \sin(p_4
(\hat{T}-1)/2) \sin(p_4/2) \right) O_T(k_4,q_4) \nonumber \\
&& + \hat{p}_4 \sin(p_4 (\hat{T}-1)/2) \sin^2(p_3 \hat{R}/2) \cos^2(k_3
\hat{R}/2) \frac{\sin^2(k_4 \hat{T}/2)}{\sin^2(k_4/2)}
\frac{\sin(q_4 \hat{T}/2)}{\tan(q_4/2)} \nonumber \\
&& + 2 \sin^2(p_3 \hat{R}/2) \sin^2(k_3 \hat{R}/2) \frac{\sin(k_4
\hat{T}/2)}{\sin(k_4/2)} \hat{p}_4 \cos(p_4/2) \sin(p_4 \hat{T}/2)
\nonumber \\ && \,\,\, \cdot \cos(q_4/2) \Sigma_T(q_4,k_4) \nonumber \\
&& - 2 \sin^2(p_3 \hat{R}/2) \hat{p}_4 \cos(p_4/2) \sin(p_4 \hat{T}/2) 
\frac{\sin(q_4 \hat{T}/2)}{\tan(q_4/2)} \Sigma_T(k_4, -k_4) \nonumber \\
&& + \sin^2(p_3 \hat{R}/2) \sin^2(q_3 \hat{R}/2) \frac{\sin(k_4
\hat{T}/2)}{\sin(k_4/2)} \nonumber \\ && \,\,\, \cdot
\hat{q}_4 \left(\cos(p_4 \hat{T}/2) - \sin(p_4
(\hat{T}-1)/2) \sin(p_4/2) \right) O_T(q_4,k_4) \nonumber \\
&& + \sin^2(p_3 \hat{R}/2) \sin^2(q_3 \hat{R}/2) \frac{\sin(k_4
\hat{T}/2)}{\sin(k_4/2)} \hat{p}_4 \sin(p_4 (\hat{T}-1)/2) \Sigma_T(q_4,k_4) 
\nonumber \\
&& - \sin^2(p_3 \hat{R}/2) \hat{p}_4 \cos(p_4/2) \sin(p_4 \hat{T}/2)
\frac{\sin(q_4 \hat{T}/2)}{\tan(q_4/2)} \Sigma_T(k_4,-k_4) \nonumber \\
&& + \hat{p}_4 \left(\sin(p_4 (\hat{T}+1)/2) + \sin(p_4 (\hat{T}-1)/2) \right)
\frac{\sin^2(k_4 \hat{T}/2)}{\sin^2(k_4/2)} \frac{\sin(q_4
\hat{T}/2)}{\tan(q_4/2)} \nonumber \\ && \,\,\, \cdot \sin^2(p_3 \hat{R}/2)
\left(\sin^2(k_3 \hat{R}/2) + \cos(k_3 \hat{R}) \right) \Bigg] \\
&&<L^{(0)} \omega^{(4E)}_{RTTT}>_{conn} \nonumber \\ &=& -
\frac{C_2(G) C_2(F)}{3} \int\limits_{BZ} \frac{d^4p}{(2\pi)^4} \int\limits_{BZ}
\frac{d^4k}{(2\pi)^4} \int\limits_{-\pi}^{\pi} \frac{dq_4}{2\pi}
\frac{1}{\hat{k}^2} \frac{1}{\hat{p}^2
\left(\hat{\vec{p}}^2 + \hat{q}_4^2 \right)} \nonumber \\ 
&& \cdot \Bigg[2 \sin(p_4 (\hat{T}+1)/2) \sin^2(p_3 \hat{R}/2)
\frac{\sin(q_4 \hat{T}/2)}{\tan(q_4/2)} \hat{T} \nonumber \\
&&- 2 \sin(p_4 (\hat{T}+1)/2) \sin^2(p_3 \hat{R}/2)
\sin^2(k_3 \hat{R}/2) \nonumber \\ && \,\,\, \cdot \frac{\sin(k_4
\hat{T}/2)}{\sin(k_4/2)} \frac{\sin((q+k)_4 \hat{T}/2)}{\sin((q+k)_4/2)}
\cos(q_4/2) \nonumber \\
&& + \sin(p_4 (\hat{T}-1)/2) \sin^2(p_3 \hat{R}/2) 
\frac{\sin(q_4 \hat{T}/2)}{\tan(q_4/2)} \hat{T} \nonumber \\
&& - \sin(p_4 (\hat{T}-1)/2) \sin^2(p_3 \hat{R}/2)
\sin^2(k_3 \hat{R}/2) \nonumber \\ && \,\,\, \cdot \frac{\sin(k_4
\hat{T}/2)}{\sin(k_4/2)}  \frac{\sin((q+k)_4 \hat{T}/2)}{\sin((q+k)_4/2)}
\cos(q_4/2) \nonumber \\ &&
+ 2 \sin^2(p_3 \hat{R}/2) \sin(p_4 \hat{T}/2)
\cos(p_4/2) \cos(q_4/2) \nonumber \\ && \,\,\, \cdot
\left(\Sigma_T(q_4,0) - \Sigma_T(k_4, q_4-k_4) \right)
\nonumber \\ && + \sin^2(p_3 \hat{R}/2) \sin(p_4 \hat{T}/2) \cos(p_4/2) 
\frac{\sin(q_4 \hat{T}/2)}{\tan(q_4/2)} \hat{T} \nonumber \\
&& + \sin^2(p_3 \hat{R}/2) \sin(p_4 \hat{T}/2) \cos(p_4/2) 
\cos(p_3 \hat{R}) \nonumber \\ && \,\,\, \cdot \frac{\sin(k_4
\hat{T}/2)}{\sin(k_4/2)}  \frac{\sin((q+k)_4 \hat{T}/2)}{\sin((q+k)_4/2)}
\cos(q_4/2) \Bigg].
\end{eqnarray}

The functions $\Sigma_{1T}$ and $\Sigma_{2T}$ are obtained from $\Sigma_1$ and
$\Sigma_2$ by replacing $\hat{R}$ with $\hat{T}$, $p_{\mu}$ with $p_4$ and
$k_{\mu}$ by $k_4$ (see section \ref{potential_NLO}). The new functions
$\Sigma_{3T}$ and $\Sigma_{4T}$ are the even parts of the following sums:
\begin{eqnarray*}
\tilde{\Sigma}_{3T} &=& e^{i(k_4-p_4-z)(\hat{T}-1)/2} \sum_{l_1 =
0}^{\hat{T}-3} \sum_{l_2 = l_1+1}^{\hat{T}-2} \sum_{l_3 = l_2+1}^{\hat{T}-1}
e^{-i k_4 l_1} e^{i p_4 l_2} e^{i q_4 l_3} \\
\tilde{\Sigma}_{4T} &=& e^{i(k_4-p_4-z)(\hat{T}-1)/2} \sum_{l_1 =
0}^{\hat{T}-3} \sum_{l_2 = l_1+1}^{\hat{T}-2} \sum_{l_3 = l_2+1}^{\hat{T}-1}
e^{i p_4 l_1} e^{-i k_4 l_2} e^{i q_4 l_3}
\end{eqnarray*}
The explicit results are:
\begin{eqnarray}
\Sigma_{3T} &=& \frac{1}{4} \frac{\sin(q_4 \hat{T}/2)}{\sin(q_4/2)}
\frac{\sin(p_4 (\hat{T}-2)/2)}{\sin(p_4/2)} \frac{\sin(k_4 (\hat{T}-2)/2)
\cos(k_4)}{\sin(k_4/2)} \nonumber \\ &&+ \frac{1}{4} \frac{\sin(q_4
\hat{T}/2)}{\sin(q_4/2)} \frac{\sin(p_4/2 - k_4)}{\sin(p_4/2)}
\frac{\sin((p-k)_4 (\hat{T}-2)/2)}{\sin((p-k)_4/2)} \nonumber \\
&&+ \frac{1}{4} \frac{\cos((p+q)_4 (\hat{T}-2)/2 + q_4/2)}{\sin(q_4/2)
\sin((p+q)_4/2)} \frac{\sin(k_4 (\hat{T}-2)/2) \cos(k_4)}{\sin(k_4/2)}
\nonumber \\ &&- \frac{1}{4} \frac{\cos(p_4/2 - k_4)}{\sin(q_4/2)
\sin((p+q)_4/2)} \frac{\sin(p+q-k)_4 (\hat{T}-2)/2)}{\sin((p+q-k)_4/2))} \\
\Sigma_{4T} &=& \frac{1}{4} \frac{\sin(q_4 \hat{T}/2)}{\sin(q_4/2)}
\frac{\sin(k_4 (\hat{T}-2)/2)}{\sin(k_4/2)} \frac{\sin(p_4 (\hat{T}-2)/2)
\cos(p_4)}{\sin(p_4/2)} \nonumber \\ && - \frac{1}{4} \frac{\sin(q_4
\hat{T}/2)}{\sin(q_4/2)} \frac{\sin(p_4 - k_4/2)}{\sin(k_4/2)}
\frac{\sin((p-k)_4 (\hat{T}-2)/2)}{\sin((p-k)_4/2)} \nonumber \\
&&+ \frac{1}{4} \frac{\cos((q-k)_4 (\hat{T}-2)/2 + q_4/2)}{\sin(q_4/2)
\sin((q-k)_4/2)} \frac{\sin(p_4 (\hat{T}-2)/2) \cos(p_4)}{\sin(p_4/2)}
\nonumber \\ && - \frac{1}{4} \frac{\cos(p_4 - k_4/2)}{\sin(q_4/2)
\sin((q-k)_4/2)} \frac{\sin(p+q-k)_4 (\hat{T}-2)/2)}{\sin((p+q-k)_4/2))}
\end{eqnarray}

\vspace{3ex}

\noindent
These results with $L^{(0)}$ above have to be compared to the expressions one
gets by inserting $S^{(0)}$ into the graphs with $\omega^{(4)}$ in the limit of
large $\hat{T}$. The graphs contributing to $<S^{(0)} \omega^{(4)}>_0$ can be
split in the same way as outlined above, and with similar arguments as before,
it can be shown that the contributions coming from $\omega^{(4)}_{RR}$,
$\omega^{(4)}_{RT}$, and $\omega^{(4)}_{RRRT}$ vanish in the limit of large
$\hat{T}$. Hence only the results for the graphs with $\omega^{(4)}_{TT}$ and
$\omega^{(4)}_{RTTT}$, which can be obtained from the sections
\ref{potential_NLO} and \ref{ASR_check_NLO}, are given here:
\begin{eqnarray}
&&<S^{(0)} \omega^{(4A)}_{TT}>_{conn} \nonumber \\ &=& - 2\frac{C_2(G)
C_2(F)}{3} \int\limits_{BZ} \frac{d^4p}{(2\pi)^4} \int\limits_{BZ}
\frac{d^4k}{(2\pi)^4} \frac{1}{\hat{p}^2}
\frac{\hat{k}^2+\hat{\vec{k}}^2}{\left(\hat{k}^2\right)^2} \frac{\sin^2(p_4
T/2)}{\sin^2(p_4/2)} \frac{\sin^2(k_4 T/2)}{\sin^2(k_4/2)} \nonumber \\ &&
\sin^2(p_3 \hat{R}/2) \sin^2(k_3 \hat{R}/2) \\
&& <S^{(0)} \omega^{(4B)}_{TT}>_{conn} \nonumber \\ &=& -\frac{C_2(G)
C_2(F)}{2} \int\limits_{BZ} \frac{d^4p}{(2\pi)^4} \int\limits_{BZ}
\frac{d^4k}{(2\pi)^4} \frac{1}{\hat{p}^2}
\frac{\hat{k}^2+\hat{\vec{k}}^2}{\left(\hat{k}^2\right)^2}
\Bigg[\frac{\sin^2(p_4 T/2)}{\sin^2(p_4/2)} \frac{\sin^2(k_4
T/2)}{\sin^2(k_4/2)} \nonumber \\ && \,\,\,
+ 4 O_T^2(p_4, k_4) \Bigg] \cdot
\left[\sin^2(p_3 \hat{R}/2) \cos^2(k_3 \hat{R}/2) + \sin^2(k_3 \hat{R}/2)
\cos^2(p_3 \hat{R}/2) \right] \nonumber \\ \\
&& <S^{(0)} \omega^{(4C)}_{TT}>_{conn} \nonumber \\ &=&
\frac{C_2(G) C_2(F)}{4} \int\limits_{BZ} \frac{d^4p}{(2\pi)^4}
\int\limits_{BZ} \frac{d^4k}{(2\pi)^4} \frac{1}{\hat{p}^2} 
\frac{\hat{k}^2+\hat{\vec{k}}^2}{\left(\hat{k}^2\right)^2} \left[4
\frac{\sin(p_4 T/2)}{\sin(p_4/2)} (\Sigma_{1T} - \Sigma_{2T}) \right.
\nonumber \\ && + \frac{\sin^2(p_4 T/2)}{\sin^2(p_4/2)}
\left(\frac{\sin^2(k_4 T/2)}{\sin^2(k_4/2)}  - T\right)  +
\frac{\sin(p_4 T/2)}{\sin(p_4/2)} \frac{\sin(k_4
T/2)}{\sin(k_4/2)} \cos(k_3 R) \nonumber \\ && \left. \quad
\left(\frac{\sin(p_4 T/2)}{\sin(p_4/2)} \frac{\sin(k_4
T/2)}{\sin(k_4/2)} -\frac{\sin((p+k)_3 T/2)}{\sin((p+k)_3/2)}
\right) \right] \cdot \sin^2(p_3 \hat{R}/2)
\\
&& <S^{(0)} \omega^{(4D)}_{TT}>_{conn} \nonumber \\ &=& \frac{C_2(G)
C_2(F)}{12} \int\limits_{BZ} \frac{d^4p}{(2\pi)^4} \int\limits_{BZ}
\frac{d^4k}{(2\pi)^4} \frac{1}{\hat{p}^2}
\frac{\hat{k}^2+\hat{\vec{k}}^2}{\left(\hat{k}^2\right)^2} \left[3
\frac{\sin^2(p_4 T/2)}{\sin^2(p_4/2)} T \right.
\nonumber \\ && + 2 \frac{\sin^2(p_4 T/2)}{\sin^2(p_4/2)}
\frac{\sin^2(k_4 T/2)}{\sin^2(k_4/2)} 
\cos^2(k_3 \hat{R}/2) + 4 \frac{\sin(p_4 T/2)}{\sin(p_4/2)}
(\Sigma_{1T} - \Sigma_{2T}) \nonumber \\
&& - \left(2  - \cos(k_3 R)\right)
\left. \frac{\sin(p_{\nu} T/2)}{\sin(p_{\nu}/2)}
\frac{\sin(k_{\nu} T/2)}{\sin(k_{\nu}/2)}
\frac{\sin((p_{\nu}+k_{\nu}) T/2)}{\sin((p_{\nu}+k_{\nu})/2)}
\right] \cdot \sin^2(p_3 \hat{R}/2) \nonumber \\ \\
&& <S^{(0)} \omega^{(4E)}_{TT}>_{conn} \nonumber \\ &=& \frac{C_2(G) C_2(F)}{3}
\int\limits_{BZ} \frac{d^4p}{(2\pi)^4} \int\limits_{BZ} \frac{d^4k}{(2\pi)^4}
\frac{1}{\hat{p}^2} \frac{\hat{k}^2+\hat{\vec{k}}^2}{\left(\hat{k}^2\right)^2}
\left[\frac{\sin^2(p_4 T/2)}{\sin^2(p_4/2)} T \sin^2(p_3 \hat{R}/2)
\right. \nonumber \\ && - \sin^2(p_3 \hat{R}/2) \sin^2(k_3 \hat{R}/2)
\left. \frac{\sin(p_4 T/2)}{\sin(p_4/2)}
\frac{\sin(k_4 T/2)}{\sin(k_4/2)} \frac{\sin((p+k)_4
T/2)}{\sin((p+k)_4/2)} \right]. \nonumber \\
\end{eqnarray}
and
\begin{eqnarray}
&&<S^{(0)} \omega^{(4A)}_{RTTT}>_{conn} \nonumber \\ &=& 0 \\
&&<S^{(0)} \omega^{(4B)}_{RTTT}>_{conn} \nonumber \\ &=& -C_2(G) C_2(F)
\int\limits_{BZ} \frac{d^4p}{(2\pi)^4} \int\limits_{BZ} \frac{d^4k}{(2\pi)^4}
\frac{1}{\hat{p}^2} \frac{\hat{k}_3 \hat{k}_4}{\left(\hat{k}^2\right)^2}
\frac{\sin^2(k_3 \hat{R}/2)}{\sin(k_3/2)}
\frac{\sin(p_4 \hat{T}/2)}{\sin(p_4/2)}
\nonumber \\
&& \cdot \cos^2(p_3 \hat{R}/2) \Bigg[\frac{\sin^2(k_4 \hat{T}/2)}{\sin(k_4/2)}
\frac{\sin(p_4 \hat{T}/2)}{\sin(p_4/2)} + 2 \cos(k_4 \hat{T}/2) O_T(k_4, p_4)
\Bigg] \nonumber \\ \\
&& <S^{(0)} \omega^{(4C)}_{RTTT}>_{conn} \nonumber \\ &=& -\frac{C_2(G)
C_2(F)}{2} \int\limits_{BZ} \frac{d^4p}{(2\pi)^4} \int\limits_{BZ}
\frac{d^4k}{(2\pi)^4} \frac{1}{\hat{p}^2} 
\frac{\hat{k}_3 \hat{k}_4}{\left(\hat{k}^2\right)^2} \nonumber \\
&& \cdot \Bigg[2 \sin^2(p_3 \hat{R}) \frac{\sin(p_4 \hat{T}/2)}{\sin(p_4/2)} 
\frac{\sin^2(k_3 \hat{R}/2)}{\sin(k_3/2)} \sin(k_4 \hat{R}/2) 
\Sigma_T(k_4, p_4) \nonumber \\
&& + 2 \sin^2(p_3 \hat{R}/2) \frac{\sin(p_4 \hat{T}/2)}{\sin(p_4/2)}
\frac{\sin^2(k_3 \hat{R}/2)}{\sin(k_3/2)} \cos(k_4 \hat{T}/2) O_T(p_4, k_4)
\nonumber \\
&& + \sin^2(p_3 \hat{R}/2) \frac{\sin(p_4 \hat{T}/2)}{\sin(p_4/2)}
\frac{\sin^2(k_3 \hat{R}/2)}{\sin(k_3/2)} \nonumber \\ && \,\,\, \cdot
\left(\sin(k_4 \hat{T}/2)\Sigma_T(p_4,k_4) + \cos(k_4 \hat{T}/2) O_T(p_4,
k_4) \right) \Bigg] \\
&& <S^{(0)} \omega^{(4D)}_{RTTT}>_{conn} \nonumber \\ &=&
\frac{C_2(G) C_2(F)}{12} \int\limits_{BZ} \frac{d^4p}{(2\pi)^4}
\int\limits_{BZ} \frac{d^4k}{(2\pi)^4} \frac{1}{\hat{p}^2}
\frac{\hat{k}_3 \hat{k}_4}{\left(\hat{k}^2\right)^2} \nonumber \\
&& \cdot \Bigg[\sin^2(p_4 \hat{T}/2) \frac{\sin(p_3
\hat{R}/2)}{\sin^2(p_3/2)} \frac{\sin^2(k_3 \hat{R}/2)}{\sin(k_3/2)}
\sin(k_4 \hat{T}/2) \Sigma_T(p_4, k_4) \nonumber \\
&& + 2 \sin^2(p_3 \hat{R}/2) \frac{\sin^2(p_4 \hat{T}/2)}{\sin^2(p_4/2)}
\frac{\sin^2(k_3 \hat{R}/2)}{\sin(k_3/2)} \frac{\sin^2(k_4
\hat{T}/2)}{\sin(k_4/2)} \nonumber \\
&& - 4 \sin^2(p_3 \hat{R}/2) \frac{\sin^2(p_4 \hat{T}/2)}{\sin^2(p_4/2)}
\frac{\sin^2(k_3 \hat{R}/2)}{\sin(k_3/2)} \cos(k_4 \hat{T}/2) O_T(p_4, k_4)
\nonumber \\
&& + 2 \sin^2(p_3 \hat{R}/2) \frac{\sin^2(p_4 \hat{T}/2)}{\sin^2(p_4/2)}
\frac{\sin^2(k_3 \hat{R}/2)}{\sin(k_3/2)} \nonumber \\ && \,\,\, \cdot
\left(\sin(k_4 \hat{T}/2) \Sigma_T(p_4, k_4) - \cos(k_4 \hat{T}/2) O_T(p_4,
k_4) \right) \Bigg] \\
&& <S^{(0)} \omega^{(4E)}_{RTTT}>_{conn} \nonumber \\ &=& -\frac{C_2(G)
C_2(F)}{2} \int\limits_{BZ} \frac{d^4p}{(2\pi)^4} \int\limits_{BZ}
\frac{d^4k}{(2\pi)^4} \frac{1}{\hat{p}^2}
\frac{\hat{k}_3 \hat{k}_4}{\left(\hat{k}^2\right)^2}
\frac{\sin(p_4 \hat{T}/2)}{\sin(p_4/2)} \sin^2(p_3 \hat{R}/2) \nonumber \\
&& \cdot \frac{\sin^2(k_3 \hat{R}/2)}{\sin(k_3/2)} \sin(k_4 \hat{T}/2) 
\frac{\sin((p+k)_4 \hat{T}/2)}{\sin((p+k)_4/2)}.
\label{end}
\end{eqnarray}

\vspace{3ex}

\noindent
Now using the expressions (\ref{start}) to (\ref{end}), it has to be shown that
\begin{equation}
\lim_{\hat{T} \to \infty} <L^{(0)} \omega^{(4)}>_{conn} 
= \lim_{\hat{T} \to \infty} \frac{1}{\hat{T}} <S^{(0)} \omega^{(4)}>_{conn}.
\end{equation}
For the parts with $\omega^{(4A)}$, this is easy to show; using
\[
\lim_{\hat{T} \to \infty} <L^{(0)} \omega^{(2)}>_{conn} 
= \lim_{\hat{T} \to \infty} \frac{1}{\hat{T}} <S^{(0)} \omega^{(2)}>_{conn}.
\]
one indeed gets:
\begin{equation}
\lim_{\hat{T} \to \infty} <L^{(0)} \omega^{(4A)}_{TT}>_{conn} 
= \lim_{\hat{T} \to \infty} \frac{1}{\hat{T}} <S^{(0)}
\omega^{(4A)}_{TT}>_{conn}
\end{equation}
and
\begin{equation}
\lim_{\hat{T} \to \infty} <L^{(0)} \omega^{(4A)}_{RTTT}>_{conn} 
= \lim_{\hat{T} \to \infty} \frac{1}{\hat{T}} <S^{(0)}
\omega^{(4A)}_{RTTT}>_{conn} = 0.
\end{equation}
All the other contributions have to be checked explicitly
by evaluating the eight- respectively nine-dimensional integrals using a
numerical integration routine like \emph{Vegas} \cite{NumericalRecipes}. This
work, which requires lots of computer time, is still in progress.

\newpage
\thispagestyle{empty}
\cleardoublepage
\chapter{Energy Sum Rule}

\label{chapter_ESR}

\renewcommand{\pw}{
  \fmfstraight
  \fmfleft{w1,w2,w3}
  \fmftop{w4,w5,w6}
  \fmfright{w9,w8,w7}
  \fmfbottom{w12,w11,w10}
  \fmf{plain}{w1,w3}
  \fmf{plain}{w3,w7}
  \fmf{plain}{w7,w9}
  \fmf{plain}{w9,w1}
  \begin{fmfsubgraph}(0.3w, 0.4h)(1cm, 0.4cm)
  \fmftop{p1,p2,p3,p4,p5}
  \fmfbottom{p10,p9,p8,p7,p6}
  \fmf{plain}{p3,p5}
  \fmf{plain}{p3,pv1}
  \fmf{plain}{pv1,p10}
  \fmf{plain}{p5,pv2}
  \fmf{plain}{pv2,p8}
  \fmf{plain}{p8,p10}
  \end{fmfsubgraph}
}

In deriving the energy sum rule in the form
\begin{equation}
\hat{V}(\hat{R},\hat{\beta}) = \lim_{\hat{T} \to \infty} 
\left[\eta_- <-\mathcal{P}'_t+\mathcal{P}'_s>_{q\bar{q}-0} 
+ \frac{1}{4} \sum_{\vec{x}} <T_{\mu\mu}(\vec{x},t)>_{q\bar{q}-0} 
\right],
\end{equation}
essentially three steps were necessary:
\begin{enumerate}
\item Taking the derivative (\ref{xiderivative}) of the potential, calculated
on an anisotropic lattice, with respect to the anisotropy parameter $\xi$, and
then returning to the isotropic lattice $\xi=1$; this led to
(\ref{anisotropicderivative}).
\item Introducing the abbreviations $\eta_{\pm}$ and using the formula derived
in \cite{Karsch2} for $\eta_+$; the result was (\ref{EnergySumRule}).
\item Taking the limit of large $\hat{T}$ and thereby restricting the sum
over all plaquettes to one fixed time slice.
\end{enumerate}

There is no problem with the second step; the formula for $\eta_+$ used there
was proven in \cite{Karsch2} analytically, as well as checked numerically. But
the first and third step are questionable; it´s not entirely clear if in the
continuum limit, the potential really becomes completely independent of $\xi$
(hence it is not clear if the derivative (\ref{xiderivative}) really is zero),
and the behaviour for large $\hat{T}$ is not clear, too.

In the first section of this chapter, some basic calculations will be done,
and the energy sum rule will be cast into an equivalent form which is easier
to check. The second section deals then with the explicit perturbative check.
This will lead to a better understanding of the various terms which contribute
to the potential energy (the energy in the electric and in the magnetic fields
and the trace anomaly). The various contributions will be classified and their
magnitudes will be compared in the third section. Finally, the limit of large
$\hat{T}$ will be examined.

\section{Preliminaries}

\label{ESR_prelim}

It will turn out that it will be helpful to write the difference between the
spatial and temporal plaquettes in the first term of (\ref{EnergySumRule}) in
the following way:
\begin{equation}
-\mathcal{P}_t + \mathcal{P}_s
= -\mathcal{P}_t - \mathcal{P}_s + 2 \mathcal{P}_s.
\end{equation}
Then the potential is given by:
\begin{eqnarray}
\hat{V} &=& \lim_{\hat{T} \to \infty} \frac{1}{\hat{T}} \left[
\eta_- <-\mathcal{P}_t - \mathcal{P}_s + 2 \mathcal{P}_s>_{q\bar{q}-0}
+ \frac{\beta_L(g_0)}{2g_0} <S>_{q\bar{q}-0} \right] \nonumber \\
&=& \lim_{\hat{T} \to \infty} \frac{1}{\hat{T}} \left[
\frac{\eta_-}{\hat{\beta}} <-S + 2 S_s>_{q\bar{q}-0}
+ \frac{\beta_L(g_0)}{2g_0} <S>_{q\bar{q}-0}
\right] \nonumber \\
&=& \! \lim_{\hat{T} \to \infty} \frac{1}{\hat{T}} \left[
- \frac{g_0^2}{2d(F)} \eta_- <S>_{q\bar{q}-0}
+ \frac{g_0^2}{d(F)} \eta_- <S_s>_{q\bar{q}-0}
+ \frac{\beta_L(g_0)}{2g_0} <S>_{q\bar{q}-0}
\right], \nonumber \\
\end{eqnarray}
where $S_s$ denotes the part of the action which comes from summing only over
the \emph{spatial} plaquettes.

Because the expectation value of the action appears in two places in the
formula above, the identity (\ref{actionidentity})
\begin{equation}
\lim_{T \to \infty} \frac{1}{T} <S>_{q\bar{q}-0, subtr} = - g_0^2
\frac{\partial}{\partial g_0^2} \hat{V}
\end{equation}
becomes very useful now. When using it, one has to pay attention that the
self-energy contributions have to be subtracted. Inserting this identity
above, one gets:
\begin{equation}
\hat{V} = \frac{g_0^4}{2d(F)} \eta_- \frac{\partial}{\partial g_0^2} \hat{V}
- \frac{\beta_L(g_0)}{2} g_0 \frac{\partial}{\partial g_0^2} \hat{V}
+ \lim_{\hat{T} \to \infty} \frac{1}{\hat{T}}
\frac{g_0^2}{d(F)} \eta_- <S_s>_{q\bar{q}-0}
\end{equation}
Until this point, everything had been exact. Now the following expansions will
be used, where the explicit expressions for $\hat{\beta}_t$ and
$\hat{\beta}_t$ were taken from \cite{Karsch2}:
\begin{eqnarray}
\eta_{\pm} &=& \frac{1}{2} \left( \frac{\partial \hat{\beta}_t}{\partial \xi}
\pm \frac{\partial \hat{\beta}_s}{\partial \xi} \right)_{\xi = 1} \nonumber \\
&=& \frac{1}{2} \left(\frac{\partial}{\partial \xi} \left (\xi \hat{\beta} + 2
d(F) \xi c_t(\xi) \pm \frac{1}{\xi} \hat{\beta} \pm \frac{2d(F)}{\xi} c_s(\xi)
\right) \right)_{\xi = 1} + O(g_0^2) \nonumber \\
&=& \frac{1}{2} \left( \hat{\beta} + 2d(F) c_t(\xi) + 2d(F) \xi \frac{\partial
c_t}{\partial \xi} \mp \frac{1}{\xi^2} \hat{\beta} \mp \frac{2d(F)}{\xi^2}
c_s(\xi) \pm \frac{2d(F)}{\xi} \frac{\partial c_s}{\partial \xi} \right)_{\xi =
1} + O(g_0^2) \nonumber \\
&=& \frac{1}{2} \left( \hat{\beta} + 2d(F) \frac{\partial
c_t}{\partial \xi} \mp \hat{\beta} \pm 2d(F) \frac{\partial c_s}{\partial \xi}
\right)_{\xi = 1} + O(g_0^2) \nonumber \\
&=& \left\{ \begin{array}{r}
d(F) \left(\frac{\partial c_t}{\partial \xi} + \frac{\partial c_s}{\partial
\xi} \right)_{\xi = 1} \\
\hat{\beta} + d(F) \left(\frac{\partial c_t}{\partial \xi} - \frac{\partial
c_s}{\partial \xi} \right)_{\xi = 1}
\end{array} \right\} + O(g_0^2)
= \left\{ \begin{array}{r}
\beta_0 d(F) \\
\frac{2d(F)}{g_0^2} + c d(F)
\end{array} \right\} + O(g_0^2)
\end{eqnarray}
Additionally one needs the leading order of the expansion of the lattice beta
function:
\begin{equation}
\beta_L(g_0) \equiv -a \frac{\partial g_0}{\partial a} = - \beta_0 g_0^3 +
O(g_0^5)
\end{equation}
with
\[
\beta_0 = \frac{11 C_2(G)}{48\pi^2}.
\]
Then one arrives at:
\begin{eqnarray}
\hat{V} &=& g_0^2 \frac{\partial}{\partial g_0^2} \hat{V}
+ \frac{1}{2} c g_0^4 \frac{\partial}{\partial g_0^2} \hat{V}
+ \frac{1}{2} \beta_0 g_0^4 \frac{\partial}{\partial g_0^2} \hat{V}
\nonumber \\ && + \lim_{\hat{T} \to \infty} \frac{1}{\hat{T}} \left[
2 <S_s>_{q\bar{q}-0} + g_0^2 c <S_s>_{q\bar{q}-0} \right] + O(g_0^6)
\nonumber \\
&=& g_0^2 \frac{\partial}{\partial g_0^2} \hat{V}
+ \left. \frac{\partial c_t}{\partial \xi} \right|_{\xi = 1}
g_0^4  \frac{\partial}{\partial g_0^2} \hat{V} \nonumber \\ &&
+ \lim_{\hat{T} \to \infty} \frac{1}{\hat{T}} \left[
2 <S_s>_{q\bar{q}-0} + g_0^2 c <S_s>_{q\bar{q}-0}\right] + O(g_0^6),
\label{ESR_NLO}
\end{eqnarray}
where
\[
c + \beta_0 = 2 \left. \frac{\partial c_t}{\partial \xi} \right|_{\xi = 1}
\]
has been used. The formula (\ref{ESR_NLO}) is equivalent to the energy
sum rule up to next-to-leading order; it will be checked now in the following
section.

All that is needed in order to do this are the expansions of the potential and
of $<S_s>_{q\bar{q}-0}$. The explicit form of the potential up to
next-to-leading order can be taken from chapter 2; for the other term, one
gets the following expansion up to next-to-leading order, which is similar to
the ones for $<SW>_{conn}$ (\ref{SW_expansion}) and $<LW>_{conn}$
(\ref{LW_expansion}):
\begin{eqnarray}
<S_s>_{q\bar{q}-0}
&=& -g_0^2 <S_s^{(0)} \omega^{(2)}>_{conn} 
 + g_0^4 <S_s^{(0)} S^{(2)} \omega^{(2)}>_{conn} \nonumber \\
&& - g_0^4 <S_s^{(0)} \frac{1}{2} (S^{(1)})^2 \omega^{(2)}>_{conn}
 + g_0^4 <S_s^{(0)} S^{(2)}_{FP} \omega^{(2)}>_{conn} \nonumber \\
&& + g_0^4 <S_s^{(0)} S^{(2)}_{meas} \omega^{(2)}>_{conn}
 + g_0^4 <S_s^{(1)} S^{(1)} \omega^{(2)}>_{conn} \nonumber \\
&& - g_0^4 <S_s^{(2)} \omega^{(2)}>_{conn}
 + g_0^4 <S_s^{(0)} S^{(1)} \omega^{(3)}>_{conn} \nonumber \\
&& - g_0^4 <S_s^{(0)} \omega^{(4)}>_{conn}
 - g_0^4 <S_s^{(1)} \omega^{(3)}>_{conn} \nonumber \\
&& - g_0^4 <S_s^{(0)} \omega^{(2)}>_{conn} <\omega^{(2)}>_0
+ O(g_0^6).
\end{eqnarray}
In contrast to (\ref{SW_expansion}) and (\ref{LW_expansion}), here a
disconnected term (the last one) appears, because of the definition
(\ref{EV_QQ}) of $<O>_{q\bar{q}-0}$.

For simplicity, but without loss of generality (because of the symmetry of
the lattice), in the following the Wilson loop will be chosen to lie in the
3-4-plane.

\section{Perturbative check}

\label{ESR_check}

\subsection{Leading order}

\label{ESR_check_LO}

In leading order, (\ref{ESR_NLO}) reduces to the following simple formula:
\begin{equation}
\hat{V} = g_0^2 \frac{\partial}{\partial g_0^2} \hat{V} - 2 g_0^2 \lim_{\hat{T}
\to \infty} \frac{1}{\hat{T}} <S_s^{(0)} \omega^{(2)}>_{conn} 
\end{equation}
where use has been made of the fact that both the expansion for
$<S_s>_{q\bar{q}-0}$ and the expansion for the potential start with $g_0^2$.
But the latter one leads also to the following identity:
\begin{equation}
\hat{V} = g_0^2 \frac{\partial}{\partial g_0^2} \hat{V} + O(g_0^4),
\end{equation}
hence in leading order, the energy sum rule is equivalent to the requirement 
\begin{equation}
\lim_{\hat{T} \to \infty} \frac{1}{\hat{T}} <S_s^{(0)}\omega^{(2)}>_{conn} = 0,
\end{equation}
i.\ e., in leading order, the expectation value of the spatial plaquettes has
to vanish in the limit of large temporal extent of the Wilson loop.

An explicit calculation for this correlator can be done by first looking at
the insertion of $S_s^{(0)}$ into an arbitrary gluon line:
\begin{equation}
<A^A_{\mu}(p) A^B_{\nu}(q) S_s^{(0)}>_{conn} = \delta^{AB}
(2 \pi)^4 \delta(p+q) \delta_{\mu}^s
\frac{\delta_{\mu\nu} \hat{\vec{p}}^2 - \hat{p}_{\mu}\hat{p}_{\nu}}
{\left(\hat{p}^2 \right)} \delta_{\nu}^s
\label{Ss_insertion}
\end{equation}
with
\begin{equation}
\delta^s_{\mu} := 1 - \delta_{\mu 4} = \left\{
\begin{array}{cc}
1 & \mu < 4 \\
0 & \mu = 4 \\
\end{array}
\right.
\end{equation}
Then the result for the correlator with the Wilson loop is:
\begin{equation}
<S_s^{(0)}\omega^{(2)}>_{conn, subtr} = -C_2(F) \int\limits_{BZ}
\frac{d^4p}{(2\pi)^4} \frac{\cos(p_4 \hat{T})}{\left(\hat{p}^2\right)^2}
\frac{\sin^2(p_3 \hat{R}/2)}{\sin^2(p_3/2)} \left(\hat{\vec{p}}^2 -
\hat{p}_3^2 \right)
\end{equation}
Taking the limit, one obviously obtains:
\begin{equation}
\lim_{\hat{T} \to \infty} \frac{1}{\hat{T}}
<S_s^{(0)}\omega^{(2)}>_{conn, subtr} = 0;
\end{equation}
the expectation value of the spatial plaquettes (the energy in the magnetic
fields) does indeed vanish in leading order, and therefore the energy sum rule
is true in leading order.

Looking at the relevant Feynman diagrams makes this a bit clearer. On the one
hand, there are the possibilities where both gluon lines end on spatial lines
of the Wilson loop:\\[1ex]
\begin{picture}(12,4)
\put(-3.5,-22.1)
{
\epsfig{file=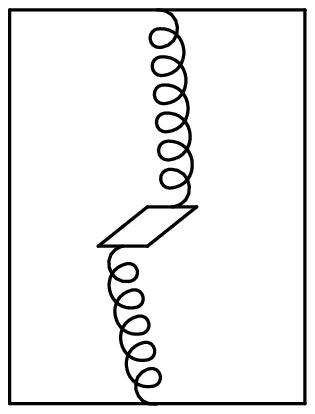, height=30cm, width=20cm}
}
\end{picture}\\
Here a sum over all spatial plaquettes is to be understood (which is denoted
by the perspective view of the plaquette), and, as usual, the end points of
the gluon lines can be attached to all four links of the plaquette. Obviously,
such diagrams can give no contribution proportional to $\hat{T}$; therefore
their contributions to $<S_s^{(0)}\omega^{(2)}>_{conn, subtr}$ vanish when
dividing by $\hat{T}$ and then taking the limit. On the other hand, the
sum over all of the diagrams of the following form:\\[1ex]
\begin{picture}(12,4)
\put(-3.5,-22.1)
{
\epsfig{file=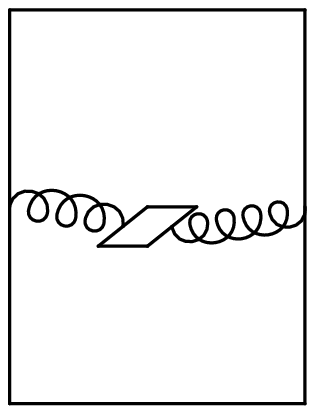, height=30cm, width=20cm}
}
\end{picture}\\
would give a contribution proportional to $\hat{T}$. But such diagrams can not
appear, because the propagator in Feynman gauge connects only links which are
parallel to each other. Only spatial plaquettes appear here, hence no gluon
line can connect the plaquettes to temporal line of the Wilson loop.

\subsection{Next-to-leading order}

\label{ESR_check_NLO}

Using the result derived above
\begin{equation}
\lim_{\hat{T} \to \infty} \frac{1}{\hat{T}}
<S_s^{(0)}\omega^{(2)}>_{conn, subtr} = 0,
\end{equation}
one sees that only the higher terms in the expansion of $<S_s>_{q\bar{q}-0}$
can give non-vanishing contributions in the limit of large $\hat{T}$. Hence
the term $g_0^2 <S_s>_{q\bar{q}-0}$ in $(\ref{ESR_NLO})$ is of order $g_0^6$
and does not contribute in next-to-leading order:
\begin{equation}
\hat{V} = g_0^2 \frac{\partial}{\partial g_0^2} \hat{V}
+ \left. \frac{\partial c_t}{\partial \xi} \right|_{\xi = 1}
g_0^4  \frac{\partial}{\partial g_0^2} \hat{V}
+ 2 \lim_{\hat{T} \to \infty} \frac{1}{\hat{T}} <S_s>_{q\bar{q}-0} + O(g_0^6).
\label{ESR_NLO2}
\end{equation}

The remaining graphs which contribute to $<S_s>_{q\bar{q}-0}$ can, as usual,
be divided into the three groups
\begin{itemize}
\item vacuum polarization graphs
\item spider graphs
\item graphs with two independent gluon lines
\end{itemize}
And again as usual, these three types of graphs will be examined separately.

\subsubsection{The vacuum polarization graphs}

\begin{figure}[t]
\begin{picture}(12,13)
\put(-3.5,-13)
{
\epsfig{file=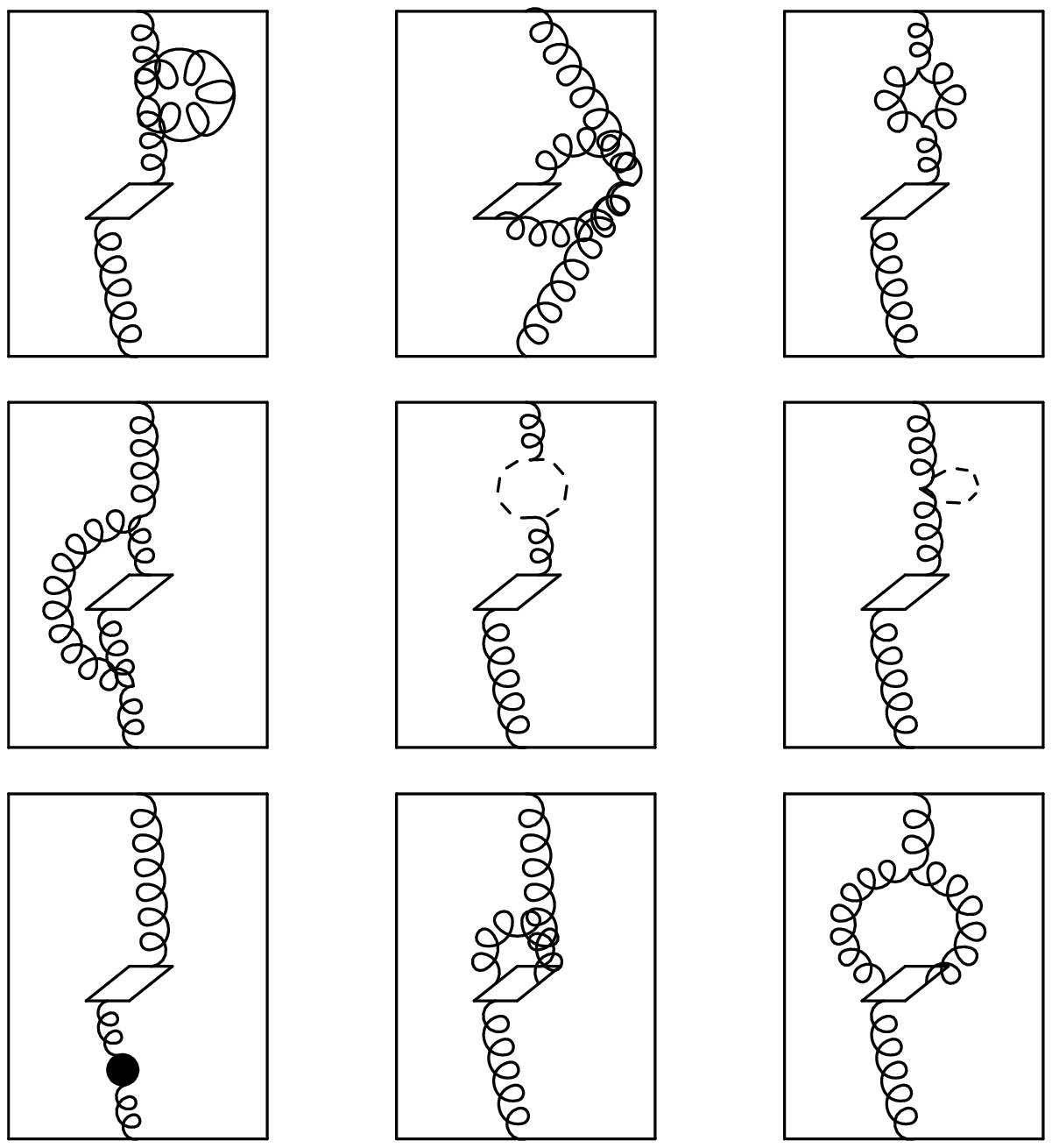, height=30cm, width=20cm}
}
\end{picture}

\caption{Vacuum polarization graphs contributing to the expectation value of
the spatial plaquettes}
\label{VP_spatial}
\end{figure}

The relevant graphs are shown in figure \ref{VP_spatial}. Using the same
argument as for the one diagram in leading order, one sees easily that most
graphs do not contribute for $\hat{T} \to \infty$; only the second and the
fourth graph survive:
\begin{eqnarray}
&&\lim_{\hat{T} \to \infty} \frac{<S_s^{(0)} S^{(2)} \omega^{(2)}>_{conn,ext}}
{\hat{T}}
= \lim_{\hat{T} \to \infty} \frac{<S_s^{(0)} \frac{1}{2} (S^{(1)})^2
\omega^{(2)}>_{conn,ext}}{\hat{T}} \nonumber \\
&=& \lim_{\hat{T} \to \infty} \frac{<S_s^{(2)}
\omega^{(2)}>_{conn}}{\hat{T}} = \lim_{\hat{T} \to \infty} \frac{<S_s^{(0)}
S^{(2)}_{FP} \omega^{(2)}>_{conn}}{\hat{T}} \nonumber \\ &=& \lim_{\hat{T} \to
\infty} \frac{<S_s^{(0)} S^{(2)}_{meas} \omega^{(2)}>_{conn}}{\hat{T}} =
\lim_{\hat{T} \to \infty} \frac{<S_s^{(1)} S^{(1)}
\omega^{(2)}>_{conn}}{\hat{T}} = 0
\end{eqnarray}
(because of the insertion of $S_s^{(0)}$, respectively $S_s^{(2)}$, only graphs
can contribute in which at least one of the external gluon lines ends on a
spatial line of the Wilson loop---but the sum over such graphs does not give
a contribution proportional to $\hat{T}$ and hence vanishes in the
limit). The subscript ''ext'' in the first two terms means that there,
only the insertions of $S_s^{(0)}$ into \emph{external} lines are considered.

The remaining graphs (number two and four) will be studied later in more
detail.

\subsubsection{The spider graphs}

As in section \ref{ASR_restriction_NLO}, there are two spider graphs which have
to be taken into account here:\\[1ex]
\begin{picture}(12,4)
\put(-3.5,-22.1)
{
\epsfig{file=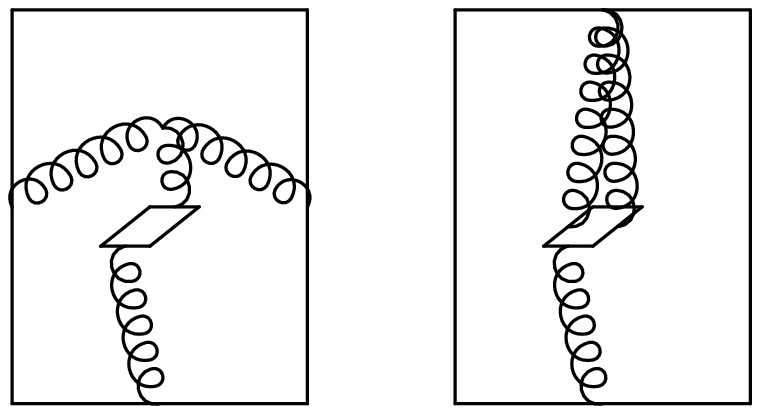, height=30cm, width=20cm}
}
\end{picture}
\\
The first graph behaves for $\hat{T} \to \infty$ like 
$<S^{(1)} \omega^{(3)}>_{conn}$, the ordinary spider graph without
plaquette insertions:
\begin{equation}
\lim_{\hat{T} \to \infty} \frac{1}{\hat{T}} <S_s^{(0)} S^{(1)}
\omega^{(3)}>_{conn} = \lim_{\hat{T} \to \infty} \frac{1}{\hat{T}} <S^{(1)}
\omega^{(3)}>_{conn} = 0
\end{equation}
Treating the second graph is even more simple; it does not contribute at all,
even for finite $\hat{T}$:
\begin{equation}
<S_s^{(1)} \omega^{(3)}>_{conn} = 0.
\end{equation}
The reason for this is that $S_s^{(1)}$ contains a three-gluon vertex. Because
this vertex only incorporates spatial plaquettes and the gluon lines are in
Feynman gauge, all three gluon lines can only be connected to the spatial
lines in the Wilson loop. But then the polarizations of all three gluons which
meet at the vertex are identical---and for this configuration, the vertex
vanishes. 

Therefore both spider graphs give no contributions. The only graphs which
are left now, beside the two vacuum polarization graphs mentioned above, are

\subsubsection{The graphs with two independent gluon lines}

Here one has to consider two types of contributions. First, as usual, there are
the contributions coming from $\omega^{(4)}$, but, as already mentioned, here
additionally a disconnected contribution appears. This is depicted in the
following two graphs:\\[1ex]
\begin{picture}(12,4)
\put(-3.5,-22.2)
{
\epsfig{file=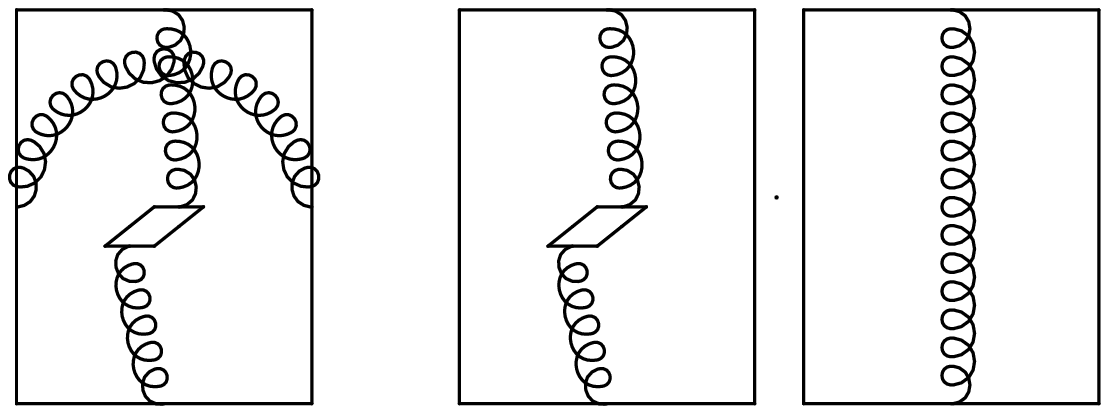, height=30cm, width=20cm}
}
\end{picture}
\\[2ex]
The first graph gives contributions proportional to $\left(C_2(F)\right)^2$
and to $C_2(F) \cdot C_2(G)$; the product of the two loops in the second
graph gives a color factor $\left(C_2(F)\right)^2$. An explicit calculation
yields:
\begin{eqnarray}
&& <S_s^{(0)} \omega^{(4A)}>_{conn} \nonumber \\ &=& 2 \left(C_2(F)\right)^2 
\int\limits_{BZ} \frac{d^4p}{(2\pi)^4}
\frac{\sin^2(p_3 \hat{R}/2)\sin^2(p_4 \hat{T}/2)}
{\hat{p}^2} \left(\frac{1}{\sin^2(p_3 /2)} +
\frac{1}{\sin^2(p_4 /2)} \right) \nonumber \\ && \cdot 
\int\limits_{BZ}
\frac{d^4p}{(2\pi)^4} \frac{\cos(p_4 \hat{T}/2)}{\left(\hat{p}^2\right)^2}
\frac{\sin^2(p_3 \hat{R}/2)}{\sin^2(p_3/2)} \left(\hat{\vec{p}}^2 -
\hat{p}_3^2 \right) \nonumber \\ && + C_2(F) C_2(G) \cdot \left( \ldots \right)
\end{eqnarray}
and
\begin{eqnarray}
&& <\omega^{(2)}>_0 <S_s^{(0)} \omega^{(2)}>_{conn} \nonumber \\
&=& \! \! -2 \left(C_2(F)\right)^2 \!
\int\limits_{BZ} \! \frac{d^4p}{(2\pi)^4}
\frac{\sin^2(p_3 \hat{R}/2)\sin^2(p_4 \hat{T}/2)}
{\hat{p}^2} \left(\frac{1}{\sin^2(p_3/2)} +
\frac{1}{\sin^2(p_4/2)} \right) \nonumber \\ && \cdot 
\int\limits_{BZ}
\frac{d^4p}{(2\pi)^4} \frac{\cos(p_4 \hat{T}/2)}{\left(\hat{p}^2\right)^2}
\frac{\sin^2(p_3 \hat{R}/2)}{\sin^2(p_3/2)} \left(\hat{\vec{p}}^2 -
\hat{p}_3^2 \right);
\end{eqnarray}
hence the contributions proportional to $\left(C_2(F)\right)^2$ cancel
exactly, and the disconnected part vanishes.

But there are still contributions left from $\omega^{(4)}$ which are
proportional to $C_2(F) \cdot C_2(G)$. These graphs can be classified into
three categories:
\begin{enumerate}
\item
Graphs in which only temporal links of the Wilson loop appear; these links
can not be connected with the spatial plaquettes using the gluon propagators
in Feynman gauge, and therefore their contributions vanish.
\item
Graphs in which only spatial links of the Wilson loop appear; these do not give
contributions proportional to $\hat{T}$ and therefore in the limit of
large $\hat{T}$, they vanish.
\item Graphs in which temporal as well as spatial links appear - these have to
be treated explicitly.
\end{enumerate}

In order to examine the contributions of the graphs of the third category,
split $\omega^{(4)}$ again into the parts $A$ to $F$ and then look only at the
graphs with spatial as well as temporal links; write for these
$\omega^{(4A)}_{RT}$ to $\omega^{(4F)}_{RT}$. The results for the various
contributions are:
\begin{eqnarray} 
&&<S_s^{(0)} \omega^{(4A)}_{RT}>_{conn} \nonumber \\ &=&
-\frac{C_2(F)C_2(G)}{6} \int\limits_{BZ} \frac{d^4p}{(2\pi)^4}
\int\limits_{BZ} \frac{d^4k}{(2\pi)^4}
\frac{\hat{\vec{p}}^2-\hat{p}_3^2}{\left(\hat{p}^2\right)^2}
\frac{1}{\hat{k}^2} \frac{\sin^2(p_3 R/2)}{\sin^2(p_3/2)}
\frac{\sin^2(k_4 T/2)}{\sin^2(k_4/2)} \nonumber \\ && \left[1 -
\cos(k_3 R) - \cos(p_4 T) + \cos(p_4 T) \cos(k_3 R) \right]
\nonumber \\ && + \left(C_2(F)\right)^2 \cdot \left( \ldots \right) \\
&&<S_s^{(0)} \omega^{(4B)}_{RT}>_{conn} \nonumber \\ &=&
-\frac{C_2(F)C_2(G)}{2} \int\limits_{BZ} \frac{d^4p}{(2\pi)^4}
\int\limits_{BZ} \frac{d^4k}{(2\pi)^4}
\frac{\hat{\vec{p}}^2-\hat{p}_3^2}{\left(\hat{p}^2\right)^2}
\frac{1}{\hat{k}^2} \frac{\sin^2(p_3 R/2)}{\sin^2(p_3/2)}
\frac{\sin^2(k_4 T/2)}{\sin^2(k_4/2)} \nonumber \\ \\
&&<S_s^{(0)} \omega^{(4C)}_{RT}>_{conn} \nonumber \\ &=& \frac{C_2(F)C_2(G)}{8}
\int\limits_{BZ} \frac{d^4p}{(2\pi)^4} \int\limits_{BZ} \frac{d^4k}{(2\pi)^4}
\frac{\hat{\vec{p}}^2-\hat{p}_3^2}{\left(\hat{p}^2\right)^2}
\frac{1}{\hat{k}^2} \nonumber \\
&& \left[3 \frac{\sin^2(p_3 R/2)}{\sin^2(p_3/2)}
\left(\frac{\sin^2(k_4 T/2)}{\sin^2(k_4/2)} - T \right) \right.
\nonumber \\ && - \frac{\sin^2(p_3 R/2)}{\sin^2(p_3/2)} \cos(p_4 T)
\left(\frac{\sin^2(k_4 T/2)}{\sin^2(k_4/2)} - 3 T \right) \nonumber \\
&& + \frac{\sin^2(k_4 T/2)}{\sin^2(k_4/2)}
\left(\frac{\sin^2(p_3 R/2)}{\sin^2(p_3/2)} - R \right) 
\left(1 - \cos(k_3 R) \right) \nonumber \\
&& \left. - 2 \frac{\sin^2(p_3 R/2)}{\sin^2(p_3/2)} \cos(p_4 T)
\frac{\sin^2(k_4 T/2)}{\sin^2(k_4/2)} \cos(k_3 R) \right] \\
&&<S_s^{(0)} \omega^{(4D)}_{RT}>_{conn} \nonumber \\ &=&
\frac{C_2(F)C_2(G)}{24} \int\limits_{BZ} \frac{d^4p}{(2\pi)^4}
\int\limits_{BZ} \frac{d^4k}{(2\pi)^4}
\frac{\hat{\vec{p}}^2-\hat{p}_3^2}{\left(\hat{p}^2\right)^2}
\frac{1}{\hat{k}^2} \nonumber \\ && \left[ \frac{\sin^2(p_3
R/2)}{\sin^2(p_3/2)} \frac{\sin^2(k_4 T/2)}{\sin^2(k_4/2)} \left(8
- 2 \cos(p_4 T) \cos(k_3 R) \right) \right. \nonumber \\
&& - 3 \frac{\sin^2(p_3 R/2)}{\sin^2(p_3/2)}
\left(\frac{\sin^2(k_4 T/2)}{\sin^2(k_4/2)} - T \right) \nonumber \\
&& - \frac{\sin^2(k_4 T/2)}{\sin^2(k_4/2)}
\left(\frac{\sin^2(p_3 R/2)}{\sin^2(p_3/2)} - R \right) \nonumber \\
&& - \frac{\sin^2(p_3 R/2)}{\sin^2(p_3/2)} \cos(p_4 T)
\left(\frac{\sin^2(k_4 T/2)}{\sin^2(k_4/2)} + 3 T \right) \nonumber \\
&& - \frac{\sin^2(k_4 T/2)}{\sin^2(k_4/2)} \cos(k_3 R)
\left(\frac{\sin^2(p_3 R/2)}{\sin^2(p_3/2)} + R \right) \\
&&<S_s^{(0)} \omega^{(4E)}_{RT}>_{conn} \nonumber \\ &=&
\frac{C_2(F)C_2(G)}{12} \int\limits_{BZ} \frac{d^4p}{(2\pi)^4}
\int\limits_{BZ} \frac{d^4k}{(2\pi)^4}
\frac{\hat{\vec{p}}^2-\hat{p}_3^2}{\left(\hat{p}^2\right)^2}
\frac{1}{\hat{k}^2} \nonumber \\ && \left[3 \frac{\sin^2(p_3
R/2)}{\sin^2(p_3/2)} T \left(1 - \cos(p_4 T) \right)
+ \frac{\sin^2(k_4 T/2)}{\sin^2(k_4/2)} R
\left(1 - \cos(k_3 R) \right) \right].
\end{eqnarray}
And finally, because of the symmetry properties in the colour indices one gets
as usual:
\begin{equation}
<S_s^{(0)} \omega^{(4F)}_{RT}>_{conn} = 0.
\end{equation} 
\\[2ex]
Adding everything, the result is:
\begin{eqnarray}
<S_s^{(0)} \omega^{(4)}_{RT}>_{conn; C_2(F)C_2(G)} 
&=& \frac{C_2(F)C_2(G)}{2} \int\limits_{BZ}
\frac{d^4p}{(2\pi)^4} \int\limits_{BZ} \frac{d^4k}{(2\pi)^4}
\frac{\hat{\vec{p}}^2-\hat{p}_3^2}{\left(\hat{p}^2\right)^2}
\frac{1}{\hat{k}^2} \nonumber \\ &&
\frac{\sin^2(p_3 R/2)}{\sin^2(p_3/2)}
\cos(p_4 T) \frac{\sin^2(k_4 T/2)}{\sin^2(k_4/2)} \cos(k_3 R) \nonumber \\
\end{eqnarray}
Divide by $\hat{T}$ and take the limit:
\begin{equation}
\lim_{\hat{T} \to \infty} \frac{<S_s^{(0)} \omega^{(4)}>_{conn; C_2(F)C_2(G)}}
{\hat{T}} 
= \lim_{\hat{T} \to \infty} \frac{<S_s^{(0)} \omega^{(4)}_{RT}>_{conn;
C_2(F)C_2(G)}} {\hat{T}}
= 0.
\end{equation}
Hence nearly all contributions vanish, and only the vacuum polarization graphs
contribute to $<S_s>_{q\bar{q}-0}$ in next-to-leading order!

\subsubsection{Check in next-to-leading order}

Making use of the fact that only the vacuum polarization graphs contribute to
the expectation value of the spatial plaquettes, the result is now simply:
\begin{eqnarray}
<S_s>_{q\bar{q}-0} \!
&=& \! g_0^4 \lim_{\hat{T} \to \infty} \frac{1}{\hat{T}} \left(<S_s^{(0)}
S^{(2)} \omega^{(2)}>_{conn} - <S_s^{(0)} \frac{1}{2} (S^{(1)})^2
\omega^{(2)}>_{conn} \right) \nonumber \\ &=& \! C_2(F) g_0^4
\int\limits_{BZ} \frac{d^3p}{(2\pi)^3} \frac{\cos(p_3
\hat{R})}{\left(\hat{\vec{p}}^2 \right)^2} \Pi^{(s)}_{44}(\vec{p}, 0).
\label{ESR_VP}
\end{eqnarray}
As usual, the self energy contributions have been subtracted here. The
vacuum polarization tensor $\Pi_{\mu\nu}^{s}$ (with the insertion of the sum
over the spatial plaquettes) is given by the sum of the following two
contributions (given for an arbitrary dimension $d$; see appendix
\ref{commonintegrals}):
\begin{eqnarray}
&&\Pi_{\mu\nu}^{Loop,s} \nonumber \\ &=& C_2(G) \int\limits_{BZ}
\frac{d^dk}{(2\pi)^d} \sum_{i,j=1}^{d-1} \frac{\hat{\vec{k}}^2 \delta_{ij} -
\hat{k}_i \hat{k}_j} {\left(\hat{k}^2\right)^2 (\widehat{p+k})^2} \nonumber \\
&& \Bigg[\delta_{ij} (\widehat{p+2k})_{\mu} (\widehat{p+2k})_{\nu} \cos(p_i/2)
\cos(p_j/2) \nonumber \\ && + \delta_{\mu\nu} (\widehat{2p+k})_i
(\widehat{2p+k})_j \cos(k_{\mu}/2) \cos(k_{\nu}/2) \nonumber \\
&& + \delta_{\mu i} \delta_{\nu j} (\widehat{p-k})^2 \cos((p+k)_i/2)
\cos((p+k)_j/2) \nonumber \\
&& + \Bigg\{ \delta_{\mu i} \Bigg((\widehat{2p+k})_j (\widehat{2p+k})_{\nu}
\cos(k_i/2) \cos(p_i/2) \nonumber \\
&& + (\widehat{p-k})_j (\widehat{p+2k})_{\nu} \cos((p+k)_i/2) \cos(p_j/2)
\nonumber \\ && + (\widehat{p-k})_{\nu} (\widehat{2p+k})_j \cos((p+k)_i/2)
\cos(k_{\nu}/2) \Bigg) \Bigg\}
+ \Bigg\{(\mu,i) \leftrightarrow (\nu,j) \Bigg\} \Bigg]
\nonumber \\ \\
&& \Pi_{\mu\nu}^{Tadpole,s} \nonumber \\ &=& -\frac{1}{2} C_2(G) \Bigg\{
\delta_{\mu\nu} \left[ \frac{2(d-2)}{d} \Delta_0 - \frac{(d+1)(d-2)}{3d(d-1)}
\Delta_0 \hat{\vec{p}}^2 \right] \nonumber \\
&& + \delta^s_{\mu\nu} \Bigg[\frac{(2d-5)(d-2)}{d-1} \Delta_0 - \frac{7}{6d}
+ \left(\frac{(2d-1)(d-2)}{4d(d-1)} \Delta_0 - \frac{1}{4d}\right)
\hat{p}_{\mu}^2 \nonumber \\ && + \left(-\frac{(5d-2)(d-2)}{6d(d-1)} \Delta_0 +
\frac{1}{d}\right) \hat{\vec{p}}^2 - \frac{d-2}{6d(d-1)} \Delta_0 \hat{p}^2
\Bigg] \nonumber \\ && + \frac{d-2}{6(d-1)} \Delta_0 \hat{p}_{\mu}
\hat{p}_{\nu} \left(\delta^s_{\mu} + \delta^s_{\nu} \right) +
\frac{(4d-1)(d-2)}{12d(d-1)} \Delta_0 \hat{p}_{\mu} \hat{p}_{\nu}
\delta^s_{\mu} \delta^s_{\nu} \Bigg\} \nonumber \\ && + \left(C_2(F) -
\frac{1}{6} C_2(G) \right) \Bigg\{ \delta_{\mu\nu} \frac{(d-2)^2}{d(d-1)}
\Delta_0 \hat{\vec{p}}^2 \nonumber \\ && + \delta^s_{\mu\nu} \Bigg[
\left(\frac{4-2d}{d} \Delta_0 + \frac{1}{d} \right) \hat{\vec{p}}^2 +
\frac{(d-2)^2}{d(d-1)} \Delta_0 \hat{p}^2 \Bigg] \nonumber \\ && -
\frac{(d-2)^2}{d(d-1)} \Delta_0 \hat{p}_{\mu} \hat{p}_{\nu}
\left(\delta^s_{\mu} + \delta^s_{\nu} \right) - \left(\frac{4-2d}{d} \Delta_0
+ \frac{1}{d} \right) \hat{p}_{\mu} \hat{p}_{\nu} \delta^s_{\mu}
\delta^s_{\nu} \Bigg\}
\end{eqnarray}
with
\begin{equation}
\delta^s_{\mu\nu} := \sum_{\mu,\nu} \delta^s_{\mu} \delta_{\mu\nu}
\delta^s_{\nu}
\end{equation}
and the usual abbreviation
\[
\Delta_0 \equiv \int\limits_{BZ} \frac{d^dk}{(2\pi)^d} \frac{1}{\hat{k}^2}
\]
(see appendix \ref{commonintegrals}).

As already explained above, vacuum polarization graphs in which the sum over
the spatial plaquettes is inserted into an external line can not contribute;
therefore the ghost graphs and the graph with the insertion of the integration
measure do not appear here, because they do not have internal gluon lines.

Only the $\mu=\nu=d=4$ components contribute in the limit of large $\hat{T}$,
and from these, only the part with $p_4=0$. Then the formulas above reduce to:
\begin{eqnarray}
\Pi_{44}^{Tadpole,s}(\vec{p}, 0) &=& C_2(G) \Delta_0 \left(\frac{1}{12}
\hat{\vec{p}}^2 - \frac{1}{2} \right) + C_2(F) \frac{1}{3}
\Delta_0 \hat{\vec{p}}^2 \\
\Pi_{44}^{Loop,s}(\vec{p}, 0) &=& C_2(G) \int\limits_{BZ} \frac{d^4k}{(2\pi)^4}
\frac{1}{\left(\hat{k}^2\right) \left((\widehat{\vec{p}+\vec{k}})^2 +
\hat{k}^2_4 \right)} \nonumber \\
&& \left[ (\widehat{2k})^2_4 \hat{\vec{k}}^2 \left(3 - \frac{1}{4}
\hat{\vec{p}}^2\right) + \hat{\vec{k}}^2 \left(1 - \frac{1}{4}
\hat{k}^2_4 \right) \left(\widehat{2 \vec{p} + \vec{k}}\right)^2 \right.
\nonumber \\ && - (\widehat{2k})^2_4  \left(\hat{\vec{k}}^2 - \frac{1}{4}
\sum_i \hat{k}_i^2 \hat{p}_i^2 \right) \nonumber \\ && - \left. \left(1 -
\frac{1}{4} \hat{k}^2_4 \right) \left(\sum_i \left(\widehat{2p+k}\right)_i
\hat{k}_i \right)^2 \right].
\end{eqnarray}
There is a source for a potential infrared divergence in the $p$-integral in
(\ref{ESR_VP}); one has to check that the vacuum polarization tensor goes to
zero for $p \to 0$. Inserting $p=0$ above, the results are:
\[
\Pi_{44}^{Tadpole,s}(0) = - \frac{1}{2} C_2(G) \Delta_0 = - \Pi_{44}^{Loop,s}(0)
\]
and therefore
\begin{equation}
\Pi_{44}^{(s)}(0) = \Pi_{44}^{Tadpole,s}(0) + \Pi_{44}^{Loop,s}(0) = 0.
\end{equation}
Hence there is no infrared divergence, and the integral can be done without
problems.

Inserting the formula (\ref{ESR_VP}) for the expectation value of the spatial
plaquettes, (\ref{ESR_NLO2}) becomes:
\begin{equation}
\hat{V} = g_0^2 \frac{\partial}{\partial g_0^2} \hat{V}
+ \left.\frac{\partial c_t}{\partial \xi}\right|_{\xi = 1}
g_0^4 \frac{\partial}{\partial g_0^2} \hat{V}
+ 2 C_2(F) g_0^4 \int\limits_{BZ} \frac{d^3p}{(2\pi)^3}
\frac{\cos(p_3 \hat{R})}{\left(\hat{\vec{p}}^2 \right)^2}
\Pi^{(s)}_{44}(\vec{p}, 0)
\end{equation}
Then the expansion for the potential
\begin{equation}
\hat{V} = g_0^2 V^{(2)} + g_0^4 V^{(4)} + O(g_0^6)
\end{equation}
yields
\begin{equation}
V^{(4)} = 2 V^{(4)}
+ \left.\frac{\partial c_t}{\partial \xi}\right|_{\xi = 1} V^{(2)}
+ 2 C_2(F) \int\limits_{BZ} \frac{d^3p}{(2\pi)^3}
\frac{\cos(p_3 \hat{R})}{\left(\hat{\vec{p}}^2 \right)^2}
\Pi^{(s)}_{44}(\vec{p}, 0).
\end{equation}
The explicit expression for the potential can be taken from chapter
\ref{potential}. Additionally, use
\begin{equation}
\left.\frac{\partial c_t}{\partial \xi} \right|_{\xi = 1} =
-\frac{1}{4} C_2(F) \cdot 0.586844 + 4 C_2(G) \cdot 0.005306,
\end{equation}
derived in \cite{Karsch2}, to obtain finally the following expression, which
is equivalent to the energy sum rule in next-to-leading order:
\begin{eqnarray}
&& \int\limits_{BZ} \frac{d^3p}{(2\pi)^3}
\frac{\cos(p_3 \hat{R})}{\left(\hat{\vec{p}}^2 \right)^2}
\Pi^{(s)}_{44}(\vec{p}, 0) \nonumber \\
&=& \frac{C_2(G)}{4 \pi \hat{R}} \left(0.023220 \ln \hat{R} + 0.057400
\right) + \frac{C_2(F)}{4 \pi \hat{R}} 0.051644.
\label{ESR_equivalent}
\end{eqnarray}

Now it is convenient to split up $\Pi^{Tadpole}$ into two parts, corresponding
to the group theoretical factors:
\begin{equation}
\Pi^{Tadpole} = C_2(G) \Pi^{Tadpole,1} + C_2(F) \Pi^{Tadpole,2};
\end{equation}
then (\ref{ESR_equivalent}) splits also into two parts:
\begin{eqnarray}
(I) && \int\limits_{BZ} \frac{d^3p}{(2\pi)^3}
\frac{\cos(p_3 \hat{R})}{\left(\hat{\vec{p}}^2 \right)^2}
\Pi^{Tadpole,s,2}_{44}(\vec{p}, 0)
= \frac{1}{4 \pi \hat{R}} 0.051644 \label{ESR_NLO_I} \\
(II) && \int\limits_{BZ} \frac{d^3p}{(2\pi)^3}
\frac{\cos(p_3 \hat{R})}{\left(\hat{\vec{p}}^2 \right)^2}
\left(\Pi^{Tadpole,s,1}_{44}(\vec{p}, 0) + \Pi^{Loop,s,1}_{44}(\vec{p},
0)\right) \nonumber \\
&=& \frac{1}{4 \pi \hat{R}} \left(0.023220 \ln \hat{R} +
0.057400 \right)
\label{ESR_NLO_II}
\end{eqnarray}

In the first part (\ref{ESR_NLO_I}), insert the explicit expression
$\Pi_{44}^{Tadpole,2}(\vec{p},0) = \frac{1}{3} \Delta_0 \, \hat{\vec{p}}^2$;
the remaining integral gives in the continuum limit (see section
\ref{potential_LO}):
\begin{equation}
\int\limits_{BZ} \frac{d^3p}{(2\pi)^3}
\frac{\cos(p_3 \hat{R})}{\hat{\vec{p}}^2 } = \frac{1}{4 \pi \hat{R}}.
\end{equation}
Hence the first part of (\ref{ESR_equivalent}) requires simply the following
formula to hold:
\begin{equation}
\frac{1}{3} \Delta_0 = 0.051644.
\end{equation}
Inserting for $\Delta_0$ the numerically calculated value (see
appendix C), one sees that this formula is indeed satisfied---to very good
accuracy!

\begin{figure}[t]
\begin{picture}(10,4)
\put(3,0){\epsfig{file=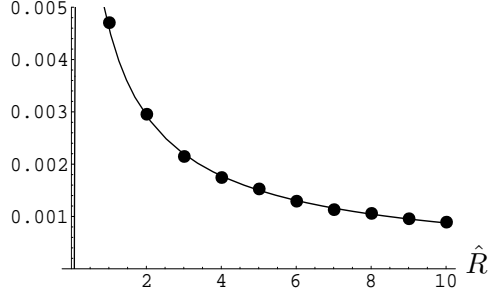, height=4cm, width=6cm}}
\put(9.1,0.3){$\hat{R}$}
\end{picture}
\caption{Comparison of the calculated values to the fitted curve; see text}
\label{Fit}
\end{figure}

The second part (\ref{ESR_NLO_II}) is a little bit more complicated. The
integral on the left hand side can be only evaluated numerically; this has
been done for $\hat{R}$ between 1 and 10 using again the routine \emph{Vegas}
from \cite{NumericalRecipes}. The resulting numbers can be compared to the
function on the right hand side.

\noindent
A fit of the calculated values to a function of the form
\[
\frac{1}{4 \pi \hat{R}} \left(a \ln \hat{R} + b \right),
\]
using Mathematica, gives:
\begin{equation}
a = 0.02268 \pm 0.00078; \quad \quad \quad
b = 0.05792 \pm 0.00132
\end{equation}
If the energy sum rule is valid, the results should be $a = 0.2322$ and $b =
0.05740$. The numbers from the fit agree very well with this in the range of
their numerical errors. Hence the second part of (\ref{ESR_equivalent}) is
also valid.

For illustrative purposes, figure \ref{Fit} gives a plot of the calculated
values (left hand side of (\ref{ESR_NLO_II})), as well as of the curve on which
they should lie if the energy sum rule is valid (right hand side of
(\ref{ESR_NLO_II})). The very good agreement between the points and the curve
is obvious.

Summarizing: Both parts, (\ref{ESR_NLO_I}) and (\ref{ESR_NLO_II}), of the
formula (\ref{ESR_equivalent}), which is equivalent to the energy sum rule in
next-to-leading order, are satisfied with good numerical accuracy, or in other
words:
\begin{center}
\emph{The energy sum rule is valid up to next-to-leading order with good
numerical accuracy.}
\end{center}

\section{Contributions to the potential}

\label{contributions_potential}

Using the validity of the energy sum rule and the results obtained for the
spatial plaquettes, one can now analyze the various contributions to the
potential:
\begin{eqnarray}
\hat{V} &=& \mbox{electric field energy} + \mbox{magnetic field energy}
\nonumber \\ && + \mbox{trace anomaly} \nonumber \\
&=& -\eta_- \lim_{\hat{T} \to \infty} \frac{1}{\hat{T}}
<\mathcal{P}_t>_{q\bar{q}-0,subtr} + \eta_- \lim_{\hat{T} \to \infty}
\frac{1}{\hat{T}} <\mathcal{P}_s>_{q\bar{q}-0,subtr}
\nonumber \\ && + \frac{1}{4} \lim_{\hat{T} \to \infty}
\frac{1}{\hat{T}} <\sum_{x, \mu} T_{\mu\mu}(x)>_{q\bar{q}-0,subtr}
\end{eqnarray}
First, look at the contribution of the magnetic field energy.
\begin{eqnarray*}
\eta_-<\mathcal{P}_s>_{q\bar{q}-0,subtr} &=& \frac{\eta_-}{\beta}
<S_s>_{q\bar{q}-0,subtr} \\
&=& <S_s>_{q\bar{q}-0,subtr} + \frac{1}{2} c g_0^2 <S_s>_{q\bar{q}-0,subtr},
\end{eqnarray*}
where the constant $c$ is again given by
\[
c = \left(\frac{\partial c_t}{\partial \xi} - \frac{\partial
c_s}{\partial \xi} \right)_{\xi = 1};
\]
the functions $c_t$ and $c_s$ can be found in \cite{Karsch2}.

As already pointed out in the last section, in the limit of large $\hat{T}$,
the leading order of $<S_s>_{q\bar{q}-0,subtr}$ vanishes, and the
lowest non-vanishing contribution is of order $g_0^4$; hence up to
next-to-leading order, the second term does not contribute here. The first term
is given by the two vacuum polarization graphs where the sum over the spatial
plaquettes has been inserted into an internal line. According to the results
obtained in the last section, one gets then for the contribution of the
magnetic field energy:
\begin{eqnarray}
&&\mbox{magnetic field energy} \nonumber \\
&=&\eta_- \lim_{\hat{T} \to \infty} \frac{1}{\hat{T}}
<\mathcal{P}_s>_{q\bar{q}-0,subtr} \nonumber \\
&=& \frac{C_2(F)}{4 \pi \hat{R}} \left(\frac{11 C_2(G)}{48 \pi^2} g_0^4
\ln\left(7.501 \hat{R}\right) + \frac{1}{2} \left.\frac{\partial
c_t}{\partial \xi}\right|_{\xi = 1} g_0^4 + \frac{1}{8} C_2(F) g_0^4 \right)
\nonumber \\
&\approx& \frac{C_2(F)}{4 \pi \hat{R}} g_0^4 \left( C_2(G) \left(0.023
\ln\hat{R} + 0.058 \right) + 0.052 \, C_2(F)\right).
\label{magneticenergy}
\end{eqnarray}
For the electric field energy, use
\begin{eqnarray*}
<\mathcal{P}_t>_{q\bar{q}-0,subtr} &=& \frac{1}{\hat{\beta}}
\left(<S>_{q\bar{q}-0,subtr} - <S_s>_{q\bar{q}-0,subtr}\right)
\end{eqnarray*}
and
\[
\lim_{\hat{T} \to \infty} \frac{1}{\hat{T}} <S>_{q\bar{q}-0,subtr} = 
\hat{\beta} \frac{\partial \hat{V}}{\partial \hat{\beta}}.
\]
Inserting the explicit form of the potential and of
$<S_s>_{q\bar{q}-0,subtr}$, one then obtains for the contribution of the
electric field energy:
\begin{eqnarray}
&& \mbox{electric field energy} \nonumber \\
&=&-\eta_- \lim_{\hat{T} \to \infty} \frac{1}{\hat{T}}
<\mathcal{P}_t>_{q\bar{q}-0,subtr} \nonumber \\
&=& - \frac{C_2(F)}{4 \pi \hat{R}} \left(g_0^2 + \frac{11 C_2(G)}{16 \pi^2}
g_0^4 \ln\left(7.501 \hat{R}\right) - \frac{1}{2} \left.\frac{\partial
c_s}{\partial \xi}\right|_{\xi = 1} g_0^4 + \frac{3}{8} C_2(F) g_0^4 \right)
\nonumber \\
&\approx& - \frac{C_2(F)}{4 \pi \hat{R}} \left(g_0^2 + g_0^4 C_2(G) \left(
0.070 \ln\hat{R} + 0.141\right) + 0.302 \, g_0^4 \, C_2(F) \right).
\label{electricenergy}
\end{eqnarray}
Finally, for the trace anomaly one can use
\[
\sum_{x, \mu} T_{\mu\mu}(x) = \frac{2\beta_L}{g_0} S
\]
and again
\[
\lim_{\hat{T} \to \infty} \frac{1}{\hat{T}} <S>_{q\bar{q}-0,subtr} = 
\hat{\beta} \frac{\partial \hat{V}}{\partial \hat{\beta}}.
\]
Then one sees that the contribution of the trace anomaly to the potential is
given by:
\begin{eqnarray}
&& \mbox{trace anomaly} \nonumber \\
&=& \frac{1}{4} \lim_{\hat{T} \to \infty}
\frac{1}{\hat{T}} <\sum_{x, \mu} T_{\mu\mu}(x)>_{q\bar{q}-0,subtr} \nonumber \\
&=&- \frac{11 C_2(F) C_2(G)}{96 \pi^2} \frac{g_0^4}{4 \pi \hat{R}} \nonumber \\
&\approx& - 0.011 C_2(F) C_2(G) \frac{g_0^4}{4 \pi \hat{R}}.
\label{traceanomaly}
\end{eqnarray}

\begin{figure}[t]
\begin{picture}(10,4)
\put(3,0){\epsfig{file=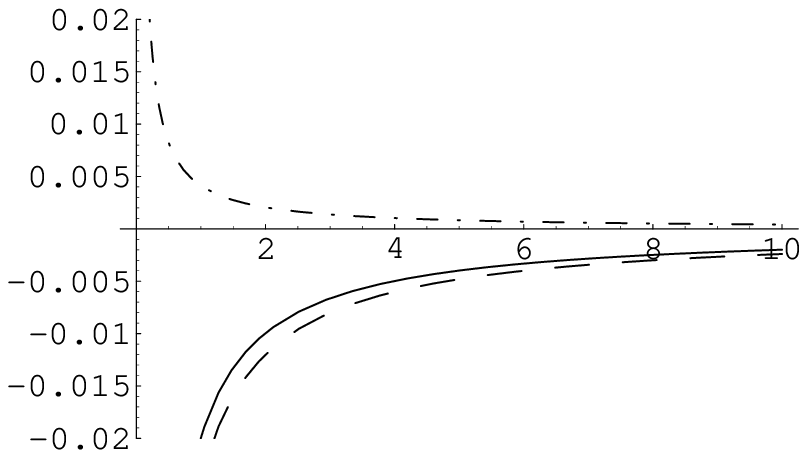, height=4cm, width=6cm}}
\put(3,4){$\hat{V}_{C_2(F)^2}/g_0^4$}
\put(9.1,1.9){$\hat{R}$}
\end{picture}
\caption{Contributions to the potential in NLO, proportional to
$\left(C_2(F)\right)^2$, divided by $g_0^4$: electric field energy (dashed),
magnetic field energy (dot-dashed), potential(solid)}
\label{potential_RR}
\end{figure}

\begin{figure}[t]
\begin{picture}(10,4)
\put(3,0){\epsfig{file=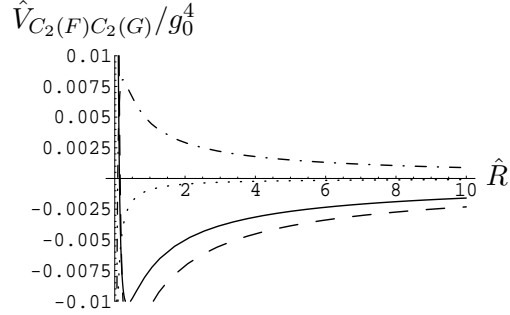, height=4cm, width=6cm}}
\put(2.8,4){$\hat{V}_{C_2(F)C_2(G)}/g_0^4$}
\put(9.1,1.9){$\hat{R}$}
\end{picture}
\caption{Contributions to the potential in NLO, proportional to
$C_2(F) C_2(G)$, divided by $g_0^4$: electric field energy (dashed), magnetic
field energy (dot-dashed), trace anomaly (dotted), potential (solid)}
\label{potential_RG}
\end{figure}

\begin{figure}[b]
\begin{picture}(10,4)
\put(3,0){\epsfig{file=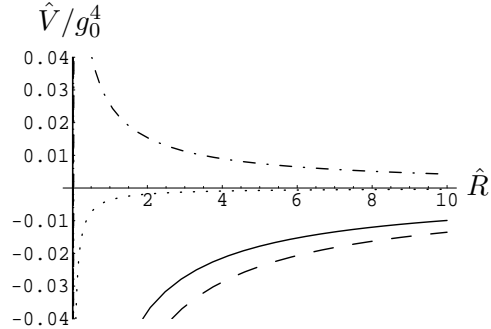, height=4cm, width=6cm}}
\put(3.4,4.1){$\hat{V}/g_0^4$}
\put(9.1,1.9){$\hat{R}$}
\end{picture}
\caption{All contributions to the potential in NLO for SU(3), divided by
$g_0^4$: electric field energy (dashed), magnetic field energy (dot-dashed),
trace anomaly (dotted), potential (solid)}
\label{potential_SU3}
\end{figure}

Looking at the three contributions (\ref{electricenergy}),
(\ref{magneticenergy}) and (\ref{traceanomaly}), one notices several things:
\begin{itemize}
\item The contribution from the electric field energy is negative and appears
already in leading order. Both group theoretical factors contribute in
next-to-leading order.
\item The contribution from the magnetic field energy is positive and appears
first in next-to-leading order; but even there, only the vacuum polarization
graphs give a non-vanishing contribution. Again, both group theoretical factors
contribute.
\item The contribution from the trace anomaly is negative and appears first
in next-to-leading order; it is proportional to $C_2(F) \cdot C_2(G)$.
\end{itemize}

Comparing the explicit numerical factors which appear in the three
contributions, or even better, looking at the plots where the magnitudes of
these terms are displayed (figures \ref{potential_RR} and \ref{potential_RG};
for the special case of $SU(3)$, see figure \ref{potential_SU3}), one
additionally sees that the energy in the magnetic fields is smaller than the
one in the electric fields, so that only a part of the electric field energy
is cancelled and the overall contribution of the energy in the fields is
negative. The trace anomaly gives an additional negative contribution, which
for large $\hat{R}$ is small compared to the energy in the fields. This is in
strong contrast with the situation for a confining potential, discussed in
section \ref{ESR_derivation}, where the trace anomaly and the energy in the
fields both gave exactly equal contributions to the potential.

\section{Restriction to a fixed time slice}

\label{ESR_restriction}

What remains to be shown is that one can restrict the sum over the plaquettes
to a fixed time slice:
\begin{eqnarray}
\lim_{\hat{T} \to \infty} \frac{1}{\hat{T}} <\mathcal{P}_t>_{q\bar{q}-0} 
&=& \lim_{\hat{T} \to \infty} <\mathcal{P}'_t(t)>_{q\bar{q}-0} \\
\lim_{\hat{T} \to \infty} \frac{1}{\hat{T}} <\mathcal{P}_s>_{q\bar{q}-0} 
&=& \lim_{\hat{T} \to \infty} <\mathcal{P}'_s(t)>_{q\bar{q}-0},
\end{eqnarray}
where $\mathcal{P}'_t(t)$ respectively $\mathcal{P}'_s(t)$ denotes the sum
over all plaquettes on the time slice $t$. Using
\begin{equation}
\mathcal{P}_s + \mathcal{P}_t = \frac{1}{\hat{\beta}} S
\end{equation}
and assuming that
\begin{equation}
\lim_{\hat{T} \to \infty} \frac{1}{\hat{T}} <S>_{q\bar{q}-0} 
= \lim_{\hat{T} \to \infty} <L(t)>_{q\bar{q}-0}
\end{equation}
(see section \ref{ESR_restriction}), it suffices to show that the restriction
to a fixed time slice works for the spatial plaquettes; this is equivalent to
showing that \begin{equation}
\lim_{\hat{T} \to \infty} \frac{1}{\hat{T}} <S_s>_{q\bar{q}-0} 
= \lim_{\hat{T} \to \infty} <L_s(t)>_{q\bar{q}-0}.
\end{equation}
The left hand side of this equation has already been discussed in section
\ref{ESR_check}; for the right hand side, one gets an analogous expansion:
\begin{eqnarray}
<L_s>_{q\bar{q}-0}
&=& -g_0^2 <L_s^{(0)} \omega^{(2)}>_{conn} 
 + g_0^4 <L_s^{(0)} S^{(2)} \omega^{(2)}>_{conn} \nonumber \\
&& - g_0^4 <L_s^{(0)} \frac{1}{2} (S^{(1)})^2 \omega^{(2)}>_{conn}
 + g_0^4 <L_s^{(0)} S^{(2)}_{FP} \omega^{(2)}>_{conn} \nonumber \\
&& + g_0^4 <L_s^{(0)} S^{(2)}_{meas} \omega^{(2)}>_{conn}
 + g_0^4 <L_s^{(1)} S^{(1)} \omega^{(2)}>_{conn} \nonumber \\
&& - g_0^4 <L_s^{(2)} \omega^{(2)}>_{conn}
 + g_0^4 <L_s^{(0)} S^{(1)} \omega^{(3)}>_{conn} \nonumber \\
&& - g_0^4 <L_s^{(0)} \omega^{(4)}>_{conn}
 - g_0^4 <L_s^{(1)} \omega^{(3)}>_{conn} \nonumber \\
&& - g_0^4 <L_s^{(0)} \omega^{(2)}>_{conn} <\omega^{(2)}>_0
+ O(g_0^6).
\end{eqnarray}

\subsection{Leading order}

\label{ESR_restriction_LO}

For calculating the correlator $<L_s^{(0)} \omega^{(2)}>_{conn}$, it is
convenient, as usual, first to look at the insertion of $L_s^{(0)}$ into an
arbitrary line. The result is very similar to the one obtained when inserting
$S_s^{(0)}$ (\ref{Ss_insertion}); the only difference is that the fourth
component of the momentum is not conserved:
\begin{equation}
<A^A_{\mu}(p) A^B_{\nu}(q) L_s^{(0)}>_{conn} = \delta^{AB}
(2 \pi)^3 \delta(\vec{p}+\vec{q}) \sum_{\mu, \nu} \delta_{\mu}^s
\frac{\delta_{\mu\nu} \hat{\vec{p}}^2 - \hat{p}_{\mu}\hat{p}_{\nu}}
{\hat{p}^2\left(\hat{\vec{p}}^2 + \hat{q}_4^2 \right)} \delta_{\nu}^s.
\label{Ls_insertion}
\end{equation}
With this, one gets the following simple result for the correlator with the
Wilson loop in leading order:
\begin{eqnarray}
&&<L_s^{(0)}(t) \omega^{(2)}>_{conn} \nonumber \\ &=& -2 C_2(F) C_2(G)
\int\limits_{BZ} \frac{d^4p}{(2\pi)^4} \int\limits_{-\pi}^{\pi}
\frac{dq_4}{2\pi} \frac{\hat{\vec{p}}^2 - \hat{p}_3^2}
{\hat{p}^2\left(\hat{\vec{p}}^2 + \hat{q}_4^2 \right)}
\frac{\sin^2(p_3 \hat{R}/2)}{\sin^2(p_3/2)} \nonumber \\
&& \cdot \sin(p_4 \hat{T}/2) \sin(q_4 \hat{T}/2) \cos((p+q)_4 (n_{c,4} - t)),
\end{eqnarray}
where again a Wilson loop lying in the 3-4-plane has been used and $n_{c,4}$ is
the fourth coordinate of the center of the loop. For $\hat{T} \to \infty$,
this correlator vanishes; for the special choice $n_{c,4} = t$ the correlator
vanishes even for every finite $\hat{T}$, because then the function under the
integral is odd.

Thus one sees that in leading order, the restriction to one time slice indeed
works:
\begin{equation}
\lim_{\hat{T} \to \infty} \frac{1}{\hat{T}} <S_s>_{q\bar{q}-0} =
\lim_{\hat{T} \to \infty} <L_s(t)>_{q\bar{q}-0} + O(g_0^4).
\end{equation}

\subsection{Next-to-leading order}

\label{ESR_restriction_NLO}

The relevant graphs for the next order can be found essentially already in
section \ref{ESR_check_NLO}: the only crucial difference is that in that
section, the plaquettes were summed over \emph{all} space-time, whereas here
they are summed only over the \emph{fixed} time slice $t$. Now, as usual, it is
convenient to look at the three different types of graphs seperately. For
simplicity, again only the special case $n_{c,4} = t = 0$ will be treated.

\subsubsection{The vacuum polarization graphs}

Here one can use arguments very similar to the ones in the sections
\ref{ESR_check_NLO} and \ref{ASR_restriction_NLO}. First look again at the
graphs where $L_s^{(0)}$ is inserted into an external line. They give a
contribution proportional to
\begin{eqnarray}
\int\limits_{BZ} \frac{d^4p}{(2\pi)^4} \int\limits_{-\pi}^{\pi}
\frac{dq_4}{2\pi} \frac{\sin^2(p_3 \hat{R}/2)}{\sin(p_3/2)} \sin(p_4 \hat{T}/2)
\sin(q_4 \hat{T}/2) \nonumber \\
\sum_{j=1}^3 \left(\frac{\Pi_{3 j}(p)}{\sin(p_3)} -
\frac{\Pi_{4 j}(p)}{\sin(p_4)}\right)
\frac{\delta_{j 3} \hat{\vec{p}}^2 - \hat{p}_j \hat{p}_3}
{\hat{p}^2 \left(\hat{\vec{p}}^2 + \hat{q}_4^2 \right)}.
\end{eqnarray}
Obviously the function under the integral is odd in $q_4$ and therefore the
integral vanishes even for every finite $\hat{T}$. This is in accordance with
the results of section \ref{ESR_check_NLO}, where it was shown that in the
limit of large $\hat{T}$, the vacuum polarisation graphs with an insertion of
$S_s^{(0)}$ into an external line do not contribute.

Next consider an insertion of $L_s^{(0)}$ into an internal line and the two
graphs coming from the vertex insertions $L_s^{(1)}$ and $L_s^{(2)}$
(corresponding to the first two graphs in figure \ref{L_graphs}, but here only
with spatial plaquettes). These give the following contribution:
\begin{eqnarray}
&&2 C_2(F) \sum_{\alpha, \beta} \int\limits_{BZ} \frac{d^4p}{(2 \pi)^4}
\int\limits_{-\pi}^{\pi} \frac{dq_4}{2 \pi}
\frac{\sin^2(p_3 \hat{R}/2) \sin(p_4 \hat{T}/2) \sin(q_4
\hat{T}/2)}{\hat{p}^2 \left(\hat{\vec{p}}^2+\hat{q}_4^2\right)} \nonumber
\\ &&
\frac{\delta_{3\alpha} - \delta_{4\alpha}}{\sin(p_{\alpha}/2)}
\frac{\delta_{3\beta} - \delta_{4\beta}}{\sin(q_{\beta}/2)}
\Pi^{L_s}_{\alpha\beta}(\vec{p},p_4,q_4),
\end{eqnarray}
where $\Pi^{L_s}$ represents the vacuum polarization tensor with $L_s^{(0)}$
inserted into one of its internal lines respectively the two contributions
with the vertex insertions $L_s^{(1)}$ or $L_s^{(2)}$. In the limit $\hat{T}
\to \infty$, this reduces to:
\begin{equation}
2 C_2(F) \sum_{\alpha, \beta} \int\limits_{BZ} \frac{d^3p}{(2 \pi)^3}
\frac{\sin^2(p_3 \hat{R}/2)}{\left(\hat{\vec{p}}^2\right)^2}
\Pi^{L_s}_{44}(\vec{p},0,0).
\end{equation}

Now first look at the gluon tadpole graph. The contribution coming from
this graph is proportional to:
\begin{eqnarray}
&& \sum_{\rho, \lambda}
\int\limits_{BZ} \frac{d^4r}{(2\pi)^4} \int\limits_{BZ} \frac{d^4s}{(2\pi)^4}
(2\pi)^4(p+q+r+s)
\Gamma_{\mu\nu\rho\lambda}^{ABCD}(p,q,r,s) \frac{(2\pi)^3
\delta(\vec{r}+\vec{s})}{\hat{r}^2\left(\hat{\vec{r}}^2 + \hat{s}_4^2
\right)} \nonumber \\ && \delta_{\rho}^s \left(\delta_{\rho\lambda} \hat{r}^2 -
\hat{r}_{\rho} \hat{r}_{\lambda} \right) \delta_{\lambda}^s,
\end{eqnarray}
where again the Delta-function coming from the four-gluon vertex has been
extraced explicitly from $\Gamma$. Now split this four-dimensional
Delta-function up into a three-dimensional spatial Delta-function and another
for the temporal components. Additionally, as explained above, it suffices to
look at the special case $p_4=q_4=0$:
\begin{eqnarray*}
&&\int\limits_{BZ} \frac{d^4r}{(2\pi)^4} \int\limits_{BZ} \frac{d^4s}{(2\pi)^4}
(2\pi)^3(\vec{p}+\vec{q}+\vec{r}+\vec{s}) (2\pi) \delta(r_4+s_4)  
\frac{(2\pi)^3 \delta(\vec{r}+\vec{s})}{\hat{r}^2\left(\hat{\vec{r}}^2
+ \hat{s}_4^2 \right)} \\ && \sum_{\rho, \lambda}
\Gamma_{\mu\nu\rho\lambda}^{ABCD}((\vec{p},0),(\vec{q},0),r,s)
\delta_{\rho}^s \left(\delta_{\rho\lambda} \hat{r}^2 - \hat{r}_{\rho}
\hat{r}_{\lambda} \right) \delta_{\lambda}^s.
\end{eqnarray*}
Carrying out the four $s$-integrations, using the second and third
Delta-function, gives:
\begin{equation}
\sum_{\rho, \lambda}
\int\limits_{BZ} \frac{d^4r}{(2\pi)^4} (2\pi)^3(\vec{p}+\vec{q})
\Gamma_{\mu\nu\rho\lambda}^{ABCD}((\vec{p},0),(\vec{q},0),r,-r)
\delta_{\rho}^s \frac{\delta_{\rho\lambda} \hat{r}^2 - \hat{r}_{\rho}
\hat{r}_{\lambda}}{\left(\hat{r}^2\right)^2} \delta_{\lambda}^s.
\end{equation}
If one compares this with the result one would have
obtained if one would have inserted the sum over \emph{all} spatial
plaquettes $S_s^{(0)}$, using (\ref{Ss_insertion}), and then looking again only
at $p_4=q_4=0$ (the only important part for $\hat{T} \to \infty$), one sees
that one gets exactly the same result, and therefore:
\begin{equation}
\lim_{\hat{T} \to \infty} <L_s^{(0)}\omega^{(2)}S^{(2)}>_{conn} =
\lim_{\hat{T} \to \infty} \frac{1}{\hat{T}}
<S_s^{(0)}\omega^{(2)}S^{(2)}>_{conn}
\end{equation}
holds, if only the insertion into the \emph{internal} line is considered. But
above it already had been shown that an insertion of $L_s^{(0}$ into an
\emph{external} line gives for $\hat{T} \to \infty$ the same result as an
insertion of $S_s^{(0)}$---therefore the result is finally that an insertion
of $L_s^{(0}$ into \emph{any} line of the gluon tadpole graph is equivalent to
an insertion of $S_s^{(0)}$ in the limit of large $\hat{T}$.

Exactly the same arguments can be made for the graph where $L_s^{(0)}$ is
inserted into any line of the gluon loop graph. There one obtains:
\begin{equation}
\lim_{\hat{T} \to \infty} <L^{(0)}\omega^{(2)}\left(S^{(1)}\right)^2>_{conn}
= \lim_{\hat{T} \to \infty} \frac{1}{\hat{T}}
<S^{(0)}\omega^{(2)}\left(S^{(1)}\right)^2>_{conn}.
\end{equation}
So, summarizing, for the vacuum polarization graphs with an insertion, the
result is:
\begin{eqnarray}
&&\lim_{\hat{T} \to \infty}
\Bigg(-<L_s^{(0)}\omega^{(2)}S^{(2)}>_{conn} +
<L_s^{(0)}\omega^{(2)}\frac{1}{2}\left(S^{(1)}\right)^2>_{conn} 
\nonumber \\ && - <L_s^{(0)}\omega^{(2)}S^{(2)}_{FP}>_{conn} -
<L_s^{(0)}\omega^{(2)}S^{(2)}_{meas}>_{conn} \Bigg) \nonumber \\
&=&\lim_{\hat{T} \to \infty} \frac{1}{\hat{T}}
\Bigg(-<S_s^{(0)}\omega^{(2)}S^{(2)}>_{conn} +
<S_s^{(0)}\omega^{(2)}\frac{1}{2}\left(S^{(1)}\right)^2>_{conn} 
\nonumber \\ && - <S_s^{(0)}\omega^{(2)}S^{(2)}_{FP}>_{conn} -
<S_s^{(0)}\omega^{(2)}S^{(2)}_{meas}>_{conn} \Bigg),
\end{eqnarray}
where now insertions into \emph{all} gluon lines are allowed.

What remains are the two additional vacuum polarization graphs, incorporating
$L_s^{(1)}$ and $L_s^{(2)}$, analogous to the first two graphs in figure
\ref{L_graphs}, but with the sum over the plaquettes restricted to a
\emph{fixed} time slice. The operators $L_s^{(1)}$ and $L_s^{(2)}$ both only
contain \emph{spatial} links, but in the limit of large $\hat{T}$, only the
$4-4-$component of the vacuum polarization tensor $\Pi$ contributes - hence
the contributions of these graphs vanish in the limit:
\begin{equation}
\lim_{\hat{T} \to \infty} <L_s^{(1)}S^{(1)}\omega^{(2)}>_{conn}
= \lim_{\hat{T} \to \infty} <L_s^{(2)}\omega^{(2)}>_{conn} = 0.
\end{equation}
Again that is identical to the results one obtains if one had inserted
$S_s$ instead of $L_s$, divided by $\hat{T}$ and taken the limit:
\begin{equation}
\lim_{\hat{T} \to \infty} \frac{<S_s^{(1)}S^{(1)}\omega^{(2)}>_{conn}}{\hat{T}}
= \lim_{\hat{T} \to \infty} \frac{<S_s^{(2)}\omega^{(2)}>_{conn}}{\hat{T}} = 0.
\end{equation}

\vspace{3ex}

Therefore one sees that for \emph{all} vacuum polarization graphs, in the
limit of large $\hat{T}$, the contributions coming from an insertion of $S_s$
(and dividing by $\hat{T}$) are exactly the same as the contributions coming
from an insertion of $L_s$, so for these graphs, the restriction to one fixed
time slice works.

\subsubsection{The spider graphs}

In section \ref{ESR_check_NLO}, it was seen that the two spider graphs do not
contribute in the limit of large $\hat{T}$. Therefore if one calculates them
with $L_s$ instead of $S_s$, their contributions should also go to zero.

For the first spider graph, corresponding to $<L_s^{(1)}
\omega^{(3)}>_{conn}$, one can use the same argument as in section
\ref{ESR_check_NLO}: $L_s^{(1)}$ contains a (slightly modified) three-gluon
vertex, which connects only gluons with a spatial polarization with each
other, and does not conserve the fourth component of the momentum. Using
Feynman gauge, the polarizations of the gluons at the vertex are the same as
the one on the Wilson loop; and because there is only one spatial direction
available on the Wilson loop, all three gluons meeting at the vertex
have the \emph{same} polarization. But for three gluons with the same
polarization, the three-gluon vertex vanishes - hence the result is simply:
\begin{equation}
<L_s^{(1)} \omega^{(3)}>_{conn} = 0
\end{equation}
- and that is equal to the result for $<S_s^{(1)} \omega^{(3)}>_{conn}$. Thus
obviously one has
\begin{equation}
\lim_{\hat{T} \to \infty} \frac{1}{\hat{T}} <S_s^{(1)} \omega^{(3)}>_{conn} =
\lim_{\hat{T} \to \infty} <L_s^{(1)} \omega^{(3)}>_{conn}.
\end{equation}

The treatment of the second spider graph is more complicated; it has to be
calculated explicitly. Fortunately one can use the results of section
\ref{ASR_restriction_NLO} for this; using that here only \emph{spatial}
plaquettes are inserted, the formula obtained there simplifies considerably,
so that in the end one gets:
\begin{eqnarray}
&& <L_s^{(0)} S^{(1)} \omega^{(3)}> \nonumber \\ &\sim&
\int\limits_{BZ} \frac{d^4p}{(2\pi)^4} \int\limits_{BZ} \frac{d^4k}{(2\pi)^4}
\int\limits_{-\pi}^{\pi} \frac{dq_4}{2\pi} \sin(p_3 \hat{R}/2) \sin(p_4
\hat{T}/2) \nonumber \\
&& \Bigg[ 2 i \frac{(\widehat{p+2k})_4 \cos(p_3/2) \sum_{j \ne
\mu} (\widehat{p+k})_j^2}{\hat{p}^2 \hat{k}^2 (\widehat{p+k})^2
\left(\left(\widehat{\vec{p}+\vec{k}}\right)^2 + \hat{q}_4^2 \right)}
\sin((k+q)_4 \hat{T}/2) O_R(p_3+k_3, -k_3) \nonumber \\
&& + 2 i \left(\frac{(\widehat{p+q})_4 \cos((p+k)_3/2) \left(\hat{\vec{k}}^2 -
\hat{k}_3^2\right) + (\widehat{2p+k})_3 \cos(q_4/2) \hat{\vec{k}}^2}
{\hat{p}^2 \hat{k}^2 \left(\hat{\vec{k}}^2 + \hat{q}_4^2
\right) \left(\left(\widehat{\vec{p}+\vec{k}}\right)^2 +
\left(\widehat{p-q}\right)_4^2 \right)} \right. \nonumber \\ &&
- \left. \frac{\sum_j (\widehat{2p+k})_j \hat{k}_j \hat{k}_3 \cos(q_4/2)}
{\hat{p}^2 \hat{k}^2 \left(\hat{\vec{k}}^2 + \hat{q}_4^2
\right) \left(\left(\widehat{\vec{p}+\vec{k}}\right)^2 +
\left(\widehat{p-q}\right)_4^2 \right)} \right) \nonumber
\\ && \cdot \frac{\sin(k_3 \hat{R}/2)}{\sin(k_3/2)} \frac{\sin((p-q)_4
\hat{T}/2)}{\sin((p-q)_4/2)} \cos((p+k)_3 \hat{R}/2) \cos(k_4 \hat{T}/2)
\nonumber \\ && + 2 i \left(\frac{(\widehat{2p+k})_3 \cos(k_3/2) \hat{k}_3^2
- \sum_j (\widehat{2p+k})_j \hat{k}_j \hat{k}_3 \cos(k_3/2)}
{\hat{p}^2 \hat{k}^2 \left(\hat{\vec{k}}^2 + \hat{q}_4^2
\right) \left(\left(\widehat{\vec{p}+\vec{k}}\right)^2 +
\left(\widehat{p-q}\right)^2_4 \right)} \right. \nonumber \\ &&
- \left. \frac{(\widehat{p-2q})_4 \cos(p_3/2)
\left(\hat{\vec{k}}^2 - \hat{k}_3^2\right)}
{\hat{p}^2 \hat{k}^2 \left(\hat{\vec{k}}^2 + \hat{q}_4^2
\right) \left(\left(\widehat{\vec{p}+\vec{k}}\right)^2 +
\left(\widehat{p-q}\right)_4^2 \right)} \right) \nonumber \\ && \cdot
\sin((k+q-p)_4 \hat{T}/2) O_R(p_3+k_3, -k_3) \nonumber \\
&& - 2 i \frac{(\widehat{k+q})_4 \cos((p+k)_3/2) \left(\hat{\vec{p}}^2
- \hat{p}_3^2 \right)}
{\hat{p}^2 \hat{k}^2 \left(\hat{\vec{p}}^2 + \hat{q}_4^2
\right) \left(\left(\widehat{\vec{p}+\vec{k}}\right)^2 +
\left(\widehat{k-q}\right)^2_4 \right)}
\frac{\sin(k_3 \hat{R}/2)}{\sin(k_3/2)} \frac{\sin((k-q)_4
\hat{T}/2)}{\sin((k-q)_4/2)} \nonumber \\ && \cdot \cos((p+k)_3 \hat{R}/2)
\cos(k_4 \hat{T}/2) \Bigg].
\end{eqnarray}
A careful analysis of the individual terms reveals that for large $\hat{T}$,
all of them go to zero, so that one indeed gets:
\begin{equation}
\lim_{\hat{T} \to \infty} \frac{1}{\hat{T}} <S_s^{(0)}
S^{(1)} \omega^{(3)}>_{conn} = \lim_{\hat{T} \to \infty} <L_s^{(0)}
S^{(1)} \omega^{(3)}>_{conn} = 0.
\end{equation}

\subsubsection{The graphs with two independent gluon lines}

There are again two graphs to consider here: first, the disconnected one,
corresponding to $<L_s^{(0)} \omega^{(2)}>_{conn} <\omega^{(2)}>_0$, and
second the one containing $\omega^{(4)}$. But using the results obtained in
leading order, where it was shown that $<L_s^{(0)} \omega^{(2)}>_{conn} = 0$,
the disconnected contribution vanishes. In the same way, one can show that
\[
<L_s^{(0)} \omega^{(4A)}>_{conn} = 0.
\]
For the remaining graphs, it is convenient to distinguish between the
following three classes again, as already in section \ref{ESR_check_NLO}. The
argumentation is also very similar:
\begin{enumerate}
\item
Graphs in which only temporal links of the Wilson loop appear; these links
can not be connected with the spatial plaquettes using the gluon propagators in
Feynman gauge, and therefore their contributions vanish.
\item
Graphs in which only spatial links of the Wilson loop appear; these will be
discussed below. The relevant parts of $\omega^{(4)}$ will be denoted by
$\omega^{(4)}_{RR}$.
\item Graphs in which temporal as well as spatial links appear - these are the
most complicated ones and will be also discussed below. The relevant parts of
$\omega^{(4)}$ will be denoted by $\omega^{(4)}_{RT}$.
\end{enumerate}
As usual, $\omega^{(4)}$ is split into the parts $B$ to $F$ (part $A$ has
already been treated above), and as usual, $\omega^{(4F)}$ gives no
contribution at all. 

Here again the results from section \ref{ASR_restriction_NLO} can be used, but 
because only spatial plaquettes are inserted, the formulas simplify
considerably. The graphs of the second category give the following result:
\begin{eqnarray}
&&<L_s^{(0)} \omega^{(4B)}_{RR}> \nonumber \\ &=& - \frac{C_2(G) C_2(F)}{2}
\int\limits_{BZ} \frac{d^4p}{(2\pi)^4} \int\limits_{BZ} \frac{d^4k}{(2\pi)^4}
\int\limits_{-\pi}^{\pi} \frac{dq_4}{2\pi} \frac{1}{\hat{k}^2}
\frac{\hat{\vec{p}}^2 - \hat{p}_3^2}{\hat{p}^2 \left(\hat{\vec{p}}^2 +
\hat{q}_4^2 \right)} \nonumber \\ &&
\cdot \Bigg[ \sin((k-p)_4 \hat{T}/2) \sin((k+q)_4 \hat{T}/2)
\frac{\sin^2(p_3 \hat{R}/2)}{\sin^2(p_3/2)}
\frac{\sin^2(k_3 \hat{R}/2)}{\sin^2(k_3/2)} \nonumber \\
&& + 4 \sin((p+k)_4 \hat{T}/2) \sin((q-k)_4 \hat{T}/2) O_R^2(p_3, k_3) 
\Bigg] \\
&&<L_s^{(0)} \omega^{(4C)}_{RR}> \nonumber \\ &=& \frac{C_2(F) C_2(G)}{2}
\int\limits_{BZ} \frac{d^4p}{(2\pi)^4} \int\limits_{BZ} \frac{d^4k}{(2\pi)^4}
\int\limits_{-\pi}^{\pi} \frac{dq_4}{2\pi} \frac{1}{\hat{k}^2}
\frac{\hat{\vec{p}}^2 - \hat{p}_3^2}{\hat{p}^2 \left(\hat{\vec{p}}^2 +
\hat{q}_4^2 \right)} \nonumber \\ &&
\cdot \sin^2(k_4 \hat{T}/2) \frac{\sin(k_3 \hat{R}/2)}{\sin(k_3/2)}
\cos(p_4 \hat{T}/2) \cos(q_4 \hat{T}/2) \\ && \cdot
\Bigg[2 \left.\left(\Sigma_1 - \Sigma_2 \right)\right|_{p_3 \leftrightarrow
k_3} + \frac{\sin(k_3 \hat{R}/2)}{\sin(k_3/2)} \Sigma_R(p_3, -p_3)
+ \frac{\sin(p_3 \hat{R}/2)}{\sin(p_3/2)} \Sigma_R(p_3,k_3) \Bigg]
\nonumber \\
&&<L_s^{(0)} \omega^{(4D)}_{RR}> \nonumber \\ &=& \frac{C_2(F) C_2(G)}{6}
\int\limits_{BZ} \frac{d^4p}{(2\pi)^4} \int\limits_{BZ} \frac{d^4k}{(2\pi)^4}
\int\limits_{-\pi}^{\pi} \frac{dq_4}{2\pi} \frac{1}{\hat{k}^2}
\frac{\hat{\vec{p}}^2 - \hat{p}_3^2}{\hat{p}^2 \left(\hat{\vec{p}}^2 +
\hat{q}_4^2 \right)} \nonumber \\ &&
\cdot \sin^2(k_4 \hat{T}/2) \frac{\sin(k_3 \hat{R}/2)}{\sin(k_3/2)}
\cos(p_4 \hat{T}/2) \cos(q_4 \hat{T}/2) \nonumber \\ && \cdot
\Bigg[2 \left.\left(\Sigma_1 - \Sigma_2 \right)\right|_{p_3 \leftrightarrow
k_3} + 2 \, \Sigma_R(0, -k_3) - 2 \, \Sigma_R(p_3-k_3,-p_3) \nonumber \\
&& + 2 \frac{\sin(k_3 \hat{R}/2)}{\sin(k_3/2)} \frac{\sin^2(p_3
\hat{R}/2)}{\sin^2(p_3/2)} - \frac{\sin(p_3 \hat{R}/2)}{\sin(p_3/2)}
\Sigma_R(p_3, k_3) \nonumber \\ && + \frac{\sin(k_3 \hat{R}/2)}{\sin(k_3/2)}
\Sigma_R(p_3, -p_3) \Bigg] \\
&&<L_s^{(0)} \omega^{(4E)}_{RR}> \nonumber \\ &=& \frac{C_2(F) C_2(G)}{6}
\int\limits_{BZ} \frac{d^4p}{(2\pi)^4} \int\limits_{BZ} \frac{d^4k}{(2\pi)^4}
\int\limits_{-\pi}^{\pi} \frac{dq_4}{2\pi} \frac{1}{\hat{k}^2}
\frac{\hat{\vec{p}}^2 - \hat{p}_3^2}{\hat{p}^2 \left(\hat{\vec{p}}^2 +
\hat{q}_4^2 \right)} \nonumber \\
&& \cdot \cos(p_4 \hat{T}/2) \cos(q_4 \hat{T}/2) \sin^2(k_4 \hat{T}/2)
\frac{\sin(k_3 \hat{R}/2)}{\sin(k_3/2)} \Bigg[
\frac{\sin(k_3 \hat{R}/2)}{\sin(k_3/2)} \hat{R} \nonumber \\
&& + \frac{\sin(k_3 \hat{R}/2)}{\sin(k_3/2)} \frac{\sin(p_3
\hat{R}/2)}{\sin(p_3/2)} \frac{\sin((p+k)_3 \hat{R}/2)}{\sin((p+k)_3/2)}
\Bigg].
\end{eqnarray}
Because of the fast oscillations of the two cosines for $\hat{T} \to \infty$,
the contributions of these integrals vanish in the limit of large $\hat{T}$.
On the other hand, the graphs of the third category give:
\begin{eqnarray}
&&<L_s^{(0)} \omega^{(4B)}_{RT}> \nonumber \\ &=& - \frac{C_2(G) C_2(F)}{2}
\int\limits_{BZ} \frac{d^4p}{(2\pi)^4} \int\limits_{BZ} \frac{d^4k}{(2\pi)^4}
\int\limits_{-\pi}^{\pi} \frac{dq_4}{2\pi} \frac{1}{\hat{k}^2}
\frac{\hat{\vec{p}}^2 - \hat{p}_3^2}{\hat{p}^2 \left(\hat{\vec{p}}^2 +
\hat{q}_4^2 \right)} \nonumber \\ && \cdot \frac{\sin^2(p_3
\hat{R}/2)}{\sin^2(p_3/2)} \frac{\sin^2(k_4 \hat{T}/2)}{\sin^2(k_4/2)}
\cos(p_4 \hat{T}/2) \cos(q_4 \hat{T}/2) \\
&& <L_s^{(0)} \omega^{(4C)}_{RT}> \nonumber \\ &=& \frac{C_2(G) C_2(F)}{2}
\int\limits_{BZ} \frac{d^4p}{(2\pi)^4} \int\limits_{BZ} \frac{d^4k}{(2\pi)^4}
\int\limits_{-\pi}^{\pi} \frac{dq_4}{2\pi} \frac{1}{\hat{k}^2}
\frac{\hat{\vec{p}}^2 - \hat{p}_3^2}{\hat{p}^2 \left(\hat{\vec{p}}^2 +
\hat{q}_4^2 \right)} \nonumber \\
&& \cdot \Bigg[\frac{\sin^2(k_4 \hat{T}/2)}{\sin^2(k_4/2)} \sin^2(k_3
\hat{R}/2) \cos(p_4 \hat{T}/2) \cos(q_4 \hat{T}/2) \Sigma_R(p_3, -p_3)
\nonumber \\ && + \frac{\sin^2(k_4 \hat{T}/2)}{\sin^2(k_4/2)} \sin^2(k_3
\hat{R}/2) \cos(p_4 \hat{T}/2) \cos(q_4 \hat{T}/2) \frac{\sin^2(p_3
\hat{R}/2)}{\sin^2(p_3/2)} \Bigg] \\
&& <L_s^{(0)} \omega^{(4D)}_{RT}> \nonumber \\ &=& \frac{C_2(G) C_2(F)}{6}
\int\limits_{BZ} \frac{d^4p}{(2\pi)^4} \int\limits_{BZ} \frac{d^4k}{(2\pi)^4}
\int\limits_{-\pi}^{\pi} \frac{dq_4}{2\pi} \frac{1}{\hat{k}^2}
\frac{\hat{\vec{p}}^2 - \hat{p}_3^2}{\hat{p}^2 \left(\hat{\vec{p}}^2 +
\hat{q}_4^2 \right)} \nonumber \\
&&\cdot \Bigg[-\frac{\sin^2(k_4 \hat{T}/2)}{\sin^2(k_4/2)} \sin^2(k_3
\hat{R}/2) \cos(p_4 \hat{T}/2) \cos(q_4 \hat{T}/2) \Sigma_R(p_3, -p_3)
\nonumber \\ && + 2 \frac{\sin^2(k_4 \hat{T}/2)}{\sin^2(k_4/2)} \sin^2(k_3
\hat{R}/2) \cos(p_4 \hat{T}/2) \cos(q_4 \hat{T}/2) \frac{\sin^2(p_3
\hat{R}/2)}{\sin^2(p_3/2)} \Bigg] \nonumber \\ \\
&& <L_s^{(0)} \omega^{(4E)}_{RT}> \nonumber \\ &=& \frac{C_2(G) C_2(F)}{6}
\int\limits_{BZ} \frac{d^4p}{(2\pi)^4} \int\limits_{BZ} \frac{d^4k}{(2\pi)^4}
\int\limits_{-\pi}^{\pi} \frac{dq_4}{2\pi} \frac{1}{\hat{k}^2}
\frac{\hat{\vec{p}}^2 - \hat{p}_3^2}{\hat{p}^2 \left(\hat{\vec{p}}^2 +
\hat{q}_4^2 \right)} \nonumber \\
&& \cdot \frac{\sin^2(k_4 \hat{T}/2)}{\sin^2(k_4/2)} \sin^2(k_3 \hat{R}/2)
\cos(p_4 \hat{T}/2) \cos(q_4 \hat{T}/2) \hat{R}.
\end{eqnarray}
As already pointed out in section \ref{ASR_restriction_NLO}, for $\hat{T} \to
\infty$, the factor $\frac{\sin^2(k_4 \hat{T}/2)}{\sin^2(k_4/2)}$ gives a
linear dependence on $\hat{T}$, but the two cosines both give factors of
$\hat{T}^{-1}$, so that in total all of these integrals go with $\hat{T}^{-1}$
in the limit and hence vanish. Thus one gets:
\begin{equation}
\lim_{\hat{T} \to \infty} <L_s^{(0)} \omega^{(4)}>
= \lim_{\hat{T} \to \infty} \frac{<S_s^{(0)} \omega^{(4)}>}{\hat{T}} = 0.
\end{equation}

\vspace{3ex}

Taking together all of the results obtained in the last subsections, one sees
that indeed
\begin{equation}
\lim_{\hat{T} \to \infty} <L_s>_{q\bar{q}-0}
= \lim_{\hat{T} \to \infty} \frac{<S_s>_{q\bar{q}-0}}{\hat{T}}
\end{equation}
is satisfied up to next-to-leading order---up to that order, the restriction
to one fixed time slice works for the spatial plaquettes. Thus the
check of the energy sum rule is now completed.

\newpage
\thispagestyle{empty}
\cleardoublepage
\chapter{Summary}

In this work, it was shown that in lattice perturbation theory, both
sum rules, (\ref{ActionSumRule}) and (\ref{EnergySumRule}), hold up to
next-to-leading order in the coupling constant. Additionally, the possibility
to restrict the expectation value of the action and the expectation
value of the magnetic field energy to the sum of the plaquettes on a fixed
time slice has been investigated. Two spin-offs of the check were a proof of
the transversality of the gluonic vacuum polarization on the lattice in
leading order and a proof of the gauge invariance of the expectation value of
the Wilson loop up to next-to-leading order. This is not completely obvious,
since gauge transformations on the lattice are implemented via unitary
transformations of the link variables.

\vspace{2ex}

The scaling behaviour of the potential which is used in the derivation of the
action sum rule was checked explicitly by using known results for the
potential \cite{Kovacs, Karsch, wrongpotential}. The crucial part of this sum
rule is the identity (\ref{potentialderivative}) respectively 
(\ref{actionidentity}). The perturbative examination of this identity yielded
methods and valuable results which could be used in the examination of the
energy sum rule.

Additionally, it opened up a way to proof the gauge invariance of the
expectation value of the Wilson loop perturbatively up to next-to-leading
order. By expressing this expectation value as a polynomial with respect to
the gauge parameter, it was possible to show that its dependance on the gauge
parameter vanishes in all orders of the gauge parameter and up to
next-to-leading order of the coupling constant. For eliminating the dependance
on the gauge parameter in first order, the action sum rule could be used.

The examination of the restriction of the expectation value of the action to
one fixed time slice (the possibility to replace this expectation value by
$\hat{T}$ times the expectation value of the Lagrangian, the sum of all
plaquettes on a fixed time slice) turned out to be much more difficult. It was
first shown that this is possible in leading order. For the graphs in
next-to-leading order, this was accomplished for the vacuum polarization
graphs, the spider graphs and some of the graphs with two independent gluon
lines; work on the rest is still in progress. The eight- and nine-dimensional
integrals have to be evaluated numerically, and that requires lots of computer
time.

\vspace{2ex}

On the other hand, all parts of the energy sum rule were shown to be true
up to next-to-leading order: first it was checked without the restriction to a
fixed time slice. It was shown that both the expectation value of the magnetic
field energy and the contribution from the trace anomaly vanish in leading
order, so that the only contribution to the potential in that order stems from
the energy in the electric fields.

In next-to-leading order it turned out that the expectation value of the
(euclidean) energy in the magnetic fields receives its only contributions from
the graphs in which the sum over the spatial plaquettes is inserted into
internal lines of the gluonic vacuum polarization.  By computing the
contributions of these graphs numerically and examining the two parts of the
potential with different group theoretical factors separately, the validity of
the energy sum rule was confirmed with good numerical accuracy.

For the expectation value of the magnetic energy it was possible to show that
it can be restricted to one fixed time slice. The leading order was again
relatively easy and could be demonstrated and explained explicitly. In the
next-to-leading order, similar arguments as before concerning the expectation
value of the action could be used to take care of the vacuum polarization
graphs. The contributions from the spider graphs and from the graphs with two
independent gluon lines were calculated explicitly and shown to vanish, as
expected.

\vspace{2ex}

Now the questions which were posed in the introduction can be addressed
using the results obtained in the checks. Unfortunately the conclusions which
were obtained here are not very helpful because they apply only if the
coupling constant is small enough to allow perturbation theory to give
sensible results. In contrast, in real physical systems like mesons, the
coupling constant is large and perturbation theory breaks down. But the
proven validity of the sum rules in the small coupling regime suggests that
they are true also in the non-perturbative region, hence it it still possible
to draw some sensible conclusions from the results.

The first interesting point to notice is that there are cancellations
between the expectation values of the energy in the electric and in the
magnetic fields: the (euclidean) electric field energy is always negative, the
magnetic field energy is positive. This is true for both contributions (with
different group theoretical factors) to the potential. But here the magnitude
of the magnetic field energy is much smaller than the magnitude of the
electric field energy, in contrast to Monte Carlo simulations in the
non-perturbative regime where they both have comparable sizes. Hence the
problem of large cancellations between these two contributions which appears
in these simulations does not appear in the weak coupling limit.

Another result is that the trace anomaly contributes only a very small part to
the potential for large quark-antiquark separations; this is in strong contrast
to the case of a confining potential, where it contributes exactly one half of
the potential energy!

\vspace{4ex}

To close, I will give an outlook to the open problems, possible extensions and
future projects. Now that the sum rules have been checked perturbatively in the
regime of a small coupling constant, one should use Monte Carlo simulations in
order to perform a check in the physical regime of large coupling constant.
There the same questions will be interesting as investigated here: the
problem of cancellations between the contributions from the magnetic and
electric field energies and the contribution of the trace anomaly, which is
expected to have the same magnitude as the energy in the fields in the
confining region. Additionally, a look at the sum rule for the glueball mass
would be interesting. 

Another promising approach is using a nonperturbative model like the
sto\-chastic vacuum model of Dosch and Sominov \cite{MSV}. A recent
calculation of the quark-antiquark potential in this model and subsequent
comparison with the predictions of the lattice sum rules gave good consistency
\cite{Frank}.

\vspace{2ex}

On the other hand, one could use the sum rules to check the consistency of the
results obtained in Monte Carlo simulations. One first step in this direction
was done already in \cite{Bali}, where a (corrected) version of Michael's
action sum rule was used.

\vspace{2ex}

A possible extension is the incorporation of dynamical fermions into the sum
rules, so that one could study the effects of string breaking more closely.
Alternatively the sum rules could be investigated for finite temperature,
which should shed some light on the phase transition to the quark-gluon plasma.

Hence there is much potential for future work on the lattice sum rules, and
lots of additional interesting results can be expected.

\newpage
\thispagestyle{empty}
\cleardoublepage
\appendix
\chapter{General SU(N) formulas}

\section{Basics}

The Lie group $G=SU(N)$ has $N^2-1$ generators. In the fundamental
representation of the group $SU(N)$, these generators, denoted by $T^A$, are
given by hermitian, traceless, complex $N \times N$ matrices. They obey the
following basic commutation and anticommutation relations:
\begin{eqnarray}
[T^A,T^B] &=& i \sum_C f_{ABC} T^C \\
\{T^A,T^B\} &=& \frac{1}{d(F)} \delta^{AB} id_{d(R)} + \sum_C d_{ABC} T^C,
\end{eqnarray}
where $d(F) = N$ is the dimension of the fundamental representation $F$
and $id_{d(F)}$ denotes the $d(F)$-dimensional identity matrix. The
real numbers $f_{ABC}$ are called \emph{structure constants}.

In the adjoint representation, the generators are denoted by $t^A$ and are
given by complex $(N^2-1)\times(N^2-1)$ matrices, whose elements are:
\begin{equation}
\left(t^A\right)_{BC} = - i f_{ABC}.
\end{equation}

\section{Traces}

For the generators $T^A$ in any representation $R$, one always has:
\begin{equation}
\mbox{Tr}(T^AT^B) = T(R) \delta^{AB}
\end{equation}
with a constant $T(R)$ depending on the representation. For the fundamental
representation, $T(R)$ is simply $\frac{1}{2}$; for the adjoint
representation, it is $N$. Using this and the commutators and anticommutators
given above, one gets:
\begin{eqnarray}
\mbox{Tr}(T^AT^B) &=& \frac{1}{2} \delta^{AB} \\
\mbox{Tr}(T^AT^BT^C) &=& \frac{1}{4} (d_{ABC} + i f_{ABC}) \\
\mbox{Tr}(T^AT^BT^CT^D) &=& \frac{1}{4d(F)} \delta_{AB}\delta_{CD}
- \frac{1}{8} \sum_E (f_{ABE} f_{CDE} - d_{ABE} d_{CDE}) \nonumber \\
&&+ \frac{i}{8} \sum_E (f_{ABE} d_{CDE} + d_{ABE} f_{CDE}) \\
\mbox{Tr}(t^At^B) &=& N \delta_{AB}.
\end{eqnarray}

\section{Sums}

For any representation $R$, the quadratic Casimir operator $C_2(R)$ is defined
by
\begin{equation}
\sum_A T^A T^A =: C_2(R) id_{d(R)}.
\end{equation}
In the special case of the fundamental representation $F$, one gets
\begin{equation}
C_2(F) = \frac{N^2-1}{2N},
\end{equation}
and for the adjoint representation $G$:
\begin{equation}
C_2(G) = N.
\end{equation}
Additionally, the following relation holds for every representation $R$:
\begin{equation}
T(R) = \frac{C_2(R) d(R)}{d(G)},
\end{equation}
where $d(G)$ is the dimension of the adjoint representation, which is equal
to the order of the group, i.\ e.\ the number of its generators. Hence for
$SU(N)$, the formula gives:
\begin{equation}
T(R) = \frac{C_2(R) d(R)}{N^2-1}.
\end{equation}
This agrees with the results for $T(R)$ given above for the fundamental as
well as the adjoint representation.

Using the formulas given above and the symmetry respectively anti-sym\-metry of
the structure constants and the $d_{ABC}$, the following sums can be evaluated:
\begin{eqnarray}
\sum_A d_{AAB} &=& 0 \\
\sum_{D,E} f_{ADE} f_{CDE} &=& C_2(G) \delta_{AC} = N \delta_{AC} \\
\sum_{D,E} d_{ADE} f_{CDE} &=& 0 \\
\sum_{D,E} d_{ADE} d_{CDE} &=& \left(4 C_2(F) - \frac{2}{d(F)} - C_2(G)\right)
\delta_{AC} = \frac{N^2-4}{N} \delta_{AC} \\
\sum_{A,B,C} f_{ABC} f_{ABC} &=& C_2(G) d(G) = N (N^2-1) \\
\sum_{A,B,C} d_{ABC} f_{ABC} &=& 0 \\
\sum_{A,B,C} d_{ABC} d_{ABC} &=& d(G) \left(4 C_2(F) - \frac{2}{d()} -
C_2(G)\right) = \frac{(N^2-1)(N^2-4)}{N}. \nonumber \\
\end{eqnarray}

\chapter{Sums along the Wilson Loop}

\label{sums}

Some sums along the Wilson loop appear so often that it is convenient to
summarize them here.

\section{Unrestricted sums}

The easiest sum is the one which contains only one vector potential:
\begin{eqnarray}
&&\sum_l A_l \nonumber \\
&=& \sum_{l=0}^{\hat{R}-1} A_{\mu}(n_0 + l \hat{\mu})
+ \sum_{l=0}^{\hat{T}-1} A_{\nu}(n_0 + \hat{R} \hat{\mu} + l \hat{\nu})
- \sum_{l=0}^{\hat{R}-1} A_{\mu}(n_0 + \hat{R} \hat{\mu} + \hat{T} \hat{\nu} -
l \hat{\mu}) \nonumber \\ &&
- \sum_{l=0}^{\hat{T}-1} A_{\nu}(n_0 + \hat{T} \hat{\nu} - l \hat{\nu})
\end{eqnarray}
Insert the Fourier representation for $A$:
\begin{equation}
A_{\alpha}(x) = \int\limits_{BZ} \frac{d^4p}{(2\pi)^4} A_{\alpha}(p)
e^{ipx+ip_{\alpha/2}}.
\end{equation}
This yields:
\begin{eqnarray}
&&\sum_l A_l \nonumber \\
&=& \sum_{\alpha} \int\limits_{BZ} \frac{d^4p}{(2\pi)^4} A_{\alpha}(p)
e^{ipn_0} \left(\delta_{\alpha\mu} \sum_{l=0}^{\hat{R}-1}
e^{ip_{\mu}l+ip_{\mu}/2} + \delta_{\alpha\nu} \sum_{l=0}^{\hat{T}-1}
e^{ip_{\mu} \hat{R} + i p_{\nu} l + i p_{\nu}/2 } \right. \nonumber \\
&& \left. - \delta_{\alpha\mu} \sum_{l=0}^{\hat{R}-1} e^{ip_{\mu}
\hat{R} + i p_{\nu} \hat{T} - i p_{\mu}l - ip_{\mu}/2}
- \delta_{\alpha\nu} \sum_{l=0}^{\hat{T}-1} e^{i p_{\nu} \hat{T} - i
p_{\nu}l - ip_{\nu}/2} \right) \nonumber \\
&=& \sum_{\alpha} \int\limits_{BZ} \frac{d^4p}{(2\pi)^4} A_{\alpha}(p)
e^{ipn_0} \left(\delta_{\alpha\mu} \frac{e^{ip_{\mu}\hat{R}}-1}{2i
\sin(p_{\mu}/2)} + \delta_{\alpha\nu} \frac{e^{ip_{\nu}\hat{T}}-1}{2i
\sin(p_{\nu}/2)} e^{i p_{\mu} \hat{R}} \right. \nonumber \\
&& \left. - \delta_{\alpha\mu} \frac{e^{-ip_{\mu}\hat{R}}-1}{-2i
\sin(p_{\mu}/2)} e^{ip_{\mu} \hat{R} + i p_{\nu} \hat{T}}
- \delta_{\alpha\nu} \frac{e^{-ip_{\nu}\hat{T}}-1}{-2i
\sin(p_{\nu}/2)} e^{i p_{\nu} \hat{T}} \right) \nonumber \\
&=& \!\! \sum_{\alpha} \int\limits_{BZ} \frac{d^4p}{(2\pi)^4} A_{\alpha}(p)
e^{ip(n_0+ \hat{\mu}\hat{R}/2+ \hat{\nu} \hat{T}/2)}
\left[\delta_{\mu\alpha} \frac{\sin(p_{\mu}
\hat{R}/2)}{\sin(p_{\mu}/2)} \left(e^{-ip_{\nu} \hat{T}/2}
- e^{ip_{\nu} \hat{T}/2} \right) \right. \nonumber \\ && \left. -
\delta_{\nu\alpha} \frac{\sin(p_{\nu} \hat{T}/2)}{\sin(p_{\nu}/2)}
\left(e^{-ip_{\mu} \hat{R}/2} - e^{ip_{\nu} \hat{R}/2} \right) \right]
\nonumber \\
&=& \!\! -2i \sum_{\alpha} \! \int\limits_{BZ}
\frac{d^4p}{(2\pi)^4} \sin(p_{\mu}\hat{R}/2) \sin(p_{\nu}\hat{T}/2)
e^{ip(n_0+\hat{\mu}\hat{R}/2+\hat{\nu}\hat{T}/2)}
\frac{A_{\alpha}(p)(\delta_{\mu\alpha} - \delta_{\nu\alpha})}
{\sin(p_{\alpha}/2)} \nonumber \\
\end{eqnarray}
Using this, one immediately gets:
\begin{eqnarray}
&&\sum_l <A^A_l A^B_{\beta}(x)>_0 \\
&=& -2i \int\limits_{BZ} \frac{d^4p}{(2\pi)^4}
\frac{\sin(p_{\mu}\hat{R}/2) \sin(p_{\nu}\hat{T}/2)}{\tilde{p}^2}
e^{ip(n_0+\hat{\mu}\hat{R}/2+\hat{\nu}\hat{T}/2-x)}
\frac{\delta_{\mu\beta} - \delta_{\nu\beta}}
{\sin(p_{\beta}/2)} \nonumber \\
&&\sum_{l_1,l_2} <A^A_{l_1} A^B_{l_2}>_0 \\
&=& +4 \delta^{AB} \int\limits_{BZ} \frac{d^4p}{(2\pi)^4}
\frac{\sin^2(p_{\mu}\hat{R}/2) \sin^2(p_{\nu}\hat{T}/2)}{\tilde{p}^2}
\left(\frac{1}{\sin^2(p_{\mu}/2)}+\frac{1}{\sin^2(p_{\nu}/2)}\right)
\nonumber
\end{eqnarray}
The last sum is the main ingredient of  $<\omega^{(2)}>_0$ and is also crucial
for calculating the vacuum polarization graphs.

\section{Restricted sums}

The following sum appears in $\omega^{(3)}$ as well as several times in
$\omega^{(4)}$ and thus is needed often:
\begin{eqnarray}
\sum_{l_1 < l_2} \left[A_{l_1}, A_{l_2}\right] &=&
\int\limits_{BZ} \frac{d^4q}{(2\pi)^4} \int\limits_{BZ} \frac{d^4k}{(2\pi)^4}
\left[A_{\beta}(q), A_{\gamma}(k)\right]
e^{i(q+k)(n_0+\hat{\mu}\hat{R}/2+\hat{\nu}\hat{T}/2)} \nonumber \\
&& \cdot \Bigg[ \Bigg\{-2 \delta_{\beta\mu} \delta_{\gamma\mu}
\sin((k+q)_{\nu}\hat{T}/2) O_T(k_{\mu}, q_{\mu}) \nonumber \\
&& + i \delta_{\beta\mu} \delta_{\gamma\mu}
\frac{\sin(k_{\mu}\hat{R}/2)\sin(q_{\mu}\hat{R}/2)}
{\sin(k_{\mu}/2)\sin(q_{\mu}/2)} \sin((k-q)_{\nu}\hat{T}/2) \nonumber \\
&& + \delta_{\beta\mu} \delta_{\gamma\nu}
\frac{\sin(k_{\nu}\hat{T}/2)\sin(q_{\mu}\hat{R}/2)}
{\sin(k_{\nu}/2)\sin(q_{\mu}/2)} \nonumber \\ && \,\,\, \cdot \left(
\cos(q_{\nu} \hat{T}/2 - k_{\mu} \hat{R}/2) +
i \sin(q_{\nu} \hat{T}/2 + k_{\mu} \hat{R}/2) \right)
\Bigg\} \nonumber \\ && -\Bigg\{\mu \leftrightarrow \nu, \hat{R}
\leftrightarrow \hat{T} \Bigg\} \Bigg]
\end{eqnarray}

\chapter{Some common integrals}

\label{commonintegrals}

In several calculations, integrals of the form
\begin{equation}
\int\limits_{BZ} \frac{d^4p}{(2\pi)^4} \frac{\hat{p}_{\mu}
\hat{p}_{\nu} \ldots}{(\hat{p}^2)^n}
\label{generalIntegral}
\end{equation}
appear. All of them can be evaluated exactly, leaving only one constant which
has to be determined numerically up to $n=2$, and an additional constant for
$n=3$. For convenience, all results here will be given for arbitrary dimension
$d$ as well as for the relevant case $d=4$. Hence the denominator is in
general given by:
\begin{equation}
\hat{p}^2 = \sum_{\mu=0}^d \hat{p}_{\mu}^2.
\end{equation}
Now first, the three most elementary integrals are:
\begin{eqnarray}
\int\limits_{BZ} \frac{d^dp}{(2\pi)^d} \frac{\hat{p}_{\mu}^2}{\hat{p}^2} &=&
\frac{1}{d} \\
\int\limits_{BZ} \frac{d^dp}{(2\pi)^d} \frac{1}{\hat{p}^2} &=:& \Delta_0 \\
\int\limits_{BZ} \frac{d^dp}{(2\pi)^d} \frac{\hat{p}_{\mu}}{\hat{p}^2} &=& 0,
\end{eqnarray}
the first following from the symmetry of the lattice, the third from the
fact that the integrand is odd, and the second simply being a
definition of the one remaining constant $\Delta_0$. This constant can be
calculated more easily by noting that the corresponding integral can be
rewritten in the following way:
\begin{eqnarray}
\int\limits_{BZ} \frac{d^dp}{(2\pi)^d} \frac{1}{\hat{p}^2} &=&
\int_0^{\infty} dt \int\limits_{BZ} \frac{d^dp}{(2\pi)^d} e^{-t \hat{p}^2} \nonumber
\\ &=& \int_0^{\infty} dt \int\limits_{BZ} \frac{d^dp}{(2\pi)^d} \exp\left(- 4 t
\sum_{\mu=0}^d \sin^2(p_{\mu}/2) \right) \nonumber \\
&=& \int_0^{\infty} dt \left(\int\limits_{-\pi}^{\pi} \frac{dp}{2\pi}
\exp\left(- 4 t \sin^2(p/2) \right) \right)^d \nonumber \\
&=& \int_0^{\infty} dt e^{-2 t d} \left(\int\limits_{-\pi}^{\pi}
\frac{dp}{2\pi} e^{ - 2 t \cos(p)} \right)^d \nonumber \\
&=& \frac{1}{2} \int_0^{\infty} dt \, e^{-t d} I_0^d(t),
\end{eqnarray}
where it has been used that the integral
\[
\int\limits_{-\pi}^{\pi} dx \, e^{-t \cos(x)}
\]
gives $2 \pi$ times the Bessel function $I_0(t)$. The remaining one-dimensional
integral can be evaluated using numerical integration routines much more
easily than the original $d$-dimensional one. In the relevant case $d=4$,
the explicit result is:
\begin{equation}
\Delta_0 \approx 0.154933.
\end{equation}

\noindent
From these three elementary integrals, one immediately gets:
\begin{equation}
\int\limits_{BZ} \frac{d^dp}{(2\pi)^d} \frac{\hat{p}_{\mu}^2}{(\hat{p}^2)^2} =
\frac{1}{d} \Delta_0.
\end{equation}
The general idea to treat the more complicated integrals can be found in
appendix B of \cite{Kawai}; they can be evaluated using partial integration.
For example:
\begin{eqnarray}
0 &=& \int\limits_{BZ} \frac{d^dp}{(2\pi)^d} \frac{\partial}{\partial p_{\mu}}
\frac{\sin(p_{\mu})}{\hat{p}^2} \nonumber \\
&=& \int\limits_{BZ} \frac{d^dp}{(2\pi)^d} \left(\frac{\cos(p_{\mu})}{\hat{p}^2}
- \frac{\sin(p_{\mu}) \cdot 4 \sin(p_{\mu}/2) \cos(p_{\mu}/2)}{(\hat{p}^2)^2}
\right) \nonumber \\
&=& \int\limits_{BZ} \frac{d^dp}{(2\pi)^d} \left(\frac{1 - 2
\sin(p_{\mu}/2)^2}{\hat{p}^2} - 8 \frac{\sin^2(p_{\mu}/2)
\cos^2(p_{\mu}/2)}{(\hat{p}^2)^2} \right) \nonumber \\
&=& \int\limits_{BZ} \frac{d^dp}{(2\pi)^d} \left(\frac{1 - \frac{1}{2}
\hat{p}_{\mu}^2}{\hat{p}^2} - 2 \frac{\hat{p}_{\mu}^2 - \frac{1}{4}
\hat{p}_{\mu}^4}{(\hat{p}^2)^2} \right)
\end{eqnarray}
and therefore, using the elementary integrals mentioned above:
\begin{equation}
\int\limits_{BZ} \frac{d^dp}{(2\pi)^d} \frac{\hat{p}_{\mu}^4}{(\hat{p}^2)^2}
= \frac{4-2d}{d} \Delta_0 + \frac{1}{d}.
\end{equation}
Now make again use of the symmetry of the lattice:
\begin{eqnarray*}
1 &=& \int\limits_{BZ} \frac{d^dp}{(2\pi)^d}
\frac{(\hat{p}^2)^2}{(\hat{p}^2)^2} \\ &=& d \int\limits_{BZ} \frac{d^dp}{(2\pi)^d}
\frac{\hat{p}_{\mu}^4}{(\hat{p}^2)^2} + (d^2-d) \int\limits_{BZ}
\frac{d^dp}{(2\pi)^d} \frac{\hat{p}_{\mu}^2 \hat{p}_{\nu}^2}{(\hat{p}^2)^2}
\end{eqnarray*}
with $\mu \ne \nu$ in the second term. This gives:
\begin{equation}
\int\limits_{BZ} \frac{d^dp}{(2\pi)^d} \frac{\hat{p}_{\mu}^2
\hat{p}_{\nu}^2}{(\hat{p}^2)^2} = \frac{2d-4}{d(d-1)} \Delta_0 \quad \quad
\mbox{for } \mu \ne \nu.
\end{equation}

Using the same methods, all integrals of the type (\ref{generalIntegral}) can
be evaluated. The results up to $n=2$ are given here:
\begin{eqnarray}
\int\limits_{BZ} \frac{d^dp}{(2\pi)^d} \frac{\hat{p}_{\mu}^2 \hat{p}_{\nu}^2}
{(\hat{p}^2)^2} &=& \delta_{\mu\nu} (\frac{4-2d}{d}\Delta_0 + \frac{1}{d}) +
(1-\delta_{\mu\nu}) \frac{2d-4}{d(d-1)} \Delta_0 \nonumber \\
&=& \delta_{\mu\nu} (-\Delta_0 + \frac{1}{4}) +
(1-\delta_{\mu\nu}) \frac{1}{3} \Delta_0 \nonumber \\
&=& (\frac{4-2d}{d-1} \Delta_0 + \frac{1}{d}) \delta_{\mu\nu} +
\frac{2d-4}{d(d-1)} \Delta_0 \nonumber \\
&=& (-\frac{4}{3} \Delta_0 + \frac{1}{4}) \delta_{\mu\nu} + \frac{1}{3}
\Delta_0 \\
\int\limits_{BZ} \frac{d^dq}{(2\pi)^d} \frac{\widehat{(p+q)}^2}{\hat{q}^2}
&=& \hat{p}^2 \left( \Delta_0 - \frac{1}{2d} \right) + 1.
\end{eqnarray}

For $n=3$, one needs an additional constant, which is defined in the following
way:
\begin{equation}
\Delta_1 := (d-4) \int\limits_{BZ} \frac{d^dp}{(2\pi)^d}
\frac{1}{\left(\hat{p}^2\right)^2}
\end{equation}
Then one gets the following results:
\begin{eqnarray}
\int\limits_{BZ} \frac{d^dp}{(2\pi)^d} \frac{\hat{p}_{\mu}^2}
{(\hat{p}^2)^3} &=& \frac{1}{d} \Delta_1 \\
\int\limits_{BZ} \frac{d^dp}{(2\pi)^d} \frac{\left(\hat{p}_{\mu}^2\right)^2}
{(\hat{p}^2)^3} &=& \frac{1}{2d} \Delta_0 - \frac{1}{d} \Delta_1 \\
\int\limits_{BZ} \frac{d^dp}{(2\pi)^d} \frac{\hat{p}_{\mu}^2 \hat{p}_{\nu}^2}
{(\hat{p}^2)^3} &=& \frac{1}{2d(d-1)} \Delta_0 + \frac{1}{d(d-1)} \Delta_1
\quad \quad \mbox{for } \mu \ne \nu \\
\int\limits_{BZ} \frac{d^dp}{(2\pi)^d} \frac{\left(\hat{p}_{\mu}^2\right)^3}
{(\hat{p}^2)^3} &=& \frac{3-2d}{d} \Delta_0 + \frac{1}{d} - \frac{4}{d}
\Delta_1 \\
\int\limits_{BZ} \frac{d^dp}{(2\pi)^d} \frac{\left(\hat{p}_{\mu}^2\right)^2
\hat{p}_{\nu}^2} {(\hat{p}^2)^3} &=& \frac{1}{d(d-1)} \Delta_0 +
\frac{4}{d(d-1)} \Delta_1 \quad \quad \mbox{for } \mu \ne \nu \\
\int\limits_{BZ} \frac{d^dp}{(2\pi)^d} \frac{\hat{p}_{\mu}^2
\hat{p}_{\nu}^2 \hat{p}_{\lambda}^2} {(\hat{p}^2)^3} &=&
\frac{2d-6}{d(d-1)(d-2)} \Delta_0 - \frac{8}{d(d-1)(d-2)} \Delta_1
\\ && \quad \quad \quad \quad \quad \quad \quad \quad \quad \quad \quad \quad
\quad \mbox{for } \mu \ne \nu \ne \lambda \nonumber
\end{eqnarray}
For $d \to 4$, the additional constant $\Delta_1$ can be calculated exactly by
extracting the infrared divergence: The integral
\[
\int\limits_{BZ} \frac{d^dp}{(2\pi)^d} \frac{1}{\left(\hat{p}^2\right)^2}
\]
is divergent for $p \to 0$. But in that limit, $\hat{p}^2$ can be replaced by
$p^2$:
\[
\lim_{d \to 4} \Delta_1 = \lim_{d \to 4} (d-4) \int\limits_{BZ}
\frac{d^dp}{(2\pi)^d} \frac{1}{\left(p^2\right)^2}.
\]
Additionally, the region of integration can be restricted to a (small) sphere
around the origin:
\[
\lim_{d \to 4} \Delta_1 = \lim_{d \to 4} (d-4) \int d\Omega_d 
\int_0^R p^{d-1} \frac{dp}{(2\pi)^d} \frac{1}{p^4}
= \lim_{d \to 4} \frac{\Omega_d}{(2\pi)^d} R^{d-4},
\]
where $\Omega_d$ is the surface of the $d$-dimensional sphere. Taking the
limit and inserting the explicit value $\Omega_4= \frac{1}{2} (2\pi)^2$,  
one finally gets:
\begin{equation}
\lim_{d \to 4} \Delta_1 = \frac{1}{2(2\pi)^2}.
\end{equation}

\chapter{Fourier transform of the potential}

\label{potentialfourier}

In section \ref{runningcoupling}, the following expression for the
quark-antiquark potential in momentum space was given:
\begin{eqnarray}
V(\vec{q}^2) &=& - \frac{g_0^2(a)}{\vec{q}^2} C_2(F) \nonumber \\
&& \cdot \left[ 1 + g_0^2(a) \left[\beta_0 \left(\ln
\frac{\pi^2}{a^2\vec{q}^2} - \gamma + \frac{31}{33} \right)
 - \frac{\bar{A}(1, 0, N)}{4 \pi} + R(N) \right] \right]. \nonumber \\
\end{eqnarray}

In order to get the dependance of the potential on $\hat{R}$, one has to
perform a Fourier transformation. In the literature usually only the result is
given; here I explain explicitly how the calculation can be done. The first
term is easy---simply use the residual theorem:
\begin{eqnarray}
&&\int \frac{d^3q}{(2\pi)^3} \frac{1}{\vec{q}^2} e^{i \vec{q} \cdot \vec{R}}
\nonumber \\ &=& \lim_{\alpha \to 0} \int \frac{d^3q}{(2\pi)^3}
\frac{1}{\vec{q}^2 + \alpha^2} e^{i \vec{q} \cdot \vec{R}}
= \lim_{\alpha \to 0} \frac{1}{(2 \pi)^2} \int\limits_0^{\infty} q^2
dq \int\limits_{-1}^1 dx \frac{1}{q^2 + \alpha^2} e^{i q R x} \nonumber \\
&=& \lim_{\alpha \to 0} \frac{1}{(2 \pi)^2 i R} \int\limits_{-\infty}^{\infty}
\frac{q}{q^2 + \alpha^2} e^{i q R} dq
= \lim_{\alpha \to 0} \frac{1}{4\pi R} e^{-\alpha r} \nonumber \\ 
&=& \frac{1}{4\pi R}.
\end{eqnarray}

In the second term, additional to the pole, there is a branch cut on the
positive imaginary axis because of the logarithm. Thus one has to use a
more complicated path of integration:
\begin{eqnarray}
\int \frac{d^3q}{(2\pi)^3}
\frac{\ln\left(\frac{\vec{q}^2}{\mu^2}\right)}{\vec{q}^2} e^{i \vec{q} \cdot
\vec{R}}
&=& \lim_{\alpha, \beta \to 0} \int \frac{d^3q}{(2\pi)^3}
\frac{\ln\left(\frac{\vec{q}^2 + \beta^2}{\mu^2}\right)}{\vec{q}^2 + \alpha^2}
e^{i \vec{q} \cdot \vec{R}} \nonumber \\
&=& \lim_{\alpha, \beta \to 0} \frac{1}{(2 \pi)^2 i R}
\int\limits_{-\infty}^{\infty} \frac{q}{q^2 + \alpha^2} \ln\left(\frac{q^2 +
\beta^2}{\mu^2}\right) e^{i q R} dq \nonumber \\
&=& \lim_{\alpha, \beta \to 0} \frac{1}{(2 \pi)^2 i R} \left[
\int\limits_{\begin{picture}(1,0.5)
\put(0,0){\epsfig{file=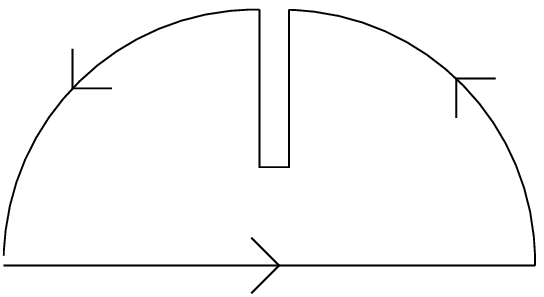, width=1cm, height=0.5cm}}
\end{picture}}
\frac{q}{q^2 + \alpha^2} \ln\left(\frac{q^2 + \beta^2}{\mu^2}\right) e^{i q R}
dq \right. \nonumber \\
&-& \int\limits_{\begin{picture}(1,0.5)
\put(0,0){\epsfig{file=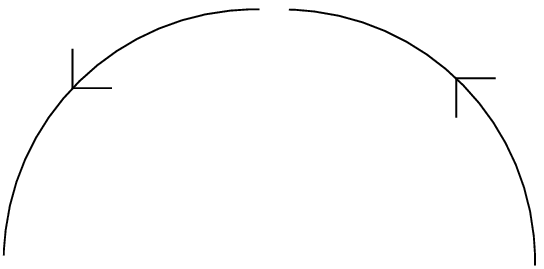, width=1cm, height=0.5cm}}
\end{picture}}
\frac{q}{q^2 + \alpha^2} \ln\left(\frac{q^2 + \beta^2}{\mu^2}\right) e^{i q R}
dq \nonumber \\
&-& \left. \int\limits_{\begin{picture}(0.1,0.5)
\put(0,0){\epsfig{file=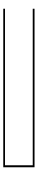, width=0.1cm, height=0.5cm}}
\end{picture}}
\frac{q}{q^2 + \alpha^2} \ln\left(\frac{q^2 + \beta^2}{\mu^2}\right) e^{i q R}
dq \right]
\end{eqnarray}
The first term can again be evaluated using the residual theorem, the second
vanishes, and the third gives a contribution from the discontinuity across
the branch cut:
\begin{eqnarray}
&=& \lim_{\beta \to 0} \frac{1}{4 \pi R} \ln\left(\frac{\beta^2}{\mu^2}\right)
- \lim_{\epsilon, \beta \to 0} \frac{1}{(2 \pi)^2 i R}
\int\limits_{i\beta-\epsilon}^{i\infty-\epsilon}
\frac{2 i \mbox{Im} \ln\left(\frac{q^2 + \beta^2}{\mu^2}\right)}{q}  e^{i q R}
dq \nonumber \\
&=& \lim_{\beta \to 0} \frac{1}{4 \pi R} \ln\left(\frac{\beta^2}{\mu^2}\right)
+ \lim_{\beta \to 0} \frac{1}{2 \pi R}
\int\limits_{i\beta}^{i\infty} \frac{1}{q} e^{i q R}
dq \nonumber \\
&=& \lim_{\beta \to 0} \frac{1}{4 \pi R} \ln\left(\frac{\beta^2}{\mu^2}\right)
+ \lim_{\beta \to 0} \frac{1}{2 \pi R}
\int\limits_{\beta R}^{\infty} \frac{1}{x} e^{- x} dx
\end{eqnarray}
Now the second term gives a modified Gamma function, which can be evaluated
exactly in the limit $\beta \to 0$:
\begin{eqnarray}
&=& \lim_{\beta \to 0} \frac{1}{4 \pi R} \ln\left(\frac{\beta^2}{\mu^2}\right)
+ \lim_{\beta \to 0} \frac{1}{2 \pi R} \Gamma(0, \beta R) \nonumber \\
&=& \lim_{\beta \to 0} \frac{1}{4 \pi R} \ln\left(\frac{\beta^2}{\mu^2}\right)
- \lim_{\beta \to 0} \frac{1}{2 \pi R} \left(\ln(\beta R) + \gamma \right)
\nonumber \\
&=& - \frac{\ln(\mu^2 R^2) + 2 \gamma}{4 \pi R}.
\end{eqnarray}
Putting everything together, the potential in coordinate space is given by:
\begin{eqnarray}
V(R) &=& - \frac{g_0^2(a)}{4 \pi R} C_2(F) \nonumber \\
&& \cdot \left[ 1 + g_0^2(a) \left[\beta_0 \left(\ln
\frac{\pi^2 R^2}{a^2} + \gamma + \frac{31}{33} \right)
 - \frac{\bar{A}(1, 0, N)}{4 \pi} + R(N) \right] \right]. \nonumber \\
\end{eqnarray}

\newpage
\thispagestyle{empty}
\cleardoublepage

\newpage
\thispagestyle{empty}
\cleardoublepage

\pagestyle{empty}

\newpage
\cleardoublepage

\begin{center}
\LARGE Acknowledgements \normalsize
\end{center}

\vspace{2ex}

First I would like to thank Prof.\ H.\ J.\ Rothe for offering this
interesting topic for my dissertation to me, for his friendly supervision and
support and the intensive discussions about crucial points of the thesis.
It was an interesting experience to check and to validate the lattice sum rules
he himself had derived.

\vspace{2ex}

Next I am very grateful to Prof.\ W.\ Wetzel, who helped me checking some of
my calculations for errors and who suggested both the proof of the
transversality of the vacuum polarization and the proof of the gauge
invariance of the expectation value of the Wilson loop to me. He invested many
hours of his time on this and gave me lots of valuable hints and suggestions.

\vspace{2ex}

Many thanks go also to Prof. U.\ Heller, who provided me with some of his
notes on his calculations of the potential on the lattice, using lattice
perturbation theory. I got several crucial clues from these notes which helped
me to understand his paper written in collaboration with Prof.\ F.\ Karsch.

\vspace{2ex}

I thank Prof.\ D.\ Gromes for his friendly interest and readiness to referee
this thesis.

\vspace{2ex}

The great working atmosphere at the Institute for Theoretical Physics in
Heidelberg, especially in the ''Westzimmer'', also has to be mentioned. I
would like to thank Tanja Robens, Markus M{\"u}ller, Christian M{\"u}ller,
Felix Schwab, Martin Pospischil, Sebastian Diehl, Kai M{\"u}ller, Hendrik
Ballhausen, Ewald Puchwein, Felix Nagel, Filipe Pacetti, Lala Adueva and Raffi
Kasarcan for interesting and enlightening discussions, cheerful conversations
and also lots of fun.

\vspace{2ex}

Special thanks go to Kai Schwenzer and Frank Steffen for
the many hours we spent together discussing and proofreading the contribution
of Prof.\ J.\ Zinn-Justin to the workshop "Topology and Geometry in Physics" of
the Graduier\-tenkolleg "Physical Systems with many Degrees of Freedom". In
these discussions, I learnt lots about toplogy, geometry and Quantum Field
Theory in general, and despite the amount of work which had to be done, in our
meetings there was always time left for fun.

\vspace{2ex}

Many thanks go to the above mentioned Graduiertenkolleg, too, for supporting
me financially during my last third months working on my thesis, and for
providing several interesting workshops. Additionally to the workshop on
topology and geometry, the ones on Biophysics and Quantum Information were
both very interesting and valuable.

\vspace{2ex}

Also I would like to thank the Landesgraduiertenf{\"o}rderung
Baden-W{\"u}rttem\-berg for financial support during most of the time I worked
on my dissertation and for an interesting evening in the Alte Aula of the
University of Heidelberg in December 2002.

\vspace{2ex}

Schlussendlich danke ich auch allen meinen Freunden innerhalb und
ausserhalb der Physik, als da w{\"a}ren Hannes Klehr, Max Urban, Wouter
Kornelis, Martina Keller, Vera Spillner und Michael Elser, f{\"u}r ihre
jahrelange Freundschaft und Unterst{\"u}tzung. Besonderer Dank geht an meine
Freundin Kristin Warnick und meine Familie f{\"ur} ihre Liebe und Geduld mit
mir.

\end{document}